\documentclass[twocolumn,pra,amsmath,amssymb,superscriptaddress,longbibliography,nofootinbib,floatfix]{revtex4-2}
\usepackage[utf8]{inputenc}
\usepackage{pgfplots}
\usepackage{bbold}
\usepackage{comment}
\usepackage[colorlinks=true, linkcolor=red, allbordercolors={white}]{hyperref}
\usepackage{algorithm}
\usepackage{algpseudocode}
\usepackage{ragged2e}
\usepackage{float}
\pgfplotsset{compat = newest}
\usepgfplotslibrary{colorbrewer}
\usepgfplotslibrary{groupplots}
\usetikzlibrary{arrows.meta}
\definecolor{color1}{HTML}{1f77b4}
\definecolor{color2}{HTML}{ff7f0e}
\definecolor{color3}{HTML}{2ca02c}
\definecolor{color4}{HTML}{d62728}
\definecolor{color5}{HTML}{9467bd}
\definecolor{color6}{HTML}{8c564b}
\definecolor{color7}{HTML}{e377c2}
\definecolor{color8}{HTML}{7f7f7f}
\definecolor{color9}{HTML}{bcbd22}
\definecolor{shadyblue}{rgb}{0.33,0.33,1}
\pgfplotsset{
  cycle list={color1\\color2\\color3\\color4\\color5\\color6\\color7\\color8\\color9\\},
}
\newcommand{\overbar}[1]{\mkern 1.5mu\overline{\mkern-1.5mu#1\mkern-1.5mu}\mkern 1.5mu}
\newcommand{\arrowIn}{
\tikz \draw[-{Stealth[length=2mm, width=1.5mm]}] (-1pt,0) -- (1pt,0);
}
\newcommand{\arrowInn}{
\tikz \draw[-{Stealth[length=7.5mm, width=5.6mm]}] (-1pt,0) -- (1pt,0);
}
\tikzset{
    new dash/.code args={on #1 off #2}{
        \csname tikz@addoption\endcsname{%
            \pgfgetpath\currentpath%
            \pgfprocessround{\currentpath}{\currentpath}%
            \csname pgf@decorate@parsesoftpath\endcsname{\currentpath}{\currentpath}%
            \pgfmathparse{\csname pgf@decorate@totalpathlength\endcsname-#1}\let\rest=\pgfmathresult%
            \pgfmathparse{#1+#2}\let\onoff=\pgfmathresult%
            \pgfmathparse{max(floor(\rest/\onoff), 1)}\let\nfullonoff=\pgfmathresult%
            \pgfmathparse{max((\rest-\onoff*\nfullonoff)/\nfullonoff+#2, #2)}\let\offexpand=\pgfmathresult%
            \pgfsetdash{{#1}{\offexpand}}{0pt}}%
    }
}

\newcommand{\pvec}[1]{\vec{#1}\mkern2mu\vphantom{#1}}

\usepackage[floatrow]{trivfloat}
\trivfloat{algorithm}

\DeclareRobustCommand{\justifyingg}{%
 \let\\\@raggedtwoe@savedcr
 \let\@gnewline\@raggedtwoe@saved@gnewline
 \leftskip\z@
 \@rightskip\z@
 \rightskip\@rightskip
 \parfillskip\JustifyingParfillskip
 \parindent\JustifyingParindent
 \@raggedtwoe@spaceskipfalse
 \@raggedtwoe@everyselectfont
}

\begin{document}

\title{Numerical simulation of non-abelian anyons}

\author{Nico Kirchner}
\affiliation{Department of Physics, TFK, Technische Universit{\"a}t M{\"u}nchen, James-Franck-Stra{\ss}e 1, D-85748 Garching, Germany}
\author{Darragh Millar}
\affiliation{Department of Theoretical Physics, National University of Ireland, Maynooth, Ireland}
\author{Babatunde M. Ayeni}
\affiliation{Department of Theoretical Physics, National University of Ireland, Maynooth, Ireland}
\author{Adam Smith}
\affiliation{School of Physics and Astronomy, University of Nottingham, Nottingham, NG7 2RD, UK}
\affiliation{Centre for the Mathematics and Theoretical Physics of Quantum Non-Equilibrium Systems, University of Nottingham, Nottingham, NG7 2RD, UK}
\author{Joost K. Slingerland}
\affiliation{Department of Theoretical Physics, National University of Ireland, Maynooth, Ireland}
\affiliation{Dublin Institute for Advanced Studies, School of Theoretical Physics, 10 Burlington Rd, Dublin, Ireland}
\author{Frank Pollmann}
\affiliation{Department of Physics, TFK, Technische Universit{\"a}t M{\"u}nchen, James-Franck-Stra{\ss}e 1, D-85748 Garching, Germany}
\affiliation{Munich Center for Quantum Science and Technology (MCQST), Schellingstr. 4, 80799 M{\"u}nchen, Germany}



\begin{abstract}

Two-dimensional systems such as quantum spin liquids or fractional quantum Hall systems exhibit anyonic excitations that possess more general statistics than bosons or fermions. This exotic statistics makes it challenging to solve even a many-body system of non-interacting anyons. We introduce an algorithm that allows to simulate anyonic tight-binding Hamiltonians on two-dimensional lattices. The algorithm is directly derived from the low energy topological quantum field theory and is suited for general abelian and non-abelian anyon models. As concrete examples, we apply the algorithm to study the energy level spacing statistics, which reveals level repulsion for free semions, Fibonacci anyons and Ising anyons. Additionally, we simulate non-equilibrium quench dynamics, where we observe that the density distribution becomes homogeneous for large times - indicating thermalization.

\end{abstract}

\maketitle

\tableofcontents

\section{Introduction}

Two-dimensional systems can support topological point-like quasiparticle excitations, so-called anyons~\cite{Leinaas1977,PhysRevLett.48.1144,PhysRevLett.49.957}, which obey statistics beyond regular bosons or fermions. These anyons can lead to novel physical phenomena and have thus attracted considerable attention over the past decades. For instance, their topological protection from local perturbations led to the idea of fault-tolerant topological quantum computing~\cite{einleitung, 1802.06176, PhysRevX.4.011036, KITAEV20032, freedman2002modular}. Experimentally, anyons are of interest since they may be realized in quantum spin liquids~\cite{Savary2016, PhysRevB.71.045110, 0904.2771, PhysRevB.65.224412, PhysRevB.69.064404} or systems exhibiting the fractional quantum Hall effect~\cite{PhysRevLett.50.1395, PhysRevLett.52.1583, PhysRevB.59.8084, STERN2008204}. While being hypothesized for quite some time~\cite{PhysRevLett.49.957}, experimental verification came only recently in form of anyon collisions for anyons featured in fractional quantum Hall systems at $\nu=1/3$ filling~\cite{doi:10.1126/science.aaz5601} and measurements of the half-integer quantization of the thermal Hall conductivity in the Kitaev material candidate $\alpha$-$\mathrm{RuCl_3}$~\cite{Kasahara2018, doi:10.1126/science.aay5551, 2104.12184}. Further, the ground state of the toric code model~\cite{KITAEV20032} was realized on a superconducting quantum computer~\cite{doi:10.1126/science.abi8378}, which was used to verify properties such as the topological entanglement entropy~\cite{PhysRevLett.96.110404} and anyonic braid statistics. In yet another work, topological string operators were measured in quantum spin liquid states probed by a programmable quantum simulator~\cite{doi:10.1126/science.abi8794}.

Theoretically, anyons may be described using the framework of topological quantum field theory (TQFT)~\cite{Witten1989, Atiyah1988}, which associates states / wave functions in the Hilbert space to two-dimensional surfaces. For many purposes, however, one can restrict oneself to unitary modular categories, which essentially form the mathematical basis of TQFT~\cite{Turaev+2016}. There are also exactly solvable microscopic models that are capable of describing systems featuring anyonic excitations~\cite{KITAEV20032, KitaevFormulae, PhysRevB.71.045110, PhysRevB.94.235136}.

It is of great interest to connect theory and experiment by finding experimentally measurable signatures or ``fingerprints'' that indicate the presence of anyonic excitations. Studying, e.g., spectroscopic properties, far-from-equilibrium dynamics and thermalization behavior may reveal such measurable signatures. For example, it was found that the spectral response of a system close to the threshold of exciting a pair of abelian anyons depends on their statistics~\cite{SpectralFunction}. Other studies focus, e.g., on anyonic systems featuring specific interactions~\cite{0902.3275, PhysRevLett.98.160409, PhysRevB.83.134439, PhysRevLett.101.050401, 10.1088/1367-2630/13/4/045014} or abelian anyons in one dimension in order to apply analytical methods~\cite{1312.4657, PhysRevLett.97.100402, PhysRevA.86.043631, PhysRevA.96.023611, Li2013, Tang_2015,PhysRevA.78.045602,  2202.06543}. As for numerical studies, simulations of anyons hopping on a square lattice for abelian anyons have been considered~\cite{PhysRevB.43.2661_fluxconvention, PhysRevB.43.10761, SemionsTorus}. The transport properties of a single abelian or non-abelian anyon on a ladder with background charges were numerically examined, which revealed ballistic transport for abelian anyons and uniformly distributed background charges and dispersive transport for non-abelian anyons~\cite{PhysRevB.90.134201, PhysRevB.89.075112}. Further, tight-binding models of non-abelian anyons on chains and ladders have been studied using exact diagonalization of small system sizes~\cite{poilblanc2013one,poilblanc2012fractionalization,soni2016effective}, and of ladders in thermodynamic limit~\cite{ayeni2018phase} using symmetric tensor network algorithms that incorporate the topological data of anyon models and anyonic diagrammatic techniques into ordinary tensor networks~\cite{PhysRevB.82.115126,konig2010anyonic,PhysRevB.89.075112,PhysRevB.93.165128}.

In this paper, we introduce an algorithm that allows for simulating both abelian and non-abelian anyons on two-dimensional lattices beyond ladders, where all anyons are mobile and are subject to an anyonic tight-binding Hamiltonian that incorporates their statistics. As an example, we utilize the algorithm to study the energy level spacing statistics and the density distribution after a quench, where we focus on semions (abelian), Fibonacci anyons and Ising anyons (both non-abelian), for concreteness.\\

The paper is structured as follows. First, in section~\ref{sec:particle_types}, we consider how to numerically simulate anyon dynamics for three concrete examples: fermions, semions and Fibonacci anyons. These examples highlight the new considerations needed for simulating general abelian and non-abelian anyons. In section~\ref{sec:formalism}, the general formalism that we will base our algorithm upon is reviewed together with the most important concepts relevant for our discussions. Then, in section~\ref{sec:basis_states}, we discuss some important aspects of the Hilbert space and choose a basis. Based on these considerations and the previously introduced formalism, we discuss how the matrix elements according to our algorithm are computed using fusion diagrams (Sec.~\ref{sec:algo}). In section~\ref{sec:results}, we discuss some results obtained from the algorithm, where we concentrate on the energy level spacing statistics and the dynamics of the density distribution after a quench in order to see the relaxation and thermalization behavior. In section~\ref{sec:conclusion}, we close the paper by giving our conclusion.

\section{Simulation of Different Anyon Types}\label{sec:particle_types}

Before introducing the general algorithm to simulate arbitrary types of anyons, we consider three concrete examples. First, fermions which are routinely simulated numerically~\cite{0101188}. Next, semions~\cite{SemionsTorus} where we have to distinguish between clockwise and counter-clockwise exchanges and need to introduce an additional degree of freedom on a torus corresponding to non-trivial topological charges associated with the boundary conditions~\cite{doi:10.1063/1.4939783}. Finally, we consider Fibonacci anyons~\cite{0902.3275} whose non-abelian statistics lead to further degrees of freedom known as ``fusion channels'', which are associated with composite particles.

\subsection{Fermions}

\begin{figure}[t]
\centering
\begin{tikzpicture}[line width=0.75pt, scale=1.2]
	\draw[black!40, line width=5pt] (0,-0.5) -- (0,3.5) node[sloped,pos=0.5,allow upside down]{\arrowInn}; ; 
	\draw[black!40, line width=5pt] (0,3.5) -- (1,-0.5) node[sloped,pos=0.5,allow upside down]{\arrowInn}; ; 
	\draw[black!40, line width=5pt] (1,-0.5) -- (1,3.5) node[sloped,pos=0.5,allow upside down]{\arrowInn}; ; 
	\draw[black!40, line width=5pt] (1,3.5) -- (2,-0.5) node[sloped,pos=0.5,allow upside down]{\arrowInn}; ; 
	\draw[black!40, line width=5pt] (2,-0.5) -- (2,1.5) node[sloped,pos=0.75,allow upside down]{\arrowInn}; ; 
	\draw[black!40, line width=5pt] (-1,3.5) -- (0,-0.5) node[sloped,pos=0.5,allow upside down]{\arrowIn}; ; 
	\draw[black] (-0.5,0) -- (3.5,0);
	\draw[black] (-0.5,1) -- (3.5,1);
	\draw[black] (-0.5,2) -- (3.5,2);
	\draw[black] (-0.5,3) -- (3.5,3);
	\draw[black] (0,-0.5) -- (0,3.5);
	\draw[black] (1,-0.5) -- (1,3.5);
	\draw[black] (2,-0.5) -- (2,3.5);
	\draw[black] (3,-0.5) -- (3,3.5);
	\filldraw [fill=white, draw=white] (-1.1,-0.5) rectangle (-0.5,3.6);
	\filldraw [fill=white, draw=white] (3.5,3.5) rectangle (-0.5,3.6);
	\filldraw [fill=white, draw=white] (-0.5,-0.5) rectangle (3.5,-0.6);
	\node[draw,circle,inner sep=1.75pt,fill,black] at (0,2) {};
	\node[draw,circle,inner sep=1.75pt,fill,black] at (0,0) {};
	\node[draw,circle,inner sep=1.75pt,fill,black] at (0,1) {};
	\node[draw,circle,inner sep=1.75pt,fill,black] at (0,3) {};
	\node[draw,circle,inner sep=1.75pt,fill,black] at (1,2) {};
	\node[draw,circle,inner sep=1.75pt,fill,black] at (1,0) {};
	\node[draw,circle,inner sep=1.75pt,fill,black] at (1,1) {};
	\node[draw,circle,inner sep=1.75pt,fill,black] at (1,3) {};
	\node[draw,circle,inner sep=1.75pt,fill,black] at (2,2) {};
	\node[draw,circle,inner sep=1.75pt,fill,black] at (2,0) {};
	\node[draw,circle,inner sep=1.75pt,fill,black] at (2,1) {};
	\node[draw,circle,inner sep=1.75pt,fill,black] at (2,3) {};
	\node[draw,circle,inner sep=1.75pt,fill,black] at (3,2) {};
	\node[draw,circle,inner sep=1.75pt,fill,black] at (3,0) {};
	\node[draw,circle,inner sep=1.75pt,fill,black] at (3,1) {};
	\node[draw,circle,inner sep=1.75pt,fill,black] at (3,3) {};
	\node[black, anchor=south west] (a) at (1-0.05,2-0.05) {$i-L_y$};
	\node[black, anchor=south west] (a) at (2,2) {$i$};
	\node[black, anchor=south west] (a) at (2,1) {$i-1$};
\end{tikzpicture}
\caption{Possible choice for the Jordan-Wigner string for site $i$. The associated contribution picks up a factor of $e^{i\pi}$ for each occupied site on the shaded string.}
\label{fig:fermion lattice}
\end{figure}
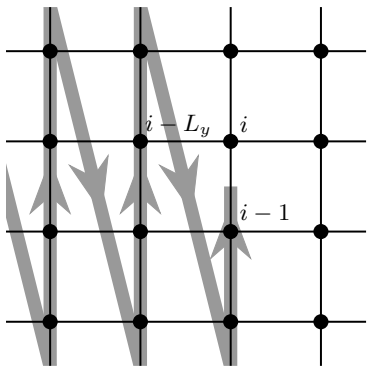
Fermions can be simulated by utilizing the so-called Jordan-Wigner transformation~\cite{Jordan1928,0101188}, where operator strings $e^{i\pi \sum_{j=0}^{i-1}n_j}$ are introduced in order to determine the signs following from the fermionic anti-commutation relations, where $n_j$ denote the site occupation operators. Using these strings, the fermionic operators can be mapped to spin-$1/2$ operators~\cite{0101188} via
\begin{align}
	c_i^{\dagger}=e^{i\pi \sum_{j=0}^{i-1}n_j}S_i^+,\quad c_i=e^{i\pi \sum_{j=0}^{i-1}n_j}S_i^-.
\end{align}
Here, $S^{\pm}_i=S^x_i\pm iS^y_i$ denote the spin raising and lowering operators, $c_i^{\dagger},c_i$ the fermionic creation and annihilation operators with $c_i^{\dagger}c_i=n_i=S^+_iS^-_i=S^z_i+1/2$ and $S_i^{x,y,z}$ the usual spin operators in $x,y,z$-direction. For a $2$D system, the strings may be chosen to be like in Fig.~\ref{fig:fermion lattice}, where we chose to number the sites column by column, starting with the left-most, as indicated by the shaded string.

In one dimension, a local Hamiltonian can be mapped via the Jordan-Wigner transformation to yet another local Hamiltonian. For higher dimensional system on the other hand, some local terms are expressed by non-local Jordan-Wigner strings, as can be seen for the convention in Fig.~\ref{fig:fermion lattice}: A local term like $c_{i}^{\dagger}c_{i-L_y}$ is mapped to the non-local term $e^{i\pi \sum_{j=i-L_y}^{i-1}(S_j^z+1/2)}S_i^+S_{i-L_y}^-$.

\subsection{Abelian Anyons: Semions}\label{sec:partcile_types_abelian}

While indistinguishable fermions acquire a phase $\pi$ under exchange, abelian anyons~\cite{PhysRevLett.49.957} may acquire a more general rational phase $\theta\in [0,2\pi)$. Importantly, we must distinguish between clockwise ($-\theta$) and counter-clockwise ($+\theta$) exchanges. Such a distinction is not possible using the simple Jordan-Wigner strings discussed above. Additionally, as will be described below, there are extra degrees of freedom if we consider periodic boundary conditions (PBC). In the following, we will focus on the case of semions with $\theta=\pi/2$~\cite{simon2020topological} to illustrate how abelian anyonic statistics may be incorporated in an algorithm.

\begin{figure}[t]
\centering
\begin{tikzpicture}[line width=0.75pt, scale=1.2]
	\draw[black] (-0.5,0) -- (3.5,0);
	\draw[black] (-0.5,1) -- (3.5,1);
	\draw[black] (-0.5,2) -- (3.5,2);
	\draw[black] (-0.5,3) -- (3.5,3);
	\draw[black] (0,-0.5) -- (0,3.5);
	\draw[black] (1,-0.5) -- (1,3.5);
	\draw[black] (2,-0.5) -- (2,3.5);
	\draw[black] (3,-0.5) -- (3,3.5);
	\draw[black][line width=2pt] (3.25,-0.75) -- (3.25,3.75);
	\node[black, anchor=west] (a) at (3.25,3.75) {Cut $B$};
	\draw[black][line width=2pt] (-0.75,-0.25) -- (3.75,-0.25);
	\node[black, anchor=west] (a) at (3.75,-0.25) {Cut $A$};
	\node[draw,circle,inner sep=1.75pt,fill,black] at (2,2) {};
	\node[draw,circle,inner sep=1.75pt,fill,black] at (2,3) {};
	\node[draw,circle,inner sep=1.75pt,fill,black] at (1,1) {};
	\node[draw,circle,inner sep=1.75pt,fill,black] at (0,2) {};
	\node[black, anchor=south east] (a) at (0,2) {$1$};
	\node[black, anchor=south east] (a) at (1,1) {$2$};
	\node[black, anchor=south east] (a) at (2,2) {$3$};
	\node[black, anchor=south east] (a) at (2,3) {$4$};
	\draw[black][line width=1.25pt] (0,2) -- (0.5,1.5);
	\draw[black][line width=1.25pt] (0.5,-0.2-0.01) -- (0.5,1.5+0.01);
	\draw[black][line width=1.25pt] (0.5-0.01,-0.2) -- (3.25,-0.2);
	\draw[black][line width=1.25pt] (1,1) -- (1.5,0.5);
	\draw[black][line width=1.25pt] (1.5,-0.15-0.01) -- (1.5,0.5+0.01);
	\draw[black][line width=1.25pt] (1.5-0.01,-0.15) -- (3.25,-0.15);
	\draw[black][line width=1.25pt] (2,2) -- (2.5,1.5);
	\draw[black][line width=1.25pt] (2.5,-0.1-0.01) -- (2.5,1.5+0.01);
	\draw[black][line width=1.25pt] (2.5-0.01,-0.1) -- (3.25,-0.1);
	\draw[black][line width=1.25pt] (2,3) -- (2.6,2.5);
	\draw[black][line width=1.25pt] (2.6,-0.05-0.01) -- (2.6,2.5+0.01);
	\draw[black][line width=1.25pt] (2.6-0.01,-0.05) -- (3.25,-0.05);
	\draw[black] (1,-0.7) -- (2,-0.7) node[sloped,pos=1,allow upside down]{\arrowIn}; ; 
	\draw[black] (-0.7,1.25) -- (-0.7,2.25) node[sloped,pos=1,allow upside down]{\arrowIn}; ; 
	\node[black, anchor=north] (a) at (1.5,-0.65) {$x$};
	\node[black, anchor=east] (a) at (-0.65,1.75) {$y$};
\end{tikzpicture}
\caption{Choice of strings associated to anyons in order to simulate abelian exchange statistics. The phases associated to the anyons' strings are $\pi/2$ for the vertical and $\pi$ for the horizontal parts for semions.}
\label{fig:abelian lattice}
\end{figure}
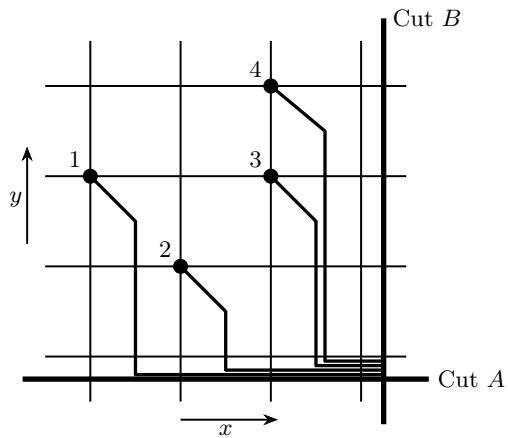

A solution to the problem of distinguishing clockwise and counter-clockwise exchanges can be found in Refs.~\cite{PhysRevB.43.10761, SemionsTorus}. New strings, as depicted in Fig.~\ref{fig:abelian lattice}, are introduced. The strings are chosen to originate from the anyons, going into the plaquettes on the lower right of the anyons' sites and then follow the negative $y$-direction until they almost reach cut $A$ that is given by $y=1/2$. Then, the strings follow the positive $x$-direction, ending when reaching cut $B$, which is given by $x=L_x+1/2$ for a $L_x\times L_y$ lattice. The two mentioned cuts can be viewed as two loops on the torus that need to be crossed in order to translate an anyon once around the system in $x$- or $y$-direction. For such translations, additional effects need to be considered, which will be done below. We first focus on translations without crossing a cut.

One key property of the anyons' strings is that they have phases associated with them. For semions, the phases associated with the vertical parts of the strings are $\pi/2$, whereas the ones associated with the horizontal parts are $\pi$. When translating anyons across the lattice, the phases corresponding to the translation processes are determined by these strings. If a semion crosses another semion's string in a direction corresponding to counter-clockwise exchange, like, e.g., for translating semion $3$ in Fig.~\ref{fig:abelian lattice} to the right or semion $4$ to the left, the wave function is multiplied by $e^{i\pi/2}$. This factor is acquired for each pair of semions (the one being translated and the one whose string is crossed) fulfilling the just mentioned condition. Similarly, the wave function is to be multiplied by $e^{-i\pi/2}$ for each time the semion being translated crosses another semion's string such that the process corresponds to a clockwise exchange, such as translating semion $1$ in Fig.~\ref{fig:abelian lattice} to the right or semion $2$ to the left. Translations in $y$-direction only yield a non-trivial factor ($e^{i\pi}$) if a horizontal string is crossed, which can only occur upon crossing cut $A$, like, e.g., for translating semion $4$ in Fig.~\ref{fig:abelian lattice} in positive $y$-direction.

The rules for translations in the bulk can be summarized by noting that the acquired factors for nearest-neighbor translations in positive $x$-direction are given by $e^{i\pi(n_{i\uparrow}-n_{j\downarrow})/2}$, where $i$ and $j$ refer to the initial and final site of the semion that is translated. The number of semions that are localized in the same column as site $i$ and possess a greater $y$-coordinate than site $i$ is denoted by $n_{i\uparrow}$, the number of semions localized in the same column as site $j$ with a smaller $y$-coordinate than site $j$ is denoted by $n_{j\downarrow}$. For translations in negative $x$-direction, the acquired phases are given by $e^{-i\pi(n_{i\uparrow}-n_{j\downarrow})/2}$.

Translations over one of the two cuts have additional effects that are more complicated. First, let us consider what happens in the system when moving a semion around the torus, as this is what translations across the cuts correspond to. Semions can be viewed as charged particles with infinitesimal flux tubes attached to them~\cite{PhysRevLett.49.957, PhysRevB.43.10761, SemionsTorus} such that exchanging two semions yields a phase of $\pi/2$, i.e., if the charges are $e$, the attached flux tubes carry magnetic fluxes of $\phi_0/2$, with the flux quantum $\phi_0=ch/e$. Thus, when translating a semion around the torus, the system effectively gains an additional flux in the respective direction, i.e., the obtained state is (despite the semion returning to its initial position) physically different from the inital one. In order to distinguish these states, we introduce wave function components, also known as ``sheets'', which reflect the topological ground state degeneracy~\cite{doi:10.1063/1.4939783}. Translating a semion twice around the torus introduces a flux in the respective direction of magnitude $\phi_0$, i.e., for particles with charge $e$, the state effectively remains unchanged since they are only affected by flux modulo $\phi_0$. We thus need precisely two sheets in order to describe semions on a torus, which doubles the size of the Hilbert space.

The presence of these two sheets is reflected in the two matrices
\begin{align}
  \begin{pmatrix}
    1 & 0\\
    0 & -1
  \end{pmatrix}\quad \text{and}\quad
  \begin{pmatrix}
    0 & 1\\
    1 & 0
  \end{pmatrix}\text{,}
  \label{eq:semion_drag}
\end{align}
which describe the transformation between the sheet components when crossing cut $A$ and $B$, respectively. The second matrix represents the sheet switching due to the introduction of an additional $\phi_0/2$ flux when translating a semion around the torus in $x$-direction. The first matrix contains the phases that are picked up when translating a semion across cut $A$ due to the flux threading the torus. When crossing one of the cuts, the respective matrix has to be applied in addition to the string rules discussed above. Note that the presence of multiple wave function components as described above is not restricted to semions. It turns out that on a torus, every abelian anyon model features multiple sheets. The total number of sheets in general depends on the number of particles and their charge and coincides with the topological ground state degeneracy~\cite{doi:10.1063/1.4939783}.

There is an important detail regarding semions that we have not mentioned yet: The total number of semions on a torus always has to be even. We explain the reason for this and how to systematically obtain the total number of sheets in the general case in Sec.~\ref{sec:basis_states} after having introduced some other important concepts.\\

It is possible to directly generalize the rules sketched above to other abelian anyon models for which counter-clockwise exchange yields a more general phase of $\theta$. The phases associated to the strings simply become $\theta$ for the vertical and $2\theta$ for the horizontal parts. The generalized matrices acquired upon crossing a cut are more complicated since their dimensions, which also agree with the number of the wave function components, depend on statistics of the anyons to be simulated. For exchange statistics corresponding to $\theta=\pi p/M$ with $p$ and $M$ being coprime integers, the total number of wave function components is $M$~\cite{PhysRevB.43.10761,StatisticsTorusFormulae}. For details regarding the general abelian case, we refer to reader to Ref.~\cite{PhysRevB.43.10761}.\\

Note that there is a generalized Jordan-Wigner transformation~\cite{PhysRevLett.86.1082} that may be used as a mapping between spins and abelian anyons in $2$D with open boundary conditions (OBC). I.e., anyonic systems featuring abelian anyons can be simulated similarly to fermionic systems by using spins. This transformation accounts for on the exchange statistics but only works for OBC, that is, it corresponds to a single wave function component.

\subsection{Non-abelian Anyons: Fibonacci Anyons}
\label{sec:partcile_types_nonabelian}

Non-abelian anyons introduce an additional complication in the form of non-unique fusion. For fermions, we know that two particles collectively behave as a composite boson. Similar to this case, one can view the composite particle of multiple abelian anyons as another anyon that is uniquely determined by the anyons forming it. For two semions for example, the composite particle carries the (neutral) vacuum charge, i.e., it is also a boson, as semions are their own antiparticles~\cite{simon2020topological}. In particular, every composite particle of abelian anyons can be considered as a single \emph{unique} anyon. This uniqueness is not present for composite particles of non-abelian anyons, which is analoguous to the composition of spins. Two spin-$1/2$s for example may form a composite spin singlet (spin-$0$) or spin triplet (spin-$1$), which may be written as $1/2\otimes 1/2 = 0 \oplus 1$. Similarly, the composition of two non-abelian anyons is not always unique. An example is given by Fibonacci anyons~\cite{0902.3275}. Two Fibonacci anyons form a composite particle that can possess either the Fibonacci anyonic charge or the vacuum charge, i.e., as for spins, the result is not unique\footnote{The analogy between spin-$1/2$s and Fibonacci anyons has even led to examining the ``golden chain''~\cite{0902.3275, PhysRevLett.98.160409}, which is essentially the analogue of the Heisenberg spin-$1/2$ chain for Fibonacci anyons.}.\\

When talking about composite particles in the context of anyons, the term ``fusion'' is usually used. For the two examples above, we thus say that two semions fuse to the vacuum charge, whereas two Fibonacci anyons fuse either to the vacuum charge or another Fibonacci anyon (or some superposition). Each of these two outcomes is refered to as a ``fusion channel''~\cite{1506.05805}. Using this terminology, the difference between abelian and non-abelian anyons in the context of fusion can be described as abelian anyons always having an unique fusion channel for each pair of anyons to fuse. For non-abelian anyons, however, there exists a pair of anyons that has multiple fusion channels~\cite{einleitung}. The fusion rules described above may be written as $s\times s=1$  ($s$ denotes the semionic charge and $1$ the vacuum charge)~\cite{1506.05805} and $\tau \times \tau = 1+\tau $ ($\tau$ denotes the Fibonacci anyonic charge)~\cite{0902.3275}.

\begin{figure}[t]
\centering
\begin{tikzpicture}[line width=0.75pt]
	\begin{scope}[shift={(0,5.25)}]
	\node[black, anchor=south east] (a) at (-0.25,0.45) {$(a)$};
	\node[draw,circle,inner sep=1.5pt,fill,black] at (0,0) {};
	\node[black] (a) at (0.25,-0.) {$\tau$};
	\node[draw,circle,inner sep=1.5pt,fill,black] at (1.5,0) {};
	\node[black] (a) at (1.75,-0.) {$\tau$};
	\node[draw,circle,inner sep=1.5pt,fill,black] at (3,0) {};
	\node[black] (a) at (3.25,-0.) {$\tau$};
	\draw (0.75,0) ellipse (1.25cm and 0.5cm);
	\draw (1.5,0) ellipse (2.1cm and 0.75cm);
	\node[black] (a) at (2,-0.35) {$1$};
	\node[black] (a) at (3.55,-0.5) {$\tau$};
	\node [black, anchor=west] (a) at (3.7,0) {
            $\begin{aligned}
                \equiv |0\rangle
            \end{aligned}$
    };
	\end{scope}
	\begin{scope}[shift={(0,3.5)}]
	\node[draw,circle,inner sep=1.5pt,fill,black] at (0,0) {};
	\node[black] (a) at (0.25,-0.) {$\tau$};
	\node[draw,circle,inner sep=1.5pt,fill,black] at (1.5,0) {};
	\node[black] (a) at (1.75,-0.) {$\tau$};
	\node[draw,circle,inner sep=1.5pt,fill,black] at (3,0) {};
	\node[black] (a) at (3.25,-0.) {$\tau$};
	\draw (0.75,0) ellipse (1.25cm and 0.5cm);
	\draw (1.5,0) ellipse (2.1cm and 0.75cm);
	\node[black] (a) at (2,-0.35) {$\tau$};
	\node[black] (a) at (3.55,-0.5) {$\tau$};
	\node [black, anchor=west] (a) at (3.7,0) {
            $\begin{aligned}
                \equiv |1\rangle
            \end{aligned}$
    };
	\end{scope}
	\begin{scope}[shift={(0,1.75)}]
	\node[draw,circle,inner sep=1.5pt,fill,black] at (0,0) {};
	\node[black] (a) at (0.25,-0.) {$\tau$};
	\node[draw,circle,inner sep=1.5pt,fill,black] at (1.5,0) {};
	\node[black] (a) at (1.75,-0.) {$\tau$};
	\node[draw,circle,inner sep=1.5pt,fill,black] at (3,0) {};
	\node[black] (a) at (3.25,-0.) {$\tau$};
	\draw (0.75,0) ellipse (1.25cm and 0.5cm);
	\draw (1.5,0) ellipse (2.1cm and 0.75cm);
	\node[black] (a) at (2,-0.35) {$\tau$};
	\node[black] (a) at (3.55,-0.5) {$1$};
	\node [black, anchor=west] (a) at (3.7,0) {
            $\begin{aligned}
                \equiv |N\rangle
            \end{aligned}$
    };
	\end{scope}
	\begin{scope}[shift={(0.5,0)}]
	\node[black, anchor=south east] (a) at (-0.75,0.25) {$(b)$};
	\node[draw,circle,inner sep=1.5pt,fill,black] at (0,0) {};
	\node[black] (a) at (0.25,-0.) {$\tau$};
	\node[draw,circle,inner sep=1.5pt,fill,black] at (1.5,0) {};
	\node[black] (a) at (1.75,-0.) {$\tau$};
	\node[draw,circle,inner sep=1.5pt,fill,black] at (3,0) {};
	\node[black] (a) at (3.25,-0.) {$\tau$};
	\draw [-Stealth][line width=1pt] (0+0.1,0-0.1) to [out=315,in=225] (1.5-0.1,0-0.1);
	\draw [-Stealth][line width=1pt] (1.5-0.1,0+0.1) to [out=135,in=45] (0+0.1,0+0.1);
	\end{scope}
	\begin{scope}[shift={(0.5,-1.5)}]
	\node[black, anchor=south east] (a) at (-0.75,0.25) {$(c)$};
	\node[draw,circle,inner sep=1.5pt,fill,black] at (0,0) {};
	\node[black] (a) at (0.25,-0.) {$\tau$};
	\node[draw,circle,inner sep=1.5pt,fill,black] at (1.5,0) {};
	\node[black] (a) at (1.75,-0.) {$\tau$};
	\node[draw,circle,inner sep=1.5pt,fill,black] at (3,0) {};
	\node[black] (a) at (3.25,-0.) {$\tau$};
	\draw [-Stealth][line width=1pt] (1.5+0.1,0-0.1) to [out=315,in=225] (3-0.1,0-0.1);
	\draw [-Stealth][line width=1pt] (3-0.1,0+0.1) to [out=135,in=45] (1.5+0.1,0+0.1);
	\end{scope}
\end{tikzpicture}
\caption{$(a)$ The three distinct ways three Fibonacci anyons can fuse. Exchanging different pairs of anyons yield different results depending on the state. The exchange in $(b)$ corresponds to Eq.~(\ref{eq:fib_example1}), whereas $(c)$ corresponds to Eq.~(\ref{eq:fib_example2}) for the first two states in $(a)$.}
\label{fig:fib_example}
\end{figure}
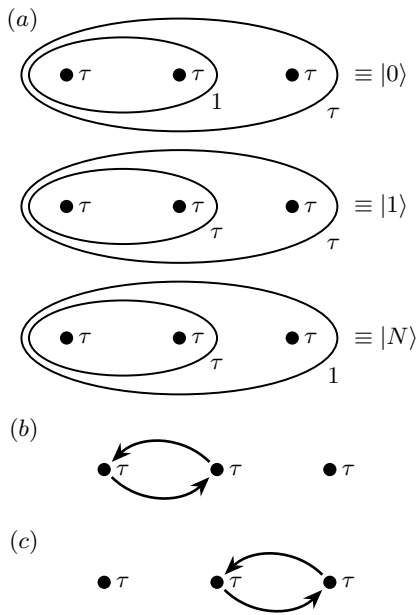
Let us take a system of three Fibonacci anyons to illustrate the different fusion channels. For three Fibonacci anyons, there are three different ways to fuse, corresponding to three different wave function components, which are depicted in Fig.~\ref{fig:fib_example}$a$, where all anyons within a loop fuse to the charge attached to it. The states are labeled $|0\rangle$, $|1\rangle$ and $|N\rangle$, as usually done in topological quantum computing~\cite{1802.06176}.

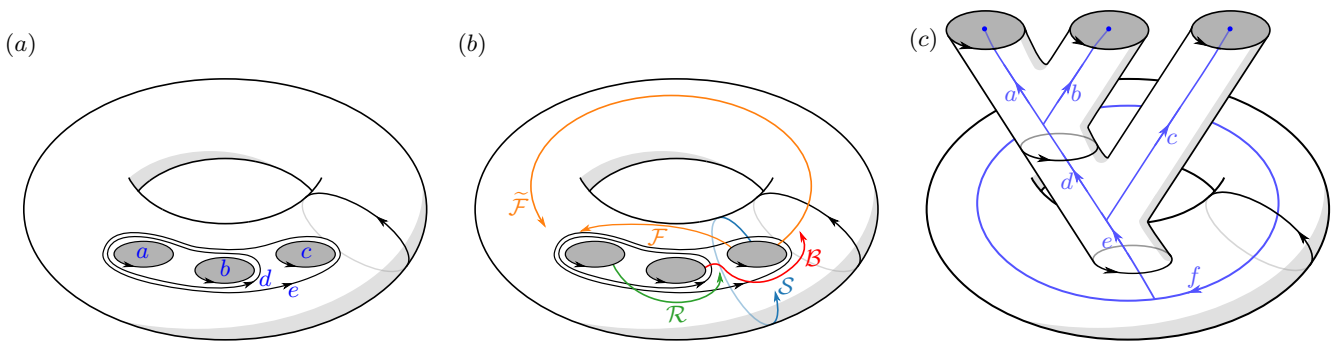
\begin{figure*}[tb]
	\centering
	\begin{tikzpicture}
	\begin{scope}[shift={(-12,0)},scale=3/4.5]
	\node[inner sep=0pt, anchor=south west] (a) at (0,0) {\includegraphics[width=0.3\textwidth]{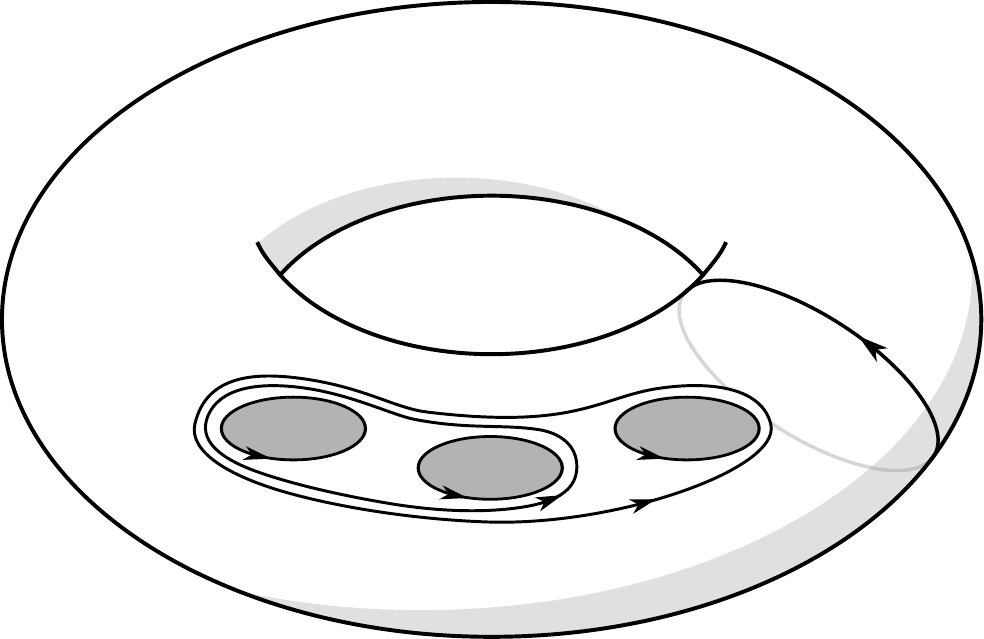}};
	\node[black, anchor=south] (a) at (0.25-0.25*4.5/4,4.75+0.75) {$(a)$};
	\node[blue, anchor=south] (a) at (2.225*1.056+0.05,1.4*1.06+0.0) {$a$};
	\node[blue, anchor=south] (a) at (3.85*1.056-0.025,1.025*1.06+0.05) {$b$};
	\node[blue, anchor=south] (a) at (5.45*1.056-0.1,1.4*1.06+0.0) {$c$};
	\node[blue, anchor=south] (a) at (3.85*1.056+0.775,1.025*1.06-0.09-0.07) {$d$};
	\node[blue, anchor=south] (a) at (5.45*1.056-0.45+0.1,1.4*1.06-0.7-0.07) {$e$};
	\end{scope}
	\begin{scope}[shift={(0,0)}, scale=3/4]
	\node[inner sep=0pt, anchor=south west] (a) at (0,0) {\includegraphics[width=0.3\textwidth]{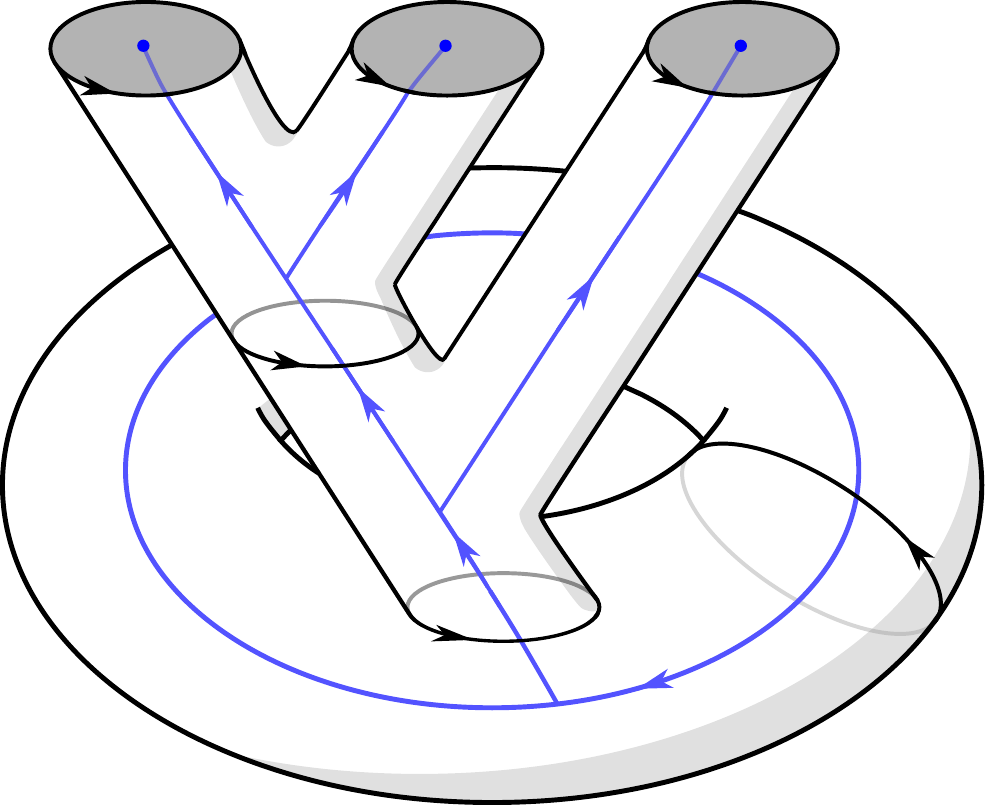}};
	\node[black, anchor=south] (a) at (0.25*4/4.5-0.25,4.75*4/4.5+0.75) {$(c)$};
	\node[shadyblue, anchor=south] (a) at (1.25*1.06+0.2,5.2*1.06-1.4) {$a$};
	\node[shadyblue, anchor=south] (a) at (2.5*1.07,5.2*1.06-1.4) {$b$};
	\node[shadyblue, anchor=south] (a) at (4.1*1.06+0.025,4.2*1.06+0.04-1.15) {$c$};
	\node[shadyblue, anchor=south] (a) at (2.2*1.06+0.18,3.15*1.06+0.05-0.85) {$d$};
	\node[shadyblue, anchor=south] (a) at (2.9*1.06+0.155,1.78*1.06+0.06-0.44) {$e$};
	\node[shadyblue, anchor=south] (a) at (4.5*1.06,0.8*1.06-0.05) {$f$};
	\end{scope}
	\begin{scope}[shift={(-6,0)}, scale=3/4.5]
\node[inner sep=0pt, anchor=south west] (a) at (0,0) {\includegraphics[width=0.3\textwidth]{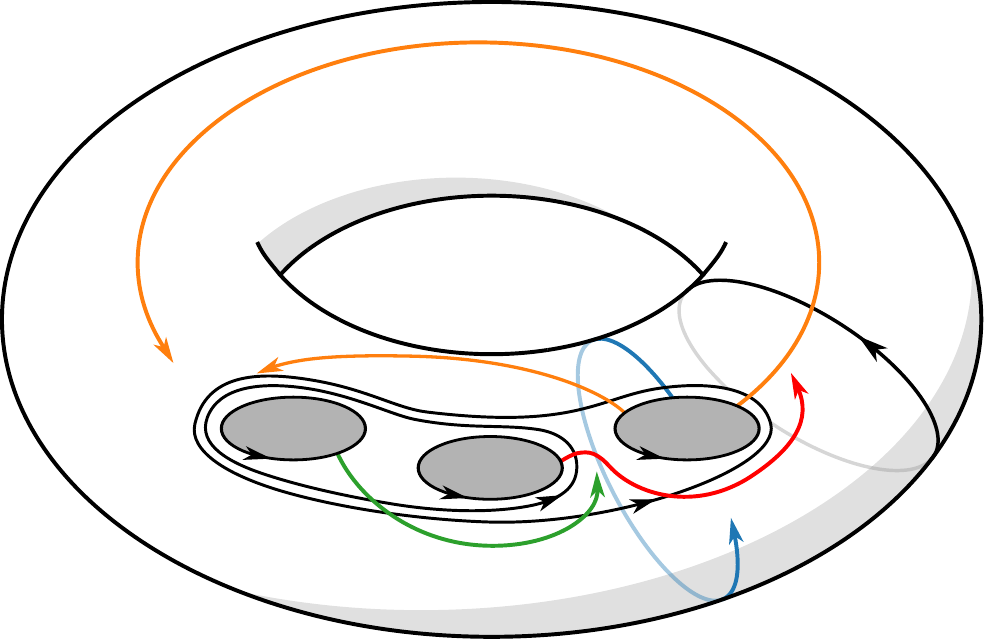}};
\node[black, anchor=south] (a) at (0.25-0.25*4.5/4,4.75+0.75) {$(b)$};
	\node[color2, anchor=south] (a) at (3.85*1.056-0.025-0.375,1.025*1.06+0.125-0.875+1.41) {$\mathcal{F}$};
	\node[color3, anchor=south] (a) at (3.85*1.056-0.025,1.025*1.06+0.125-1) {$\mathcal{R}$};
	\node[red, anchor=south] (a) at (3.85*1.056+2.7,1.025*1.06+0.125+0.125) {$\mathcal{B}$};
	\node[color2, anchor=south] (a) at (3.85*1.056-3.175,1.025*1.06+0.125+1.125) {$\mathcal{\widetilde{F}}$};
	\node[color1, anchor=south] (a) at (3.85*1.056+2.17,1.025*1.06+0.125-0.445) {$\mathcal{S}$};
	\end{scope}
	\end{tikzpicture}
	\caption{$(a)$ Torus with anyons associated with the three punctures. These are, similar to their fusion products, measured by the black oriented lines. $(b)$ Five examples for operations that need to be considered for anyons hopping on a lattice with PBC: $\mathcal{F}$, $\mathcal{R}$ and $\mathcal{B}$ correspond to translations of anyons that can be computed using the $R$- and $F$-moves. Translations of anyons around the torus in $x$-direction ($\mathcal{\widetilde{F}}$) can also be calculated with the $F$-moves. Evaluating translations around the torus in $y$-direction ($\mathcal{S}$) additionally requires the punctured torus $S$-matrix $S^{(z)}$. $(c)$ Fusion diagram associated with the punctured torus. The final fusion product (charge $e$) fuses with the anyonic charge $f$ moving along a non-contractible loop, which follows the negative $x$-direction; the black loop being threaded by $f$ follows the (positive) $y$-direction. The fusion order of the charges indicated here is used as the conventional fusion order for braiding.}
	\label{fig:torus1}
\end{figure*}

It turns out that exchanging (or more generally, braiding) a pair of anyons may have different effects on the state, depending on the fusion products. That is, exchanges are in general associated with unitary matrices rather than simple phases. To illustrate this, we consider the three states $|0\rangle$, $|1\rangle$ and $|N\rangle$ in Fig.~\ref{fig:fib_example}$a$. If the first and second Fibonacci anyon are exchanged counter-clockwise (Fig.~\ref{fig:fib_example}$b$), the transformation between the three wave function components corresponding to the three states is described by the matrix~\cite{1802.06176}
\begin{align}
  \begin{pmatrix}
    e^{-4\pi i/5} & 0 & 0\\
    0 & e^{3\pi i/5} & 0\\
    0 & 0 & e^{3\pi i/5}
  \end{pmatrix}\text{.}
  \label{eq:fib_example1}
\end{align}
In the second case, i.e., counter-clockwise exchanging the second and the third anyon (Fig.~\ref{fig:fib_example}$c$), the corresponding matrix is~\cite{1802.06176}
\begin{align}
  \begin{pmatrix}
    \phi^{-1}e^{4\pi i/5} & \phi^{-1/2}e^{-3\pi i/5} & 0\\
    \phi^{-1/2}e^{-3\pi i/5} & -\phi^{-1} & 0\\
    0 & 0 & e^{3\pi i/5}
  \end{pmatrix}\text{,}
  \label{eq:fib_example2}
\end{align}
where $\phi=(1+\sqrt{5})/2$ is the golden ratio. The off-diagonal entries in this matrix imply that if the initial state is $|0\rangle$ or $|1\rangle$, the result of swapping the second an the third anyon is a superposition of the two states. In both scenarios, we can see that the exchanges do not simply correspond to phases as in the abelian case but to unitary matrices.\\

Using the above example, it can be seen that it is not sufficient to simply determine how many anyons are exchanged in a certain process (and whether or not this exchange is clockwise) as in the abelian case. We need to develop a more sophisticated algorithm that is able to incorporate the non-abelian nature of the anyons. It needs to keep track of the states' fusion products and compute the effects of exchanging anyons based on this information.

\section{Formalism}\label{sec:formalism}

In this section, we review the most important aspects of the formalism~\cite{1506.05805, preskill1999lecture, simon2020topological, 0707.4206, bonderson_2007, doi:10.1063/1.4939783} used for describing abelian and non-abelian anyons. This formalism allows us to connect the physical picture of anyons on a torus (Sec.~\ref{sec:phys_pic}) to fusion diagrams (Sec.~\ref{sec:diagrams}), which is crucial for the discussion of the basis states later on in Sec.~\ref{sec:basis_states}. Considering the different effects translations of anyons may have on the states, we discuss braiding in the fusion diagrams in terms of the $F$-moves (Sec.~\ref{sec:Fmoves}), the $R$-moves (Sec.~\ref{sec:Rmoves}) and the braid operators (Sec.~\ref{sec:braid_op}). For the special case of translating anyons around the torus in $y$-direction, we introduce the punctured torus $S$-matrix in Sec.~\ref{sec:dual} and show how it can be utilized to transform diagrams to their dual versions in order to deal with such translations. We note that due to the formalism being used for both abelian and non-abelian anyons, the resulting algorithm discussed later on reproduces the algorithm sketched in Sec.~\ref{sec:partcile_types_abelian} for the special case of abelian anyon models.

\subsection{Physical Picture}\label{sec:phys_pic}

Physically, we can associate states in the Hilbert space with surfaces, as done in topological quantum field theory (TQFT)~\cite{Witten1989}. In this context, localized anyonic excitations on top of a ground state can be thought of as punctures in the system with which the corresponding topological charges are associated. Here, when talking about ``ground states'', we refer to systems without anyonic excitations. In other contexts, we can refer to states containing anyonic excitations as ground states of the corresponding \emph{punctured} system since anyonic charges are associated with each boundary component~\cite{doi:10.1063/1.4939783}. For PBC, the resulting physical picture is thus a torus with punctures for each anyon, as depicted in Fig.~\ref{fig:torus1}$a$ for three anyons labeled $a$, $b$ and $c$ that are measured\footnote{Anyonic charges are in general measured with respect to oriented loops. This concept will be introduced in Sec.~\ref{sec:diagrams} when discussing fusion diagrams.} with respect to the black oriented loops. Similar to what is depicted in Fig.~\ref{fig:fib_example}$a$, there are also loops that measure the fusion of multiple anyons. In this case, charges $a$ and $b$ fuse to $d$ and $d$ and $c$ (or alternatively, $a$, $b$ and $c$) fuse to charge $e$. Translations of anyons thus correspond to translations of punctures, as depicted in Fig.~\ref{fig:torus1}$b$, where five different translations of punctures that need to be considered are indicated. These include counter-clockwise exchange of the first two anyons' fusion product with the third anyon ($\mathcal{F}$), counter-clockwise exchange of the first and second anyon ($\mathcal{R}$), counter-clockwise exchange of the second and third anyon ($\mathcal{B}$), translation of an anyon around the torus in $x$-direction ($\mathcal{\widetilde{F}}$) and translation of an anyon around the torus in $y$-direction ($\mathcal{S}$). After introducing the representation of anyons using fusion diagrams, we discuss the corresponding operations on them that are needed to compute the effects of all the different translations in the following sections step by step.

In Sec.~\ref{sec:partcile_types_abelian}, we mentioned that translating a semion around the torus changes the physical state. In order to distinguish these different states, we introduce another black loop in Fig.~\ref{fig:torus1}$a$ that measures such charges. Note that the process of translating an anyon around the torus is independent of other anyons in the system since even in the absence of any localized anyons, we can create an anyon-antianyon pair, move one anyon around the torus and annihilate the pair of anyons again, which in general changes the state. The additional black loop is thus important to distinguish different wave function components corresponding to different anyonic charges being translated around the torus. The physical intuition behind this loop will become clearer after the introduction of fusion diagrams below, which will lead us to Fig.~\ref{fig:torus1}$c$.

\subsection{Diagrammatic Representation}\label{sec:diagrams}

Let us start by discussing some basics about anyon models and fusion and introducing the general diagrammatics of anyon models. For this and all the other discussions regarding diagrammatics, we follow Refs.~\cite{simon2020topological},~\cite{0707.4206} and~\cite{2102.05677}. For additional details, we refer the reader to Refs.~\cite{1506.05805, preskill1999lecture, simon2020topological, 0707.4206, bonderson_2007, doi:10.1063/1.4939783,2102.05677}.

Each anyon model contains a set $\mathcal{C}$ of anyonic / topological charges, which are conserved quantum numbers obeying the commutative and associative fusion algebra
\begin{align}
	a\times b=\sum_c N_{ab}^cc,
	\label{eq:fusion_def}
\end{align}
where $N_{ab}^c$ are the so-called fusion multiplicities. The fusion multiplicities are non-negative integers describing the number of distinct ways in which two anyonic charges $a$ and $b$ can fuse to charge $c$. In section~\ref{sec:partcile_types_nonabelian}, the concept of fusion was already introduced for semions ($s\times s=1$, i.e., $N_{ss}^1$) and Fibonacci anyons ($\tau \times \tau = 1+\tau$, i.e., $N_{\tau\tau}^1=N_{\tau\tau}^{\tau}=1$). Equation~(\ref{eq:fusion_def}) simply represents the generalization of these fusion rules. As for the set of anyonic charges $\mathcal{C}$, $\mathcal{C}=\lbrace 1,s \rbrace$ for semions and $\mathcal{C}=\lbrace 1,\tau \rbrace$ for Fibonacci anyons, where $s,\tau$ and $1$ again denote the semionic charge, Fibonacci anyonic charge and vacuum charge, respectively. For Ising anyons, $\mathcal{C}=\lbrace 1,\sigma , \psi \rbrace$ and $N_{\sigma \sigma}^1=N_{\sigma \sigma}^{\psi}=N_{\psi \psi}^1=N_{\sigma \psi}^{\sigma}=1$, where $\sigma$ are Ising anyons and $\psi$ fermions. The vacuum charge $1$ is special in the sense that each anyon model contains a unique charge $1\in \mathcal{C}$ which obeys the fusion rules $N_{a1}^c=\delta_{ac}$ and $N_{ab}^1=\delta_{\overbar{a}b}$, where $\overbar{a}$ refers to the conjugate charge (``antiparticle'') of charge $a$, which is also contained in $\mathcal{C}$: $\overbar{a} \in \mathcal{C}$, with $\overbar{1}=1$. This means that fusion with the vacuum charge is trivial.

The difference between abelian and non-abelian anyon models can now be described as non-abelian anyon models possessing the property of there being at least a single pair of charges $a$ and $b$ for which $\sum_c N_{ab}^c>1$, i.e., there are multiple fusion channels for these two charges. Abelian anyon models on the other hand possess the property that there is a unique fusion product for each pair of charges, i.e., $\sum_cN_{ab}^c=1$ for all $a,b\in\mathcal{C}$.

With the concept of fusion, we can introduce the diagrammatic notation for anyons, where the anyonic charges are represented by oriented lines with the corresponding charge labels attached to them. The following diagram is an example, in which charges $a$ and $b$ fuse to charge $c$:
\begin{align}
\begin{split}
\begin{tikzpicture}[line width=0.75pt, scale=0.75]
	\draw [black,domain=1:3, samples=10] plot ({\x*cos(120)}, {\x*sin(120)+0.75});
	\draw [black,domain=0:3-0.866/sin(60)*2, samples=10] plot ({\x*cos(60)-0.5*2}, {\x*sin(60)+0.75+0.866*2});	
	\node[black, anchor=south] (a) at (-1.5,3.3) {$a$};
	\node[black, anchor=south] (a) at (-1.5+1,3.3) {$b$};
	\node[black, anchor=north east] (a) at (-0.75+0.1,0.75+1.299+0.1) {$c$};
	\draw[black] (-1.25+0.005,0.75+2.165-0.00866) -- (-1.25,0.75+2.165) node[sloped,pos=1,allow upside down]{\arrowIn}; ; 
	\draw[black] (-0.75+0.005,0.75+1.299-0.00866) -- (-0.75,0.75+1.299) node[sloped,pos=1,allow upside down]{\arrowIn}; ; 
	\draw[black] (-0.75-0.005,0.75+2.165-0.00866) -- (-0.75,0.75+2.165) node[sloped,pos=1,allow upside down]{\arrowIn}; ; 
	\node [black, anchor=west] (a) at (-0.45,0.75+1.299+0.433) {
            $\begin{aligned}
                .
            \end{aligned}$
    };
\end{tikzpicture}
\end{split}
\label{eq:fusionabc}
\end{align}
This is only allowed if $N_{ab}^c\neq 0$ and corresponds to the diagram
\begin{align}
\begin{split}
\begin{tikzpicture}[line width=0.75pt]
	\begin{scope}[shift={(0,5.25)}]
	\node[draw,circle,inner sep=1.5pt,fill,black] at (0,0) {};
	\node[black] (a) at (0.25,-0.) {$a$};
	\node[draw,circle,inner sep=1.5pt,fill,black] at (1.5,0) {};
	\node[black] (a) at (1.75,-0.) {$b$};
	\draw (0.75,0) ellipse (1.25cm and 0.5cm);
	\draw[black] (0.75+1.25*0.5,-0.5*0.866) -- (0.75+1.25*0.5+0.00866*1.25,-0.5*0.866+0.005*0.5) node[sloped,pos=1,allow upside down]{\arrowIn}; ; 
	\node[black] (a) at (1.5,-0.6) {$c$};
	\end{scope}
\end{tikzpicture}
\end{split}
\label{eq:fusionabc_loop}
\end{align}
in the notation introduced previously in Fig.~\ref{fig:fib_example}$a$ for Fibonacci anyons. Oriented loops, such as the one encircling charges $a$ and $b$, are associated with projectors onto anyonic charges~\cite{doi:10.1063/1.4939783, arxiv.1306.2379}. In notations displaying such loops, like in, e.g., Eq.~(\ref{eq:fusionabc_loop}), we only associate anyonic charges consistent with the fusion rules with these loops since otherwise, the state is annihilated by the projection. Such loops can be thought of as measuring the total charge of the enclosed anyons. In particular, we can also associate oriented loops with the punctures of the surface, such that these loops measure the anyonic charges of the excitations, as depicted for the punctures in Fig.~\ref{fig:torus1}$a$. Note that the orientation of the loop measuring charge $c$ in Eq.~(\ref{eq:fusionabc_loop}) is of importance since reversing its orientation corresponds to measuring the conjugated charge $\overline{c}$. Similar to reversing the orientation of the loops, one can also reverse the orientations of the lines in the fusion diagrams (Eq.~(\ref{eq:fusionabc})). This corresponds to replacing an anyonic charge by its conjugate charge and may be interpreted as a particle moving forward in time being identical to its antiparticle moving backwards in time.

In Eqs.~(\ref{eq:fusionabc}) and (\ref{eq:fusionabc_loop}), one would in principle need to introduce an additional label taking values $1,...,N_{ab}^c$ in order to distinguish the different ways in which the charges $a$ and $b$ can combine to form charge $c$. For convenience, we assume that $N_{ab}^c \in \lbrace 0,1 \rbrace$ such that we can ignore such labels and simplify the notation. Diagrams such as the one in Eq.~(\ref{eq:fusionabc}) can be thought of as states in the corresponding ``fusion space''. We denote the state represented by this diagram as $|(ab)_c \rangle$, where the charges in the brackets fuse to the charge in brackets' index.

In order to connect the fusion diagrams to physical states like the one depicted in Fig.~\ref{fig:torus1}$a$, we deform the torus containing the anyons such that it resembles the diagrams. This deformation is done in such a way that the resulting state does not differ from the initial one in a topological sense, i.e., the punctures and the handle (and the associated charges) remain unchanged. The final result can be seen in Fig.~\ref{fig:torus1}$c$, where the fusion diagram is depicted inside the torus. The diagram contains the additional anyon $f$, which moves along a non-contractible loop in $x$-direction. This anyon corresponds to the flux that we used in Sec.~\ref{sec:partcile_types_abelian} to distinguish the two wave function components for semions, which can be now described by $f=1$ and $f=s$. Note that anyon $f$ is not associated with any puncture and does not represent an excitation; it is also present in the system's ground states~\cite{doi:10.1063/1.4939783} and corresponds to the charge $f$ moving around the torus. It is essential in order to distinguish the different physical states, as indicated above for the semionic case. Looking at Fig.~\ref{fig:torus1}$c$, it can be seen that the state can represented by the diagram

\begin{align}
\begin{split}
\begin{tikzpicture}[line width=0.75pt, scale=0.8]
	\draw (0,0) ellipse (1cm and 0.75cm);
	\draw (0,0) ellipse (0.25cm and 0.25cm);
	\node[black, anchor=north west] (a) at (0.01,0.0) {$y$};
	\draw[black] (0.177,0.177) -- (-0.177,-0.177);
	\draw[black] (0.177,-0.177) -- (-0.177,0.177);
	\node[black, anchor=south] (a) at (-1.5,3.3) {$a$};
	\node[black, anchor=south] (a) at (-1.5/3,3.3) {$b$};
	\node[black, anchor=south] (a) at (1.5/3,3.3) {$c$};
	\node[black, anchor=north] (a) at (-1.5/3+0.05,0.75+2.598/3-0.3) {$e$};
	\node[black, anchor=north] (a) at (-1.5/3*2,0.75+2.598/3*2-0.2) {$d$};
	\node[black] (a) at (0.91+0.05,0.75) {$f$};
	\draw[black] (0,0.75) -- (-1.5/3,0.75+2.598/3) node[sloped,pos=0.50,allow upside down]{\arrowIn}; ; 
	\draw[black] (-1.5/3,0.75+2.598/3) -- (-1.5/3*2,0.75+2.598/3*2) node[sloped,pos=0.50,allow upside down]{\arrowIn}; ; 
	\draw[black] (-1.5/3*2,0.75+2.598/3*2) -- (-1.5,0.75+2.598) node[sloped,pos=0.50,allow upside down]{\arrowIn}; ; 
	\draw[black] (-1.5/3,0.75+2.598/3) -- (1.5/3,0.75+2.598) node[sloped,pos=0.50,allow upside down]{\arrowIn}; ; 
	\draw[black] (-1.5/3*2,0.75+2.598/3*2) -- (-1.5/3,0.75+2.598) node[sloped,pos=0.50,allow upside down]{\arrowIn}; ; 	
	\draw[black, line width=0.05pt] (0.707+0.707*0.001,0.707*0.75-0.707*0.75*0.001) -- (0.707,0.707*0.75) node[sloped,pos=0.50,allow upside down]{\arrowIn}; ; 
    \node[black, anchor=west] (a) at (1.05,2.96/2-0.2) {,};
\end{tikzpicture}
\end{split}
\label{eq:torus_repr}
\end{align}
where anyons $a$ and $b$ fuse to $d$ and $d$ and $c$ to $e$. The symbol $\otimes$ represents a non-contractible loop and the attached index $y$ is used to represent its direction. This loop is thus complementary to the one along which charge $f$ moves. The index is of importance as both directions need to be considered in the algorithm. Similar to before, we may choose to write the state depicted in Eq.~(\ref{eq:torus_repr}) as $|((ab)_dc)_e;f\rangle$.

In Fig.~\ref{fig:torus1}$c$ and Eq.~(\ref{eq:torus_repr}), it can be seen that the fusion product of all anyons (charge $e$) is connected to charge $f$~\cite{InsideOutsideBases}. This implies that for the state in Fig.~\ref{fig:torus1}$c$ to exist, $N_{ef}^f \neq 0$ has to be fulfilled. Fig.~\ref{fig:torus1}$c$ also indicates the convention for the fusion order that will be used later on for the basis states: First, the two ``left'' anyons $a$ and $b$ fuse, then their fusion product fuses with the anyon associated with the third puncture. This fusion product then fuses with the anyon threading the torus. This fusion order can be generalized to more anyons by letting them fuse one after another, starting from the left. The generalized diagram for $N$ anyons then reads
\begin{align}
\begin{split}
\begin{tikzpicture}[line width=0.75pt]
	\draw (0,0.15) ellipse (0.8cm and 0.6cm);
	\draw (0,0.15) ellipse (0.2cm and 0.2cm);
	\node[black, anchor=north west] (a) at (0.01,0.15) {$y$};
	\draw[black] (0.177*0.8,0.177*0.8+0.15) -- (-0.177*0.8,-0.177*0.8+0.15);
	\draw[black] (0.177*0.8,-0.177*0.8+0.15) -- (-0.177*0.8,0.177*0.8+0.15);
	\draw [black,domain=0:2.1, samples=10] plot ({\x*cos(120)}, {\x*sin(120)+0.75});
	\draw [black,domain=2.7:4, samples=10] plot ({\x*cos(120)}, {\x*sin(120)+0.75});
	\draw [black,domain=0:4-0.6928/sin(60), samples=10] plot ({\x*cos(60)-0.4}, {\x*sin(60)+0.75+0.6928});
	\draw [black,domain=0:4-0.6928/sin(60)*2, samples=10] plot ({\x*cos(60)-0.4*2}, {\x*sin(60)+0.75+0.6928*2});
	\draw [black,domain=0:4-0.6928/sin(60)*4, samples=10] plot ({\x*cos(60)-0.4*4}, {\x*sin(60)+0.75+0.6928*4});
	\node[black, anchor=south] (a) at (-2,4.15) {$\alpha_1$};
	\node[black, anchor=south] (a) at (-2+0.8,4.15) {$\alpha_2$};
	\node[black, anchor=south] (a) at (-2+0.8*3-0.05,4.15) {$\alpha_{N-1}$};
	\node[black, anchor=south] (a) at (-2+0.8*4+0.05,4.15) {$\alpha_{N}$};
	\node[black, anchor=north] (a) at (-2+0.3,4-0.6928+0.1) {$f_1$};
	\node[black, anchor=north] (a) at (-2+0.4*2-0.1,4-0.6928*2-0.175) {$f_{N-3}$};
	\node[black, anchor=north] (a) at (-2+0.4*3-0.2,4-0.6928*3) {$f_{N-2}$};
	\node[black, anchor=north] (a) at (-2+0.4*4-0.2,4-0.6928*4+0.05) {$f_{N-1}$};
	\node[black] (a) at (0.9*0.8,0.7*0.8+0.15+0.1) {$f_N$};
	\draw[black] (-0.5*4*9/10+0.005,0.75+0.866*4*9/10-0.00866) -- (-0.5*4*9/10,0.75+0.866*4*9/10) node[sloped,pos=1,allow upside down]{\arrowIn}; ; 
	\draw[black] (-0.5*4*7.5/10+0.005,0.75+0.866*4*7.5/10-0.00866) -- (-0.5*4*7.5/10,0.75+0.866*4*7.5/10) node[sloped,pos=1,allow upside down]{\arrowIn}; ; 
	\draw[black] (-0.5*4*4.5/10+0.005,0.75+0.866*4*4.5/10-0.00866) -- (-0.5*4*4.5/10,0.75+0.866*4*4.5/10) node[sloped,pos=1,allow upside down]{\arrowIn}; ; 
	\draw[black] (-0.5*4*3/10+0.005,0.75+0.866*4*3/10-0.00866) -- (-0.5*4*3/10,0.75+0.866*4*3/10) node[sloped,pos=1,allow upside down]{\arrowIn}; ; 
	\draw[black] (-0.5*4*1/10+0.005,0.75+0.866*4*1/10-0.00866) -- (-0.5*4*1/10,0.75+0.866*4*1/10) node[sloped,pos=1,allow upside down]{\arrowIn}; ; 
	\draw[black] (0.8/2-0.005,5.657/2-0.00866) -- (0.8/2,5.657/2) node[sloped,pos=1,allow upside down]{\arrowIn}; ; 
	\draw[black] (-0.4/2-0.005,6.35/2-0.00866) -- (-0.4/2,6.35/2) node[sloped,pos=1,allow upside down]{\arrowIn}; ; 
	\draw[black] (-2.8/2-0.005,7.735/2-0.00866) -- (-2.8/2,7.735/2) node[sloped,pos=1,allow upside down]{\arrowIn}; ; 
	\draw[black, line width=0.05pt] (0.707*0.8+0.707*0.001*0.8,0.707*0.75*0.8-0.707*0.75*0.001*0.8+0.15) -- (0.707*0.8,0.707*0.75*0.8+0.15) node[sloped,pos=0.50,allow upside down]{\arrowIn}; ; 
	\draw [line width=2pt, line cap=round, new dash=on 0pt off 5pt] (-1.05*4/5-1.35/5,2.57*4/5+3.09/5) -- (-1.05/5-1.35*4/5,2.57/5+3.09*4/5);
	\draw [line width=2pt, line cap=round, new dash=on 0pt off 5pt] (-2+0.8+0.6-0.1,4.15+0.3) -- (-2+0.8*3-0.6-0.1,4.15+0.3);
	\draw [line width=2pt, line cap=round, new dash=on 0pt off 5pt] (-0.2*0.65-1.4*0.35,6.35/2*0.65+7.735/2*0.35) -- (-0.2*0.35-1.4*0.65,6.35/2*0.35+7.735/2*0.65);
	\node [black, anchor=west] (a) at (-2+0.8*4+0.15,3.824/2) {
            $\begin{aligned}
                ,
            \end{aligned}$
    };
\end{tikzpicture}
\end{split}
\label{eq:general_state}
\end{align}
where $\alpha_i$ are the anyonic charges associated with the punctures of the torus, $f_i$ their fusion products according to the defined fusion order with $i=1,\ldots,N-1$ and $f_N$ the anyon moving along the non-contractible loop. The corresponding state is denoted by $| (\ldots((\alpha_1\alpha_2)_{f_1}\alpha_3)_{f_2}\ldots \alpha_N)_{f_{N-1}};f_N\rangle$. The diagrams in Eq.~(\ref{eq:general_state}) define the canonical form for the fusion diagrams that will be used for the basis states to be introduced in Sec.~\ref{sec:basis_states}. They will be utilized for all braid operations to be performed in our algorithm and will be connected to lattice configurations of anyons via a fusion order on the lattice. Apart from the fusion consistency conditions in all other vertices that also have to be satisfied, the generalized consistency condition involving the final fusion product of all the anyons and the anyon moving along the non-contractible loop reads
\begin{align}
	N^{f_N}_{f_{N-1}f_N} \neq 0.
	\label{eq:fusion_consistency}
\end{align}
Note that for abelian anyon models, Eq.~(\ref{eq:fusion_consistency}) can only be fulfilled if $f_{N-1}=1$ due to the relations $\sum_cN_{ab}^c=1$ and $N_{1a}^b=\delta_{ab}$. I.e., abelian anyons must fuse to the vacuum charge in order to be able to exist on a torus, which implies that semions must appear in even numbers.

\subsection{The \texorpdfstring{$F$}{F}-moves}\label{sec:Fmoves}

Using the diagrammatic notation, we can now take a closer look at translation processes and their effects on the fusion diagrams by utilizing the five examples depicted in Fig.~\ref{fig:torus1}$b$. Let us start by considering $\mathcal{F}$, which translates anyon $c$ to the left such that the resulting fusion diagram is not in canonical form since the first anyons to fuse are still $a$ and $b$, which are now on the right side of the diagram. We can relate the obtained non-conventional fusion order to the canonical one using the $F$-moves:
\begin{align}
\begin{split}
\begin{tikzpicture}[line width=0.75pt, scale=0.75]
	\draw [black,domain=0:3, samples=10] plot ({\x*cos(120)}, {\x*sin(120)+0.75});
	\draw [black,domain=0:3-0.866/sin(60), samples=10] plot ({\x*cos(60)-0.5}, {\x*sin(60)+0.75+0.866});
	\draw [black,domain=0:3-0.866/sin(60)*2, samples=10] plot ({\x*cos(60)-0.5*2}, {\x*sin(60)+0.75+0.866*2});	
	\node[black, anchor=south] (a) at (-1.5,3.3) {$c$};
	\node[black, anchor=south] (a) at (-1.5+1,3.3) {$a$};
	\node[black, anchor=south] (a) at (-1.5+2,3.3) {$b$};
	\node[black, anchor=north east] (a) at (-0.25+0.1,0.75+0.433+0.1) {$e$};
	\node[black, anchor=north east] (a) at (-0.75+0.1,0.75+1.299+0.1) {$f$};
	\draw[black] (-1.25+0.005,0.75+2.165-0.00866) -- (-1.25,0.75+2.165) node[sloped,pos=1,allow upside down]{\arrowIn}; ; 
	\draw[black] (-0.75+0.005,0.75+1.299-0.00866) -- (-0.75,0.75+1.299) node[sloped,pos=1,allow upside down]{\arrowIn}; ; 
	\draw[black] (-0.25+0.005,0.75+0.433-0.00866) -- (-0.25,0.75+0.433) node[sloped,pos=1,allow upside down]{\arrowIn}; ; 
	\draw[black] (0-0.005,0.75+1.732-0.00866) -- (0,0.75+1.732) node[sloped,pos=1,allow upside down]{\arrowIn}; ; 
	\draw[black] (-0.75-0.005,0.75+2.165-0.00866) -- (-0.75,0.75+2.165) node[sloped,pos=1,allow upside down]{\arrowIn}; ; 
	\begin{scope}[shift={(5.0,0)}, xscale=-1]
	\node [black, anchor=east] (a) at (0.65,0.75+1.299) {
            $\begin{aligned}
                =\sum_{d}\left[F^{cab}_{e}\right]_{fd}
            \end{aligned}$
    };
	\draw [black,domain=0:3, samples=10] plot ({\x*cos(120)}, {\x*sin(120)+0.75});
	\draw [black,domain=0:3-0.866/sin(60), samples=10] plot ({\x*cos(60)-0.5}, {\x*sin(60)+0.75+0.866});
	\draw [black,domain=0:3-0.866/sin(60)*2, samples=10] plot ({\x*cos(60)-0.5*2}, {\x*sin(60)+0.75+0.866*2});
	\node[black, anchor=south] (a) at (-1.5,3.3) {$b$};
	\node[black, anchor=south] (a) at (-1.5+1,3.3) {$a$};
	\node[black, anchor=south] (a) at (-1.5+2,3.3) {$c$};
	\node[black, anchor=north west] (a) at (-0.25+0.1,0.75+0.433+0.1) {$e$};
	\node[black, anchor=north west] (a) at (-0.75+0.1,0.75+1.299+0.1) {$d$};
	\draw[black] (-1.25-0.005,0.75+2.165-0.00866) -- (-1.25,0.75+2.165) node[sloped,pos=1,allow upside down]{\arrowIn}; ; 
	\draw[black] (-0.75-0.005,0.75+1.299-0.00866) -- (-0.75,0.75+1.299) node[sloped,pos=1,allow upside down]{\arrowIn}; ; 
	\draw[black] (-0.25-0.005,0.75+0.433-0.00866) -- (-0.25,0.75+0.433) node[sloped,pos=1,allow upside down]{\arrowIn}; ; 
	\draw[black] (0+0.005,0.75+1.732-0.00866) -- (0,0.75+1.732) node[sloped,pos=1,allow upside down]{\arrowIn}; ; 
	\draw[black] (-0.75+0.005,0.75+2.165-0.00866) -- (-0.75,0.75+2.165) node[sloped,pos=1,allow upside down]{\arrowIn}; ; 
	\node[black, anchor=west] (a) at (-1.65,0.75+1.299) {,};
	\end{scope}
\end{tikzpicture}
\end{split}
\label{eq:Fmoves}
\end{align}
where we omitted the parts of the diagrams that are not affected by the $F$-moves. In the considered scenario (operation $\mathcal{F}$ in Fig.~\ref{fig:torus1}$b$), one would thus have use the inverse $F$-moves to bring the diagram back to its canonical form. Due to the unitarity of the $F$-moves, the inverse is given by
\begin{align}
	\left[F^{abc}_{d}\right]^{-1}_{ef}=\left[F^{abc}_{d}\right]^{\dagger}_{ef} = \left[F^{abc}_{d}\right]_{fe}^*.
\end{align}
Further, the $F$-moves are $F^{abc}_d=\mathbb{1}$ if any of the charges $a$, $b$ or $c$ is the trivial charge $1$ and the fusion is allowed by the fusion rules. As for the anyon models to be studied in this paper, the only non-trivial $F$-move for semions is $\left[F^{sss}_s\right]_{11}=-1$~\cite{1506.05805}. For Fibonacci anyons, the non-trivial $F$-moves are~\cite{1802.06176}
\begin{align}
	F^{\tau\tau\tau}_{\tau}=\left( \begin{array}{rrr}
\phi^{-1} & \phi^{-1/2} \\ 
\phi^{-1/2} & -\phi^{-1} \\
\end{array} \right),
\label{eq:Fmove_Fib}
\end{align}
where the first (second) entry in each row / column corresponds to charge $1$ ($\tau$) and $\phi=(1+\sqrt{5})/2$ is again the golden ratio. For Ising anyons, the non-trivial $F$-moves~\cite{bonderson_2007} are $[F^{\sigma \psi \sigma}_{\psi}]_{\sigma \sigma}=[F^{\psi \sigma \psi}_{\sigma}]_{\sigma \sigma}=-1$ and
\begin{align}
	F^{\sigma \sigma \sigma}_{\sigma}=\frac{1}{\sqrt{2}}\left( \begin{array}{rrr}
1 & 1 \\ 
1 & -1 \\
\end{array} \right),
\label{eq:Fmove_Ising}
\end{align}
where the first and second entries in each row / column correspond to $1$ and $\psi$, respectively.

It turns out that the $F$-moves are even sufficient to obtain the canonical form of the braided fusion diagrams for processes that translate anyons around the torus in $x$-direction, like $\mathcal{\widetilde{F}}$ in Fig.~\ref{fig:torus1}$b$. A calculation showing the general idea of how this works is given later in Fig.~\ref{eq:drag_x_2p} in Sec.~\ref{sec:algo_cuts} for the case of two anyons. The generalization to $N$ anyons can be found in App.~\ref{app_example_computations_sheet_hopping}. Note that above, we only considered the change in fusion order due to the process $\mathcal{F}$ in Fig.~\ref{fig:torus1}$b$. We ignored that one first has to exchange anyon $c$ with $d$ (the fusion product of anyons $a$ and $b$) to arrive at this state. Such exchanges are considered in the following section.

\subsection{The \texorpdfstring{$R$}{R}-moves}\label{sec:Rmoves}

Exchanges of two anyons that fuse with each other, such as indicated by $\mathcal{R}$ in Fig.~\ref{fig:torus1}$b$, may be represented in the fusion diagrams as
\begin{align}
\begin{split}
\begin{tikzpicture}[line width=0.75pt, scale=0.75]
	\draw [black,domain=1:2, samples=10] plot ({\x*cos(120)}, {\x*sin(120)+0.75});
	\draw[black] (-1,2.48) arc (-60:47.5:0.5);
	\draw[black] (-1,2.48+0.866) arc (120:240:0.5);
	\draw [black,domain=0:0.577, samples=10] plot ({\x*cos(30)-1}, {\x*sin(30)+2.48+0.866});
	\draw [black,domain=0.125:0.577, samples=10] plot ({-1*\x*cos(30)-1}, {\x*sin(30)+2.48+0.866});
	\node[black, anchor=south] (a) at (-1.5,3.58) {$b$};
	\node[black, anchor=south] (a) at (-1.5+1,3.58) {$a$};
	\node[black, anchor=north east] (a) at (-0.75+0.1,0.75+1.299+0.1) {$d$};
	\draw[black] (-1.25+0.00866,3.49-0.005) -- (-1.25,3.49) node[sloped,pos=1,allow upside down]{\arrowIn}; ; 
	\draw[black] (-0.75+0.005,0.75+1.299-0.00866) -- (-0.75,0.75+1.299) node[sloped,pos=1,allow upside down]{\arrowIn}; ; 
	\draw[black] (-0.75-0.00866,3.49-0.005) -- (-0.75,3.49) node[sloped,pos=1,allow upside down]{\arrowIn}; ; 
	\begin{scope}[shift={(3.3,0.14)}]
	\node [black, anchor=east] (a) at (-1.65,0.75+1.299+0.433) {
            $\begin{aligned}
                =R^{ab}_{d}
            \end{aligned}$
    };
	\draw [black,domain=1:3, samples=10] plot ({\x*cos(120)}, {\x*sin(120)+0.75});
	\draw [black,domain=0:3-0.866/sin(60)*2, samples=10] plot ({\x*cos(60)-0.5*2}, {\x*sin(60)+0.75+0.866*2});	
	\node[black, anchor=south] (a) at (-1.5,3.3) {$b$};
	\node[black, anchor=south] (a) at (-1.5+1,3.3) {$a$};
	\node[black, anchor=north east] (a) at (-0.75+0.1,0.75+1.299+0.1) {$d$};
	\draw[black] (-1.25+0.005,0.75+2.165-0.00866) -- (-1.25,0.75+2.165) node[sloped,pos=1,allow upside down]{\arrowIn}; ; 
	\draw[black] (-0.75+0.005,0.75+1.299-0.00866) -- (-0.75,0.75+1.299) node[sloped,pos=1,allow upside down]{\arrowIn}; ; 
	\draw[black] (-0.75-0.005,0.75+2.165-0.00866) -- (-0.75,0.75+2.165) node[sloped,pos=1,allow upside down]{\arrowIn}; ; 
	\node [black, anchor=west] (a) at (-0.45,0.75+1.299+0.433) {
            $\begin{aligned}
                ,
            \end{aligned}$
    };
	\end{scope}
\end{tikzpicture}
\end{split}
\label{eq:Rmove}
\end{align}
where it was used that the braided diagram can be written in terms of the unbraided diagram using $R$-moves. In Eq.~(\ref{eq:Rmove}), we omitted the parts of the diagrams that are unaffected by the $R$-moves. The $R$-moves are unitary, i.e., they obey $\left(R^{ab}_d\right)^{-1}=\left(R^{ab}_d\right)^{\dagger}=\left(R^{ba}_d\right)^*$, where $R^{ab}_d$ corresponds to counter-clockwise and $\left(R^{ab}_d\right)^{-1}$ to clockwise exchange. Further, braiding with the vacuum charge is trivial, $R^{1b}_d=R^{a1}_d=1$. Note that the same exchange as above was already introduced in Fig.~\ref{fig:fib_example}$b$ for Fibonacci anyons. The corresponding $R$-moves are given by~\cite{1802.06176}
\begin{align}
	R^{\tau\tau}_1=e^{-4\pi i/5} \quad \text{and}\quad R^{\tau\tau}_{\tau}=e^{3\pi i/5}.
\label{eq:Rmove_Fib}
\end{align}
With this, it is easy to see that the exchange in Fig.~\ref{fig:fib_example}$b$ indeed corresponds to the matrix in Eq.~(\ref{eq:fib_example1}). For semions, the only non-trivial $R$-move is $R^{ss}_1=i$~\cite{simon2020topological}. The non-trivial $R$-moves for Ising anyons are $R^{\sigma \sigma}_{1}=e^{-i\frac{\pi}{8}}$, $R^{\sigma \sigma}_{\psi}=e^{i\frac{3\pi}{8}}$, $R^{\sigma \psi}_{\sigma}=R^{\psi \sigma}_{\sigma}=e^{-i\frac{\pi}{2}}$ and $R^{\psi \psi}_{1}=-1$~\cite{bonderson_2007}.

\subsection{The Braid Operator}\label{sec:braid_op}

After having briefly discussed how exchanges of anyons that fuse with each other affect the diagrams, it is natural to go one step further and consider exchanges of anyons that do not directly fuse with each other. An example for such an exchange process is given by $\mathcal{B}$ in Fig.~\ref{fig:torus1}$b$, where anyon $b$ is exchanged with anyon $c$. The idea how to resolve such braids is not complicated: We know from Sec.~\ref{sec:Fmoves} that we can use $F$-moves to change the order in which the anyons fuse with each other. We can thus simply apply $F$-moves until the anyons to be exchanged directly fuse with each other, then use the $R$-moves introduced in the previous section and finally transform the diagram back to its canonical form using $F$-moves again. For the counter-clockwise exchange corresponding to $\mathcal{B}$ in Fig.~\ref{fig:torus1}$b$, this is written in diagrams as
\begin{align}
\begin{split}
\begin{tikzpicture}[line width=0.75pt, scale=0.75]
	\draw [black,domain=0:3, samples=10] plot ({\x*cos(120)}, {\x*sin(120)+0.75});
	\draw [black,domain=0:1.5-0.13, samples=10] plot ({\x*cos(60)-0.5}, {\x*sin(60)+0.75+0.866});
	\draw [black,domain=1.5+0.13:3-0.866/sin(60), samples=10] plot ({\x*cos(60)-0.5}, {\x*sin(60)+0.75+0.866});
	\draw [black,domain=0:(3-0.866/sin(60)*2)/2, samples=10] plot ({\x*cos(60)-0.5*2}, {\x*sin(60)+0.75+0.866*2});
	\draw [black,domain=(3-0.866/sin(60)*2)/2:3-0.866/sin(60)*2, samples=10] plot ({\x*cos(60)+1}, {\x*sin(60)+0.75+0.866*2});
	\draw [black,domain=0-0.01:2+0.01, samples=10] plot ({\x-0.75}, {2.915});
	\node[black, anchor=south] (a) at (-1.5,3.3) {$a$};
	\node[black, anchor=south] (a) at (-1.5+3,3.3) {$c$};
	\node[black, anchor=south] (a) at (-1.5+2,3.3) {$b$};
	\node[black, anchor=north east] (a) at (-0.25+0.1,0.75+0.433+0.1) {$e$};
	\node[black, anchor=north east] (a) at (-0.75+0.1,0.75+1.299+0.1) {$d$};
	\draw[black] (-1.25+0.005,0.75+2.165-0.00866) -- (-1.25,0.75+2.165) node[sloped,pos=1,allow upside down]{\arrowIn}; ; 
	\draw[black] (-0.75+0.005,0.75+1.299-0.00866) -- (-0.75,0.75+1.299) node[sloped,pos=1,allow upside down]{\arrowIn}; ; 
	\draw[black] (-0.25+0.005,0.75+0.433-0.00866) -- (-0.25,0.75+0.433) node[sloped,pos=1,allow upside down]{\arrowIn}; ; 
	\draw[black] (0-0.005,0.75+1.732-0.00866) -- (0,0.75+1.732) node[sloped,pos=1,allow upside down]{\arrowIn}; ; 
	\draw[black] (0.025-0.005,2.915) -- (0.025,2.915) node[sloped,pos=1,allow upside down]{\arrowIn}; ; 
	\begin{scope}[shift={(7,0)}]
	\node [black, anchor=east] (a) at (-1.65,0.75+1.299) {
            $\begin{aligned}
                =\sum_{f}\left[B^{abc}_{e}\right]_{df}
            \end{aligned}$
    };
	\draw [black,domain=0:3, samples=10] plot ({\x*cos(120)}, {\x*sin(120)+0.75});
	\draw [black,domain=0:3-0.866/sin(60), samples=10] plot ({\x*cos(60)-0.5}, {\x*sin(60)+0.75+0.866});
	\draw [black,domain=0:3-0.866/sin(60)*2, samples=10] plot ({\x*cos(60)-0.5*2}, {\x*sin(60)+0.75+0.866*2});
	\node[black, anchor=south] (a) at (-1.5,3.3) {$a$};
	\node[black, anchor=south] (a) at (-1.5+1,3.3) {$b$};
	\node[black, anchor=south] (a) at (-1.5+2,3.3) {$c$};
	\node[black, anchor=north east] (a) at (-0.25+0.1,0.75+0.433+0.1) {$e$};
	\node[black, anchor=north east] (a) at (-0.75+0.1,0.75+1.299+0.1) {$f$};
	\draw[black] (-1.25+0.005,0.75+2.165-0.00866) -- (-1.25,0.75+2.165) node[sloped,pos=1,allow upside down]{\arrowIn}; ; 
	\draw[black] (-0.75+0.005,0.75+1.299-0.00866) -- (-0.75,0.75+1.299) node[sloped,pos=1,allow upside down]{\arrowIn}; ; 
	\draw[black] (-0.25+0.005,0.75+0.433-0.00866) -- (-0.25,0.75+0.433) node[sloped,pos=1,allow upside down]{\arrowIn}; ; 
	\draw[black] (0-0.005,0.75+1.732-0.00866) -- (0,0.75+1.732) node[sloped,pos=1,allow upside down]{\arrowIn}; ; 
	\draw[black] (-0.75-0.005,0.75+2.165-0.00866) -- (-0.75,0.75+2.165) node[sloped,pos=1,allow upside down]{\arrowIn}; ; 
	\end{scope}
	\end{tikzpicture}
	\end{split}\label{eq:defB}
\end{align}
where the unitary braid operator $B$ is
\begin{align}
	&\left[ B^{abc}_e \right]_{df} = \sum_g \left[ F^{acb}_e \right]_{dg} R^{cb}_g  \left[ F^{abc}_e \right]^{-1}_{gf}.
	\label{eq:defB_formula}
\end{align}
Its inverse $\left[ B^{abc}_e \right]_{df}^{-1}=\left[ B^{abc}_e \right]_{df}^{\dagger}$ can be used to compute clockwise exchanges of anyons. Note that the exchange associated with $\mathcal{B}$ in Fig.~\ref{fig:torus1}$b$ is the same as the one introduced in Fig.~\ref{fig:fib_example}$c$ for Fibonacci anyons. Using Eq.~(\ref{eq:defB_formula}) together with Eqs.~(\ref{eq:Fmove_Fib}) and (\ref{eq:Rmove_Fib}), it is straight forward to verify that the exchange in Fig.~\ref{fig:fib_example}$c$ indeed corresponds to the matrix in Eq.~(\ref{eq:fib_example2}).

In general, when dealing with more than three anyons, exchanges can be computed in a similar way by using multiple $F$-moves until the $R$-moves can be applied and then transforming the diagram back to the canonical form. With this, one can in principle compute any exchanges of anyons, no matter how many anyons are involved and which anyons are to be exchanged. The remaining translation processes that need to be considered involve translations of anyons around the torus in $y$-direction, as indicated by $\mathcal{S}$ in Fig.~\ref{fig:torus1}$b$. For such translations, we need to introduce the punctured torus $S$-matrix.

\subsection{Dual States and the Punctured Torus \texorpdfstring{$S$}{S}-matrix}\label{sec:dual}

Our approach of dealing with translation processes that involve anyons moving around the torus in $y$-direction, such as $\mathcal{S}$ in Fig.~\ref{fig:torus1}$b$, is to transform the corresponding fusion diagrams such that the translations resemble translation processes of anyons around the torus in $x$-direction, like $\mathcal{\widetilde{F}}$ in Fig.~\ref{fig:torus1}$b$, since we already know how to deal with these translations. The sought-for transformation changes the direction of the non-contractible loops in the fusion diagrams and can be achieved using the punctured torus $S$-matrix~\cite{2102.05677, InsideOutsideBases}, as we will see below. The punctured torus $S$-matrix is given by

\begin{align}
\begin{split}
&\begin{tikzpicture}[line width=0.75pt, scale=0.75]
	\draw (1.25,-0.75) ellipse (1.25cm and 0.5cm);
	\draw (0.625+1.25,0.75) ellipse (0.625cm and 0.5cm);
	\filldraw [fill=white, draw=white] (0,-0.725) rectangle (2.5,-0.2);
	\filldraw [fill=white, draw=white] (1.25,0.225) rectangle (2.5,0.725);
	\draw[black] (-0.75,0) arc (180:45:0.75);
	\draw[black] (-0.75,0) arc (180:385:0.75);
	\draw[black] (0.5,0) arc (180:-135:0.75);
	\draw[black] (0.5,0) arc (180:205:0.75);
	\draw[black] (-0.75,-0.001) -- (-0.75,0) node[sloped,pos=1,allow upside down]{\arrowIn}; ; 
	\draw[black] (0.5,-0.001) -- (0.5,0) node[sloped,pos=1,allow upside down]{\arrowIn}; ; 
	\node[black, anchor=east] (a) at (-0.7,0) {$a$};
	\node[black, anchor=east] (a) at (0.55,0) {$b$};
	\node[black, anchor=west] (a) at (2.4,0) {$z$};
	\draw[black] (2.5,-0.75) -- (2.5,0.75) node[sloped,pos=0.5,allow upside down]{\arrowIn}; ;
    \node [black, anchor=east] (a) at (-1.1,0) {
            $\begin{aligned}
                S^{(z)}_{ab}=\frac{1}{\mathcal{D}\sqrt{d_z}}
            \end{aligned}$
    };
\end{tikzpicture}
\end{split}\label{eq:Sz}\\
&=\frac{1}{\mathcal{D}}\sum_c d_ad_b\frac{\theta_c}{\theta_a\theta_b}\left[F^{b\overline{b}a}_a\right]_{1c} \left[F^{b\overline{b}a}_a\right]^{-1}_{c\overline{z}} \left[F^{\overline{z}a\overline{a}}_{\overline{z}}\right]_{a1},\nonumber
\end{align}
with $N_{a\overline{a}}^z\neq 0$ and $N_{b\overline{b}}^z\neq 0$; $d_a$ denotes the quantum dimension of charge $a$, $\theta_a$ its topological spin and $\mathcal{D}$ the total quantum dimension. These quanities are given by
\begin{align}
\begin{split}
\begin{tikzpicture}[line width=0.75pt, scale=0.75]
	\draw (0,0) ellipse (0.75cm and 0.75cm);
	\draw[black] (-0.75,-0.005) -- (-0.75,0) node[sloped,pos=1,allow upside down]{\arrowIn}; ; 
	\node[black, anchor=east] (a) at (-0.7,0) {$a$};
	\node [black, anchor=west] (a) at (0.8,0) {
            $\begin{aligned}
                =\left|\left[F^{a\overline{a}a}_a\right]_{11}\right|^{-1},
            \end{aligned}$
    };
    \node [black, anchor=east] (a) at (-1.0,0) {
            $\begin{aligned}
                d_a=
            \end{aligned}$
    };
\end{tikzpicture}
\end{split}\\
\begin{split}
\begin{tikzpicture}[line width=0.75pt, scale=0.75]
	\draw (0,0) ellipse (0.75cm and 0.75cm);
	\draw (2,0) ellipse (0.75cm and 0.75cm);
	\filldraw [fill=white, draw=white] (0.55,-0.75) rectangle (1.45,0.75);
	\draw[black] (0.5,-0.559) -- (1.5,0.559) node[sloped,pos=0.25,allow upside down]{\arrowIn}; ; 
	\draw [black,domain=0:0.35, samples=10] plot ({0.5+\x}, {0.559-2*0.559*\x});
	\draw [black,domain=0.65:1, samples=10] plot ({0.5+\x}, {0.559-2*0.559*\x});
	\node[black, anchor=north west] (a) at (0.625,-0.559/2+0.15) {$a$};
	\node [black, anchor=west] (a) at (2.85,0) {
            $\begin{aligned}
                =\sum_c \frac{d_c}{d_a}R^{aa}_c
            \end{aligned}$
    };
    \node [black, anchor=east] (a) at (-0.7,0) {
            $\begin{aligned}
                \theta_a=\frac{1}{d_a}
            \end{aligned}$
    };
\end{tikzpicture}
\end{split}\\
\begin{split}
\text{and}\quad \mathcal{D}=\sqrt{\sum_ad_a^2}.
\end{split}
\end{align}

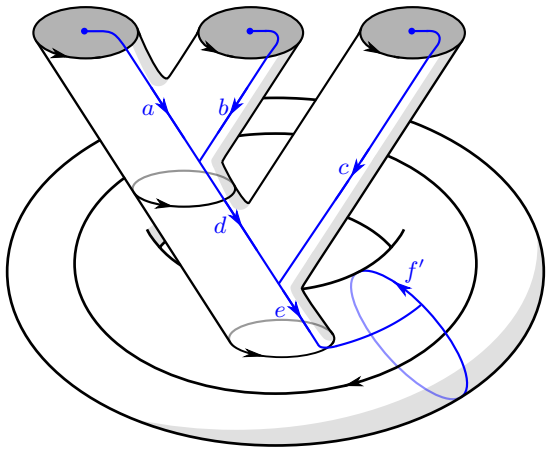
\begin{figure}[tb]
	\centering
	\begin{tikzpicture}
\node[inner sep=0pt, anchor=south west] (a) at (0,0) {\includegraphics[width=0.4\textwidth]{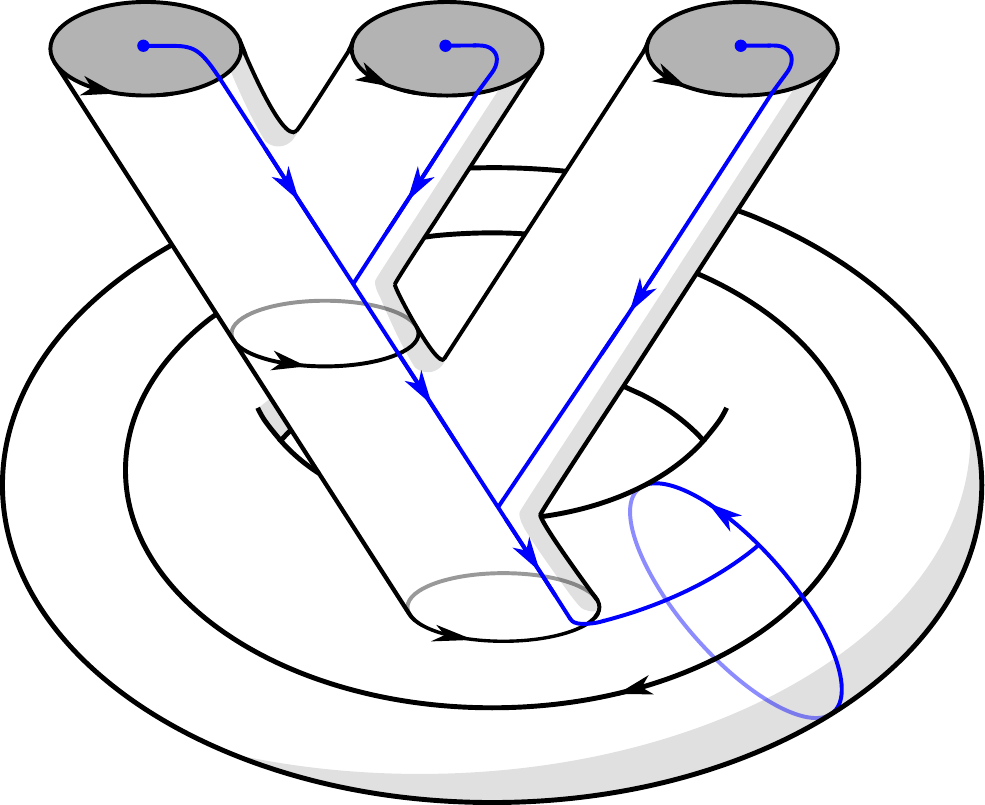}};
	\node[blue, anchor=south] (a) at (1.25*1.06+0.575,5.2*1.06-1.2) {$a$};
	\node[blue, anchor=south] (a) at (2.5*1.06+0.25,5.2*1.06-1.2) {$b$};
	\node[blue, anchor=south] (a) at (4.1*1.06+0.15,4.2*1.06+0.04-0.975) {$c$};
	\node[blue, anchor=south] (a) at (2.2*1.06+0.525,3.15*1.06+0.05-0.65) {$d$};
	\node[blue, anchor=south] (a) at (2.9*1.06+0.58,1.78*1.06+0.06-0.325) {$e$};
	\node[blue, anchor=south] (a) at (4.5*1.06+0.675,0.8*1.06+0.05+1.175) {$f'$};
	\end{tikzpicture}
	\caption{Physical picture of the dual state in the so-called ``outside basis''~\cite{InsideOutsideBases} obtained by transforming with the punctured torus $S$-matrix. The charge $f'$ moves along a loop complementary to the one of charge $f$ in the initial state in Fig.~\ref{fig:torus1}$c$ and the lines corresponding to the charges associated with the punctures are outside the torus rather than inside.}
	\label{fig:torus_outsidebasis}
\end{figure}

In many other contexts, $S^{(1)}$ is simply refered to as ``the $S$-matrix''. $S^{(z)}$ can be used to transform a state to its dual representation. In the dual basis, the anyon moving along the non-contractible loop is replaced by a different anyon moving along another, complementary non-contractible loop, which may be illustrated using Fig.~\ref{fig:torus1}$c$: When transforming with $S^{(e)}$, anyon $f$ is replaced by another anyon $f'$ moving along a complementary non-contractible loop, similar to the black one measuring the anyonic charge $f$ in the initial state. This black loop is then also transformed to a non-contractible loop complementary to its initial one after the transformation, similar to the initial loop of charge $f$. The resulting physical picture is illustrated in Fig.~\ref{fig:torus_outsidebasis}. The lines corresponding to the charges associated with the punctures are also slightly different: Rather than inside the torus, they are now located outside, similar to the non-contractible loop of charge $f'$; the orientation of the lines is also reversed\footnote{This can be seen by computing the inner product~\cite{kirillov1996inner} between a fusion diagram and its transformed counterpart, which corresponds to $S^{(z)}$ in Eq.~(\ref{eq:Sz}). Note that the inner product we refer to here corresponds to the trace of the inner product as defined in Ref.~\cite{InsideOutsideBases}.}.

For $S^{(z)}$ to actually be the correct transformation, all the charges associated with the punctures on the torus need to fuse to charge $z$. The two different basis choices are also refered to as ``inside basis'' and ``outside basis''~\cite{InsideOutsideBases}; Fig.~\ref{fig:torus1}$c$ corresponds to the inside basis and Fig.~\ref{fig:torus_outsidebasis} to the outside basis.

In the diagrammatic notation, the basis change using the punctured torus $S$-matrix reads
\begin{align}
\begin{split}
\begin{tikzpicture}[line width=0.75pt, scale=0.8]
	\draw (0,0) ellipse (1cm and 0.75cm);
	\draw (0,0) ellipse (0.25cm and 0.25cm);
	\node[black, anchor=north west] (a) at (0.01,0.0) {$y$};
	\draw[black] (0.177,0.177) -- (-0.177,-0.177);
	\draw[black] (0.177,-0.177) -- (-0.177,0.177);
	\node[black, anchor=south] (a) at (-1.5,3.3) {$a$};
	\node[black, anchor=south] (a) at (-1.5/3,3.3) {$b$};
	\node[black, anchor=south] (a) at (1.5/3,3.3) {$c$};
	\node[black, anchor=north] (a) at (-1.5/3+0.05,0.75+2.598/3-0.3) {$e$};
	\node[black, anchor=north] (a) at (-1.5/3*2,0.75+2.598/3*2-0.2) {$d$};
	\node[black] (a) at (0.91+0.05,0.75) {$f$};
	\draw[black] (0,0.75) -- (-1.5/3,0.75+2.598/3) node[sloped,pos=0.50,allow upside down]{\arrowIn}; ; 
	\draw[black] (-1.5/3,0.75+2.598/3) -- (-1.5/3*2,0.75+2.598/3*2) node[sloped,pos=0.50,allow upside down]{\arrowIn}; ; 
	\draw[black] (-1.5/3*2,0.75+2.598/3*2) -- (-1.5,0.75+2.598) node[sloped,pos=0.50,allow upside down]{\arrowIn}; ; 
	\draw[black] (-1.5/3,0.75+2.598/3) -- (1.5/3,0.75+2.598) node[sloped,pos=0.50,allow upside down]{\arrowIn}; ; 
	\draw[black] (-1.5/3*2,0.75+2.598/3*2) -- (-1.5/3,0.75+2.598) node[sloped,pos=0.50,allow upside down]{\arrowIn}; ; 	
	\draw[black, line width=0.05pt] (0.707+0.707*0.001,0.707*0.75-0.707*0.75*0.001) -- (0.707,0.707*0.75) node[sloped,pos=0.50,allow upside down]{\arrowIn}; ; 
	\begin{scope}[shift={(5.4+0.5,0)}]
	\node [black, anchor=east] (a) at (-1.65,2.96/2) {
            $\begin{aligned}
                \xrightarrow{\text{dual}}\sum_{f'}S_{ff'}^{(e)\dagger}
            \end{aligned}$
    };
    \node[black, anchor=west] (a) at (1.1,2.96/2-0.1) {,};
	\draw (0,0) ellipse (1cm and 0.75cm);
	\draw (0,0) ellipse (0.25cm and 0.25cm);
	\node[black, anchor=north west] (a) at (0.01,0.0) {$x$};
	\draw[black] (0.177,0.177) -- (-0.177,-0.177);
	\draw[black] (0.177,-0.177) -- (-0.177,0.177);
	\node[black, anchor=south] (a) at (-1.5,3.3) {$a$};
	\node[black, anchor=south] (a) at (-1.5/3,3.3) {$b$};
	\node[black, anchor=south] (a) at (1.5/3,3.3) {$c$};
	\node[black, anchor=north] (a) at (-1.5/3+0.05,0.75+2.598/3-0.3) {$e$};
	\node[black, anchor=north] (a) at (-1.5/3*2,0.75+2.598/3*2-0.2) {$d$};
	\node[black] (a) at (0.975,0.75) {$f'$};
	\draw[black] (-1.5/3,0.75+2.598/3) -- (0,0.75) node[sloped,pos=0.50,allow upside down]{\arrowIn}; ; 
	\draw[black] (-1.5/3*2,0.75+2.598/3*2) -- (-1.5/3,0.75+2.598/3) node[sloped,pos=0.50,allow upside down]{\arrowIn}; ; 
	\draw[black] (-1.5,0.75+2.598) -- (-1.5/3*2,0.75+2.598/3*2) node[sloped,pos=0.50,allow upside down]{\arrowIn}; ; 
	\draw[black] (1.5/3,0.75+2.598) -- (-1.5/3,0.75+2.598/3) node[sloped,pos=0.50,allow upside down]{\arrowIn}; ; 
	\draw[black] (-1.5/3,0.75+2.598) -- (-1.5/3*2,0.75+2.598/3*2) node[sloped,pos=0.50,allow upside down]{\arrowIn}; ; 	
	\draw[black, line width=0.05pt] (0.707,0.707*0.75) -- (0.707+0.707*0.001,0.707*0.75-0.707*0.75*0.001) node[sloped,pos=0.50,allow upside down]{\arrowIn}; ; 
	\end{scope}
\end{tikzpicture}
\end{split}
\label{eq:transform_example}
\end{align}
where we denote the change to the dual basis by $\xrightarrow{\text{dual}}$ and $\otimes_x$ represents a non-contractible loop complementary to the loop in $y$-direction along which $f'$ moves. The sum runs over all these anyonic charges $f'$ and the orientation of all lines associated with the punctures is reversed. The orientation of the line corresponding to the anyon moving along the non-contractible loop is also reversed upon changing the basis, which follows from the definition in Eq.~(\ref{eq:Sz}). Diagrams corresponding to states in the dual space are commonly depicted upside down~\cite{bonderson_2007, simon2020topological}. We will not do so here for convenience. Whether a diagram corresponds to a state in the dual space or not can always be infered from the index $x$ or $y$ attached to $\otimes$ and the orientation of the lines. Note that the transformation in Eq.~(\ref{eq:transform_example}) is done using $S^{(e)\dagger}$ rather than $S^{(e)}$. This is a detail arising from the way the crossings of the non-contractible loops in Eq.~(\ref{eq:Sz}) are oriented with respect to each other~\cite{KitaevFormulae}.

\section{Basis States on the Lattice}\label{sec:basis_states}

Before discussing how to use the diagrams introduced in the previous section to construct the anyonic tight-binding Hamiltonian, we need to introduce a basis of the Hilbert space on the lattice and make a few remarks. In this section, we introduce the basis states by connecting anyon configurations to fusion diagrams. Then, in Sec.~\ref{sec:sheets}, we discuss how to systematically determine the wave function components, which correspond to different fusion products in the fusion diagrams in Eq.~(\ref{eq:general_state}). For systems featuring anyonic excitations of distinct charges, i.e., for distinguishable anyons, additional aspects regarding the wave function components need to be considered. These aspects are discussed in Sec.~\ref{sec:anyonorderings} and may be omitted if all anyonic excitations in the system have the same charge. In particular, this is the case for the semion and the Fibonacci anyon model. Even for the Ising anyon model, we will focus on the case where all excitations are Ising anyons such that the details of Sec.~\ref{sec:anyonorderings} are not needed to reproduce our results presented in Sec.~\ref{sec:results}.

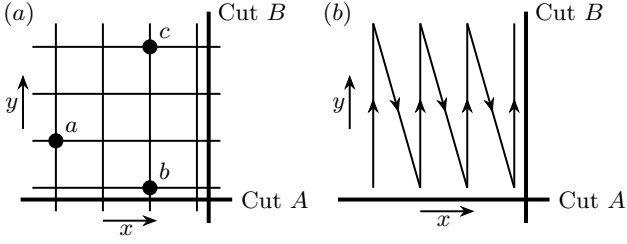
\begin{figure}[t]
\centering
\begin{tikzpicture}[line width=0.75pt, scale=0.625]
	\draw[black] (-0.5,0) -- (3.5,0);
	\draw[black] (-0.5,1) -- (3.5,1);
	\draw[black] (-0.5,2) -- (3.5,2);
	\draw[black] (-0.5,3) -- (3.5,3);
	\draw[black] (0,-0.5) -- (0,3.5);
	\draw[black] (1,-0.5) -- (1,3.5);
	\draw[black] (2,-0.5) -- (2,3.5);
	\draw[black] (3,-0.5) -- (3,3.5);
	\draw[black][line width=1.5pt] (3.25,-0.75) -- (3.25,3.75);
	\node[black, anchor=west] (a) at (3.25,3.75) {Cut $B$};
	\draw[black][line width=1.5pt] (-0.75,-0.25) -- (3.75,-0.25);
	\node[black, anchor=west] (a) at (3.75,-0.25) {Cut $A$};
	\node[black, anchor=south east] (a) at (-0.25,3.25) {$(a)$};
	\node[draw,circle,inner sep=1.75pt,fill,black] at (2,0) {};
	\node[draw,circle,inner sep=1.75pt,fill,black] at (2,3) {};
	\node[draw,circle,inner sep=1.75pt,fill,black] at (0,1) {};
	\node[black, anchor=south west] (a) at (2,0) {$b$};
	\node[black, anchor=south west] (a) at (2,3) {$c$};
	\node[black, anchor=south west] (a) at (0,1) {$a$};
	\draw[black] (1,-0.7) -- (2,-0.7) node[sloped,pos=1,allow upside down]{\arrowIn}; ; 
	\draw[black] (-0.7,1.25) -- (-0.7,2.25) node[sloped,pos=1,allow upside down]{\arrowIn}; ; 
	\node[black, anchor=north] (a) at (1.5,-0.6) {$x$};
	\node[black, anchor=east] (a) at (-0.6,1.75) {$y$};
	\begin{scope}[shift={(6.75,0)}]
	\node[black, anchor=south east] (a) at (-0.25,3.25) {$(b)$};
	\draw[black] (0,0) -- (0,3.5) node[sloped,pos=0.5,allow upside down]{\arrowIn}; ; 
	\draw[black] (0,3.5) -- (1,0) node[sloped,pos=0.5,allow upside down]{\arrowIn}; ; 
	\draw[black] (1,0) -- (1,3.5) node[sloped,pos=0.5,allow upside down]{\arrowIn}; ; 
	\draw[black] (1,3.5) -- (2,0) node[sloped,pos=0.5,allow upside down]{\arrowIn}; ; 
	\draw[black] (2,0) -- (2,3.5) node[sloped,pos=0.5,allow upside down]{\arrowIn}; ; 
	\draw[black] (2,3.5) -- (3,0) node[sloped,pos=0.5,allow upside down]{\arrowIn}; ; 
	\draw[black] (3,0) -- (3,3.5) node[sloped,pos=0.5,allow upside down]{\arrowIn}; ; 
	\draw[black] (1,-0.5) -- (2,-0.5) node[sloped,pos=1,allow upside down]{\arrowIn}; ; 
	\draw[black] (-0.5,1.25) -- (-0.5,2.25) node[sloped,pos=1,allow upside down]{\arrowIn}; ; 
	\node[black, anchor=north] (a) at (1.5,-0.4) {$x$};
	\node[black, anchor=east] (a) at (-0.4,1.75) {$y$};
	\draw[black][line width=1.5pt] (3.25,-0.75) -- (3.25,3.75);
	\node[black, anchor=west] (a) at (3.25,3.75) {Cut $B$};
	\draw[black][line width=1.5pt] (-0.75,-0.25) -- (3.75,-0.25);
	\node[black, anchor=west] (a) at (3.75,-0.25) {Cut $A$};
	\end{scope}
\end{tikzpicture}
\caption{$(a)$ Choice of cuts $A$ and $B$ on a lattice containing three anyons corresponding to the diagram in Eq.~(\ref{eq:torus_repr}) if the fusion products are chosen accordingly. $(b)$ Choice of fusion order. Anyons with small $x$-coordinates fuse first; among those with the same $x$-coordinate, anyons with smaller $y$-coordinates fuse first.}
\label{fig:lattice}
\end{figure}

The first step towards choosing the basis states is to connect anyon configurations on the lattice with the fusion diagrams. This is done as depicted in Fig.~\ref{fig:lattice}. By noting that the fusion order convention is indicated in Fig.~\ref{fig:lattice}$b$, the first anyons to fuse in the corresponding diagrams are the ones with the smallest $x$-coordinates. Among those with the same $x$-coordinates, the anyons with smaller $y$-coordinates fuse before those with larger $y$-coordinates. This means that the anyon configuration depicted in Fig.~\ref{fig:lattice}$a$ may correspond to the diagram in Eq.~(\ref{eq:torus_repr}). However, we cannot tell from the anyon configuration itself whether or not this is the case since it only shows the charges of the localized anyons and not of their fusion products. That is, in terms of the previously introduced notation, the configuration in Fig.~\ref{fig:lattice}$a$ corresponds to some superposition of states $|((ab)_dc)_e;f\rangle$, where $d$, $e$ and $f$ are not determined by the configuration itself. As this information is of fundamental importance for the underlying physics, we introduce tuples $\mathbf{f}$ that contain the fusion products and the anyon moving along the non-contractible loop. The tuple corresponding to the state $|((ab)_dc)_e;f\rangle$ is $\mathbf{f}=(d,e,f)$. Combining anyon configurations and tuples $\mathbf{f}$ thus fixes the topological charges in the fusion diagrams, i.e., different fusion products in the diagrams correspond to different tuples. The tuples therefore correspond to the wave function components of a physical state.

With this consideration, a basis of the Hilbert space for $N$ anyons associated with the diagrams in Eq.~(\ref{eq:general_state}) is given by the states
\begin{align}
	\begin{split}
	&|(\alpha_k,\vec{r}_k)_{k=1}^N, \mathbf{f}\rangle \label{eq:realspacebasis} \\
	&\equiv | (\alpha_k,\vec{r}_k)_{k=1}^N \rangle\otimes | (\ldots((\alpha_1\alpha_2)_{f_1}\alpha_3)_{f_2}\ldots \alpha_N)_{f_{N-1}};f_n\rangle,
	\end{split}
\end{align}
where $(\alpha_k,\vec{r}_k)$ denote tuples containing the charge $ \alpha_k$ and position $ \vec{r}_k $ of the $k$-th anyon; $(\alpha_k,\vec{r}_k)_{k=1}^N$ denotes all such tuples ordered according to the fusion order in Fig.~\ref{fig:lattice}$b$. The tuples $\mathbf{f}$ contain the fusion products of the fusion diagram in Eq.~(\ref{eq:general_state}) associated with the state, i.e., $\mathbf{f}=(f_1,f_2,\ldots,f_{N})$. We discuss how to systematically obtain all the tuples that are consistent with the fusion rules in Sec.~\ref{sec:sheets}. The overlap between two states $|(\alpha_k,\vec{r}_k)_{k=1}^N, \mathbf{f}\rangle$  and $|(\alpha'_k,\pvec{r}'_k)_{k=1}^N, \mathbf{f'}\rangle$ is given by
\begin{align}
\begin{split}
	&\langle(\alpha'_k,\pvec{r}'_k)_{k=1}^N,\mathbf{f'}|(\alpha_k,\vec{r}_k)_{k=1}^N, \mathbf{f}\rangle \\ &=  \left( \prod_{k=1}^N \delta_{\alpha'_k\alpha_k}\delta_{\pvec{r}'_k \vec{r}_k} \right) \times  \delta_{\mathbf{f'}\mathbf{f}},
\end{split}
\end{align}
where $\delta_{x'x}=1$ if $x'=x$ and $\delta_{x'x}=0$ otherwise; this holds for topological charges, (position) vectors and tuples.\\

Similar to the abelian case sketched in Sec.~\ref{sec:partcile_types_abelian}, two cuts, one in $x$- and one in $y$-direction named cuts $A$ and $B$, were introduced to the lattice depicted in Fig.~\ref{fig:lattice}. Cut $A$ is chosen to have a $y$-coordinate equal to $y=1/2$, such that translating an anyon with a $y$-coordinate of $y=L_y$ in positive $y$-direction translates it over the cut, where $y=L_y$ corresponds to $y=0$ due to the PBC. Similarly, cut $B$ is chosen to have an $x$-coordinate of $x=1/2$, with $x=L_x$ corresponding to $x=0$. The two cuts may be viewed as two loops that need to be passed when translating an anyon around the torus. They are needed in order to keep track of the topological charge associated with the PBCs and thus play a special role when constructing the Hamiltonian later on, as they do in the algorithm sketched in Sec.~\ref{sec:partcile_types_abelian}.

\subsection{Determining the Wave Function Components}\label{sec:sheets}

We have seen above that the tuples $\mathbf{f}$ are essential for assigning an anyon configuration to a fusion diagram / wave function component. It is thus important to discuss a systematic way to determine the tuples corresponding to actual physical states, i.e., determining fusion products that agree with the fusion rules of the anyons in the system. Before considering the general case, let us take a look at the semion, the Fibonacci anyon and the Ising anyon model for concreteness.

For semions, with the fusion rules $s\times s=1$ and $s\times 1 =s$~\cite{1506.05805}, there are either two or zero wave function components. If there is an even number of semions in the system, the anyons always fuse to the vacuum charge. Then, the consistency condition~(\ref{eq:fusion_consistency}) is fulfilled for both $f_N=1$ and $f_N=s$ and thus, there are two different wave function components corresponding to the tuples $(1,s,1,s,\ldots,1,1)$ and $(1,s,1,s,\ldots,1,s)$. If the number of semions is odd, the consistency condition~(\ref{eq:fusion_consistency}) can never be fulfilled, i.e., such a system can not exist on a torus, which corresponds to zero wave function components / tuples $\mathbf{f}$. This argument agrees with the constraint $e^{i2\theta N}=1$ for abelian anyons~\cite{PhysRevB.43.10761} (with $\theta=\pi/2$ for semions) and also confirms that semions have two wave function components on a torus, as stated in Sec.~\ref{sec:partcile_types_abelian}.

For Fibonacci anyons, the consistency condition~(\ref{eq:fusion_consistency}) can be fulfilled for an arbitrary (positive) number of $\tau$-anyons. The reason for this is the non-abelian fusion rule $\tau \times \tau = 1+\tau $~\cite{0902.3275}, which makes sure that both cases $f_{N-1}=1$ and $f_{N-1}=\tau$ are consistent with $f_N=\tau$. Thus, any number of Fibonacci anyons may exist on a torus, which is in sharp contrast to the semion case or more generally the abelian case, for which $f_{N-1}=1$ has to be fulfilled. For, e.g., a single Fibonacci anyon, the tuple corresponding to the only wave function component is $(\tau)$, which only contains the anyon associated with the non-contractible loop. For two Fibonacci anyons, the tuples are $(1,1)$, $(1,\tau)$ and $(\tau ,\tau)$, i.e., there are three wave function components. More generally, the total number of wave function components $n_c(N)$ for $N>0$ Fibonacci anyons is
\begin{align}
	n_{c}(N) = 2F_{N-1}+F_N,
\end{align}
with the Fibonacci series $F_{N+1}=F_N+F_{N-1}$, $F_0=0$ and $F_1=1$. This formula is based on the fact that if the $N$ $\tau$-anyons fuse to $1$, both $f_N=1$ and $f_N=\tau$ are consistent with Eq.~(\ref{eq:fusion_consistency}), whereas if they fuse to $\tau$, only $f_N=\tau$ is allowed. Using that the number of different ways of $N$ $\tau$s fusing to $\tau$ is given by $F_N$~\cite{0902.3275}, one arrives at the above relation as due to the fusion rules, the only way for $N$ $\tau$s to fuse to $1$ is for the first $N-1$ $\tau$s to fuse to $\tau$.

For Ising anyons, we can see from the fusion rules $\sigma \times \sigma = 1+\psi$, $\sigma \times 1 = \sigma$ and $\sigma \times \psi = \sigma$~\cite{bonderson_2007} that for an odd number of $\sigma$s, that is, for $N$ odd, $f_{N-1}=\sigma$. This implies that the consistency condition~(\ref{eq:fusion_consistency}) cannot be fulfilled, i.e., only an even number of Ising anyons can exist on tori, which is similar to the semionic case. From the just mentioned fusion rules and $\psi \times \psi =1$, we also see that $f_N=1,\sigma,\psi$ is consistent with $f_{N-1}=1$, whereas only $f_N=\sigma$ is consistent with $f_{N-1}=\psi$. Overall, taking the intermediate fusion products into account, the total number of wave function components $n_c(N)$ for $N>0$ Ising anyons can be written as
\begin{align}
	n_c(N) = \begin{cases}
	\sqrt{2}^{N+2},\quad &\text{for $N$ even}\\ 
	0,\quad &\text{for $N$ odd.} 
	\end{cases}
\end{align}
\\

In the general case, the total number of wave function components can be obtained by iterating over all combinations of fusion products $\lbrace f_i \rbrace$ in Eq.~(\ref{eq:general_state}) and checking the fusion multiplicites in each of the $N$ vertices. If there is a vertex that is associated with a multiplicity $N_{ab}^c=0$, the corresponding state is unphysical. The total number of tuples $\mathbf{f}=(f_1,f_2,\ldots,f_{N})$ associated with physical states thus corresponds to the number of wave function components\footnote{Note again that we focus on $N_{ab}^c\in \lbrace 0,1 \rbrace$. For $N_{ab}^c >1$, there may be multiple wave function components associated with the same tuple $\mathbf{f}$, which may be distinguished by additional labels.}.

\begin{algorithm}[t]
\caption{\justifyingg{Construction of the set of tuples $\mathbf{f}$ whose entries $f_i$ are the fusion products of $N$ anyons $\lbrace \alpha_i\rbrace$ as defined by the canonical form of the fusion diagrams in Eq.~(\ref{eq:general_state}) such that fusion is consistent with the fusion multiplicities $\lbrace N_{ab}^c\rbrace$. Input $\mathcal{C}$ contains all topological charges of the anyon model.}}
\begin{algorithmic}[1]
\Function{GetTuples}{$\lbrace \alpha_i\rbrace, \lbrace N_{ab}^c\rbrace, \mathcal{C}$}

\State $\lbrace \mathbf{f}\rbrace \gets$ set containing all tuples $(f)$ fulfilling $N_{\alpha_{1}\alpha_{2}}^{f}>0$

\For{$i\gets 1$ to $N-2$}

\For{$ \mathbf{f}$ in $ \lbrace \mathbf{f}\rbrace$ and $\mathbf{for}$ $f$ in $\mathcal{C}$}

\If{$N_{f_{i}\alpha_{i+2}}^{f}>0$}
\State add tuple $(f_{1},f_{2},\ldots,f_{i},f)$ to $ \lbrace \mathbf{f}\rbrace$
\EndIf

\EndFor
\State remove all tuples containing $i$ entries from $ \lbrace \mathbf{f}\rbrace$
\EndFor

\For{$ \mathbf{f}$ in $ \lbrace \mathbf{f}\rbrace$ and $\mathbf{for}$ $f$ in $\mathcal{C}$}
\If{$N_{f_{N-1}f}^{f}>0$}
\State add tuple $(f_{1},f_{2},\ldots,f_{N-1},f)$ to $ \lbrace \mathbf{f}\rbrace$
\EndIf
\EndFor

\State remove all tuples containing $N-1$ entries from $ \lbrace \mathbf{f}\rbrace$

\State \Return $ \lbrace \mathbf{f}\rbrace$
\EndFunction
\end{algorithmic}
\label{algo:tuples}
\end{algorithm}

The method for determining the wave function components and tuples described above is a brute force approach since it requires to iterate over all configurations of $\lbrace f_i \rbrace$ and checking the corresponding fusion multiplicities. By noting that the number of distinct configurations of charges in the tuples scale exponentially with the number of anyons localized in the system, it is clear that we would like to employ a more efficient method. This can be done by building the tuples $\mathbf{f}$ iteratively, as shown in form of pseudocode in Alg.~\ref{algo:tuples}. We start with a set including all tuples that contain an anyon consistent the fusion of the first and second anyon. From this point, we successively check the fusion multiplicities in the vertices in Eq.~(\ref{eq:general_state}) and extend the tuples by consistent anyonic charges until we end up with tuples $\mathbf{f}$ that contain $N$ entries. These final tuples correspond to the wave function components; the total number of wave function components is thus $|\lbrace \mathbf{f}\rbrace|$.

Using semions as an example, it can be seen that the above method of finding the tuples is more efficient: For $N$ semions, we have to check $2N$ fusion multiplicities and carry out the operations needed to iteratively build the tuples. For the brute force approach on the other hand, we need to check up to $2^NN$ fusion multiplicities. In both cases, we end up with the tuples $(1,s,1,s,\ldots,1,1)$ and $(1,s,1,s,\ldots,1,s)$ for an even number of semions.

\subsection{Distinguishable Anyons}\label{sec:anyonorderings}

Up to now, we only considered the fusion products for a given anyon configuration when talking about the wave function components and the corresponding tuples $\mathbf{f}$. This is sufficient as long as we focus on systems whose anyonic excitations ($\lbrace \alpha_i\rbrace$ in Eq.~(\ref{eq:general_state})) are of the same charge. If this is no longer the case, i.e., if there are distinguishable anyonic excitations, different orderings of the anyons $\lbrace\alpha_i\rbrace$ may result in different tuples $\mathbf{f}$ due to different intermediate fusion products. Let us illustrate this by considering the toric code~\cite{KITAEV20032}, which features three topological charges apart from the trivial vacuum charges: electrical charges $e$, magnetic vortices $m$ and fermions $f$. These charges obey the fusion rules~\cite{1506.05805}
\begin{equation}
\begin{aligned}
	e \times e &= 1 \quad &m\times m &=1 \quad &f\times f &= 1\\
	e\times m &= f \quad &e\times f &=m \quad &m\times f &= e.
\end{aligned}
\end{equation}
All other fusion processes are trivial as they involve the vacuum charge $1$. If we now consider a system featuring two electric ($e$) and two magnetic ($m$) excitations, it can be seen from the convention of the fusion order on the lattice in Fig.~\ref{fig:lattice}$b$ that there are anyon configurations for which, e.g., $\alpha_1=\alpha_2=e$ and $\alpha_3=\alpha_4=m$ in the fusion diagrams in Eq.~(\ref{eq:general_state}). In this case, the set of tuples $\lbrace \mathbf{f}\rbrace$ is given by
\begin{align}
	\lbrace \mathbf{f}\rbrace=\lbrace (1,&m,1,1),(1,m,1,e),(1,m,1,m),(1,m,1,f) \rbrace\nonumber \\
	&\text{for}\quad \alpha_1=\alpha_2=e,\quad \alpha_3=\alpha_4=m.
	\label{eq:anyonordering1}
\end{align}
On the other hand, there are also anyon configurations corresponding to $\alpha_1=\alpha_2=m$ and $\alpha_3=\alpha_4=e$ for the same system, i.e., the set of tuples $\lbrace \mathbf{f'}\rbrace$ is
\begin{align}
	\lbrace \mathbf{f'}\rbrace=\lbrace (&1,e,1,1),(1,e,1,e),(1,e,1,m),(1,e,1,f) \rbrace\nonumber \\
	&\text{for}\quad \alpha_1=\alpha_2=m,\quad \alpha_3=\alpha_4=e.
	\label{eq:anyonordering2}
\end{align}
It can be seen from Eqs.~(\ref{eq:anyonordering1}) and (\ref{eq:anyonordering2}) that the tuples corresponding to the two anyon orderings do not agree with each other. The components of the wave function, which correspond to the different fusion diagrams, thus depend on the anyon ordering. One key observation in this context is that the number of tuples in the set $\lbrace \mathbf{f}\rbrace$ does not depend on the anyon ordering, that is, $|\lbrace \mathbf{f}\rbrace|=|\lbrace \mathbf{f'}\rbrace|$ for two different anyon orderings. This is shown in App.~\ref{app_distinguishable_anyons} and implies that the total number of wave function components does not change upon exchanging anyons despite the fact that the tuples associated with some components may change.

In practice, when dealing with distinguishable anyonic excitations, we can determine the tuples $\mathbf{f}$ associated with the wave function components by utilizing the iterative method described by Alg.~\ref{algo:tuples} in the previous section. The only difference is that we have to apply this method for all different anyon orderings, i.e., for all \emph{distinct permutations} of the anyonic charges $\lbrace \alpha_i\rbrace$, and associate them with the wave function components accordingly. A pseudocode implementing this procedure is given in Alg.~\ref{algo:tuples2}. It shows that we associate a set $\lbrace \mathbf{f}\rbrace$ of tuples that corresponds to the different wave function components with each anyon ordering, i.e., we make the wave function components dependent on the ordering. Note that it is possible to make this algorithm more efficient by realizing that tuples obtained for a certain anyon ordering do not change upon permuting the first and the second anyon (fusion is commutative). One can thus use the result of the permuted anyon ordering if the tuples have already been computed for this case.

\begin{algorithm}[t]
\caption{\justifyingg{Construction of the tuples $\mathbf{f}$ for all distinct anyon orderings using Alg.~\ref{algo:tuples}. The tuples correspond to the wave function components and depend on the anyon ordering, they are stored in the dictionary $WFCtuples$ (``wave function component tuples'').}}
\begin{algorithmic}[1]
\Function{GetTuplesForOrderings}{$\lbrace \alpha_i\rbrace$, $\lbrace N_{ab}^c\rbrace$, $\mathcal{C}$}
\State $permutations \gets$ list of distinct permutations of the anyons $\lbrace \alpha_i\rbrace$

\State $WFCtuples \gets$ empty dictionary

\For{$ordering$ in $permutations$}

\State $\lbrace \mathbf{f}\rbrace \gets \textsc{GetTuples}(ordering, \lbrace N_{ab}^c\rbrace, \mathcal{C}$)
\State $WFCtuples[\text{key}=ordering] \gets \lbrace \mathbf{f}\rbrace$
\EndFor

\State \Return $ WFCtuples$
\EndFunction
\end{algorithmic}
\label{algo:tuples2}
\end{algorithm}

\begin{figure*}[t]
\centering
\begin{tikzpicture}[line width=0.75pt]
	\begin{scope}[shift={(-5.7,4.375-1.2)}, scale=0.8]
	\draw[black] (-0.5,0) -- (3.5,0);
	\draw[black] (-0.5,1) -- (3.5,1);
	\draw[black] (-0.5,2) -- (3.5,2);
	\draw[black] (-0.5,3) -- (3.5,3);
	\draw[black] (0,-0.5) -- (0,3.5);
	\draw[black] (1,-0.5) -- (1,3.5);
	\draw[black] (2,-0.5) -- (2,3.5);
	\draw[black] (3,-0.5) -- (3,3.5);
	\node[draw,circle,inner sep=1.75pt,fill,black] at (0,2) {};
	\node[draw,circle,inner sep=1.75pt,fill,black] at (1,3) {};
	\node[draw,circle,inner sep=1.75pt,fill,black] at (1,1) {};
	\node[black, anchor=south east] (a) at (-0.25,3.25) {$(a)$};
	\node[black, anchor=north west] (a) at (0,2) {$\tau$};
	\node[black, anchor=north west] (a) at (1,1) {$\tau$};
	\node[black, anchor=north west] (a) at (1,3) {$\tau$};
	\draw [-Stealth][line width=1pt] (1+0.05,1+0.05) to [out=45,in=135] (2-0.05,1+0.05);
	\end{scope}
	\begin{scope}[shift={(0.75,5.25)}]
	\node[black, anchor=south east] (a) at (-2.,3.25*0.8+4.375-1.2-5.25) {$(b)$};
	\node [black, anchor=east] (a) at (-0.55,0) {
            $\begin{aligned}
                T_{\vec{r}_2,\vec{r}_2+\vec{e}_x}
            \end{aligned}$
    };
	\node[draw,circle,inner sep=1.5pt,fill,black] at (0,0) {};
	\node[black] (a) at (0.25,-0.) {$\tau$};
	\node[draw,circle,inner sep=1.5pt,fill,black] at (1.5,0) {};
	\node[black] (a) at (1.75,-0.) {$\tau$};
	\node[draw,circle,inner sep=1.5pt,fill,black] at (3,0) {};
	\node[black] (a) at (3.25,-0.) {$\tau$};
	\draw (0.75,0) ellipse (1.25cm and 0.5cm);
	\draw (1.5,0) ellipse (2.1cm and 0.75cm);
	\node[black] (a) at (2,-0.35) {$1$};
	\node[black] (a) at (3.55,-0.5) {$\tau$};
	\end{scope}
	\begin{scope}[shift={(5.55,5.25)}]
	\node [black, anchor=east] (a) at (-0.65,0) {
            $\begin{aligned}
                =
            \end{aligned}$
    };
	\node[draw,circle,inner sep=1.5pt,fill,black] at (0,0) {};
	\node[black] (a) at (0.25,-0.) {$\tau$};
	\node[draw,circle,inner sep=1.5pt,fill,black] at (1.5,0) {};
	\node[black] (a) at (1.75,-0.) {$\tau$};
	\node[draw,circle,inner sep=1.5pt,fill,black] at (3,0) {};
	\node[black] (a) at (3.25,-0.) {$\tau$};
	\draw [black,domain=0:180] plot ({0.5*cos(\x)}, {0.35*sin(\x)});
	\draw [black,domain=0:180] plot ({0.5*cos(\x)+3}, {0.35*sin(\x)});
	\draw [black,domain=180:360] plot ({2*cos(\x)+1.5}, {0.65*sin(\x)});
	\draw [black,domain=180:360] plot ({1*cos(\x)+1.5}, {0.35*sin(\x)});
	\draw (1.5,0) ellipse (2.1cm and 0.75cm);
	\node[black] (a) at (2.4,0.25) {$1$};
	\node[black] (a) at (3.55,-0.5) {$\tau$};
	\end{scope}
	\begin{scope}[shift={(3.05,3.5)}]
	\node [black, anchor=east] (a) at (-0.55,-0.25) {
            $\begin{aligned}
                =\sum_{f_1'\in\lbrace 1,\tau \rbrace} -t^{-1} \mathcal{H}_{s's}
            \end{aligned}$
    };
	\node[draw,circle,inner sep=1.5pt,fill,black] at (0,0) {};
	\node[black] (a) at (0.25,-0.) {$\tau$};
	\node[draw,circle,inner sep=1.5pt,fill,black] at (1.5,0) {};
	\node[black] (a) at (1.75,-0.) {$\tau$};
	\node[draw,circle,inner sep=1.5pt,fill,black] at (3,0) {};
	\node[black] (a) at (3.25,-0.) {$\tau$};
	\draw (0.75,0) ellipse (1.25cm and 0.5cm);
	\draw (1.5,0) ellipse (2.1cm and 0.75cm);
	\node[black] (a) at (2,-0.35) {$f_1'$};
	\node[black] (a) at (3.55,-0.5) {$\tau$};
	\end{scope}
\end{tikzpicture}
\caption{$(a)$ The indicated translation of the second anyon using $T_{\vec{r}_2,\vec{r}_2+\vec{e}_x}$ leads to fusion in the final state that does not agree with the convention for the basis states. $(b)$ The translated state can be expressed in terms of the basis states using the Hamiltonian's matrix elements $\mathcal{H}_{s's}$, which depend on $f_1'$. This translation is equivalent to the translation process $\mathcal{B}$ in the torus picture in Fig.~\ref{fig:torus1}$b$.}
\label{fig:fib_translation_example}
\end{figure*}
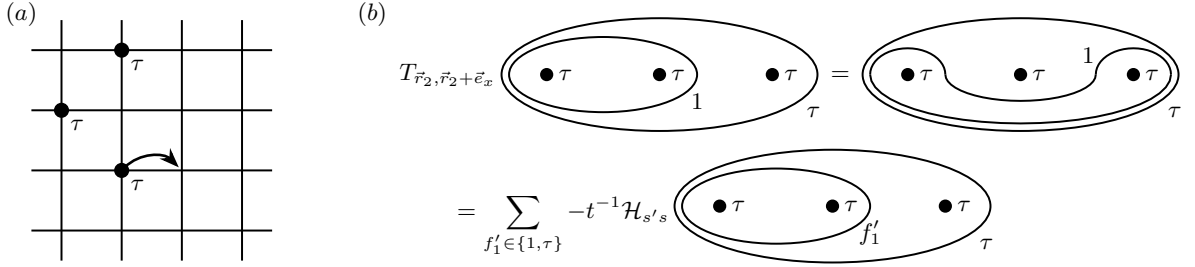

\section{Computation of the Matrix Elements}\label{sec:algo}

The goal of our algorithm is to simulate anyons hopping on a $2$D $L_x\times L_y$ square lattice according to a tight-binding Hamiltonian $\mathcal{H}$ for anyons. This Hamiltonian can be written as
\begin{align}
	\mathcal{H}=-t\sum_{\langle ij \rangle} \left(T_{\vec{r}_i,\vec{r}_j}+ \mathrm{H.c.}\right) ,
	\label{eq:hamiltonian}
\end{align}
where the sum runs over all nearest-neighbor lattice site pairs $i$ and $j$. The hopping amplitude is denoted by $t$ and the translation operator translating the anyon located at position $\vec{r}_i$ (corresponding to site $i$) to position $\vec{r}_j$ (corresponding to site $j$) by $T_{\vec{r}_i,\vec{r}_j}$.

For bosonic or fermionic systems, we can rewrite the translation operators in terms of creation and annihilation operators, i.e., $T_{\vec{r}_i,\vec{r}_j}=b_j^{\dagger}b_i$ and $T_{\vec{r}_i,\vec{r}_j}=c_j^{\dagger}c_i$, where $b_i^{\dagger}(b_i)$ and $c_i^{\dagger}(c_i)$ denote the bosonic and fermionic creation (annihilation) operators. For anyonic systems on the other hand, such a notation is in general not possible. The reason for this can be seen when considering semions. We know from the considerations in the previous section that only an even number of semions can exist on a torus, which implies that anyonic systems do in general not have the standard Fock spaces bosonic or fermionic systems feature. We thus have to rely on the translation operators $T_{\vec{r}_i,\vec{r}_j}$ rather than on creation and annihilation operators.

Note that we focus on hard-core anyons that do not allow multiple anyons to be localized at the same site. Information on how to generalize the algorithm is provided in App.~\ref{app_commutation}. Since we consider square lattices with PBC, we can also write the Hamiltonian in Eq.~(\ref{eq:hamiltonian}) as
\begin{align}
	\mathcal{H}=-t\sum_{i} \left( T_{\vec{r}_i,\vec{r}_i+\vec{e}_x}+T_{\vec{r}_i,\vec{r}_i+\vec{e}_y}+ \mathrm{H.c.}\right) ,
	\label{eq:hamiltonian_square_lattice}
\end{align}
where $\vec{e}_x$ ($\vec{e}_y$) denotes the unit vector in positive $x$-direction ($y$-direction); the lattice spacing is set to unity. The states given by Eq.~(\ref{eq:realspacebasis}) will be used as basis to explain the construction of the anyonic tight-binding Hamiltonian $\mathcal{H}$ for an arbitrary anyon model by computing the matrix elements with respect to these basis states. The matrix elements are given by
\begin{align}
\begin{split}
	&\quad \,\,\, \langle(\alpha'_k,\pvec{r}'_k)_{k=1}^N, \mathbf{f'}|\mathcal{H}|(\alpha_k,\vec{r}_k)_{k=1}^N, \mathbf{f}\rangle \equiv \mathcal{H}_{s's},
\end{split}
\label{eq:H_projection}
\end{align}
where we used the shorthand notation $|s \rangle \equiv|(\alpha_k,\vec{r}_k)_{k=1}^N, \mathbf{f}\rangle$ and $|s' \rangle \equiv|(\alpha'_k,\pvec{r}'_k)_{k=1}^N, \mathbf{f'}\rangle$ for more convenient indices. From the form of the Hamiltonian in Eqs.~(\ref{eq:hamiltonian}) or (\ref{eq:hamiltonian_square_lattice}), it can be seen that $\mathcal{H}_{s's}$ is only non-zero if the anyon configuration associated with $|s'\rangle$ can be obtained by translating a single anyon in the configuration of $|s\rangle$ to a nearest neighboring site. In the following discussions, we will thus assume that $|s'\rangle$ denotes such a state.\\

\begin{figure*}[t]
	\centering
\begin{tikzpicture}[line width=0.75pt]
	\begin{scope}[shift={(-7.75,3.16/2-1.5*0.8)}, scale=0.8]
	\draw[black] (-0.5,0) -- (3.5,0);
	\draw[black] (-0.5,1) -- (3.5,1);
	\draw[black] (-0.5,2) -- (3.5,2);
	\draw[black] (-0.5,3) -- (3.5,3);
	\draw[black] (0,-0.5) -- (0,3.5);
	\draw[black] (1,-0.5) -- (1,3.5);
	\draw[black] (2,-0.5) -- (2,3.5);
	\draw[black] (3,-0.5) -- (3,3.5);
	\node[draw,circle,inner sep=1.75pt,fill,black] at (0,2) {};
	\node[draw,circle,inner sep=1.75pt,fill,black] at (1,2) {};
	\node[draw,circle,inner sep=1.75pt,fill,black] at (1,3) {};
	\node[draw,circle,inner sep=1.75pt,fill,black] at (1,1) {};
	\node[draw,circle,inner sep=1.75pt,fill,black] at (2,0) {};
	\node[black, anchor=south east] (a) at (-0.25,3.25) {$(a)$};
	\node[black, anchor=north west] (a) at (0,2) {$a$};
	\node[black, anchor=north west] (a) at (1,1) {$b$};
	\node[black, anchor=north west] (a) at (1,2) {$c$};
	\node[black, anchor=north west] (a) at (1,3) {$d$};
	\node[black, anchor=north west] (a) at (2,0) {$e$};
	\draw [-Stealth][line width=1pt] (1+0.05,1+0.05) to [out=45,in=135] (2-0.05,1+0.05);
	\end{scope}
	\node [black, anchor=east] (a) at (-1.65,3.16/2) {
            $\begin{aligned}
                T_{\vec{r}_b,\vec{r}_b+\vec{e}_x}
            \end{aligned}$
    };
	\draw (0,0.15) ellipse (1*0.8 and 0.75*0.8);
	\draw (0,0.15) ellipse (0.25*0.8 and 0.25*0.8);
	\node[black, anchor=south east] (a) at (-3,3.25*0.8+3.16*0.5-1.5*0.8) {$(b)$};
	\node[black, anchor=north west] (a) at (0.03,0.15) {$y$};
	\draw[black] (0.177*0.8,0.177*0.8+0.15) -- (-0.177*0.8,-0.177*0.8+0.15);
	\draw[black] (0.177*0.8,-0.177*0.8+0.15) -- (-0.177*0.8,0.177*0.8+0.15);
	\node[black, anchor=south] (a) at (-1.5,3.3) {$a$};
	\node[black, anchor=south] (a) at (-1.5+0.6,3.3) {$b$};
	\node[black, anchor=south] (a) at (-1.5+0.6*2,3.3) {$c$};
	\node[black, anchor=south] (a) at (-1.5+0.6*3,3.3) {$d$};
	\node[black, anchor=south] (a) at (-1.5+0.6*4,3.3) {$e$};
	\node[black, anchor=north] (a) at (-1.5+0.3,3.25-0.5196) {$f_1$};
	\node[black, anchor=north] (a) at (-1.5+0.3*2,3.25-0.5196*2) {$f_2$};
	\node[black, anchor=north] (a) at (-1.5+0.3*3,3.25-0.5196*3) {$f_3$};
	\node[black, anchor=north] (a) at (-1.5+0.3*4-0.1*0.5,3.25-0.5196*4+0.1732*0.5) {$f_4$};
	\node[black] (a) at (0.9*0.8,0.7*0.8+0.15+0.075) {$f_5$};
	\draw[black] (0,0.75) -- (-1.5,0.75+2.598) node[sloped,pos=0.90,allow upside down]{\arrowIn}; ; 
	\draw[black] (-1.5*0.69,0.75+2.598*0.69) -- (-1.5*0.7,0.75+2.598*0.7) node[sloped,pos=1,allow upside down]{\arrowIn}; ; 
	\draw[black] (-1.5*0.49,0.75+2.598*0.49) -- (-1.5*0.5,0.75+2.598*0.5) node[sloped,pos=1,allow upside down]{\arrowIn}; ; 
	\draw[black] (-1.5*0.29,0.75+2.598*0.29) -- (-1.5*0.3,0.75+2.598*0.3) node[sloped,pos=1,allow upside down]{\arrowIn}; ; 
	\draw[black] (-1.5*0.09,0.75+2.598*0.09) -- (-1.5*0.1,0.75+2.598*0.1) node[sloped,pos=1,allow upside down]{\arrowIn}; ; 
	\draw[black] (-1.2,0.75+0.5196*4) -- (-1.5+0.6,0.75+2.598) node[sloped,pos=0.50,allow upside down]{\arrowIn}; ; 
	\draw[black] (-1.2+0.3,0.75+0.5196*3) -- (-1.5+0.6*2,0.75+2.598) node[sloped,pos=0.50,allow upside down]{\arrowIn}; ; 
	\draw[black] (-1.2+0.3*2,0.75+0.5196*2) -- (-1.5+0.6*3,0.75+2.598) node[sloped,pos=0.50,allow upside down]{\arrowIn}; ; 
	\draw[black] (-1.2+0.3*3,0.75+0.5196*1) -- (-1.5+0.6*4,0.75+2.598) node[sloped,pos=0.50,allow upside down]{\arrowIn}; ; 
	\draw[black, line width=0.05pt] (0.707*0.8+0.707*0.8*0.001,0.707*0.8*0.75-0.707*0.8*0.75*0.001+0.15) -- (0.707*0.8,0.707*0.8*0.75+0.15) node[sloped,pos=0.50,allow upside down]{\arrowIn}; ; 	
	\begin{scope}[shift={(3.5,0)}]
	\node [black, anchor=east] (a) at (-1.65,3.16/2) {
            $\begin{aligned}
                =
            \end{aligned}$
    };
	\draw (0,0.15) ellipse (1*0.8 and 0.75*0.8);
	\draw (0,0.15) ellipse (0.25*0.8 and 0.25*0.8);
	\node[black, anchor=north west] (a) at (0.03,0.15) {$y$};
	\draw[black] (0.177*0.8,0.177*0.8+0.15) -- (-0.177*0.8,-0.177*0.8+0.15);
	\draw[black] (0.177*0.8,-0.177*0.8+0.15) -- (-0.177*0.8,0.177*0.8+0.15);
	\draw [black,domain=0:3, samples=10] plot ({\x*cos(120)}, {\x*sin(120)+0.75});
	\draw [black,domain=0:3-0.5196/sin(60), samples=10] plot ({\x*cos(60)-0.3}, {\x*sin(60)+0.75+0.5196});
	\draw [black,domain=0:3-0.5196/sin(60)*2-0.4-0.03, samples=10] plot ({\x*cos(60)-0.3*2}, {\x*sin(60)+0.75+0.5196*2});
	\draw [black,domain=3-0.5196/sin(60)*2-0.2+0.03:3-0.5196/sin(60)*2, samples=10] plot ({\x*cos(60)-0.3*2}, {\x*sin(60)+0.75+0.5196*2});
	\draw [black,domain=0:3-0.5196/sin(60)*3-0.4-0.03, samples=10] plot ({\x*cos(60)-0.3*3}, {\x*sin(60)+0.75+0.5196*3});
	\draw [black,domain=3-0.5196/sin(60)*3-0.2+0.03:3-0.5196/sin(60)*3, samples=10] plot ({\x*cos(60)-0.3*3}, {\x*sin(60)+0.75+0.5196*3});
	\draw [black,domain=0:3-0.5196/sin(60)*4-0.3, samples=10] plot ({\x*cos(60)-0.3*4}, {\x*sin(60)+0.75+0.5196*4});
	\draw [black,domain=3-0.5196/sin(60)*4-0.3:3-0.5196/sin(60)*4, samples=10] plot ({\x*cos(60)+1.2}, {\x*sin(60)+0.75+0.5196*4});
	\draw [black,domain=-1.05-0.01:1-0.33-0.05, samples=10] plot ({\x}, {3.088});
	\draw [black,domain=1.175-0.33+0.05:1.35+0.01, samples=10] plot ({\x}, {3.088});
	\node[black, anchor=south] (a) at (-1.5,3.3) {$a$};
	\node[black, anchor=south] (a) at (-1.5+0.6*5,3.3) {$b$};
	\node[black, anchor=south] (a) at (-1.5+0.6*2,3.3) {$c$};
	\node[black, anchor=south] (a) at (-1.5+0.6*3,3.3) {$d$};
	\node[black, anchor=south] (a) at (-1.5+0.6*4,3.3) {$e$};
	\node[black, anchor=north] (a) at (-1.5+0.3,3.25-0.5196) {$f_1$};
	\node[black, anchor=north] (a) at (-1.5+0.3*2,3.25-0.5196*2) {$f_2$};
	\node[black, anchor=north] (a) at (-1.5+0.3*3,3.25-0.5196*3) {$f_3$};
	\node[black, anchor=north] (a) at (-1.5+0.3*4-0.1*0.5,3.25-0.5196*4+0.1732*0.5) {$f_4$};
	\node[black] (a) at (0.9*0.8,0.7*0.8+0.15+0.075) {$f_5$};
	\draw[black] (0,0.75) -- (-1.5,0.75+2.598) node[sloped,pos=0.90,allow upside down]{\arrowIn}; ; 
	\draw[black] (-1.5*0.69,0.75+2.598*0.69) -- (-1.5*0.7,0.75+2.598*0.7) node[sloped,pos=1,allow upside down]{\arrowIn}; ; 
	\draw[black] (-1.5*0.49,0.75+2.598*0.49) -- (-1.5*0.5,0.75+2.598*0.5) node[sloped,pos=1,allow upside down]{\arrowIn}; ; 
	\draw[black] (-1.5*0.29,0.75+2.598*0.29) -- (-1.5*0.3,0.75+2.598*0.3) node[sloped,pos=1,allow upside down]{\arrowIn}; ; 
	\draw[black] (-1.5*0.09,0.75+2.598*0.09) -- (-1.5*0.1,0.75+2.598*0.1) node[sloped,pos=1,allow upside down]{\arrowIn}; ; 
	\draw[black] (-0.16,3.088) -- (-0.15,3.088) node[sloped,pos=1,allow upside down]{\arrowIn}; ; 
	\draw[black] (-1.2+0.3,0.75+0.5196*3) -- (-0.6,0.75+2.598/2+0.5196*1.5) node[sloped,pos=1,allow upside down]{\arrowIn}; ; 
	\draw[black] (-1.2+0.3*2,0.75+0.5196*2) -- (-0.15,0.75+2.598/2+0.5196*1) node[sloped,pos=1,allow upside down]{\arrowIn}; ; 
	\draw[black] (-1.2+0.3*3,0.75+0.5196*1) -- (0.3,0.75+2.598/2+0.5196*0.5) node[sloped,pos=1,allow upside down]{\arrowIn}; ; 
	\draw[black, line width=0.05pt] (0.707*0.8+0.707*0.8*0.001,0.707*0.8*0.75-0.707*0.8*0.75*0.001+0.15) -- (0.707*0.8,0.707*0.8*0.75+0.15) node[sloped,pos=0.50,allow upside down]{\arrowIn}; ; 	
	\end{scope}
\end{tikzpicture}
	\caption{Illustration of the braid rules for translating anyons in positive $x$-direction without crossing cut $B$. $(a)$ Lattice with an anyon configuration and indication which anyon is to be translated. $(b)$ Effect of the translation on the fusion diagram.}
	\label{eq:rule1}
\end{figure*}
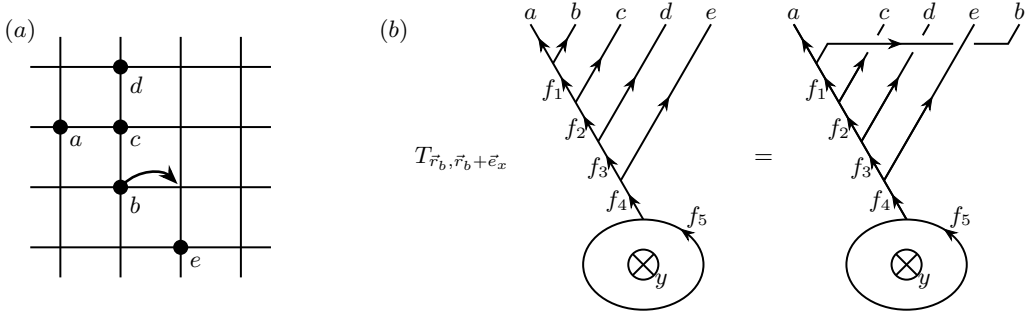

Let us first illustrate the general idea by considering the anyon configuration depicted in Fig.~\ref{fig:fib_translation_example}$a$, which features Fibonacci anyons. The ``second'' anyon of the initial state ($|0\rangle$ from Fig.~\ref{fig:fib_example}$a$) is translated by $T_{\vec{r}_2,\vec{r}_2+\vec{e}_x}$ in such a way the anyon ordering of the final state in Fig.~\ref{fig:fib_translation_example}$b$ is not canonical (see Fig.~\ref{fig:lattice}$b$). This is actually the same scenario as indicated by the translation process $\mathcal{B}$ in the torus picture in Fig.~\ref{fig:torus1}$b$. Due to the definition of the Hamiltonian $\mathcal{H}$, its matrix elements $\mathcal{H}_{s's}$ can be used to rewrite the translated state in terms of the basis states. In the case considered in Fig.~\ref{fig:fib_translation_example}, this is done by summing over all possible fusion products $f_1'$ of the first two anyons, where we suppressed in the notation that the states $|s'\rangle$ and thus $\mathcal{H}_{s's}$ depend on $f_1'$. More generally, rewriting the translation operators using the Hamiltonian's matrix elements requires summing over multiple fusion products. Note that for convenience, the diagrams in Fig.~\ref{fig:fib_translation_example}$b$ do not contain the anyon associated with the non-contractible loop ($f_N$ in Eq.~(\ref{eq:general_state})).

With this general idea, we discuss the rules to actually compute the matrix elements of the Hamiltonian $\mathcal{H}$ using fusion diagrams in the two following sections. To simplify this discussion, we note that it is sufficient to consider translations of anyons only in positive $x$- and $y$-direction since translations along the respective negative directions can be obtained by hermitian conjugation. Further, by looking at the fusion order in Fig.~\ref{fig:lattice}$b$, we can see that we need to distinguish four different cases: Translations in the bulk in $x$-direction, translations in the bulk in $y$-direction, translations over cut $A$ and translation over cut $B$.

\subsection{Translations in the Bulk}\label{sec:algo_bulk}

We start by considering translations that do not cross cuts $A$ or $B$. Although there is no physical boundary due to the PBC, we refer to such translations as being ``in the bulk''. The cuts represent loops on the torus that need be crossed when translating an anyon around the system. Such processes lead to additional effects due to different braids in the fusion diagrams and are discussed in Sec.~\ref{sec:algo_cuts}.

Let us first discuss translations of anyons in $y$-direction, which is the easiest case to be considered. From the fusion order in Fig.~\ref{fig:lattice}$b$, it can be seen that translating an anyon in the bulk in $y$-direction does not affect the fusion order, i.e., the order of the anyons in the basis states corresponding to the initial and final state is the same. Therefore, the corresponding matrix elements are $\mathcal{H}_{s's}=-t$. Here, we assumed that the anyon configuration of $|s'\rangle$ can be obtained from the one of $|s\rangle$ by translating an anyon in $y$-direction to a neighboring site; the matrix element is zero otherwise.\\

A scenario in which braiding is involved can occur when translating an anyon in $x$-direction. This can be seen using the anyon configuration depicted in Fig.~\ref{eq:rule1}$a$ and the fusion order in Fig.~\ref{fig:lattice}$b$. Initially, the anyons are in alphabetical order, whereas after the translation, $b$ comes after $c$, $d$ and $e$ and is now the last anyon. This example shows that the fusion order is changed when translating an anyon in positive $x$-direction and there is another anyon with an $x$-coordinate equal to the $x$-coordinate of the former anyon before (after) the translation and a $y$-coordinate larger (smaller) than the one of the anyon to be translated.

The effects of the above translation process on the fusion diagrams can be seen in Fig.~\ref{eq:rule1}$b$, where $\vec{r}_b$ denotes the position of anyon $b$ and $f_1$ to $f_5$ the remaining anyonic charges. Anyon $b$ is moved to come after $e$ such that the resulting fusion diagram is not in canonical form since the line associated with the translated anyon moves across the lines of the other anyons that are affected by the changed anyon order, that is, the corresponding lines are braided. In general, there are two different ways a braid can occur. The line associated with the translated anyon can either move ``in front of'' or ``behind'' the lines of the anyon it is braided with. In the case considered in Fig.~\ref{eq:rule1}$b$, anyon $b$ moves in front of $c$ and $d$ and behind $e$. Here, moving a line in front of another one corresponds to counter-clockwise exchange, which can be seen using the anyon configuration in Fig.~\ref{eq:rule1}$a$ since in this example, the indicated translation corresponds to counter-clockwise exchange of the two anyon pairs $b$ and $c$ and $b$ and $d$. Similarly, this translation also corresponds to exchanging anyons $b$ and $e$ clockwise, which is depicted in the fusion diagrams by the line of anyon $b$ moving behind the one of anyon $e$. Using this method, the braids in the fusion diagrams can be generalized to arbitrary anyon configurations; Fig.~\ref{eq:rule1} serves as a summary that covers every case that may be encountered.

\begin{figure}[t]
	\centering
\begin{tikzpicture}[line width=0.75pt]	
	\begin{scope}[shift={(0,-0.25+0.25*0.9659*0.5-7)}, scale=1, xscale=1]
	\draw[black!50, line width=0.4pt, densely dashed] (0+1*0.2588,1*0.9659*0.5) -- (0+1*0.2588,-0.5*0.9659*0.5);
	\draw[black!50, line width=0.4pt, densely dashed] (1+1*0.2588,1*0.9659*0.5) -- (1+1*0.2588,-0.5*0.9659*0.5);
	\draw[black!50, line width=0.4pt, densely dashed] (1+3*0.2588,3*0.9659*0.5) -- (1+3*0.2588,-0.5*0.9659*0.5);
	\draw[black!50, line width=0.4pt, densely dashed] (2,0) -- (2,-0.5*0.9659*0.5);
	\draw[black!50, line width=0.4pt, densely dashed] (3+2*0.2588,2*0.9659*0.5) -- (3+2*0.2588,-0.5*0.9659*0.5);
	
	\draw[black] (-0.5,0) -- (3.5,0);
	\draw[black] (-0.5+1*0.2588,1*0.9659*0.5) -- (3.5+1*0.2588,1*0.9659*0.5);
	\draw[black] (-0.5+2*0.2588,2*0.9659*0.5) -- (3.5+2*0.2588,2*0.9659*0.5);
	\draw[black] (-0.5+3*0.2588,3*0.9659*0.5) -- (3.5+3*0.2588,3*0.9659*0.5);
	\draw[black] (0-0.5*0.2588,-0.5*0.9659*0.5) -- (3.5*0.2588,3.5*0.9659*0.5);
	\draw[black] (1-0.5*0.2588,-0.5*0.9659*0.5) -- (1+3.5*0.2588,3.5*0.9659*0.5);
	\draw[black] (2-0.5*0.2588,-0.5*0.9659*0.5) -- (2+3.5*0.2588,3.5*0.9659*0.5);
	\draw[black] (3-0.5*0.2588,-0.5*0.9659*0.5) -- (3+3.5*0.2588,3.5*0.9659*0.5);
	\draw[black][line width=1.5pt] (-0.75-0.25*0.2588,-0.25*0.9659*0.5) -- (3.75-0.25*0.2588,-0.25*0.9659*0.5);
	\node[black, anchor=west] (a) at (3.75-0.25*0.2588,-0.25*0.9659*0.5) {Cut $A$};
	\draw[black][line width=1.5pt] (3.25-0.75*0.2588,-0.75*0.9659*0.5) -- (3.25+3.75*0.2588,3.75*0.9659*0.5);
	\node[black, anchor=west] (a) at (3.25+3.75*0.2588,3.75*0.9659*0.5) {Cut $B$};
	\node[draw,circle,inner sep=1.75pt,fill,black] at (0+1*0.2588,1*0.9659*0.5) {};
	\node[draw,circle,inner sep=1.75pt,fill,black] at (1+1*0.2588,1*0.9659*0.5) {};
	\node[draw,circle,inner sep=1.75pt,fill,black] at (1+3*0.2588,3*0.9659*0.5) {};
	\node[draw,circle,inner sep=1.75pt,fill,black] at (2,0) {};
	\node[draw,circle,inner sep=1.75pt,fill,black] at (3+2*0.2588,2*0.9659*0.5) {};
	\node[black, anchor=north west] (a) at (0+1*0.2588,1*0.9659*0.5) {$a$};
	\node[black, anchor=north west] (a) at (1+1*0.2588,1*0.9659*0.5) {$b$};
	\node[black, anchor=north west] (a) at (1+3*0.2588,3*0.9659*0.5) {$c$};
	\node[black, anchor=south west] (a) at (2+0.05,0) {$d$};
	\node[black, anchor=north east] (a) at (3+2*0.2588-0.05,2*0.9659*0.5) {$e$};
	\draw[black] (-1.5-0.3,-0.7+0.7-0.25*0.9659*0.5) -- (-0.75-0.3,-0.7+0.7-0.25*0.9659*0.5) node[sloped,pos=1,allow upside down]{\arrowIn}; ; 
	\draw[black] (-1.5-0.3,-0.7+0.7-0.25*0.9659*0.5) -- (-1.5-0.3,-0.7+0.75+0.7-0.25*0.9659*0.5) node[sloped,pos=1,allow upside down]{\arrowIn}; ; 
	\draw[black] (-1.5-0.3,-0.7+0.7-0.25*0.9659*0.5) -- (-1.5-0.3+0.75*0.2588,-0.7+0.75*0.9659*0.5+0.7-0.25*0.9659*0.5) node[sloped,pos=1,allow upside down]{\arrowIn}; ; 
	\node[black, anchor=north] (a) at (-1.125-0.3,-0.65+0.7-0.25*0.9659*0.5) {$x$};
	\node[black, anchor=east] (a) at (-0.95-0.3,-0.45+0.7-0.25*0.9659*0.5) {$y$};
	\node[black, anchor=east] (a) at (-1.5-0.3+0.05,-0.7+0.75*0.5+0.7-0.25*0.9659*0.5) {$t$};
	\draw[black, line width=0.4pt] (0+1*0.2588,1*0.9659*0.5) -- (0+1*0.2588,1*0.9659*0.5+0.7);
	\draw[black, line width=0.4pt] (0+1*0.2588,-0.35) -- (0+1*0.2588,-0.5*0.9659*0.5);
	\draw[black, line width=0.4pt] (1+1*0.2588,1*0.9659*0.5) -- (1+1*0.2588,1*0.9659*0.5+0.7);
	\draw[black, line width=0.4pt] (1+1*0.2588,-0.35) -- (1+1*0.2588,-0.5*0.9659*0.5);
	\draw[black, line width=0.4pt] (1+3*0.2588,3*0.9659*0.5) -- (1+3*0.2588,3*0.9659*0.5+0.7);
	\draw[black, line width=0.4pt] (1+3*0.2588,-0.35) -- (1+3*0.2588,-0.5*0.9659*0.5);
	\draw[black, line width=0.4pt] (2,0) -- (2,0+0.7);
	\draw[black, line width=0.4pt] (2,-0.35) -- (2,-0.5*0.9659*0.5);
	\draw[black, line width=0.4pt] (3+2*0.2588,2*0.9659*0.5) -- (3+2*0.2588,2*0.9659*0.5+0.7);
	\draw[black, line width=0.4pt] (3+2*0.2588,-0.35) -- (3+2*0.2588,-0.5*0.9659*0.5);
	\end{scope}

	\begin{scope}[shift={(1.2*1.2+0.5+1*0.2588,-7-2.598*0.8-2.0+0.45)}, yscale=1, xscale=1, line width=0.4pt]
	\draw (0,0.15) ellipse (1*0.8 and 0.75*0.8);
	\draw (0,0.15) ellipse (0.25*0.8 and 0.25*0.8);
	\node[black, anchor=north west] (a) at (0.03,0.15) {$y$};
	\draw[black] (0.177*0.8,0.177*0.8+0.15) -- (-0.177*0.8,-0.177*0.8+0.15);
	\draw[black] (0.177*0.8,-0.177*0.8+0.15) -- (-0.177*0.8,0.177*0.8+0.15);
	\draw[black, line width=0.05pt] (0.707*0.8+0.707*0.8*0.001,0.707*0.8*0.75-0.707*0.8*0.75*0.001+0.15) -- (0.707*0.8,0.707*0.8*0.75+0.15) node[sloped,pos=0.50,allow upside down]{\arrowIn}; ; 	
	\node[black] (a) at (0.9*0.8,0.7*0.8+0.15+0.075) {$f_5$};
	
	\draw [-Stealth][black] (-1.5*0.8*1.2,2.598*0.8*0.8+0.75) to [out=131,in=270] (-1.2*1.2-0.5,-0.25+0.25*0.9659*0.5+2.598*0.8+2.0-0.6+0.02-0.2);
	\draw [-Stealth][black] (-1.5*0.8*1.2,2.598*0.8*0.8+0.75) to [out=49,in=270] (-1.2*1.2+0.5,-0.25+0.25*0.9659*0.5+2.598*0.8+2.0-0.6+0.02-0.2);
	\draw [-Stealth][black] (-1.5*0.8*1.2*0.75,2.598*0.8*0.8*0.75+0.75) to [out=49,in=270] (-1.2*1.2+0.5+2*0.2588,-0.25+0.25*0.9659*0.5+2.598*0.8+2.0-0.6+0.02-0.2);
	\draw [-Stealth][black] (-1.5*0.8*1.2*0.5,2.598*0.8*0.8*0.5+0.75) to [out=49,in=270] (-1.2*1.2+1.5-1*0.2588,-0.25+0.25*0.9659*0.5+2.598*0.8+2.0-0.6+0.02-0.2);
	\draw [-Stealth][black] (-1.5*0.8*1.2*0.25,2.598*0.8*0.8*0.25+0.75) to [out=49,in=270] (-1.2*1.2+2.5+1*0.2588,-0.25+0.25*0.9659*0.5+2.598*0.8+2.0-0.6+0.02-0.2);
	
	\begin{scope}[shift={(0,0.75-0.75*0.8)}, yscale=0.8, xscale=1.2, line width=0.4pt]
	\node[black, anchor=north] (a) at (-1.5+0.3,3.25-0.5196) {$f_1$};
	\node[black, anchor=north] (a) at (-1.5+0.3*2,3.25-0.5196*2) {$f_2$};
	\node[black, anchor=north] (a) at (-1.5+0.3*3,3.25-0.5196*3) {$f_3$};
	\node[black, anchor=north] (a) at (-1.5+0.3*4-0.1*0.5,3.25-0.5196*4+0.1732*0.5) {$f_4$};
	\draw[black] (0,0.75) -- (-1.5*0.8,0.75+2.598*0.8); 
	\draw[black] (-1.5*0.7+0.01,0.75+2.598*0.7-0.0115) -- (-1.5*0.7,0.75+2.598*0.7) node[sloped,pos=1,allow upside down]{\arrowIn}; ; 
	\draw[black] (-1.5*0.5+0.01,0.75+2.598*0.5-0.0115) -- (-1.5*0.5,0.75+2.598*0.5) node[sloped,pos=1,allow upside down]{\arrowIn}; ; 
	\draw[black] (-1.5*0.3+0.01,0.75+2.598*0.3-0.0115) -- (-1.5*0.3,0.75+2.598*0.3) node[sloped,pos=1,allow upside down]{\arrowIn}; ; 
	\draw[black] (-1.5*0.1+0.01,0.75+2.598*0.1-0.0115) -- (-1.5*0.1,0.75+2.598*0.1) node[sloped,pos=1,allow upside down]{\arrowIn}; ; 
	\end{scope}
	\end{scope}
\end{tikzpicture}
	\caption{Merging fusion diagrams and anyon configurations into a single picture explains the braids associated with translations. Braids correspond to line crossings that are obtained upon tranlating an anyon to a neighboring site.}
	\label{fig:intuitive_picture}
\end{figure}
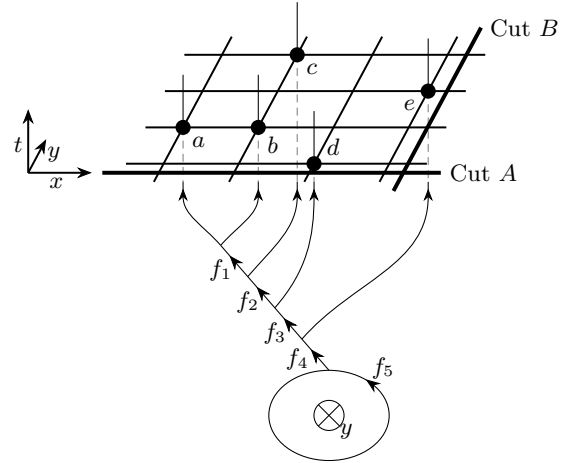

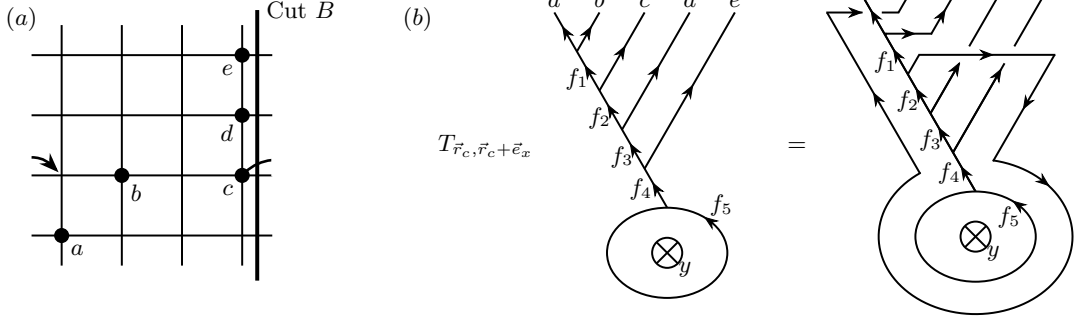
\begin{figure*}[t]
	\centering
\begin{tikzpicture}[line width=0.75pt]
	\node [black, anchor=east] (a) at (-1.65,3.16/2) {
            $\begin{aligned}
                T_{\vec{r}_c,\vec{r}_c+\vec{e}_x}
            \end{aligned}$
    };
    \node[black, anchor=south east] (a) at (-3,3.25*0.8+3.16*0.5-1.5*0.8) {$(b)$};
	\draw (0,0.15) ellipse (1*0.8 and 0.75*0.8);
	\draw (0,0.15) ellipse (0.25*0.8 and 0.25*0.8);
	\node[black, anchor=north west] (a) at (0.03,0.15) {$y$};
	\draw[black] (0.177*0.8,0.177*0.8+0.15) -- (-0.177*0.8,-0.177*0.8+0.15);
	\draw[black] (0.177*0.8,-0.177*0.8+0.15) -- (-0.177*0.8,0.177*0.8+0.15);
	\node[black, anchor=south] (a) at (-1.5,3.3) {$a$};
	\node[black, anchor=south] (a) at (-1.5+0.6,3.3) {$b$};
	\node[black, anchor=south] (a) at (-1.5+0.6*2,3.3) {$c$};
	\node[black, anchor=south] (a) at (-1.5+0.6*3,3.3) {$d$};
	\node[black, anchor=south] (a) at (-1.5+0.6*4,3.3) {$e$};
	\node[black, anchor=north] (a) at (-1.5+0.3,3.25-0.5196) {$f_1$};
	\node[black, anchor=north] (a) at (-1.5+0.3*2,3.25-0.5196*2) {$f_2$};
	\node[black, anchor=north] (a) at (-1.5+0.3*3,3.25-0.5196*3) {$f_3$};
	\node[black, anchor=north] (a) at (-1.5+0.3*4-0.1*0.5,3.25-0.5196*4+0.1732*0.5) {$f_4$};
	\node[black] (a) at (0.9*0.8,0.7*0.8+0.15+0.075) {$f_5$};
	\draw[black] (0,0.75) -- (-1.5,0.75+2.598) node[sloped,pos=0.90,allow upside down]{\arrowIn}; ; 
	\draw[black] (-1.5*0.69,0.75+2.598*0.69) -- (-1.5*0.7,0.75+2.598*0.7) node[sloped,pos=1,allow upside down]{\arrowIn}; ; 
	\draw[black] (-1.5*0.49,0.75+2.598*0.49) -- (-1.5*0.5,0.75+2.598*0.5) node[sloped,pos=1,allow upside down]{\arrowIn}; ; 
	\draw[black] (-1.5*0.29,0.75+2.598*0.29) -- (-1.5*0.3,0.75+2.598*0.3) node[sloped,pos=1,allow upside down]{\arrowIn}; ; 
	\draw[black] (-1.5*0.09,0.75+2.598*0.09) -- (-1.5*0.1,0.75+2.598*0.1) node[sloped,pos=1,allow upside down]{\arrowIn}; ; 
	\draw[black] (-1.2,0.75+0.5196*4) -- (-1.5+0.6,0.75+2.598) node[sloped,pos=0.50,allow upside down]{\arrowIn}; ; 
	\draw[black] (-1.2+0.3,0.75+0.5196*3) -- (-1.5+0.6*2,0.75+2.598) node[sloped,pos=0.50,allow upside down]{\arrowIn}; ; 
	\draw[black] (-1.2+0.3*2,0.75+0.5196*2) -- (-1.5+0.6*3,0.75+2.598) node[sloped,pos=0.50,allow upside down]{\arrowIn}; ; 
	\draw[black] (-1.2+0.3*3,0.75+0.5196*1) -- (-1.5+0.6*4,0.75+2.598) node[sloped,pos=0.50,allow upside down]{\arrowIn}; ; 
	\draw[black, line width=0.05pt] (0.707*0.8+0.707*0.8*0.001,0.707*0.8*0.75-0.707*0.8*0.75*0.001+0.15) -- (0.707*0.8,0.707*0.8*0.75+0.15) node[sloped,pos=0.50,allow upside down]{\arrowIn}; ; 	
	\begin{scope}[shift={(4.1,3.16/2-2.729/2)}]
	\node [black, anchor=east] (a) at (-1.975-0.15,2.729/2) {
            $\begin{aligned}
                =
            \end{aligned}$
    };
	\draw (0,0.15) ellipse (1*0.8 and 0.75*0.8);
	\draw (0,0.15) ellipse (0.25*0.8 and 0.25*0.8);
	\node[black, anchor=north west] (a) at (0.03,0.15) {$y$};
	\draw[black] (0.177*0.8,0.177*0.8+0.15) -- (-0.177*0.8,-0.177*0.8+0.15);
	\draw[black] (0.177*0.8,-0.177*0.8+0.15) -- (-0.177*0.8,0.177*0.8+0.15);
	\draw [black,domain=0:3, samples=10] plot ({\x*cos(120)}, {\x*sin(120)+0.75});
	\draw [black,domain=0:2.1-0.13-0.5196/sin(60), samples=10] plot ({\x*cos(60)-0.3}, {\x*sin(60)+0.75+0.5196});
	\draw [black,domain=2.1+0.13-0.5196/sin(60):3-0.5196/sin(60), samples=10] plot ({\x*cos(60)-0.3}, {\x*sin(60)+0.75+0.5196});
	\draw [black,domain=0:2.1-0.13-0.5196/sin(60)*2, samples=10] plot ({\x*cos(60)-0.3*2}, {\x*sin(60)+0.75+0.5196*2});
	\draw [black,domain=2.1+0.13-0.5196/sin(60)*2:3-0.5196/sin(60)*2, samples=10] plot ({\x*cos(60)-0.3*2}, {\x*sin(60)+0.75+0.5196*2});
	\draw [black,domain=0:3-0.5196/sin(60)*3-0.9, samples=10] plot ({\x*cos(60)-0.3*3}, {\x*sin(60)+0.75+0.5196*3});
	\draw [black,domain=0:2.425-0.5196/sin(60)*4+0.01, samples=10] plot ({\x*cos(60)-0.3*4}, {\x*sin(60)+0.75+0.5196*4});
	\draw [black,domain=2.425-0.5196/sin(60)*3-0.01:3-0.5196/sin(60)*3, samples=10] plot ({\x*cos(60)-0.3*3}, {\x*sin(60)+0.75+0.5196*3});
	\draw [black,domain=-1.1875-0.02:-0.5875+0.01, samples=10] plot ({\x}, {2.85});
	\draw[black] (-0.8876,2.85) -- (-0.8875,2.85) node[sloped,pos=1,allow upside down]{\arrowIn}; ; 
	\draw[black] (-0.444-0.005,3.099-0.00866) -- (-0.444,3.099) node[sloped,pos=1,allow upside down]{\arrowIn}; ; 
	\draw [black,domain=-0.45-0.01-0.3:1.35+0.01-0.3, samples=10] plot ({\x}, {3.088-0.5196});
	\draw [black,domain=0-0.01-0.125:3-0.5196/sin(60)-0.895, samples=10] plot ({\x*cos(60)+0.3}, {\x*sin(60)+0.75+0.5196});
	\draw [black,domain=0.419-0.14:2.75, samples=10] plot ({\x*cos(120)-0.6}, {\x*sin(120)+0.75});
	\draw [black,domain=2.75-0.5196/sin(60)*4:3-0.5196/sin(60)*4, samples=10] plot ({\x*cos(60)-0.3*4}, {\x*sin(60)+0.75+0.5196*4});
	\draw [black,domain=-1.975-0.01:-1.375-0.138+0.01, samples=10] plot ({\x}, {3.1316});
	\draw [black,domain=-1.375+0.138-0.01:-1.025+0.01, samples=10] plot ({\x}, {3.1316});
	\node[black, anchor=south] (a) at (-1.5,3.3) {$a$};
	\node[black, anchor=south] (a) at (-1.5+0.6,3.3) {$c$};
	\node[black, anchor=south] (a) at (-1.5+0.6*2,3.3) {$b$};
	\node[black, anchor=south] (a) at (-1.5+0.6*3,3.3) {$d$};
	\node[black, anchor=south] (a) at (-1.5+0.6*4,3.3) {$e$};
	\node[black, anchor=north] (a) at (-1.5+0.3,3.25-0.5196) {$f_1$};
	\node[black, anchor=north] (a) at (-1.5+0.3*2,3.25-0.5196*2) {$f_2$};
	\node[black, anchor=north] (a) at (-1.5+0.3*3,3.25-0.5196*3) {$f_3$};
	\node[black, anchor=north] (a) at (-1.5+0.3*4-0.1*0.5,3.25-0.5196*4+0.1732*0.5) {$f_4$};
	\node[black, anchor=north west] (a) at (0.17,0.67) {$f_5$};
	\draw[black] (0,0.75) -- (-1.5,0.75+2.598) node[sloped,pos=0.90,allow upside down]{\arrowIn}; ; 
	\draw[black] (-1.5*0.69,0.75+2.598*0.69) -- (-1.5*0.7,0.75+2.598*0.7) node[sloped,pos=1,allow upside down]{\arrowIn}; ; 
	\draw[black] (-1.5*0.49,0.75+2.598*0.49) -- (-1.5*0.5,0.75+2.598*0.5) node[sloped,pos=1,allow upside down]{\arrowIn}; ; 
	\draw[black] (-1.5*0.29,0.75+2.598*0.29) -- (-1.5*0.3,0.75+2.598*0.3) node[sloped,pos=1,allow upside down]{\arrowIn}; ; 
	\draw[black] (-1.5*0.09,0.75+2.598*0.09) -- (-1.5*0.1,0.75+2.598*0.1) node[sloped,pos=1,allow upside down]{\arrowIn}; ; 
	\draw[black] (-1.392+0.5*0.15+0.005,2.122-0.866*0.15-0.00866) -- (-1.392+0.5*0.15,2.122-0.866*0.15) node[sloped,pos=1,allow upside down]{\arrowIn}; ; 
	\draw[black] (-1.975+0.474-0.095,3.1316) -- (-1.975+0.475-0.095,3.1316) node[sloped,pos=1,allow upside down]{\arrowIn}; ; 
	\draw[black] (0.449-0.3,3.088-0.5196) -- (0.45-0.3,3.088-0.5196) node[sloped,pos=1,allow upside down]{\arrowIn}; ; 
	\draw[black] (-0.6,1.789) -- (-0.215*9/10-0.6/10,2.46*9/10+1.789/10) node[sloped,pos=1,allow upside down]{\arrowIn}; ; 
	\draw[black] (-1.2+0.3*3,0.75+0.5196*1) -- (0.3,0.75+2.598/2+0.5196*0.5) node[sloped,pos=1,allow upside down]{\arrowIn}; ; 
	\draw[black] (0.675+0.005,1.919+0.00866) -- (0.675,1.919) node[sloped,pos=1,allow upside down]{\arrowIn}; ; 
	\draw[black, line width=0.05pt] (1.5*0.707*1.074*0.8,1.2*0.707*1.074*0.8+0.15) -- (1.5*0.707*1.074*0.8+1.5*0.707*0.001,1.2*0.707*1.074*0.8-1.2*0.707*0.001+0.15) node[sloped,pos=0.50,allow upside down]{\arrowIn}; ; 
	\draw[black, line width=0.05pt] (0.707*0.8+0.707*0.8*0.001,0.707*0.8*0.75-0.707*0.8*0.75*0.001+0.15) -- (0.707*0.8,0.707*0.8*0.75+0.15) node[sloped,pos=0.50,allow upside down]{\arrowIn}; ; 	
	\draw [black,domain=-240+4.5:79.5+0.3, samples=100] plot ({1.5*cos(\x)*1.074*0.8}, {1.2*sin(\x)*1.074*0.8+0.15});
	\end{scope}
	\begin{scope}[shift={(-8.05,3.16/2-1.5*0.8)}, scale=0.8]
	\draw[black] (-0.5,0) -- (3.5,0);
	\draw[black] (-0.5,1) -- (3.5,1);
	\draw[black] (-0.5,2) -- (3.5,2);
	\draw[black] (-0.5,3) -- (3.5,3);
	\draw[black] (0,-0.5) -- (0,3.5);
	\draw[black] (1,-0.5) -- (1,3.5);
	\draw[black] (2,-0.5) -- (2,3.5);
	\draw[black] (3,-0.5) -- (3,3.5);
	\draw[black][line width=1.5pt] (3.25,-0.75) -- (3.25,3.75);
	\node[black, anchor=south east] (a) at (-0.25,3.25) {$(a)$};
	\node[black, anchor=west] (a) at (3.25,3.75) {Cut $B$};
	\node[draw,circle,inner sep=1.75pt,fill,black] at (0,0) {};
	\node[draw,circle,inner sep=1.75pt,fill,black] at (1,1) {};
	\node[draw,circle,inner sep=1.75pt,fill,black] at (3,1) {};
	\node[draw,circle,inner sep=1.75pt,fill,black] at (3,3) {};
	\node[draw,circle,inner sep=1.75pt,fill,black] at (3,2) {};
	\node[black, anchor=north west] (a) at (0,0) {$a$};
	\node[black, anchor=north west] (a) at (1,1) {$b$};
	\node[black, anchor=north east] (a) at (3,1) {$c$};
	\node[black, anchor=north east] (a) at (3,3) {$e$};
	\node[black, anchor=north east] (a) at (3,2) {$d$};
	\draw [-Stealth][line width=1pt] (3+0.05,1+0.05) to [out=45,in=135] (4-0.05,1+0.05);
	\draw [-Stealth][line width=1pt] (-0.5,1+0.25+0.05) to [out=0,in=135] (-0.05,1+0.05);
	\filldraw[draw=white,fill=white] (3.5,1) rectangle (4.05,1.5);
	\end{scope}
\end{tikzpicture}
	\caption{Illustration of the braid rules for translating anyons in positive $x$-direction across cut $B$. $(a)$ Lattice with an anyon configuration and indication which anyon is to be translated. $(b)$ Effect of the translation on the fusion diagram.}
	\label{eq:translate_B}
\end{figure*}

Alternatively, the braids associated with the translation process in Fig.~\ref{eq:rule1}$a$ can be explained by merging the fusion diagram and the anyon configuration into a single picture, as done in Fig.~\ref{fig:intuitive_picture} for a different anyon configuration. In this picture, translating an anyon to a neighboring site results in moving the associated line correspondingly from the initial site to the final one at the given point in time. If anyon $b$ in Fig.~\ref{fig:intuitive_picture} is translated in positive $x$-direction, the corresponding line moves in front of the line associated with $c$ and behind the line associated with $d$, as expected from the rules discussed above using Fig.~\ref{eq:rule1}. This 3D picture can thus explain the braids introduced by translations in $x$-direction. It also shows that translations in the bulk in $y$-direction are trivial since such hoppings can never lead to line crossings. Further, as we will see in the next section, Fig.~\ref{fig:intuitive_picture} also explains the braids associated with translations across cuts $A$ and $B$.\\

Using the rules above, the matrix elements $\mathcal{H}_{s's}$ can be computed as follows. First, the braided fusion diagram corresponding to the translation of state $|s\rangle$ (see Fig.~\ref{eq:rule1}$b$) needs to be obtained. Then, braid operators as discussed in Sec.~\ref{sec:formalism} have to be applied on the diagram until it is expressed as superposition of fusion diagrams in canonical form~(\ref{eq:general_state}). For each fusion diagram in this superposition, the corresponding tuple can be extracted and compared to the tuple associated with the fusion diagram of the final state $|s'\rangle$. If both tuples agree, the matrix element $\mathcal{H}_{s's}$ is given by the prefactor of the corresponding fusion diagram in the superposition of diagrams multiplied by $-t$. We refer to App.~\ref{app_example_computations_bulk} for more details, where we explicitely consider the example in Fig.~\ref{eq:rule1}$b$.

\subsection{Translations over the Cuts}\label{sec:algo_cuts}

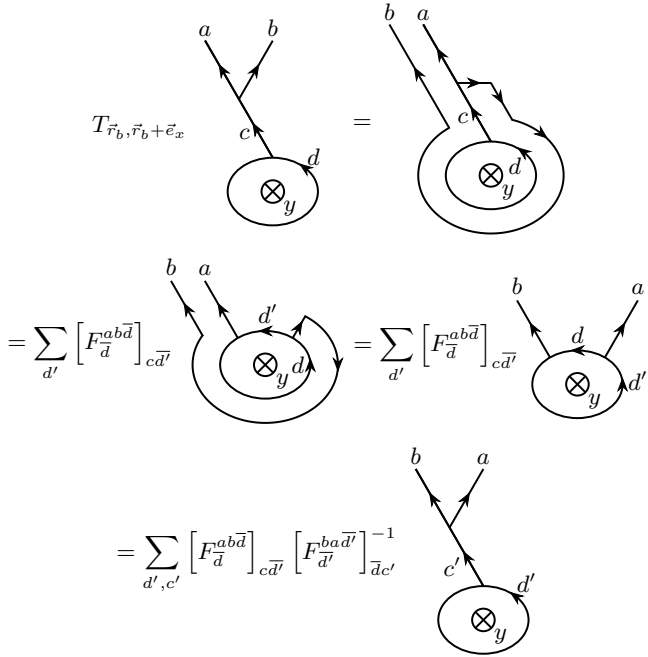
\begin{figure}[t]
\centering
\begin{tikzpicture}[line width=0.75pt]
\begin{scope}[shift={(1,0)}, scale=0.75]
	\node [black, anchor=east] (a) at (-1.35,2.61/2) {
            $\begin{aligned}
                T_{\vec{r}_b,\vec{r}_b+\vec{e}_x}
            \end{aligned}$
    };
	\draw (0,0.15) ellipse (1*0.8 and 0.75*0.8);
	\draw (0,0.15) ellipse (0.25*0.8 and 0.25*0.8);
	\node[black, anchor=north west] (a) at (0.03,0.15) {$y$};
	\draw[black] (0.177*0.8,0.177*0.8+0.15) -- (-0.177*0.8,-0.177*0.8+0.15);
	\draw[black] (0.177*0.8,-0.177*0.8+0.15) -- (-0.177*0.8,0.177*0.8+0.15);
	\node[black, anchor=south] (a) at (-1.2,2.75) {$a$};
	\node[black, anchor=south] (a) at (-0.0006,2.75) {$b$};
	\node[black, anchor=north] (a) at (-1.5+0.3*3.5-0.05,3.25-0.5196*3.5) {$c$};
	\node[black] (a) at (0.9*0.8,0.7*0.8+0.15+0.05) {$d$};
	\draw[black] (0,0.75) -- (-1.5*0.8,0.75+2.598*0.8) node[sloped,pos=0.75,allow upside down]{\arrowIn}; ; 
	\draw[black] (0,0.75) -- (-1.5*0.8,0.75+2.598*0.8) node[sloped,pos=0.25,allow upside down]{\arrowIn}; ; 
	\draw[black] (-1.2+0.3*2,0.75+0.5196*2) -- (-0.6+0.9*0.666,0.75+0.5196*2+1.5588*0.666) node[sloped,pos=0.50,allow upside down]{\arrowIn}; ; 
	\draw[black, line width=0.05pt] (0.707*0.8+0.707*0.001,0.707*0.75*0.8-0.707*0.75*0.001+0.15) --(0.707*0.8,0.707*0.75*0.8+0.15) node[sloped,pos=0.50,allow upside down]{\arrowIn}; ; 
	\end{scope}
	\begin{scope}[shift={(2.9+1,2.61/2-2.18/2)}, scale=0.75]
	\node [black, anchor=east] (a) at (-1.7995-0.15,2.18/2) {
            $\begin{aligned}
                =
            \end{aligned}$
    };
	\draw (0,0.15) ellipse (1*0.8 and 0.75*0.8);
	\draw (0,0.15) ellipse (0.25*0.8 and 0.25*0.8);
	\node[black, anchor=north west] (a) at (0.03,0.15) {$y$};
	\draw[black] (0.177*0.8,0.177*0.8+0.15) -- (-0.177*0.8,-0.177*0.8+0.15);
	\draw[black] (0.177*0.8,-0.177*0.8+0.15) -- (-0.177*0.8,0.177*0.8+0.15);
	\draw [black,domain=0.419-0.14:2.399, samples=10] plot ({\x*cos(120)-0.6}, {\x*sin(120)+0.75});
	\draw [black,domain=0.582-0.14:1.8*0.5+0.582*0.5+0.01, samples=10] plot ({\x*cos(120)+0.6}, {\x*sin(120)+0.75});
	\node[black, anchor=south] (a) at (-1.2,2.75) {$a$};
	\node[black, anchor=south] (a) at (-1.7995,2.75) {$b$};
	\node[black, anchor=north] (a) at (-1.5+0.3*3.5-0.05,3.25-0.5196*3.5) {$c$};
	\node[black, anchor=west] (a) at (0.17-0.02,0.42-0.1) {$d$};
	\draw[black] (0,0.75) -- (-1.5*0.8,0.75+2.598*0.8) node[sloped,pos=0.75,allow upside down]{\arrowIn}; ; 
	\draw[black] (0,0.75) -- (-1.5*0.8,0.75+2.598*0.8) node[sloped,pos=0.25,allow upside down]{\arrowIn}; ; 
	\draw[black] (0.1568-0.005,1.518+0.00866) -- (0.1568,1.518) node[sloped,pos=1,allow upside down]{\arrowIn}; ; 
	\draw[black] (-1.2+0.3*2,0.75+0.5196*2) -- (-0.6+0.9*0.666+0.01,0.75+0.5196*2) node[sloped,pos=0.50,allow upside down]{\arrowIn}; ; 
	\draw[black] (-1.3045+0.005,1.97-0.00866) -- (-1.3045,1.97) node[sloped,pos=1,allow upside down]{\arrowIn}; ; 
	\draw[black, line width=0.05pt] (0.707*0.8+0.707*0.001,0.707*0.75*0.8-0.707*0.75*0.001+0.15) --(0.707*0.8,0.707*0.75*0.8+0.15) node[sloped,pos=0.50,allow upside down]{\arrowIn}; ; 
	\draw[black, line width=0.05pt] (0.707*1.5*1.074*0.8,0.707*1.2*1.074*0.8+0.15) -- (0.707*1.5*1.074*0.8+0.707*1.5*1.074*0.001,0.707*1.2*1.074*0.8-0.707*1.2*1.074*0.001+0.15) node[sloped,pos=0.50,allow upside down]{\arrowIn}; ; 
	\draw [black,domain=-235.5:73.5, samples=100] plot ({1.5*cos(\x)*1.074*0.8}, {1.2*sin(\x)*1.074*0.8+0.15});
	\end{scope}
	\begin{scope}[shift={(0.9,-3.2+2.61/2-1.02/2+0.1)}, scale=0.75]
	\node [black, anchor=east] (a) at (-1.47-0.0,1.02/2-0.12) {
            $\begin{aligned}
                =\sum_{d'}\left[F^{ab\overline{d}}_{\overline{d}}\right]_{c\overbar{d'}}
            \end{aligned}$
    };
	\draw (0,0.15) ellipse (1*0.8 and 0.75*0.8);
	\draw (0,0.15) ellipse (0.25*0.8 and 0.25*0.8);
	\node[black, anchor=north west] (a) at (0.03,0.15) {$y$};
	\draw[black] (0.177*0.8,0.177*0.8+0.15) -- (-0.177*0.8,-0.177*0.8+0.15);
	\draw[black] (0.177*0.8,-0.177*0.8+0.15) -- (-0.177*0.8,0.177*0.8+0.15);
	\node[black, anchor=south] (a) at (-1.071,1.5885) {$a$};
	\node[black, anchor=south] (a) at (-1.671,1.5885) {$b$};
	\node[black, anchor=east] (a) at (0.8+0.05,0.15) {$d$};
	\draw[black, line width=0.05pt] (0.8,0.149) -- (0.8,0.15) node[sloped,pos=0.50,allow upside down]{\arrowIn}; ; 
	\draw[black, line width=0.05pt] (1.5*1.074*0.8,0.001+0.15) -- (1.5*1.074*0.8,0.15) node[sloped,pos=0.50,allow upside down]{\arrowIn}; ; 
	\draw[black, line width=0.05pt] (0.001,0.75) -- (0,0.75) node[sloped,pos=0.50,allow upside down]{\arrowIn}; ; 
	\node[black, anchor=south] (a) at (0.05,0.725) {$d'$};
	\draw [black,domain=-211:57, samples=100] plot ({1.5*cos(\x)*1.074*0.8}, {1.2*sin(\x)*1.074*0.8+0.15});
	\draw [black,domain=0-0.03:0.571*2, samples=10] plot ({\x*cos(120)-0.5}, {\x*sin(120)+0.6495});
	\draw[black, line width=0.05pt] (-0.7855+0.005,1.144-0.00866) -- (-0.7855,1.144) node[sloped,pos=1,allow upside down]{\arrowIn}; ; 
	\draw [black,domain=0-0.03:0.571-0.14, samples=10] plot ({\x*cos(60)+0.5}, {\x*sin(60)+0.6495});
	\draw[black, line width=0.05pt] (0.6002-0.005,0.8231-0.00886) -- (0.6002,0.8231) node[sloped,pos=1,allow upside down]{\arrowIn}; ; 
	\draw [black,domain=0.21-0.2:0.571*2, samples=10] plot ({\x*cos(120)-1.1}, {\x*sin(120)+0.6495});
	\draw[black, line width=0.05pt] (-1.388+0.005,1.1483-0.00886) -- (-1.388,1.1483) node[sloped,pos=1,allow upside down]{\arrowIn}; ; 
	\end{scope}
	\begin{scope}[shift={(0.9+4.15,-3+0.7112/2-1.45/2+0.805)}, scale=0.75]
	\node [black, anchor=east] (a) at (-0.571-0.3,1.45/2-0.12+0.1) {
            $\begin{aligned}
                =\sum_{d'}\left[F^{ab\overline{d}}_{\overline{d}}\right]_{c\overbar{d'}}
            \end{aligned}$
    };
	\draw (0,0.15) ellipse (1*0.8 and 0.75*0.8);
	\draw (0,0.15) ellipse (0.25*0.8 and 0.25*0.8);
	\node[black, anchor=north west] (a) at (0.03,0.15) {$y$};
	\draw[black] (0.177*0.8,0.177*0.8+0.15) -- (-0.177*0.8,-0.177*0.8+0.15);
	\draw[black] (0.177*0.8,-0.177*0.8+0.15) -- (-0.177*0.8,0.177*0.8+0.15);
	\node[black, anchor=south] (a) at (-1.071,1.5885) {$b$};
	\node[black, anchor=south] (a) at (1.071,1.5885) {$a$};
	\node[black, anchor=east] (a) at (1.41,0.22) {$d'$};
	\draw[black, line width=0.05pt] (0.8,-0.001+0.15) -- (0.8,0.15) node[sloped,pos=0.50,allow upside down]{\arrowIn}; ; 
	\draw[black, line width=0.05pt] (0.001,0.75) -- (0,0.75) node[sloped,pos=0.50,allow upside down]{\arrowIn}; ; 
	\node[black, anchor=south] (a) at (0,0.725) {$d$};
	\draw [black,domain=0-0.02:0.571*2, samples=10] plot ({\x*cos(120)-0.5}, {\x*sin(120)+0.6495});
	\draw[black, line width=0.05pt] (-0.7855+0.005,1.144-0.00866) -- (-0.7855,1.144) node[sloped,pos=1,allow upside down]{\arrowIn}; ; 
	\draw [black,domain=0-0.02:0.571*2, samples=10] plot ({\x*cos(60)+0.5}, {\x*sin(60)+0.6495});
	\draw[black, line width=0.05pt] (0.7855-0.005,1.144-0.00866) -- (0.7855,1.144) node[sloped,pos=1,allow upside down]{\arrowIn}; ; 
	\end{scope}
	\begin{scope}[shift={(2.25+0.1+1.45,-5.7+0.7112/2-2.61/2+0.65+0.3)}, scale=0.75]
	\node [black, anchor=east] (a) at (-1.35,2.61/2-0.1) {
            $\begin{aligned}
                =\sum_{d',c'}\left[F^{ab\overline{d}}_{\overline{d}}\right]_{c\overbar{d'}}\left[F^{ba\overbar{d'}}_{\overbar{d'}}\right]_{\overline{d}c'}^{-1}
            \end{aligned}$
    };
	\draw (0,0.15) ellipse (1*0.8 and 0.75*0.8);
	\draw (0,0.15) ellipse (0.25*0.8 and 0.25*0.8);
	\node[black, anchor=north west] (a) at (0.03,0.15) {$y$};
	\draw[black] (0.177*0.8,0.177*0.8+0.15) -- (-0.177*0.8,-0.177*0.8+0.15);
	\draw[black] (0.177*0.8,-0.177*0.8+0.15) -- (-0.177*0.8,0.177*0.8+0.15);
	\node[black, anchor=south] (a) at (-1.2,2.75) {$b$};
	\node[black, anchor=south] (a) at (-0.0006,2.75) {$a$};
	\node[black, anchor=north] (a) at (-1.5+0.3*3.5-0.1,3.25-0.5196*3.5) {$c'$};
	\node[black] (a) at (0.9*0.8+0.05,0.7*0.8+0.15+0.05+0.05) {$d'$};
	\draw[black] (0,0.75) -- (-1.5*0.8,0.75+2.598*0.8) node[sloped,pos=0.75,allow upside down]{\arrowIn}; ; 
	\draw[black] (0,0.75) -- (-1.5*0.8,0.75+2.598*0.8) node[sloped,pos=0.25,allow upside down]{\arrowIn}; ; 
	\draw[black] (-1.2+0.3*2,0.75+0.5196*2) -- (-0.6+0.9*0.666,0.75+0.5196*2+1.5588*0.666) node[sloped,pos=0.50,allow upside down]{\arrowIn}; ; 
	\draw[black, line width=0.05pt] (0.707*0.8+0.707*0.001,0.707*0.75*0.8-0.707*0.75*0.001+0.15) -- (0.707*0.8,0.707*0.75*0.8+0.15) node[sloped,pos=0.50,allow upside down]{\arrowIn}; ; 
	\end{scope}
\end{tikzpicture}
\caption{Translating anyon $b$ across cut $B$ corresponds to translating it around the non-contractible loop. The resulting fusion diagram can be expressed as sum over diagrams in canonical form using $F$-moves.}
\label{eq:drag_x_2p}
\end{figure}

Finally, let us consider translations of anyons across the two cuts, which corresponds to translating them around the torus in the respective directions. Before looking at the example in Fig.~\ref{eq:translate_B} that can be used to generalize the occuring braids for translations in $x$-direction, we start by illustrating how translations around the torus are dealt with diagrammatically using the simplest example of two anyons $a$ and $b$ fusing to anyon $c$, which is depicted in Fig.~\ref{eq:drag_x_2p}; the anyon moving along the non-contratible loop has charge $d$. Anyon $b$ is translated in positive $x$-direction over cut $B$, which is indicated by the translation operator $T_{\vec{r}_b,\vec{r}_b+\vec{e}_x}$ acting on the initial fusion diagram. This translation corresponds to translating the respective anyon around the non-contractible loop\footnote{The process of translating an anyon around the non-contractible loop is very similar to what can be found in Ref.~\cite{PhysRevB.86.155111} for a chain of anyons. But in contrast to them, we do not include any twist/phase factor associated with the translation of anyons around the torus. We give an argument why we chose to exclude non-trivial phase factors for these processes in App.~\ref{app:Verlinde}.} in the fusion diagram and by using $F$-moves, this diagram can be expressed in terms of fusion diagrams in canonical form again. This calculation can be generalized to the corresponding translation in arbitrary fusion diagrams, which can be expressed as accordingly weighted superpositions of diagrams in canonical form. The weights are strings of $F$-moves and are introduced in App.~\ref{app_example_computations_sheet_hopping}.\\

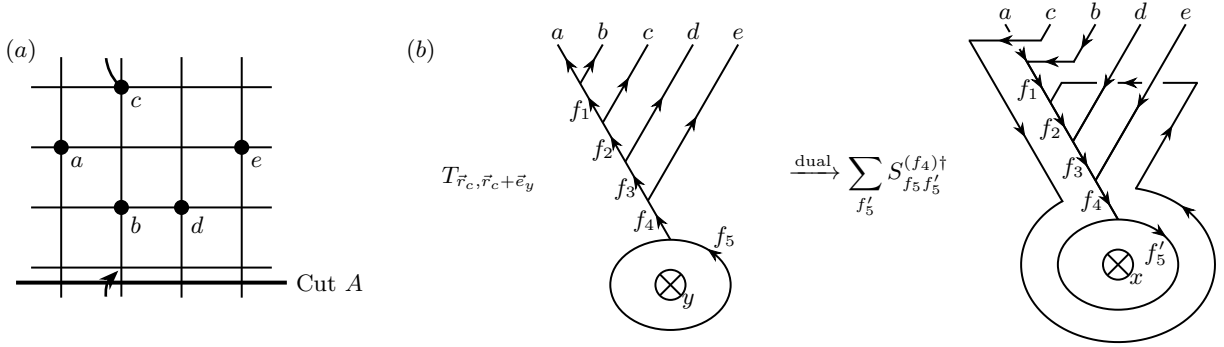
\begin{figure*}[t]
	\centering
\begin{tikzpicture}[line width=0.75pt]
	\node [black, anchor=east] (a) at (-1.65,3.16/2) {
            $\begin{aligned}
                T_{\vec{r}_c,\vec{r}_c+\vec{e}_y}
            \end{aligned}$
    };
    \node[black, anchor=south east] (a) at (-3,3.25*0.8+3.16*0.5-1.5*0.8) {$(b)$};
	\draw (0,0.15) ellipse (1*0.8 and 0.75*0.8);
	\draw (0,0.15) ellipse (0.25*0.8 and 0.25*0.8);
	\node[black, anchor=north west] (a) at (0.03,0.15) {$y$};
	\draw[black] (0.177*0.8,0.177*0.8+0.15) -- (-0.177*0.8,-0.177*0.8+0.15);
	\draw[black] (0.177*0.8,-0.177*0.8+0.15) -- (-0.177*0.8,0.177*0.8+0.15);
	\node[black, anchor=south] (a) at (-1.5,3.3) {$a$};
	\node[black, anchor=south] (a) at (-1.5+0.6,3.3) {$b$};
	\node[black, anchor=south] (a) at (-1.5+0.6*2,3.3) {$c$};
	\node[black, anchor=south] (a) at (-1.5+0.6*3,3.3) {$d$};
	\node[black, anchor=south] (a) at (-1.5+0.6*4,3.3) {$e$};
	\node[black, anchor=north] (a) at (-1.5+0.3,3.25-0.5196) {$f_1$};
	\node[black, anchor=north] (a) at (-1.5+0.3*2,3.25-0.5196*2) {$f_2$};
	\node[black, anchor=north] (a) at (-1.5+0.3*3,3.25-0.5196*3) {$f_3$};
	\node[black, anchor=north] (a) at (-1.5+0.3*4-0.1*0.5,3.25-0.5196*4+0.1732*0.5) {$f_4$};
	\node[black] (a) at (0.9*0.8,0.7*0.8+0.15+0.075) {$f_5$};
	\draw[black] (0,0.75) -- (-1.5,0.75+2.598) node[sloped,pos=0.90,allow upside down]{\arrowIn}; ; 
	\draw[black] (-1.5*0.69,0.75+2.598*0.69) -- (-1.5*0.7,0.75+2.598*0.7) node[sloped,pos=1,allow upside down]{\arrowIn}; ; 
	\draw[black] (-1.5*0.49,0.75+2.598*0.49) -- (-1.5*0.5,0.75+2.598*0.5) node[sloped,pos=1,allow upside down]{\arrowIn}; ; 
	\draw[black] (-1.5*0.29,0.75+2.598*0.29) -- (-1.5*0.3,0.75+2.598*0.3) node[sloped,pos=1,allow upside down]{\arrowIn}; ; 
	\draw[black] (-1.5*0.09,0.75+2.598*0.09) -- (-1.5*0.1,0.75+2.598*0.1) node[sloped,pos=1,allow upside down]{\arrowIn}; ; 
	\draw[black] (-1.2,0.75+0.5196*4) -- (-1.5+0.6,0.75+2.598) node[sloped,pos=0.50,allow upside down]{\arrowIn}; ; 
	\draw[black] (-1.2+0.3,0.75+0.5196*3) -- (-1.5+0.6*2,0.75+2.598) node[sloped,pos=0.50,allow upside down]{\arrowIn}; ; 
	\draw[black] (-1.2+0.3*2,0.75+0.5196*2) -- (-1.5+0.6*3,0.75+2.598) node[sloped,pos=0.50,allow upside down]{\arrowIn}; ; 
	\draw[black] (-1.2+0.3*3,0.75+0.5196*1) -- (-1.5+0.6*4,0.75+2.598) node[sloped,pos=0.50,allow upside down]{\arrowIn}; ; 
	\draw[black, line width=0.05pt] (0.707*0.8+0.707*0.8*0.001,0.707*0.8*0.75-0.707*0.8*0.75*0.001+0.15) -- (0.707*0.8,0.707*0.8*0.75+0.15) node[sloped,pos=0.50,allow upside down]{\arrowIn}; ; 	
	\begin{scope}[shift={(5.45+0.5,2.96/2-2.4212/2)}]
	\node [black, anchor=east] (a) at (-1.975-0.15,2.4212/2) {
            $\begin{aligned}
                \xrightarrow{\text{dual}}\sum_{f_5'}S^{(f_4)\dagger}_{f_5f_5'}
            \end{aligned}$
    };
	\draw (0,0.15) ellipse (1*0.8 and 0.75*0.8);
	\draw (0,0.15) ellipse (0.25*0.8 and 0.25*0.8);
	\node[black, anchor=north west] (a) at (0.03,0.15) {$x$};
	\draw[black] (0.177*0.8,0.177*0.8+0.15) -- (-0.177*0.8,-0.177*0.8+0.15);
	\draw[black] (0.177*0.8,-0.177*0.8+0.15) -- (-0.177*0.8,0.177*0.8+0.15);
	\draw [black,domain=0:2.75-0.13, samples=10] plot ({\x*cos(120)}, {\x*sin(120)+0.75});
	\draw [black,domain=2.75+0.13:3, samples=10] plot ({\x*cos(120)}, {\x*sin(120)+0.75});
	\draw [black,domain=0:3-0.5196/sin(60), samples=10] plot ({\x*cos(60)-0.3}, {\x*sin(60)+0.75+0.5196});
	\draw [black,domain=0:3-0.5196/sin(60)*2, samples=10] plot ({\x*cos(60)-0.3*2}, {\x*sin(60)+0.75+0.5196*2});
	\draw [black,domain=0:3-0.5196/sin(60)*3-0.9, samples=10] plot ({\x*cos(60)-0.3*3}, {\x*sin(60)+0.75+0.5196*3});
	\draw [black,domain=0:2.425-0.5196/sin(60)*4+0.01, samples=10] plot ({\x*cos(60)-0.3*4}, {\x*sin(60)+0.75+0.5196*4});
	\draw [black,domain=2.425-0.5196/sin(60)*3-0.01:3-0.5196/sin(60)*3, samples=10] plot ({\x*cos(60)-0.3*3}, {\x*sin(60)+0.75+0.5196*3});
	\draw [black,domain=-1.1875-0.02:-0.5875+0.01, samples=10] plot ({\x}, {2.85});
	\draw[black] (-0.8875,2.85) -- (-0.8876,2.85) node[sloped,pos=0,allow upside down]{\arrowIn}; ; 
	\draw[black] (-0.444,3.099) -- (-0.444-0.005,3.099-0.00866) node[sloped,pos=0,allow upside down]{\arrowIn}; ; 
	\draw [black,domain=-0.45-0.01-0.3:1-0.93-0.3-0.05, samples=10] plot ({\x}, {3.088-0.5196});
	\draw [black,domain=1.175-0.93-0.3+0.05:1-0.33-0.3-0.05, samples=10] plot ({\x}, {3.088-0.5196});
	\draw [black,domain=1.175-0.33-0.3+0.05:1.35+0.01-0.3, samples=10] plot ({\x}, {3.088-0.5196});
	\draw [black,domain=0-0.01-0.12:3-0.5196/sin(60)-0.9+0.003, samples=10] plot ({\x*cos(60)+0.3}, {\x*sin(60)+0.75+0.5196});
	\draw [black,domain=0.419-0.15:2.75, samples=10] plot ({\x*cos(120)-0.6}, {\x*sin(120)+0.75});
	\draw [black,domain=2.75-0.5196/sin(60)*4:3-0.5196/sin(60)*4, samples=10] plot ({\x*cos(60)-0.3*4}, {\x*sin(60)+0.75+0.5196*4});
	\draw [black,domain=-1.975-0.01:-1.025+0.01, samples=10] plot ({\x}, {3.1316});
	\node[black, anchor=south] (a) at (-1.5,3.3) {$a$};
	\node[black, anchor=south] (a) at (-1.5+0.6,3.3) {$c$};
	\node[black, anchor=south] (a) at (-1.5+0.6*2,3.3) {$b$};
	\node[black, anchor=south] (a) at (-1.5+0.6*3,3.3) {$d$};
	\node[black, anchor=south] (a) at (-1.5+0.6*4,3.3) {$e$};
	\node[black, anchor=north] (a) at (-1.5+0.3,3.25-0.5196) {$f_1$};
	\node[black, anchor=north] (a) at (-1.5+0.3*2,3.25-0.5196*2) {$f_2$};
	\node[black, anchor=north] (a) at (-1.5+0.3*3,3.25-0.5196*3) {$f_3$};
	\node[black, anchor=north] (a) at (-1.5+0.3*4-0.1*0.5,3.25-0.5196*4+0.1732*0.5) {$f_4$};
	\node[black] (a) at (0.5,0.3) {$f_5'$};
	\draw[black] (-1.5*0.85,0.75+2.598*0.85) -- (0,0.75) node[sloped,pos=0.01,allow upside down]{\arrowIn}; ; 
	\draw[black] (-1.5*0.7,0.75+2.598*0.7) -- (-1.5*0.69,0.75+2.598*0.69) node[sloped,pos=0,allow upside down]{\arrowIn}; ; 
	\draw[black] (-1.5*0.5,0.75+2.598*0.5) -- (-1.5*0.49,0.75+2.598*0.49) node[sloped,pos=0,allow upside down]{\arrowIn}; ; 
	\draw[black] (-1.5*0.3,0.75+2.598*0.3) -- (-1.5*0.29,0.75+2.598*0.29) node[sloped,pos=0,allow upside down]{\arrowIn}; ; 
	\draw[black] (-1.5*0.1,0.75+2.598*0.1) -- (-1.5*0.09,0.75+2.598*0.09) node[sloped,pos=0,allow upside down]{\arrowIn}; ; 
	\draw[black] (-1.355+0.5*0.15,2.057-0.866*0.15) -- (-1.355+0.5*0.15+0.005,2.057-0.866*0.15-0.00866) node[sloped,pos=0,allow upside down]{\arrowIn}; ; 
	\draw[black] (-1.975+0.475,3.1316) -- (-1.975+0.474,3.1316) node[sloped,pos=0,allow upside down]{\arrowIn}; ; 
	\draw[black] (0.4575-0.3,3.088-0.5196) -- (0.45-0.3,3.088-0.5196) node[sloped,pos=0,allow upside down]{\arrowIn}; ; 
	\draw[black] (-0.15,0.75+2.598/2+0.5196*1) -- (-1.2+0.3*2,0.75+0.5196*2) node[sloped,pos=0,allow upside down]{\arrowIn}; ; 
	\draw[black] (0.3,0.75+2.598/2+0.5196*0.5) -- (-1.2+0.3*3,0.75+0.5196*1) node[sloped,pos=0,allow upside down]{\arrowIn}; ; 
	\draw[black] (0.675,1.919) -- (0.675+0.005,1.919+0.00866) node[sloped,pos=0,allow upside down]{\arrowIn}; ; 
	\draw[black, line width=0.05pt] (0.707*0.8,0.707*0.75*0.8+0.15)--(0.707*0.8+0.707*0.001,0.707*0.75*0.8-0.707*0.75*0.001+0.15) node[sloped,pos=0.50,allow upside down]{\arrowIn}; ; 
	\draw[black, line width=0.05pt] (0.707*1.5*1.074*0.8+0.707*1.5*1.074*0.001,0.707*1.2*1.074*0.8-0.707*1.2*1.074*0.001+0.15) -- (0.707*1.5*1.074*0.8,0.707*1.2*1.074*0.8+0.15) node[sloped,pos=0.50,allow upside down]{\arrowIn}; ; 
	\draw [black,domain=-235.5:79.8, samples=100] plot ({1.5*cos(\x)*1.074*0.8}, {1.2*sin(\x)*1.074*0.8+0.15});
	\end{scope}
	\begin{scope}[shift={(-8.1,3.16/2-1.5*0.8)}, scale=0.8]
	\draw[black] (-0.5,0) -- (3.5,0);
	\draw[black] (-0.5,1) -- (3.5,1);
	\draw[black] (-0.5,2) -- (3.5,2);
	\draw[black] (-0.5,3) -- (3.5,3);
	\draw[black] (0,-0.5) -- (0,3.5);
	\draw[black] (1,-0.5) -- (1,3.5);
	\draw[black] (2,-0.5) -- (2,3.5);
	\draw[black] (3,-0.5) -- (3,3.5);
	\draw[black][line width=1.5pt] (-0.75,-0.25) -- (3.75,-0.25);
	\node[black, anchor=south east] (a) at (-0.25,3.25) {$(a)$};
	\node[black, anchor=west] (a) at (3.75,-0.25) {Cut $A$};
	\node[draw,circle,inner sep=1.75pt,fill,black] at (0,2) {};
	\node[draw,circle,inner sep=1.75pt,fill,black] at (1,3) {};
	\node[draw,circle,inner sep=1.75pt,fill,black] at (2,1) {};
	\node[draw,circle,inner sep=1.75pt,fill,black] at (1,1) {};
	\node[draw,circle,inner sep=1.75pt,fill,black] at (3,2) {};
	\node[black, anchor=north west] (a) at (0,2) {$a$};
	\node[black, anchor=north west] (a) at (1,1) {$b$};
	\node[black, anchor=north west] (a) at (1,3) {$c$};
	\node[black, anchor=north west] (a) at (2,1) {$d$};
	\node[black, anchor=north west] (a) at (3,2) {$e$};
	\draw [-Stealth][line width=1pt] (1-0.05,3+0.05) to [out=135,in=225] (1-0.05,4-0.05);
	\draw [-Stealth][line width=1pt] (1-0.05,-1+0.05) to [out=135,in=225] (1-0.05,0-0.05);
	\filldraw[draw=white,fill=white] (0.5,3.5) rectangle (1.5,4);
	\filldraw[draw=white,fill=white] (0.5,-1) rectangle (1.5,-0.5);
	\end{scope}
\end{tikzpicture}
	\caption{Illustration of the braid rules for translating anyons in positive $y$-direction across cut $A$. $(a)$ Lattice with an anyon configuration and indication which anyon is to be translated. $(b)$ Effect of the translation on the fusion diagram.}
	\label{eq:translate_A}
\end{figure*}

Using this consideration, let us now look at the more general example depicted in Fig.~\ref{eq:translate_B}, where the anyon configuration is shown in Fig.~\ref{eq:translate_B}$a$. Initially, the anyons are in alphabetical order and after translating anyon $c$, the order becomes $a,c,b,d,e$. The braids among the anyons in the diagram in Fig.~\ref{eq:translate_B}$b$ follow the same rules as before. The line associated with anyon $c$ moves in front of the lines of $d$ and $e$ and behind the one of $a$, in accordance with the picture introduced in Fig.~\ref{fig:intuitive_picture}. The only difference to the case discussed before is that upon crossing cut $B$, i.e., in between the braids, the translated anyon moves around the torus. We can thus use Fig.~\ref{fig:intuitive_picture} also in this case to determine the braids as long as we keep in mind how to move anyons crossing cut $B$ around the non-contractible loop in the fusion diagrams; Fig.~\ref{eq:translate_B} contains all information needed to apply the braids to any case that may be encountered.

The Hamiltonian's matrix elements $\mathcal{H}_{s's}$ are computed in the same way as for translations in the bulk. First, we apply the above rules to obtain the braided fusion diagram of the translated state, as done in Fig.~\ref{eq:translate_B}$b$. Then, we apply braid operators until the braided diagram is expressed as superposition of fusion diagrams in canonical form and read off the corresponding contributions that have to be multiplied by $-t$. Details, including an explicit treatment of the example in Fig.~\ref{eq:translate_B}$b$, are provided in App.~\ref{app_example_computations_sheet_hopping}.\\

The final case to be considered involves translations of anyons across cut $A$, which corresponds to translating them around the torus in $y$-direction. In contrast to translations in $y$-direction in the bulk, braiding is involved in this case, which is illustrated by the example depicted in Fig.~\ref{eq:translate_A}. The initial anyon configuration (Fig.~\ref{eq:translate_A}$a$) is again chosen such that the anyons are in alphabetical order. After translating anyon $c$, the order becomes $a,c,b,d,e$, which implies that braids need to be involved in the fusion diagrams. The first step towards understanding how the braids in Fig.~\ref{eq:translate_A}$b$ arise is to note that we need to transform the initial fusion diagram using the punctured torus $S$-matrix introduced in Sec.~\ref{sec:dual}. Since going to the dual space corresponds to changing the direction of the non-contractible loop of anyon $g$ in Fig.~\ref{eq:translate_A}$b$, we can treat translations around the torus in $y$-direction in the transformed fusion diagrams in the same way as translations around the torus in $x$-direction in canonical form. That is, the anyon crossing cut $A$ moves clockwise around the non-contractible loop of the transformed diagram. Finally, we have to determine the braids among the anyons. This can be done using the 3D picture in Fig.~\ref{fig:intuitive_picture}: When translating an anyon ($c$ in Fig.~\ref{fig:intuitive_picture}) across cut $A$, it braids with all anyons possessing a larger $x$-coordinate as its line has to move behind the lines of such anyons. Similarly, the translated anyon braids with all anyons featuring smaller $x$-coordinates by moving in front of them. The only anyons not involved in any braids are those in the same column as the anyon being translated. We can use Fig.~\ref{eq:translate_A} as summary for translations across cut $A$; it contains all information needed to construct the braided fusion diagrams for all cases that may be encountered.

The Hamiltonian's matrix elements are again computed in the same way as for the other cases. We need to apply the braids to the diagrams and then resolve them using braid operators. The final superposition of fusion diagrams in canonical form reveals the matrix elements. An explicit treatment of the case depicted in Fig.~\ref{eq:translate_A}$b$ can be found in App.~\ref{app_example_computations_sheet_hopping}.\\

Overall, the general structure used for computing the Hamiltonian $\mathcal{H}$ can be summarized in the pseudocode in Alg.~\ref{algo:Hamiltonian}, which assumes that the basis states were already constructed using Alg.~\ref{algo:tuples} or \ref{algo:tuples2}. By interpreting the $i$-th basis state as unit vector in the $i$-th direction, the Hamiltonian becomes a matrix whose columns correspond to the vectors associated with the superpositions obtained after applying the Hamiltonian on the basis states. As discussed above, we split the translations into the four different cases and focus on translations in the positive directions since their counterparts in the negative directions can be obtained via hermitian conjugation. The functions defined to treat the four cases are introduced as pseudocodes in App.~\ref{app_example_computations}.

\begin{algorithm}[t]
\caption{\justifyingg{Computation of the Hamiltonian using the basis states $\lbrace |(\alpha_k,\vec{r}_k)_{k=1}^N,\mathbf{f}\rangle \rbrace$. The hopping amplitude is denoted by $t$; all other parameters, in particular those defining the anyon model, are contained in $\lbrace \ldots \rbrace$ for convenience. The two lattice vectors are $\vec{e}_x$ and $\vec{e}_y$.}}
\begin{algorithmic}[1]

\Function{ComputeHamiltonian}{$\,\,\,\,t$,$\,\,\,\,\lbrace |(\alpha_k,\vec{r}_k)_{k=1}^N,\mathbf{f}\rangle \rbrace$,\newline$\lbrace \ldots \rbrace$}

\State $basis \gets$ list of all basis states $|s\rangle = |(\alpha_k,\vec{r}_k)_{k=1}^N, \mathbf{f}\rangle$; the $i$-th entry $|s_i\rangle$ is represented by $\vec{e}_i$
\State $H \gets$ empty matrix whose dimensions equal the basis size

\For{$|s_i\rangle$ in $basis$}
\For{$k\gets 1$ to $N$}

\If{$\vec{r}_k\cdot \vec{e}_x < L_x-1$}
\State $\vec{v}\gets \textsc{TranslationBulkX}(\vec{e}_i,k,\lbrace \ldots \rbrace)$
\Else
\State $\vec{v}\gets \textsc{TranslationCutB}(\vec{e}_i,k,\lbrace \ldots \rbrace)$
\EndIf

\If{$\vec{r}_k\cdot \vec{e}_y < L_y-1$}
\State $\pvec{v}'\gets \textsc{TranslationBulkY}(\vec{e}_i,k)$
\Else
\State $\pvec{v}'\gets \textsc{TranslationCutA}(\vec{e}_i,k,\lbrace \ldots \rbrace)$
\EndIf

\For{$j\gets 1$ to $\mathrm{dim}(\vec{v})$}
\State $H_{ji} = -t(\vec{v}+\pvec{v}')\cdot \vec{e}_j$
\EndFor

\EndFor
\EndFor

\State \Return $H+H^{\dagger}$
\EndFunction
\end{algorithmic}
\label{algo:Hamiltonian}
\end{algorithm}

For the case of abelian anyon models, the algorithm described above is related to the algorithm introduced in Ref.~\cite{PhysRevB.43.10761} and sketched in Sec.~\ref{sec:partcile_types_abelian}. This relation is shown in App.~\ref{app_equi} and reveals that our algorithm and the one in Ref.~\cite{PhysRevB.43.10761} use different conventions for the external fluxes $\Phi_x$ and $\Phi_y$ that have not been introduced yet. For our algoirthm, we may choose to incorporate external fluxes as additional factors of $e^{2\pi i\Phi_x/\phi_0}$ and $e^{2\pi i\Phi_y/\phi_0}$ that are acquired when translating an anyon over cut $B$ or $A$ in positive $x$- or $y$-direction, respectively, where $\phi_0=hc/e$ denotes the flux quantum~\cite{PhysRevB.43.10761, SemionsTorus}.

It is additionally possible to construct momentum states that block diagonalize the Hamiltonian $\mathcal{H}$, which allows us to numerically simulate larger systems. The basic idea is to construct eigenstates to translation operators that translate all anyons simultaneously in $x$- or $y$-direction. A detailed discussion of the momentum states and their construction can be found in App.~\ref{app_mom}.

Finally, we also note that the above algorithm is not restricted to square lattices. An algorithm for other $2$D lattice forms may be obtained by establishing a fusion order, similar to Fig.~\ref{fig:lattice}$b$. The hopping of anyons can then be expressed as discussed above; the rules determining the braids remain unchanged.

\section{Simulation Results}
\label{sec:results}

To demonstrate the algorithm discussed in the previous sections, we apply it to study spectral properties and non-equilibrium dynamics. Specifically, we start with a light discussion of the energy eigenvalues of systems containing one and two anyons, showing the similarities and differences with systems of bosons and fermions. Next, we focus on systems of multiple anyons, where we discuss the statistical distribution of the energy levels and on the density distribution following a quench. We consider semions, Fibonacci anyons and Ising anyons as simple representations of abelian and non-abelian anyons, respectively, and compare them to hard-core bosons (HCBs). For the quenches, we further compare them to fermions. 
In our simulations, we use translation invariance to construct momentum states and split the Hamiltonian into disconnected momentum sectors, as explained in App.~\ref{app_mom}.

\subsection{One- and Two-Particle Energies and Energy Level Spacing Statistics}\label{sec:levelspacing}

The tight-binding model of a single anyon on a torus has the same spectrum as a single fermion/boson in 2D with PBC. We confirmed that our algorithm works even in this simple case for models that permit having a single anyon on a torus, like the Fibonacci anyon model.

Within the bulk of the lattice, hopping of an anyon is trivial. Further, the topological amplitudes associated with hopping processes around the torus turn also out to be trivial. Therefore, the behavior of a single anyon on a torus is like a single fermion/boson on a 2D lattice with periodic boundary conditions, and hence they have the same spectrum.

When the number of anyons is two or more, some of the rich peculiarities of braiding statistics start to manifest if the particles are allowed to experience their statistics through braiding. In particular, unlike fermions and bosons, the energies are no longer sums of energies of individual anyons, as illustrated in Table~\ref{table:GSEnergies}.\\

\begin{table}
\centering
\begin{tabular}{c | c | c}
$ 2 \times E^{(1)}_{\mathrm{Fib}}$ &  $E^{(2)}_{\mathrm{Fib}}$ & $(n_x, n_y) $ \\[2pt]
\hline
$ -8	$& 	$-7.7314$ 	& 	$(0,0)  $\\
$ -6.494$	& $-7.2514$	& 	$(0,-1), (0,1), (-1,0), (1,0) $\\
$-4.9879$	& $-6.923$	& 	$(-1,-1), (-1,1), (1,-1), (1,1)  $  \\
$ -3.1099$ & $-6.2969$	& 	$(0,-2), (0,2), (-2,0), (2,0)  $
\end{tabular}
\caption{The energy of a collection of anyons is not the sum of the individual particles' energies. Here, we show a sample of the ground state energies in different momentum sectors for systems of Fibonacci anyons on $7 \times 7$ lattices. The first column shows twice the energies of one particle, $E^{(1)}_{\mathrm{Fib}}$, the second column the energies for two particles, $E^{(2)}_{\mathrm{Fib}}$.
The momenta are quantized as $k_x = \frac{2 \pi}{7} n_x$ and $k_y = \frac{2 \pi}{7} n_y$. Note that we only show a subset of the full spectra.}
\label{table:GSEnergies}
\end{table}

Let us now use a more sophisticated method for analyzing the energy spectra by studying the spectral statistics of the Hamiltonian. In particular, consider the ratios $r_n$~\cite{PhysRevB.75.155111, Kollath_2010} of consecutive energy level spacings with
\begin{align}
	r_n=\frac{\min\lbrace \delta_n,\delta_{n-1} \rbrace}{\max\lbrace \delta_n,\delta_{n-1} \rbrace}, \quad \delta_n=E_{n+1}-E_n,
\end{align}
where the energy levels $E_n$ are ordered ascendingly, i.e., $\delta_n\geq 0$. It is expected that the level spacing statistics of integrable Hamiltonians follow the Poisson distribution. For non-integrable systems on the other hand, the statistics are expected to behave like the Wigner-Dyson distribution of the Gaussian orthogonal ensemble (GOE) for time-reversal (TR) invariant systems and like the Gaussian unitary ensemble (GUE) for systems breaking TR symmetry~\cite{doi:10.1080/00018732.2016.1198134}. The Poisson (Poi), GOE and GUE predictions for the distribution of $r$ are given by~\cite{PhysRevLett.110.084101}
\begin{align}
	P_{\mathrm{Poi}}(r)&=\frac{2}{(1+r)^2}, \\ P_{\mathrm{GOE}}(r)&=\frac{27}{4}\frac{r+r^2}{(1+r+r^2)^{5/2}} \quad\text{and}\\ P_{\mathrm{GUE}}(r)&=\frac{81\sqrt{3}}{2\pi}\frac{(r+r^2)^2}{(1+r+r^2)^{4}}.
\end{align}
When looking at the distribution of $r$, it is important to consider the symmetries of the system. For example, due to the Hamiltonian's translation invariance, momentum states with distinct momentum quantum numbers are decoupled and thus, the corresponding expectation values are conserved. Similarly, the spatial reflection symmetries may lead to further conserved quantities. In general, the presence of such conserved quantities is expected to lead to Poisson statistics. If conserved quantities are absent, the energy levels are expected to show level repulsion, which corresponds to Wigner-Dyson statistics. Therefore, if one analyzes the level spacing statistics without restriction to a certain sector of the quantum numbers, a non-integrable, interacting system may seem to be integrable and non-interacting due to the lack of level repulsion between the energy levels in different sectors~\cite{doi:10.1080/00018732.2016.1198134}. It is thus important to only look at the statistics of energy levels associated with states that share the same quantum numbers. Here, we choose to look at three different momentum sectors.

\begin{figure}[t]
	\centering
	\includegraphics{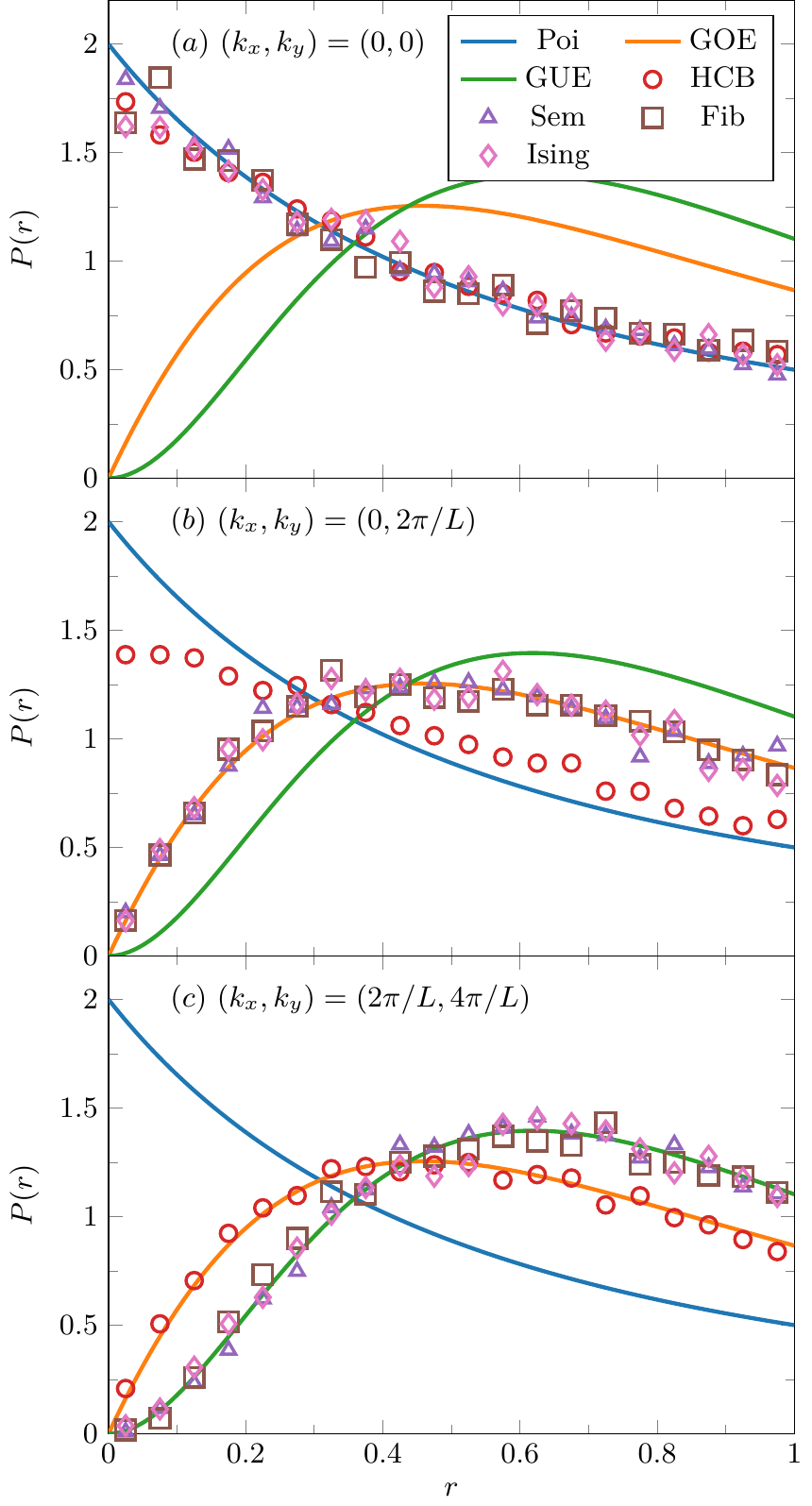}
	\caption{$P(r)$ for HCBs on a $9\times 9$ lattice with column coupling $r_{col}=0.5$, semions on a $8\times 8$ lattice, Fibonacci anyons on a $6\times 6$ and Ising anyons on a $6\times 6$ lattice for $N=4$ particles in the $(a)$: $k_x=k_y=0$, $(b)$: $k_x=0$, $k_y=2\pi/L$ and $(c)$: $k_x=2\pi/L$, $k_y=4\pi/L$ momentum sector together with the Poisson, GOE and GUE predictions.}
	\label{fig:levelstat}
\end{figure}

In Fig.~\ref{fig:levelstat}, the Poisson, GOE and GUE predictions of $r$ are plotted together with the distributions $P(r)$ obtained from exact diagonalization (ED)~\cite{lin1993, Weisse2008, Zhang_2010} for semions, Fibonacci anyons and Ising anyons on $L\times L$ square lattices in the three momentum sectors $k_x=k_y=0$ (Fig.~\ref{fig:levelstat}$a$), $k_x=0$, $k_y=2\pi /L$ (Fig.~\ref{fig:levelstat}$b$) and $k_x=2\pi /L$, $k_y=4\pi /L$ (Fig.~\ref{fig:levelstat}$c$). The corresponding distributions for HCBs are also shown for comparison\footnote{We choose to forgo showing the distributions $P(r)$ for fermions in Fig.~\ref{fig:levelstat} since due to fermions being non-interacting, their energy level spacings are expected to follow the Poissonian prediction in every momentum sector.}. For all simulations, the particle number is set to $N=4$. Due to the total number of wave function components depending on the particle type, different lattice sizes are used: $L=9$ is chosen for HCBs, $L=8$ for semions and $L=6$ for both Fibonacci anyons and Ising anyons.

From the results, it can be seen that $P(r)$ follows the Poisson distribution in the $k_x=k_y=0$ momentum sector for all particle types. This is expected since there are four spatial reflection symmetries that we do not account for in our simulations. Hence, as explained above, the distributions of $r$ follow the Poissonian prediction. For HCBs, it can be seen that $P(r)$ follows the GOE prediction in the $k_x=2\pi /L$, $k_y=4\pi /L$ momentum sector due to the absence of further symmetries, HCBs being interacting particles and the system being TR symmetric. In the $k_x=0$, $k_y=2\pi /L$ sector, the distribution is neither Poissonian nor GOE-like but in between. This is characteristic for spectra containing two symmetry blocks~\cite{PhysRevX.12.011006}, which is the case in the given situation due to the presence of the reflection symmetry in $x$-direction. For more than two symmetry blocks in the energy spectrum, the distribution of $r$ tends towards the Poisson prediction more strongly (for $k_x=k_y=0$, there are $2^4$ unaccounted symmetry sectors in total, which leads to $P(r)$ following the Poissonian prediction). For semions, Fibonacci anyons and Ising anyons, $P(r)$ agrees with the GUE prediction for $k_x=2\pi /L$, $k_y=4\pi /L$. The reason is that the studied anyonic systems break TR symmetry. This can be seen by noting that under TR, anyons are mapped to their TR partners that feature the opposite exchange statistics~\cite{PhysRevB.94.115139}. The anyonic systems further break the individual spatial reflection symmetries since the corresponding operators map counter-clockwise exchanges to clockwise exchanges and vice versa. The combination of a reflection symmetry and TR symmetry is however conserved (i.e., by exchanging anyons with their TR partners and reversing braid directions, we effectively recover the initial system). This leads to $P(r)$ following the GOE distribution in the $k_x=0$, $k_y=2\pi /L$ momentum sector where the combination of TR symmetry and reflection symmetry in $x$-direction is present\footnote{We mentioned earlier that the presence of TR symmetry leads to GOE statistics. This statement is not quite precise: the actual condition to obtain GOE statistics is the presence of \emph{any} anti-unitary symmetry~\cite{MRobnik_1986}, which is exactly what we observe for the anyonic systems studied here.}. The presence of further combinations of reflection symmetries and TR symmetry in the $k_x=k_y=0$ momentum sector leads to $P(r)$ following the Poisson distribution. The above results also suggest that for the considered anyonic systems, there are no additional symmetries to be exploited in every momentum sector in order to further block diagonalize the Hamiltonian.

It is important to note that there is a fundamental difference between HCBs and the anyons that goes beyond the presence or absence of TR symmetry: For HCBs, a hard-core potential has to be introduced in order to avoid multiple bosons being localized on the same site, which may be absorbed into the on-site commutation relations by making them anti-commutation relations. For the considered anyons however, the localization of multiple anyons on a single site is prohibited by their exchange statistics, similar to fermions. It follows that despite the anyons also obeying the Pauli exclusion principle, their statistics and non-local properties make them behave like interacting particles.

\subsection{Quench Dynamics}\label{sec:quench}

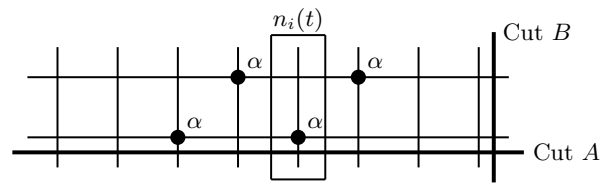
\begin{figure}[t]
\centering
\begin{tikzpicture}[line width=0.75pt, scale=0.8]
	\draw[black] (-4.5,0) -- (3.5,0);
	\draw[black] (-4.5,1) -- (3.5,1);
	\draw[black] (-4,-0.5) -- (-4,1.5);
	\draw[black] (-3,-0.5) -- (-3,1.5);
	\draw[black] (-2,-0.5) -- (-2,1.5);
	\draw[black] (-1,-0.5) -- (-1,1.5);
	\draw[black] (0,-0.5) -- (0,1.5);
	\draw[black] (1,-0.5) -- (1,1.5);
	\draw[black] (2,-0.5) -- (2,1.5);
	\draw[black] (3,-0.5) -- (3,1.5);
	\draw[black][line width=1.5pt] (3.25,-0.75) -- (3.25,1.75);
	\node[black, anchor=west] (a) at (3.25,1.75) {Cut $B$};
	\draw[black][line width=1.5pt] (-4.75,-0.25) -- (3.75,-0.25);
	\node[black, anchor=west] (a) at (3.75,-0.25) {Cut $A$};
	\node[draw,circle,inner sep=1.75pt,fill,black] at (-1,1) {};
	\node[draw,circle,inner sep=1.75pt,fill,black] at (-2,0) {};
	\node[draw,circle,inner sep=1.75pt,fill,black] at (0,0) {};
	\node[draw,circle,inner sep=1.75pt,fill,black] at (1,1) {};
	\node[black, anchor=south west] (a) at (-1,1) {$\alpha$};
	\node[black, anchor=south west] (a) at (-2,0) {$\alpha$};
	\node[black, anchor=south west] (a) at (0,0) {$\alpha$};
	\node[black, anchor=south west] (a) at (1,1) {$\alpha$};
	\draw[black] (0.45,-0.7) -- (0.45,1.7);
	\draw[black] (-0.45,-0.7) -- (-0.45,1.7);
	\draw[black] (-0.45,1.7) -- (0.45,1.7);
	\draw[black] (-0.45,-0.7) -- (0.45,-0.7);
	\node[black, anchor=south] (a) at (0,1.6) {$n_i(t)$};
\end{tikzpicture}
\caption{Initial configuration for the quench simulations on periodic lattices with $L_y=2$: $N=4$ particles of type $\alpha$ are localized in the middle of the lattice in a zigzag pattern. The operator $n_i(t)$ measures the number of particles localized in the $i$-th column at time $t$.}
\label{fig:ladder}
\end{figure}

Lastly, let us consider non-equilibrium dynamics following a quench, where the density distribution over time is monitored using ED. The lattice is now chosen to be periodic with $L_y=2$, as depicted in Fig.~\ref{fig:ladder}. This figure also shows the initial state, in which $N=4$ particles are localized in the middle of the lattice in a zigzag pattern such that there is one particle per column and no two particles are initially on neighboring sites. The zigzag pattern is chosen since two anyons are not allowed to be on the same site. It should thus reduce the blocking of the initial dynamics. Further, we choose an equal superposition of all wave function components / fusion diagrams for the initial state, i.e., the initial state can be written as $|\lbrace \mathbf{f} \rbrace|^{-1/2}\sum_{\mathbf{f}}|(\alpha_k,\vec{r}_k)_{k=1}^N, \mathbf{f}\rangle$, where $(\alpha_k,\vec{r}_k)_{k=1}^N$ denotes the anyon configuration described above and depicted in Fig.~\ref{fig:ladder} and $\mathbf{f}$ all tuples that are consistent with the fusion rules; $\lbrace \mathbf{f} \rbrace$ is the set containing all these tuples $\mathbf{f}$.

\begin{figure*}[t]
\centering
\includegraphics{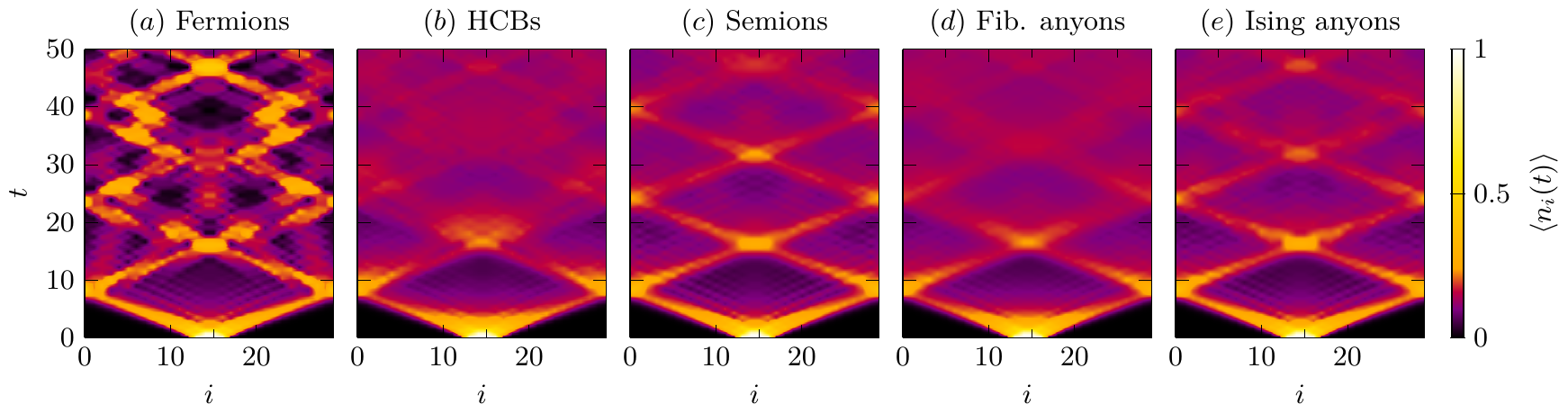}
\caption{Particle density per column $\langle n_{i}(t) \rangle$ for $(a)$ fermions, $(b)$ HCBs, $(c)$ semions, $(d)$ Fibonacci anyons and $(e)$ Ising anyons on a $30\times 2$ ladder. Initially, the $N=4$ particles are localized in the middle of the ladder in a zigzag pattern (see Fig. \ref{fig:ladder}). The initial state further features an equal superposition of all wave function components. For HCBs, the column coupling is set to $r_{col}=0.5$.}
\label{fig:density_plots}
\end{figure*}

The particle density is measured for each column rather than for each site since we are interested in the time dependence of the density along the $x$-direction. The operator measuring the density in the $i$-th column at time $t$ is denoted by $n_i(t)$, as indicated in Fig.~\ref{fig:ladder}. In order to obtain results that allow for better comparison, the lattice is chosen to have the same size, $30\times2$, for all particle types to be considered. In Fig.~\ref{fig:density_plots}, the time dependent particle density $\langle n_i(t) \rangle$ is plotted for fermions, HCBs, semions, Fibonacci anyons and Ising anyons. For HCBs, we slightly modify the Hamiltonian in Eq.~(\ref{eq:hamiltonian_square_lattice}). When hopping from one site to another in $y$-direction, hopping in the ``bulk'' is equivalent to hopping across cut $A$ due to the trivial statistics and $L_y=2$, see Fig.~\ref{fig:ladder}. This means that the coupling between two sites connected by a single hopping process in $y$-direction is effectively doubled. In order to make the system isotropic and avoid unintended effects, we modify the $y$-coupling by introducing a ``column coupling'' $r_{col}$ (i.e., $T_{\vec{r}_i,\vec{r}_i+\vec{e}_y}\rightarrow r_{col} T_{\vec{r}_i,\vec{r}_i+\vec{e}_y}$ in the Hamiltonian~(\ref{eq:hamiltonian_square_lattice})) and set it to $r_{col}=0.5$ such that the mentioned effect is compensated. This issue is unique to bosons since all other particle types feature non-trivial statistics, i.e., translations over cut $A$ differ from translations in the bulk.

For the fermionic case, the non-interacting behavior can be seen by the way the density spreads over time. When two wave packets collide, they simply pass through each other due to the lack of interactions. This leads to interference effects that do not decay with increasing time. I.e., even in the limit of infinite times, fluctuations in the density distribution can be observed. This is indicated in Fig.~\ref{fig:density_plots}$a$, where the fluctuations are prominent at all times. Quite the opposite is observed for HCBs in Fig.~\ref{fig:density_plots}$b$. Due to their interacting behavior, wave packages can not fully pass through each other, which leads to decaying interference effects and the density distribution becoming more homogeneous with increasing time. For semions (Fig.~\ref{fig:density_plots}$c$), Fibonacci anyons (Fig.~\ref{fig:density_plots}$d$) and Ising anyons (Fig.~\ref{fig:density_plots}$e$), the observed behavior is very similar to the one obtained for HCBs, i.e., the density distribution becomes more homogeneous at larger times.

We note that results very similar to those in Fig.~\ref{fig:density_plots} are obtained for other superpositions of the wave function components in the initial state, i.e., the above observations seem to be general for a quench with localized anyons. It is thus concluded that free semions, Fibonacci anyons and Ising anyons on a 2D lattice behave similarly to interacting particles if their dynamics are governed by a tight-binding Hamiltonian that merely accounts for their statistics, as their energy level spacing statistics show level repulsion and their density distributions after a quench seem to become homogeneous in the limit of large times.\\

A similarly interesting observation was made for 1D systems of hard-core abelian anyons of arbitrary statistics governed by the anyonic tight-binding Hamiltonian~\cite{1312.4657}. It was found that for quenches, one-body observables relax to the predictions of the generalized Gibbs ensemble for all abelian statistics (including HCBs) except for fermionic ones, suggesting that abelian anyons behave like interacting particles. This observation for 1D chains is consistent with what we found for free semions on quasi 2D lattices.

\section{Conclusion}\label{sec:conclusion}

We have developed an algorithm that is capable of simulating arbitrary abelian and non-abelian anyons subject to an anyonic tight-binding Hamiltonian that incorporates the anyons' statistics in two dimensions, where we focused on periodic boundary conditions. The algorithm can also be generalized to other, non-periodic, boundary conditions, which may feature anyonic charges on the boundaries. In the algorithm, the effects of anyonic statistics are expressed as braids in fusion diagrams. We also introduce momentum states in App.~\ref{app_mom} in order to block diagonalize the Hamiltonian. The main differences to other algorithms~\cite{PhysRevB.43.2661_fluxconvention, PhysRevB.43.10761, SemionsTorus,PhysRevB.89.075112} is that the presented algorithm is designed to deal with non-abelian anyon models, where all anyons are mobile on the lattice.

Our simulation results indicate thermalizing behavior for semions, Fibonacci anyons and Ising anyons: The statistical distributions of the energy levels feature level repulsion within the momentum sectors and the density distributions after a quench seem to converge to homogeneous distributions.

These results are only a first demonstration of the algorithm. In future, it can help to find new signatures that may be used to distinguish different anyonic charges as, e.g., done in one dimension for abelian anyons using the momentum distribution of the ground state~\cite{Li2013} or in two dimensions by measuring the spectral response of a system close to the threshold of exciting a pair of abelian anyons~\cite{SpectralFunction}. It would be particularly useful to identify differences between abelian and non-abelian anyons, as done for the transport properties of a single anyon on a ladder with background charges~\cite{PhysRevB.89.075112,PhysRevB.90.134201}. For the latter goal, one might suggest to study systems of three or more anyons in greater detail since exchanging two non-abelian anyons in this case does in general no longer correspond to simple $R$-moves like for two anyons. As for numerical methods, it may be beneficial to implement the presented algorithm using matrix product states or tensor product states as done in, e.g., Refs. \cite{PhysRevB.89.075112, PhysRevB.93.165128, konig2010anyonic, PhysRevB.92.115135, PhysRevB.82.115126} for other Hamiltonians. In particular, interactions and constraints on fusion products can be easily incorporated into our algorithm.

\section*{Acknowledgments}

NK, AS and FP were financially supported by the European Research Council (ERC) under the European Union’s Horizon 2020 research and innovation program under grant agreement No.~771537. AS was supported by a Research Fellowship from the Royal Commission for the Exhibition of 1851. FP acknowledges the support of the Deutsche Forschungsgemeinschaft (DFG, German Research Foundation) under Germany’s Excellence Strategy EXC-2111-390814868. FP’s research is part of the Munich Quantum Valley, which is supported by the Bavarian state government with funds from the Hightech Agenda Bayern Plus. The research of DM, BMA and JKS was financially supported by Science Foundation Ireland (SFI) through Principal Investigator Award 16/IA/4524.

The two groups of authors (NK, AS, FP and DM, BMA, JKS) had been working independently on substantially similar projects and parts of their work have been made public separately in~\cite{nico_kirchner_firstVersion} and~\cite{darragh_thesis}. On learning of each other's work, we decided to collaborate on the topic, resulting in the present manuscript where our different methods and algorithms have been cross-checked and harmonized.

\section*{Data Availability}

Access~\cite{nico_kirchner_2022_6777951} to the data plotted in Sec.~\ref{sec:results} and an example code for simulating the discussed models (without implementation of momentum states) may be granted upon reasonable request.

\appendix

\section{Examples for Braiding Diagrams and Pseudocodes for the Different Translation Cases}
\label{app_example_computations}

\begin{figure*}[t]
	\centering
\begin{tikzpicture}[line width=0.75pt]
\begin{scope}[shift={(-3.4,-5)}]
	\draw (0,0.15) ellipse (1*0.8 and 0.75*0.8);
	\draw (0,0.15) ellipse (0.25*0.8 and 0.25*0.8);
	\node[black, anchor=north west] (a) at (0.03,0.15) {$y$};
	\draw[black] (0.177*0.8,0.177*0.8+0.15) -- (-0.177*0.8,-0.177*0.8+0.15);
	\draw[black] (0.177*0.8,-0.177*0.8+0.15) -- (-0.177*0.8,0.177*0.8+0.15);
	\draw [black,domain=0:3, samples=10] plot ({\x*cos(120)}, {\x*sin(120)+0.75});
	\draw [black,domain=0:3-0.5196/sin(60), samples=10] plot ({\x*cos(60)-0.3}, {\x*sin(60)+0.75+0.5196});
	\draw [black,domain=0:3-0.5196/sin(60)*2-0.4-0.03, samples=10] plot ({\x*cos(60)-0.3*2}, {\x*sin(60)+0.75+0.5196*2});
	\draw [black,domain=3-0.5196/sin(60)*2-0.2+0.03:3-0.5196/sin(60)*2, samples=10] plot ({\x*cos(60)-0.3*2}, {\x*sin(60)+0.75+0.5196*2});
	\draw [black,domain=0:3-0.5196/sin(60)*3-0.4-0.03, samples=10] plot ({\x*cos(60)-0.3*3}, {\x*sin(60)+0.75+0.5196*3});
	\draw [black,domain=3-0.5196/sin(60)*3-0.2+0.03:3-0.5196/sin(60)*3, samples=10] plot ({\x*cos(60)-0.3*3}, {\x*sin(60)+0.75+0.5196*3});
	\draw [black,domain=0:3-0.5196/sin(60)*4-0.3, samples=10] plot ({\x*cos(60)-0.3*4}, {\x*sin(60)+0.75+0.5196*4});
	\draw [black,domain=3-0.5196/sin(60)*4-0.3:3-0.5196/sin(60)*4, samples=10] plot ({\x*cos(60)+1.2}, {\x*sin(60)+0.75+0.5196*4});
	\draw [black,domain=-1.05-0.01:1-0.33-0.05, samples=10] plot ({\x}, {3.088});
	\draw [black,domain=1.175-0.33+0.05:1.35+0.01, samples=10] plot ({\x}, {3.088});
	\node[black, anchor=south] (a) at (-1.5,3.3) {$a$};
	\node[black, anchor=south] (a) at (-1.5+0.6*5,3.3) {$b$};
	\node[black, anchor=south] (a) at (-1.5+0.6*2,3.3) {$c$};
	\node[black, anchor=south] (a) at (-1.5+0.6*3,3.3) {$d$};
	\node[black, anchor=south] (a) at (-1.5+0.6*4,3.3) {$e$};
	\node[black, anchor=north] (a) at (-1.5+0.3,3.25-0.5196) {$f_1$};
	\node[black, anchor=north] (a) at (-1.5+0.3*2,3.25-0.5196*2) {$f_2$};
	\node[black, anchor=north] (a) at (-1.5+0.3*3,3.25-0.5196*3) {$f_3$};
	\node[black, anchor=north] (a) at (-1.5+0.3*4-0.1*0.5,3.25-0.5196*4+0.1732*0.5) {$f_4$};
	\node[black] (a) at (0.9*0.8,0.7*0.8+0.15+0.075) {$f_5$};
	\draw[black] (0,0.75) -- (-1.5,0.75+2.598) node[sloped,pos=0.90,allow upside down]{\arrowIn}; ; 
	\draw[black] (-1.5*0.69,0.75+2.598*0.69) -- (-1.5*0.7,0.75+2.598*0.7) node[sloped,pos=1,allow upside down]{\arrowIn}; ; 
	\draw[black] (-1.5*0.49,0.75+2.598*0.49) -- (-1.5*0.5,0.75+2.598*0.5) node[sloped,pos=1,allow upside down]{\arrowIn}; ; 
	\draw[black] (-1.5*0.29,0.75+2.598*0.29) -- (-1.5*0.3,0.75+2.598*0.3) node[sloped,pos=1,allow upside down]{\arrowIn}; ; 
	\draw[black] (-1.5*0.09,0.75+2.598*0.09) -- (-1.5*0.1,0.75+2.598*0.1) node[sloped,pos=1,allow upside down]{\arrowIn}; ; 
	\draw[black] (-0.16,3.088) -- (-0.15,3.088) node[sloped,pos=1,allow upside down]{\arrowIn}; ; 
	\draw[black] (-1.2+0.3,0.75+0.5196*3) -- (-0.6,0.75+2.598/2+0.5196*1.5) node[sloped,pos=1,allow upside down]{\arrowIn}; ; 
	\draw[black] (-1.2+0.3*2,0.75+0.5196*2) -- (-0.15,0.75+2.598/2+0.5196*1) node[sloped,pos=1,allow upside down]{\arrowIn}; ; 
	\draw[black] (-1.2+0.3*3,0.75+0.5196*1) -- (0.3,0.75+2.598/2+0.5196*0.5) node[sloped,pos=1,allow upside down]{\arrowIn}; ; 
	\draw[black, line width=0.05pt] (0.707*0.8+0.707*0.8*0.001,0.707*0.8*0.75-0.707*0.8*0.75*0.001+0.15) -- (0.707*0.8,0.707*0.8*0.75+0.15) node[sloped,pos=0.50,allow upside down]{\arrowIn}; ; 	
	\end{scope}
	\begin{scope}[shift={(7.0,-5)}]
	\node [black, anchor=east] (a) at (-1.65,3.16/2-0.17) {
            $\begin{aligned}
                =\sum_{f_1',f_2',f_3'}\left[B^{acb}_{f_2}\right]_{f_1f_1'}\left[B^{f_1'db}_{f_3}\right]_{f_2f_2'}\left[B'^{f_2'eb}_{f_4}\right]_{f_3f_3'}
            \end{aligned}$
    };
	\draw (0,0.15) ellipse (1*0.8 and 0.75*0.8);
	\draw (0,0.15) ellipse (0.25*0.8 and 0.25*0.8);
	\node[black, anchor=north west] (a) at (0.03,0.15) {$y$};
	\draw[black] (0.177*0.8,0.177*0.8+0.15) -- (-0.177*0.8,-0.177*0.8+0.15);
	\draw[black] (0.177*0.8,-0.177*0.8+0.15) -- (-0.177*0.8,0.177*0.8+0.15);
	\node[black, anchor=south] (a) at (-1.5,3.3) {$a$};
	\node[black, anchor=south] (a) at (-1.5+0.6,3.3) {$c$};
	\node[black, anchor=south] (a) at (-1.5+0.6*2,3.3) {$d$};
	\node[black, anchor=south] (a) at (-1.5+0.6*3,3.3) {$e$};
	\node[black, anchor=south] (a) at (-1.5+0.6*4,3.3) {$b$};
	\node[black, anchor=north] (a) at (-1.5+0.3-0.075,3.25-0.5196+0.04) {$f_1'$};
	\node[black, anchor=north] (a) at (-1.5+0.3*2-0.075,3.25-0.5196*2+0.04) {$f_2'$};
	\node[black, anchor=north] (a) at (-1.5+0.3*3-0.075,3.25-0.5196*3+0.04) {$f_3'$};
	\node[black, anchor=north] (a) at (-1.5+0.3*4-0.1*0.5-0.075,3.25-0.5196*4+0.1732*0.5+0.045) {$f_4$};
	\node[black] (a) at (0.9*0.8,0.7*0.8+0.15+0.075) {$f_5$};
	\draw[black] (0,0.75) -- (-1.5,0.75+2.598) node[sloped,pos=0.90,allow upside down]{\arrowIn}; ; 
	\draw[black] (-1.5*0.69,0.75+2.598*0.69) -- (-1.5*0.7,0.75+2.598*0.7) node[sloped,pos=1,allow upside down]{\arrowIn}; ; 
	\draw[black] (-1.5*0.49,0.75+2.598*0.49) -- (-1.5*0.5,0.75+2.598*0.5) node[sloped,pos=1,allow upside down]{\arrowIn}; ; 
	\draw[black] (-1.5*0.29,0.75+2.598*0.29) -- (-1.5*0.3,0.75+2.598*0.3) node[sloped,pos=1,allow upside down]{\arrowIn}; ; 
	\draw[black] (-1.5*0.09,0.75+2.598*0.09) -- (-1.5*0.1,0.75+2.598*0.1) node[sloped,pos=1,allow upside down]{\arrowIn}; ; 
	\draw[black] (-1.2,0.75+0.5196*4) -- (-1.5+0.6,0.75+2.598) node[sloped,pos=0.50,allow upside down]{\arrowIn}; ; 
	\draw[black] (-1.2+0.3,0.75+0.5196*3) -- (-1.5+0.6*2,0.75+2.598) node[sloped,pos=0.50,allow upside down]{\arrowIn}; ; 
	\draw[black] (-1.2+0.3*2,0.75+0.5196*2) -- (-1.5+0.6*3,0.75+2.598) node[sloped,pos=0.50,allow upside down]{\arrowIn}; ; 
	\draw[black] (-1.2+0.3*3,0.75+0.5196*1) -- (-1.5+0.6*4,0.75+2.598) node[sloped,pos=0.50,allow upside down]{\arrowIn}; ; 
	\draw[black, line width=0.05pt] (0.707*0.8+0.707*0.8*0.001,0.707*0.8*0.75-0.707*0.8*0.75*0.001+0.15) -- (0.707*0.8,0.707*0.8*0.75+0.15) node[sloped,pos=0.50,allow upside down]{\arrowIn}; ; 
	\end{scope}
\end{tikzpicture}
\caption{Example showing how to express the braided fusion diagram in Fig.~\ref{eq:rule1}$b$ in terms of diagrams in canonical form.}
\label{eq:translation_bulk_example}
\end{figure*}
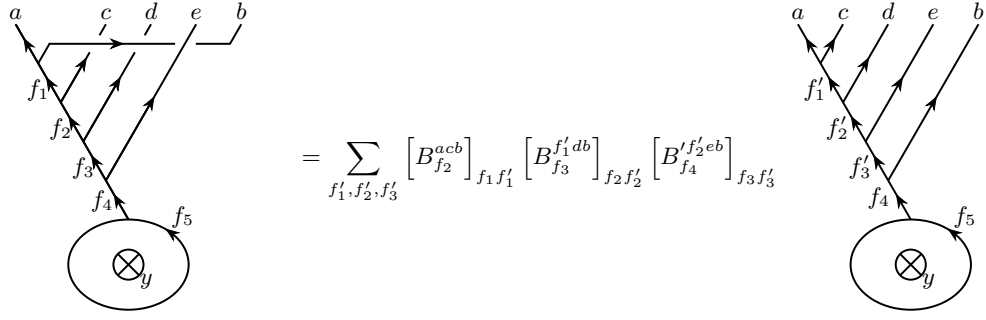

In this section, we illustrate how braided fusion diagrams can be expressed as superposition of diagrams in canonical form~(\ref{eq:general_state}) using some examples. We use this opportunity to introduce pseudocodes that implement functions treating the four different translations discussed in Sec.~\ref{sec:algo}; these functions were refered to in Alg.~\ref{algo:Hamiltonian}. Let us start by defining a second braid operator, $B'$, which is very similar to the braid operator $B$ defined in Eqs.~(\ref{eq:defB}) and (\ref{eq:defB_formula}):
\begin{align}
	\begin{split}
	\begin{tikzpicture}[line width=0.75pt, scale=0.75]
	\begin{scope}[shift={(0,-3.75)}]
	\draw [black,domain=0:3, samples=10] plot ({\x*cos(120)}, {\x*sin(120)+0.75});
	\draw [black,domain=0:3-0.866/sin(60), samples=10] plot ({\x*cos(60)-0.5}, {\x*sin(60)+0.75+0.866});
	\draw [black,domain=0:(3-0.866/sin(60)*2)/2, samples=10] plot ({\x*cos(60)-0.5*2}, {\x*sin(60)+0.75+0.866*2});
	\draw [black,domain=(3-0.866/sin(60)*2)/2:3-0.866/sin(60)*2, samples=10] plot ({\x*cos(60)+1}, {\x*sin(60)+0.75+0.866*2});
	\draw [black,domain=0-0.01:1-0.138, samples=10] plot ({\x-0.75}, {2.915});
	\draw [black,domain=1+0.138:2+0.01, samples=10] plot ({\x-0.75}, {2.915});
	\node[black, anchor=south] (a) at (-1.5,3.3) {$a$};
	\node[black, anchor=south] (a) at (-1.5+3,3.3) {$c$};
	\node[black, anchor=south] (a) at (-1.5+2,3.3) {$b$};
	\node[black, anchor=north east] (a) at (-0.25+0.1,0.75+0.433+0.1) {$d$};
	\node[black, anchor=north east] (a) at (-0.75+0.1,0.75+1.299+0.1) {$e$};
	\draw[black] (-1.25+0.005,0.75+2.165-0.00866) -- (-1.25,0.75+2.165) node[sloped,pos=1,allow upside down]{\arrowIn}; ; 
	\draw[black] (-0.75+0.005,0.75+1.299-0.00866) -- (-0.75,0.75+1.299) node[sloped,pos=1,allow upside down]{\arrowIn}; ; 
	\draw[black] (-0.25+0.005,0.75+0.433-0.00866) -- (-0.25,0.75+0.433) node[sloped,pos=1,allow upside down]{\arrowIn}; ; 
	\draw[black] (0-0.005,0.75+1.732-0.00866) -- (0,0.75+1.732) node[sloped,pos=1,allow upside down]{\arrowIn}; ; 
	\draw[black] (0.025-0.005,2.915) -- (0.025,2.915) node[sloped,pos=1,allow upside down]{\arrowIn}; ; 
	\end{scope}
	\begin{scope}[shift={(7.0,-3.75)}]
	\node [black, anchor=east] (a) at (-1.65,0.75+1.299) {
            $\begin{aligned}
                =\sum_{f}\left[B'^{abc}_{d}\right]_{ef}
            \end{aligned}$
    };
	\draw [black,domain=0:3, samples=10] plot ({\x*cos(120)}, {\x*sin(120)+0.75});
	\draw [black,domain=0:3-0.866/sin(60), samples=10] plot ({\x*cos(60)-0.5}, {\x*sin(60)+0.75+0.866});
	\draw [black,domain=0:3-0.866/sin(60)*2, samples=10] plot ({\x*cos(60)-0.5*2}, {\x*sin(60)+0.75+0.866*2});
	\node[black, anchor=south] (a) at (-1.5,3.3) {$a$};
	\node[black, anchor=south] (a) at (-1.5+1,3.3) {$b$};
	\node[black, anchor=south] (a) at (-1.5+2,3.3) {$c$};
	\node[black, anchor=north east] (a) at (-0.25+0.1,0.75+0.433+0.1) {$d$};
	\node[black, anchor=north east] (a) at (-0.75+0.1,0.75+1.299+0.1) {$f$};
	\draw[black] (-1.25+0.005,0.75+2.165-0.00866) -- (-1.25,0.75+2.165) node[sloped,pos=1,allow upside down]{\arrowIn}; ; 
	\draw[black] (-0.75+0.005,0.75+1.299-0.00866) -- (-0.75,0.75+1.299) node[sloped,pos=1,allow upside down]{\arrowIn}; ; 
	\draw[black] (-0.25+0.005,0.75+0.433-0.00866) -- (-0.25,0.75+0.433) node[sloped,pos=1,allow upside down]{\arrowIn}; ; 
	\draw[black] (0-0.005,0.75+1.732-0.00866) -- (0,0.75+1.732) node[sloped,pos=1,allow upside down]{\arrowIn}; ; 
	\draw[black] (-0.75-0.005,0.75+2.165-0.00866) -- (-0.75,0.75+2.165) node[sloped,pos=1,allow upside down]{\arrowIn}; ; 
	\node[black, anchor=west] (a) at (0.65,0.75+1.299) {,};
	\end{scope}
\end{tikzpicture}
\end{split}
\label{eq:defBprime}
\end{align}
where 
\begin{align}
	\left[ B'^{abc}_d \right]_{ef} = \sum_g \left[ F^{acb}_d \right]_{eg} \left(R^{cb}_g\right)^{-1}  \left[ F^{abc}_d \right]^{-1}_{gf}.
\end{align}
Using the unitarity of the $F$- and $R$-moves, it can be seen that $B$ and $B'$ are related by (note the reversed order of charges $b$ and $c$)
\begin{align}
	\left[ B^{abc}_d \right]^{-1}_{ef}=\left[ B^{abc}_d \right]^{\dagger}_{ef}=\left[ B'^{acb}_d \right]_{ef}.
\end{align}
This means that one of those operators would be sufficient in order to resolve braids. Nevertheless, both operators will be used as doing so provides more clarity. With this, we discuss in the following how to resolve the braids for the three examples given in Sec.~\ref{sec:algo} and introduce the corresponding pseudocodes; Sec.~\ref{app_example_computations_sheet_hopping} further includes the generalization of the operation depicted in Fig.~\ref{eq:drag_x_2p}, which corresponds to the translation of an anyon around the torus.

\subsection{Translations in the Bulk}
\label{app_example_computations_bulk}

\begin{algorithm}[t]
\caption{\justifyingg{Computation of the final state after translating the $k$-th anyon by one lattice spacing in positive $y$-direction in the bulk. The initial state $|(\alpha_k,\vec{r}_k)_{k=1}^N, \mathbf{f}\rangle$ is represented by $\vec{e}_i$.}}
\begin{algorithmic}[1]
\Function{TranslationBulkY}{$\vec{e}_i,k$}

\State $\vec{v}\gets$ vector representing $|(\alpha_{k'},\vec{r}_{k'}+\delta_{kk'}\vec{e}_y)_{k'=1}^N, \mathbf{f}\rangle$

\State \Return $\vec{v}$
\EndFunction
\end{algorithmic}
\label{algo:Hamiltonian_y_bulk}
\end{algorithm}

Lets us start by considering translations in the bulk in $y$-direction. Since this case does not involve braiding among the anyons, the corresponding pseudocode, which is shown in Alg.~\ref{algo:Hamiltonian_y_bulk}, merely needs to change the position vector of the anyon to be translated.

For translations of anyons in $x$-direction, the rules summarized in Fig.~\ref{eq:rule1} are to be applied to the fusion diagrams. Braided diagrams that are obtained by these rules, like the one in Fig.~\ref{eq:rule1}$b$, can be expressed as superposition of diagrams in canonical form using the braid operators defined in Eqs.~(\ref{eq:defB}) and (\ref{eq:defBprime}). Concretely, for the example depicted in Fig.~\ref{eq:rule1}$b$, we obtain the expression in Fig.~\ref{eq:translation_bulk_example}. Using this result, the Hamiltonian's matrix element $\mathcal{H}_{s's}$ in Eq.~(\ref{eq:H_projection}) can be computed by comparing the superposition of fusion diagrams in canonical form to the fusion diagram of the final state $|s'\rangle$. It is obtained by multiplying the contribution in the superposition that corresponds to the same fusion diagram as $|s'\rangle$ by $-t$. For the case depicted in Fig.~\ref{eq:translation_bulk_example}, the matrix element is
\begin{align}
	\mathcal{H}_{s's}=-t\left[B^{acb}_{f_2}\right]_{f_1f_1'}\left[B^{f_1'db}_{f_3}\right]_{f_2f_2'}\left[B'^{f_2'eb}_{f_4}\right]_{f_3f_3'}\delta_{f_4f_4'}\delta_{f_5f_5'},
\end{align}
where we assumed that the tuple associated with the fusion diagram of the final state $|s'\rangle$ is $\mathbf{f'}=(f_1',f_2',f_3',f_4',f_5')$. The operator string can be generalized by applying a $B$($B'$)-operator each time the translated anyon moves in front of (behind) another anyon. Each application of these operators changes an intermediate fusion product on which the subsequent operator depends. Summing over all these intermediate fusion products yields the desired superposition of fusion diagrams in canonical form. From the discussion in Sec.~\ref{sec:algo_bulk}, it follows that $B$-operators are always applied before $B'$-operators. This generalization is summarized as pseudocode in Alg.~\ref{algo:Hamiltonian_x_bulk}.

\begin{algorithm}[t]
\caption{\justifyingg{Computation of the final state after translating the $k$-th anyon by one lattice spacing in positive $x$-direction in the bulk. The initial state $|(\alpha_k,\vec{r}_k)_{k=1}^N, \mathbf{f}\rangle$ is represented by $\vec{e}_i$. The matrices corresponding to braid operators only act on the subspace defined by the fusion degrees of freedom, i.e., on the second part in the tensor product $|(\alpha_k,\vec{r}_k)_{k=1}^N\rangle \otimes |\mathbf{f}\rangle$.}}
\begin{algorithmic}[1]
\Function{TranslationBulkX}{$\vec{e}_i,k,\lbrace \ldots \rbrace$}

\State $n_{ccw}, n_{cw} \gets$ number of anyons with which $\alpha_k$ is braided counter-clockwise and clockwise, respectively
\State $\vec{v}\gets$ vector representing $|(\alpha_{k'},\vec{r}_{k'}+\delta_{kk'}\vec{e}_x)_{k'=1}^N, \mathbf{f}\rangle$

\For{$i\gets 1$ to $n_{ccw}$}
\State $B\gets$ matrix corresponding to $B^{ f_{k-3+i} \alpha_{k+i} \alpha_k }_{f_{k-1+i}}$, where $\lbrace f_k \rbrace$ are the fusion products associated with $\vec{v}$; $f_{-1}\equiv 1,f_0\equiv \alpha_1$
\State $\vec{v}\gets B\cdot \vec{v}$
\EndFor

\For{$i\gets n_{ccw}+1$ to $n_{ccw}+n_{cw}$}
\State $B'\gets$ matrix corresponding to $B'^{ f_{k-3+i} \alpha_{k+i} \alpha_k }_{f_{k-1+i}}$
\State $\vec{v}\gets B'\cdot \vec{v}$
\EndFor

\State \Return $\vec{v}$
\EndFunction
\end{algorithmic}
\label{algo:Hamiltonian_x_bulk}
\end{algorithm}

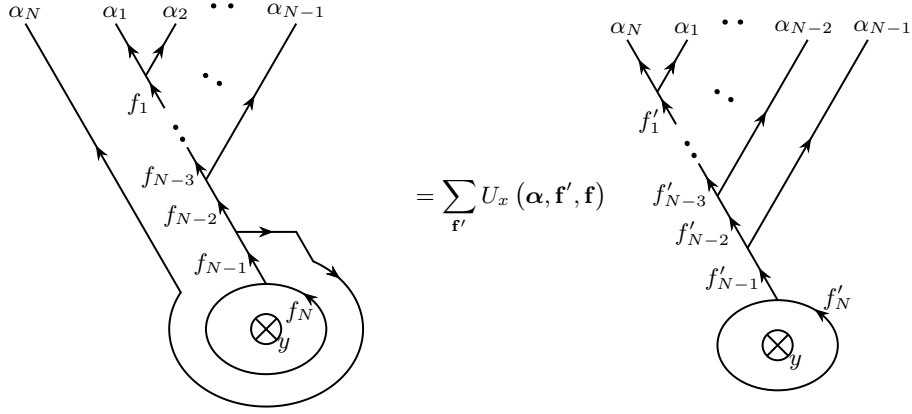
\begin{figure*}[t]
	\centering
\begin{tikzpicture}[line width=0.75pt]
	\begin{scope}[shift={(0+0.4,4.01/2-3.58/2)}]
	\draw (0,0.15) ellipse (1*0.8 and 0.75*0.8);
	\draw (0,0.15) ellipse (0.25*0.8 and 0.25*0.8);
	\node[black, anchor=north west] (a) at (0.03,0.15) {$y$};
	\draw[black] (0.177*0.8,0.177*0.8+0.15) -- (-0.177*0.8,-0.177*0.8+0.15);
	\draw[black] (0.177*0.8,-0.177*0.8+0.15) -- (-0.177*0.8,0.177*0.8+0.15);
	\draw [black,domain=0:2.1, samples=10] plot ({\x*cos(120)}, {\x*sin(120)+0.75});
	\draw [black,domain=2.7:4, samples=10] plot ({\x*cos(120)}, {\x*sin(120)+0.75});
	\draw [black,domain=0:4-0.6928/sin(60)*2, samples=10] plot ({\x*cos(60)-0.4*2}, {\x*sin(60)+0.75+0.6928*2});
	\draw [black,domain=0:4-0.6928/sin(60)*4, samples=10] plot ({\x*cos(60)-0.4*4}, {\x*sin(60)+0.75+0.6928*4});
	\node[black, anchor=south] (a) at (-2,4.15) {$\alpha_1$};
	\node[black, anchor=south] (a) at (-2+0.8,4.15) {$\alpha_2$};
	\node[black, anchor=south] (a) at (-2+0.8*3,4.15) {$\alpha_{N-1}$};
	\node[black, anchor=south] (a) at (-2-1.2,4.15) {$\alpha_N$};
	\node[black, anchor=north] (a) at (-2+0.3,4-0.6928+0.1) {$f_1$};
	\node[black, anchor=north] (a) at (-2+0.4*2-0.1,4-0.6928*2-0.175) {$f_{N-3}$};
	\node[black, anchor=north] (a) at (-2+0.4*3-0.2,4-0.6928*3) {$f_{N-2}$};
	\node[black, anchor=north] (a) at (-2+0.4*4-0.2,4-0.6928*4+0.05) {$f_{N-1}$};
	\node[black, anchor=west] (a) at (0.17-0.04,0.42-0.04) {$f_N$};
	\draw[black] (-0.5*4*9/10+0.005,0.75+0.866*4*9/10-0.00866) -- (-0.5*4*9/10,0.75+0.866*4*9/10) node[sloped,pos=1,allow upside down]{\arrowIn}; ; 
	\draw[black] (-0.5*4*7.5/10+0.005,0.75+0.866*4*7.5/10-0.00866) -- (-0.5*4*7.5/10,0.75+0.866*4*7.5/10) node[sloped,pos=1,allow upside down]{\arrowIn}; ; 
	\draw[black] (-0.5*4*4.5/10+0.005,0.75+0.866*4*4.5/10-0.00866) -- (-0.5*4*4.5/10,0.75+0.866*4*4.5/10) node[sloped,pos=1,allow upside down]{\arrowIn}; ; 
	\draw[black] (-0.5*4*3/10+0.005,0.75+0.866*4*3/10-0.00866) -- (-0.5*4*3/10,0.75+0.866*4*3/10) node[sloped,pos=1,allow upside down]{\arrowIn}; ; 
	\draw[black] (-0.5*4*1/10+0.005,0.75+0.866*4*1/10-0.00866) -- (-0.5*4*1/10,0.75+0.866*4*1/10) node[sloped,pos=1,allow upside down]{\arrowIn}; ; 
	\draw[black] (-0.4/2-0.005,6.35/2-0.00866) -- (-0.4/2,6.35/2) node[sloped,pos=1,allow upside down]{\arrowIn}; ; 
	\draw[black] (-2.8/2-0.005,7.735/2-0.00866) -- (-2.8/2,7.735/2) node[sloped,pos=1,allow upside down]{\arrowIn}; ; 
	\draw[black, line width=0.05pt] (0.707*0.8+0.707*0.001*0.8,0.707*0.75*0.8-0.707*0.75*0.001*0.8+0.15) -- (0.707*0.8,0.707*0.75*0.8+0.15) node[sloped,pos=0.50,allow upside down]{\arrowIn}; ; 
	\draw [line width=2pt, line cap=round, new dash=on 0pt off 5pt] (-1.05*4/5-1.35/5,2.57*4/5+3.09/5) -- (-1.05/5-1.35*4/5,2.57/5+3.09*4/5);
	\draw [line width=2pt, line cap=round, new dash=on 0pt off 5pt] (-2+0.8+0.6-0.1,4.15+0.3) -- (-2+0.8*3-0.55-0.1,4.15+0.3);
	\draw [line width=2pt, line cap=round, new dash=on 0pt off 5pt] (-0.2*0.65-1.4*0.35,6.35/2*0.65+7.735/2*0.35) -- (-0.2*0.35-1.4*0.65,6.35/2*0.35+7.735/2*0.65);
	\draw[black, line width=0.05pt] (0.707*1.5*1.074*0.8,0.707*1.2*1.074*0.8+0.15) -- (0.707*1.5*1.074*0.8+0.707*1.5*1.074*0.001,0.707*1.2*1.074*0.8-0.707*1.2*1.074*0.001+0.15) node[sloped,pos=0.50,allow upside down]{\arrowIn}; ; 
	\draw [black,domain=-208.5:61, samples=100] plot ({1.5*cos(\x)*1.074*0.8}, {1.2*sin(\x)*1.074*0.8+0.15});
	\draw [black,domain=0.07-0.21:4, samples=10] plot ({\x*cos(120)-1.2}, {\x*sin(120)+0.75});
	\draw [black,domain=0.53-0.19:0.8, samples=10] plot ({\x*cos(120)+0.8}, {\x*sin(120)+0.75});
	\draw [black,domain=0:0.81, samples=10] plot ({\x-0.4}, {0.75+0.6928});
	\draw[black] (0.005-0.005,0.75+0.6928) -- (0.005,0.75+0.6928) node[sloped,pos=1,allow upside down]{\arrowIn}; ; 
	\draw[black] (-2.2175+0.005,2.51-0.00866) -- (-2.2175,2.51) node[sloped,pos=1,allow upside down]{\arrowIn}; ; 
	\end{scope}
	\begin{scope}[shift={(7.2,0)}]
	\node [black, anchor=east] (a) at (-2.15,4.01/2) {
            $\begin{aligned}
                =\sum_{\mathbf{f'}}U_x\left( \text{\boldmath$\alpha$},\mathbf{f'}, \mathbf{f} \right)
            \end{aligned}$
    };
	\draw (0,0.15) ellipse (1*0.8 and 0.75*0.8);
	\draw (0,0.15) ellipse (0.25*0.8 and 0.25*0.8);
	\node[black, anchor=north west] (a) at (0.03,0.15) {$y$};
	\draw[black] (0.177*0.8,0.177*0.8+0.15) -- (-0.177*0.8,-0.177*0.8+0.15);
	\draw[black] (0.177*0.8,-0.177*0.8+0.15) -- (-0.177*0.8,0.177*0.8+0.15);
	\draw [black,domain=0:2.1, samples=10] plot ({\x*cos(120)}, {\x*sin(120)+0.75});
	\draw [black,domain=2.7:4, samples=10] plot ({\x*cos(120)}, {\x*sin(120)+0.75});
	\draw [black,domain=0:4-0.6928/sin(60), samples=10] plot ({\x*cos(60)-0.4}, {\x*sin(60)+0.75+0.6928});
	\draw [black,domain=0:4-0.6928/sin(60)*2, samples=10] plot ({\x*cos(60)-0.4*2}, {\x*sin(60)+0.75+0.6928*2});
	\draw [black,domain=0:4-0.6928/sin(60)*4, samples=10] plot ({\x*cos(60)-0.4*4}, {\x*sin(60)+0.75+0.6928*4});
	\node[black, anchor=south] (a) at (-2,4.15) {$\alpha_N$};
	\node[black, anchor=south] (a) at (-2+0.8,4.15) {$\alpha_1$};
	\node[black, anchor=south] (a) at (-2+0.8*3-0.05,4.15) {$\alpha_{N-2}$};
	\node[black, anchor=south] (a) at (-2+0.8*4+0.2,4.15) {$\alpha_{N-1}$};
	\node[black, anchor=north] (a) at (-2+0.3,4-0.6928+0.1) {$f'_1$};
	\node[black, anchor=north] (a) at (-2+0.4*2-0.1,4-0.6928*2-0.175) {$f'_{N-3}$};
	\node[black, anchor=north] (a) at (-2+0.4*3-0.2,4-0.6928*3) {$f'_{N-2}$};
	\node[black, anchor=north] (a) at (-2+0.4*4-0.2,4-0.6928*4+0.1) {$f'_{N-1}$};
	\node[black] (a) at (0.9*0.8+0.08,0.7*0.8+0.15+0.075+0.0) {$f_N'$};
	\draw[black] (-0.5*4*9/10+0.005,0.75+0.866*4*9/10-0.00866) -- (-0.5*4*9/10,0.75+0.866*4*9/10) node[sloped,pos=1,allow upside down]{\arrowIn}; ; 
	\draw[black] (-0.5*4*7.5/10+0.005,0.75+0.866*4*7.5/10-0.00866) -- (-0.5*4*7.5/10,0.75+0.866*4*7.5/10) node[sloped,pos=1,allow upside down]{\arrowIn}; ; 
	\draw[black] (-0.5*4*4.5/10+0.005,0.75+0.866*4*4.5/10-0.00866) -- (-0.5*4*4.5/10,0.75+0.866*4*4.5/10) node[sloped,pos=1,allow upside down]{\arrowIn}; ; 
	\draw[black] (-0.5*4*3/10+0.005,0.75+0.866*4*3/10-0.00866) -- (-0.5*4*3/10,0.75+0.866*4*3/10) node[sloped,pos=1,allow upside down]{\arrowIn}; ; 
	\draw[black] (-0.5*4*1/10+0.005,0.75+0.866*4*1/10-0.00866) -- (-0.5*4*1/10,0.75+0.866*4*1/10) node[sloped,pos=1,allow upside down]{\arrowIn}; ; 
	\draw[black] (0.8/2-0.005,5.657/2-0.00866) -- (0.8/2,5.657/2) node[sloped,pos=1,allow upside down]{\arrowIn}; ; 
	\draw[black] (-0.4/2-0.005,6.35/2-0.00866) -- (-0.4/2,6.35/2) node[sloped,pos=1,allow upside down]{\arrowIn}; ; 
	\draw[black] (-2.8/2-0.005,7.735/2-0.00866) -- (-2.8/2,7.735/2) node[sloped,pos=1,allow upside down]{\arrowIn}; ; 
	\draw[black, line width=0.05pt] (0.707*0.8+0.707*0.001*0.8,0.707*0.75*0.8-0.707*0.75*0.001*0.8+0.15) -- (0.707*0.8,0.707*0.75*0.8+0.15) node[sloped,pos=0.50,allow upside down]{\arrowIn}; ; 
	\draw [line width=2pt, line cap=round, new dash=on 0pt off 5pt] (-1.05*4/5-1.35/5,2.57*4/5+3.09/5) -- (-1.05/5-1.35*4/5,2.57/5+3.09*4/5);
	\draw [line width=2pt, line cap=round, new dash=on 0pt off 5pt] (-2+0.8+0.6-0.1,4.15+0.3) -- (-2+0.8*3-0.6-0.1,4.15+0.3);
	\draw [line width=2pt, line cap=round, new dash=on 0pt off 5pt] (-0.2*0.65-1.4*0.35,6.35/2*0.65+7.735/2*0.35) -- (-0.2*0.35-1.4*0.65,6.35/2*0.35+7.735/2*0.65);
	\end{scope}	
\end{tikzpicture}
\caption{A fusion diagram whose final anyon moves along a non-contractible loop can be expressed in terms of fusion diagrams in canonical form using $U_x\left( \text{\boldmath$\alpha$},\mathbf{f'}, \mathbf{f} \right)$, which is given by Eq.~(\ref{eq:sheet_hopping_def}), where $\mathbf{f}$ and $\mathbf{f'}$ are the tuples associated with the initial and final diagrams and $\text{\boldmath$\alpha$}=(\alpha_1,\alpha_2,\ldots,\alpha_N)$ contains the anyonic charges of the unbraided initial fusion diagram.}
\label{eq:diagram_sheethopping}
\end{figure*}

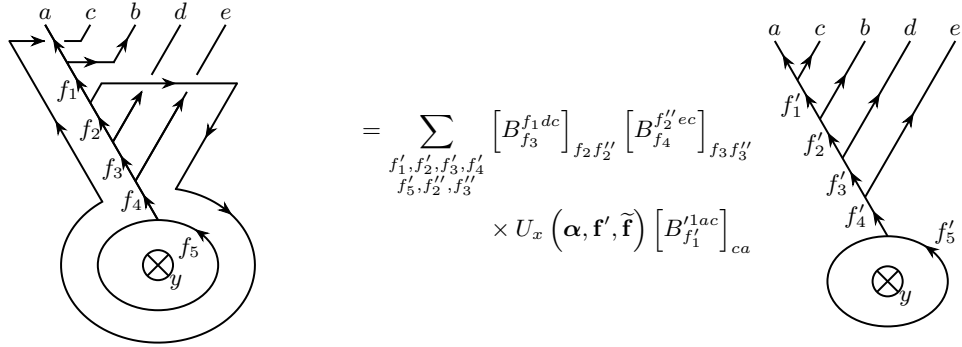
\begin{figure*}[t]
	\centering
\begin{tikzpicture}[line width=0.75pt]
\begin{scope}[shift={(-9.7,3.16/2-2.729/2)}]
	\draw (0,0.15) ellipse (1*0.8 and 0.75*0.8);
	\draw (0,0.15) ellipse (0.25*0.8 and 0.25*0.8);
	\node[black, anchor=north west] (a) at (0.03,0.15) {$y$};
	\draw[black] (0.177*0.8,0.177*0.8+0.15) -- (-0.177*0.8,-0.177*0.8+0.15);
	\draw[black] (0.177*0.8,-0.177*0.8+0.15) -- (-0.177*0.8,0.177*0.8+0.15);
	\draw [black,domain=0:3, samples=10] plot ({\x*cos(120)}, {\x*sin(120)+0.75});
	\draw [black,domain=0:2.1-0.13-0.5196/sin(60), samples=10] plot ({\x*cos(60)-0.3}, {\x*sin(60)+0.75+0.5196});
	\draw [black,domain=2.1+0.13-0.5196/sin(60):3-0.5196/sin(60), samples=10] plot ({\x*cos(60)-0.3}, {\x*sin(60)+0.75+0.5196});
	\draw [black,domain=0:2.1-0.13-0.5196/sin(60)*2, samples=10] plot ({\x*cos(60)-0.3*2}, {\x*sin(60)+0.75+0.5196*2});
	\draw [black,domain=2.1+0.13-0.5196/sin(60)*2:3-0.5196/sin(60)*2, samples=10] plot ({\x*cos(60)-0.3*2}, {\x*sin(60)+0.75+0.5196*2});
	\draw [black,domain=0:3-0.5196/sin(60)*3-0.9, samples=10] plot ({\x*cos(60)-0.3*3}, {\x*sin(60)+0.75+0.5196*3});
	\draw [black,domain=0:2.425-0.5196/sin(60)*4+0.01, samples=10] plot ({\x*cos(60)-0.3*4}, {\x*sin(60)+0.75+0.5196*4});
	\draw [black,domain=2.425-0.5196/sin(60)*3-0.01:3-0.5196/sin(60)*3, samples=10] plot ({\x*cos(60)-0.3*3}, {\x*sin(60)+0.75+0.5196*3});
	\draw [black,domain=-1.1875-0.02:-0.5875+0.01, samples=10] plot ({\x}, {2.85});
	\draw[black] (-0.8876,2.85) -- (-0.8875,2.85) node[sloped,pos=1,allow upside down]{\arrowIn}; ; 
	\draw[black] (-0.444-0.005,3.099-0.00866) -- (-0.444,3.099) node[sloped,pos=1,allow upside down]{\arrowIn}; ; 
	\draw [black,domain=-0.45-0.01-0.3:1.35+0.01-0.3, samples=10] plot ({\x}, {3.088-0.5196});
	\draw [black,domain=0-0.01-0.125:3-0.5196/sin(60)-0.895, samples=10] plot ({\x*cos(60)+0.3}, {\x*sin(60)+0.75+0.5196});
	\draw [black,domain=0.419-0.14:2.75, samples=10] plot ({\x*cos(120)-0.6}, {\x*sin(120)+0.75});
	\draw [black,domain=2.75-0.5196/sin(60)*4:3-0.5196/sin(60)*4, samples=10] plot ({\x*cos(60)-0.3*4}, {\x*sin(60)+0.75+0.5196*4});
	\draw [black,domain=-1.975-0.01:-1.375-0.138+0.01, samples=10] plot ({\x}, {3.1316});
	\draw [black,domain=-1.375+0.138-0.01:-1.025+0.01, samples=10] plot ({\x}, {3.1316});
	\node[black, anchor=south] (a) at (-1.5,3.3) {$a$};
	\node[black, anchor=south] (a) at (-1.5+0.6,3.3) {$c$};
	\node[black, anchor=south] (a) at (-1.5+0.6*2,3.3) {$b$};
	\node[black, anchor=south] (a) at (-1.5+0.6*3,3.3) {$d$};
	\node[black, anchor=south] (a) at (-1.5+0.6*4,3.3) {$e$};
	\node[black, anchor=north] (a) at (-1.5+0.3,3.25-0.5196) {$f_1$};
	\node[black, anchor=north] (a) at (-1.5+0.3*2,3.25-0.5196*2) {$f_2$};
	\node[black, anchor=north] (a) at (-1.5+0.3*3,3.25-0.5196*3) {$f_3$};
	\node[black, anchor=north] (a) at (-1.5+0.3*4-0.1*0.5,3.25-0.5196*4+0.1732*0.5) {$f_4$};
	\node[black, anchor=north west] (a) at (0.17-0.03,0.67-0.03) {$f_5$};
	\draw[black] (0,0.75) -- (-1.5,0.75+2.598) node[sloped,pos=0.90,allow upside down]{\arrowIn}; ; 
	\draw[black] (-1.5*0.69,0.75+2.598*0.69) -- (-1.5*0.7,0.75+2.598*0.7) node[sloped,pos=1,allow upside down]{\arrowIn}; ; 
	\draw[black] (-1.5*0.49,0.75+2.598*0.49) -- (-1.5*0.5,0.75+2.598*0.5) node[sloped,pos=1,allow upside down]{\arrowIn}; ; 
	\draw[black] (-1.5*0.29,0.75+2.598*0.29) -- (-1.5*0.3,0.75+2.598*0.3) node[sloped,pos=1,allow upside down]{\arrowIn}; ; 
	\draw[black] (-1.5*0.09,0.75+2.598*0.09) -- (-1.5*0.1,0.75+2.598*0.1) node[sloped,pos=1,allow upside down]{\arrowIn}; ; 
	\draw[black] (-1.392+0.5*0.15+0.005,2.122-0.866*0.15-0.00866) -- (-1.392+0.5*0.15,2.122-0.866*0.15) node[sloped,pos=1,allow upside down]{\arrowIn}; ; 
	\draw[black] (-1.975+0.474-0.095,3.1316) -- (-1.975+0.475-0.095,3.1316) node[sloped,pos=1,allow upside down]{\arrowIn}; ; 
	\draw[black] (0.449-0.3,3.088-0.5196) -- (0.45-0.3,3.088-0.5196) node[sloped,pos=1,allow upside down]{\arrowIn}; ; 
	\draw[black] (-0.6,1.789) -- (-0.215*9/10-0.6/10,2.46*9/10+1.789/10) node[sloped,pos=1,allow upside down]{\arrowIn}; ; 
	\draw[black] (-1.2+0.3*3,0.75+0.5196*1) -- (0.3,0.75+2.598/2+0.5196*0.5) node[sloped,pos=1,allow upside down]{\arrowIn}; ; 
	\draw[black] (0.675+0.005,1.919+0.00866) -- (0.675,1.919) node[sloped,pos=1,allow upside down]{\arrowIn}; ; 
	\draw[black, line width=0.05pt] (1.5*0.707*1.074*0.8,1.2*0.707*1.074*0.8+0.15) -- (1.5*0.707*1.074*0.8+1.5*0.707*0.001,1.2*0.707*1.074*0.8-1.2*0.707*0.001+0.15) node[sloped,pos=0.50,allow upside down]{\arrowIn}; ; 
	\draw[black, line width=0.05pt] (0.707*0.8+0.707*0.8*0.001,0.707*0.8*0.75-0.707*0.8*0.75*0.001+0.15) -- (0.707*0.8,0.707*0.8*0.75+0.15) node[sloped,pos=0.50,allow upside down]{\arrowIn}; ; 	
	\draw [black,domain=-240+4.5:79.5+0.3, samples=100] plot ({1.5*cos(\x)*1.074*0.8}, {1.2*sin(\x)*1.074*0.8+0.15});
	\end{scope}
	\node [black, anchor=east] (a) at (-1.65,3.16/2-0.1) {
            $\begin{aligned}
                =\sum_{\substack{f_1',f_2',f_3',f_4'\\f_5',f_2'',f_3''}}&\left[B^{f_1dc}_{f_3} \right]_{f_2f_2''}\left[B^{f_2''ec}_{f_4}  \right]_{f_3f_3''}\\ &\times U_x\left( \text{\boldmath $\alpha $},\mathbf{f'}, \widetilde{\mathbf{f}} \right) \left[ B'^{1ac}_{f_1'} \right]_{ca}
            \end{aligned}$
    };
	\draw (0,0.15) ellipse (1*0.8 and 0.75*0.8);
	\draw (0,0.15) ellipse (0.25*0.8 and 0.25*0.8);
	\node[black, anchor=north west] (a) at (0.03,0.15) {$y$};
	\draw[black] (0.177*0.8,0.177*0.8+0.15) -- (-0.177*0.8,-0.177*0.8+0.15);
	\draw[black] (0.177*0.8,-0.177*0.8+0.15) -- (-0.177*0.8,0.177*0.8+0.15);
	\node[black, anchor=south] (a) at (-1.5,3.3) {$a$};
	\node[black, anchor=south] (a) at (-1.5+0.6,3.3) {$c$};
	\node[black, anchor=south] (a) at (-1.5+0.6*2,3.3) {$b$};
	\node[black, anchor=south] (a) at (-1.5+0.6*3,3.3) {$d$};
	\node[black, anchor=south] (a) at (-1.5+0.6*4,3.3) {$e$};
	\node[black, anchor=north] (a) at (-1.5+0.3-0.075,3.25-0.5196+0.04) {$f_1'$};
	\node[black, anchor=north] (a) at (-1.5+0.3*2-0.075,3.25-0.5196*2+0.04) {$f_2'$};
	\node[black, anchor=north] (a) at (-1.5+0.3*3-0.075,3.25-0.5196*3+0.04) {$f_3'$};
	\node[black, anchor=north] (a) at (-1.5+0.3*4-0.1*0.5-0.075,3.25-0.5196*4+0.1732*0.5+0.04) {$f_4'$};
	\node[black] (a) at (0.9*0.8+0.05,0.7*0.8+0.15+0.1) {$f_5'$};
	\draw[black] (0,0.75) -- (-1.5,0.75+2.598) node[sloped,pos=0.90,allow upside down]{\arrowIn}; ; 
	\draw[black] (-1.5*0.69,0.75+2.598*0.69) -- (-1.5*0.7,0.75+2.598*0.7) node[sloped,pos=1,allow upside down]{\arrowIn}; ; 
	\draw[black] (-1.5*0.49,0.75+2.598*0.49) -- (-1.5*0.5,0.75+2.598*0.5) node[sloped,pos=1,allow upside down]{\arrowIn}; ; 
	\draw[black] (-1.5*0.29,0.75+2.598*0.29) -- (-1.5*0.3,0.75+2.598*0.3) node[sloped,pos=1,allow upside down]{\arrowIn}; ; 
	\draw[black] (-1.5*0.09,0.75+2.598*0.09) -- (-1.5*0.1,0.75+2.598*0.1) node[sloped,pos=1,allow upside down]{\arrowIn}; ; 
	\draw[black] (-1.2,0.75+0.5196*4) -- (-1.5+0.6,0.75+2.598) node[sloped,pos=0.50,allow upside down]{\arrowIn}; ; 
	\draw[black] (-1.2+0.3,0.75+0.5196*3) -- (-1.5+0.6*2,0.75+2.598) node[sloped,pos=0.50,allow upside down]{\arrowIn}; ; 
	\draw[black] (-1.2+0.3*2,0.75+0.5196*2) -- (-1.5+0.6*3,0.75+2.598) node[sloped,pos=0.50,allow upside down]{\arrowIn}; ; 
	\draw[black] (-1.2+0.3*3,0.75+0.5196*1) -- (-1.5+0.6*4,0.75+2.598) node[sloped,pos=0.50,allow upside down]{\arrowIn}; ; 
	\draw[black, line width=0.05pt] (0.707*0.8+0.707*0.8*0.001,0.707*0.8*0.75-0.707*0.8*0.75*0.001+0.15) -- (0.707*0.8,0.707*0.8*0.75+0.15) node[sloped,pos=0.50,allow upside down]{\arrowIn}; ; 
\end{tikzpicture}
\caption{Example showing how to express the braided fusion diagram in Fig.~\ref{eq:translate_B}$b$ in terms of diagrams in canonical form, where $ \text{\boldmath$\alpha$}=(a,b,d,e,c)$, $\widetilde{\mathbf{f}}=(f_1,f_2'',f_3'',f_4,f_5)$ and $\mathbf{f'}=(f_1',f_2',f_3',f_4',f_5')$.}
\label{eq:translation_cut_example}
\end{figure*}

\subsection{Translations over the Cuts}
\label{app_example_computations_sheet_hopping}

According to the rules discussed in Sec.~\ref{sec:algo_cuts}, each anyon crossing either of the two cuts moves around the torus. We sketched how such a process can be expressed in terms of fusion diagrams in canonical form for a system of two anyons in Fig.~\ref{eq:drag_x_2p}. Here, we generalize this to arbitrary fusion diagrams and introduce the new quantity $U_x\left( \text{\boldmath$\alpha$},\mathbf{f'}, \mathbf{f} \right)$ that is defined by the relation in Fig.~\ref{eq:diagram_sheethopping}. Here, $\text{\boldmath$\alpha$}=(\alpha_1,\alpha_2,\ldots,\alpha_N)$ denotes the tuple containing the anyonic charges of the unbraided initial fusion diagram (i.e., before any translations were performed). The tuples $\mathbf{f}$ and $\mathbf{f'}$ are the tuples associated with the initial diagram and the diagrams in the superposition, respectively. The sum runs over all tuples $\mathbf{f'}$ that are consisitent with the fusion rules.


As indicated, $U_x\left( \text{\boldmath$\alpha$},\mathbf{f'}, \mathbf{f} \right)$ depends on the anyons $\lbrace \alpha_k \rbrace$ and their order, the fusion diagram associated with the initial state and the one associated with respective contribution to the final state. It is given by
\begin{widetext}
\begin{align}
\begin{split}
	U_x\left( \text{\boldmath$\alpha$},\mathbf{f'}, \mathbf{f} \right) =  \begin{cases}
	\left[F^{f_{N-2}\alpha_N\overbar{f_N}}_{\overbar{f_N}}\right]_{f_{N-1}\overbar{f_N'}}\left[F^{\alpha_Nf_{N-2}\overbar{f_N'}}_{\overbar{f_N'}}\right]^{-1}_{\overbar{f_N}f'_{N-1}}\left[F^{\alpha_N\alpha_1\alpha_2}_{f'_2}\right]^{-1}_{f_1f'_1}&\\
                \qquad\quad\times\prod_{j=0}^{N-4}\left[F^{\alpha_Nf_{N-3-j}\alpha_{N-1-j}}_{f'_{N-1-j}}\right]^{-1}_{f_{N-2-j}f'_{N-2-j}}, \quad &\text{for $N>2$}\\
    \left[F^{\alpha_1 \alpha_2\overbar{f_2}}_{\overbar{f_2}}\right]_{f_1\overbar{f_2'}}\left[F^{\alpha_2 \alpha_1 \overbar{f_2'}}_{\overbar{f_2'}}\right]_{\overbar{f_2}f_1'}^{-1}, \quad\quad &\text{for $N=2$}
	\end{cases}
\end{split}
\label{eq:sheet_hopping_def}
\end{align}
\end{widetext}
and is trivial for $N=1$. By taking another look at Fig.~\ref{eq:diagram_sheethopping}, it can be seen that moving all $N$ anyons one after another around the torus does not change the fusion diagram since doing so recovers the initial order of anyons without braiding them. Here, the lines associated with the fusion products may also be moved clockwise around the torus to make the equivalence to the fusion diagram of the initial state in Eq.~(\ref{eq:general_state}) more apparent. Equivalently, this observation can be expressed in terms of $U_x\left( \text{\boldmath$\alpha$},\mathbf{f'}, \mathbf{f} \right)$ as
\begin{widetext}
\begin{align}
	\sum_{\mathbf{f}^{(1)},\mathbf{f}^{(2)},\ldots,\mathbf{f}^{(N-1)}}  U_x\left( \sigma^{N-1}(\text{\boldmath$\alpha$}),\mathbf{f}^{(N)}, \mathbf{f}^{(N-1)} \right)   \ldots U_x\left( \sigma(\text{\boldmath$\alpha$}),\mathbf{f}^{(2)}, \mathbf{f}^{(1)} \right)   U_x\left( \text{\boldmath$\alpha$},\mathbf{f}^{(1)}, \mathbf{f}^{(0)} \right) = \delta_{\mathbf{f}^{(0)}\mathbf{f}^{(N)}},
	\label{eq:sheet_hopping_identity_tuples}
\end{align}
\end{widetext}
where $\sigma(1,2,\ldots,N) = (N,1,2,\ldots,N-1)$. Note that the order of the anyonic charges in the arguments of $U_x\left( \text{\boldmath$\alpha$},\mathbf{f'}, \mathbf{f} \right)$ in Eq.~(\ref{eq:sheet_hopping_identity_tuples}) is crucial. We can now express the braided fusion diagram depicted in Fig.~\ref{eq:translate_B}$b$ in terms of canonical diagrams, as done in Fig.~\ref{eq:translation_cut_example}, where $ \text{\boldmath$\alpha$}=(a,b,d,e,c)$, $\widetilde{\mathbf{f}}=(f_1,f_2'',f_3'',f_4,f_5)$ and $\mathbf{f'}=(f_1',f_2',f_3',f_4',f_5')$. The final $B'$-operator can be replaced by the appropriate $R$-move. The corresponding matrix elements $\mathcal{H}_{s's}$ can be computed in the same way as described at the end of the previous section (Sec.~\ref{app_example_computations_bulk}). A pseudocode for computing the final state after such translations is Alg.~\ref{algo:Hamiltonian_x_cutB}, which clearly shows parallels to Alg.~\ref{algo:Hamiltonian_x_bulk} when it comes to braids among the anyons.\\

\begin{algorithm}[t]
\caption{\justifyingg{Computation of the final state after translating the $k$-th anyon by one lattice spacing in positive $x$-direction across cut $B$. The initial state $|(\alpha_k,\vec{r}_k)_{k=1}^N, \mathbf{f}\rangle$ is represented by $\vec{e}_i$. The matrices corresponding to braid operators only act on the subspace defined by the fusion degrees of freedom, i.e., on the second part in the tensor product $|(\alpha_k,\vec{r}_k)_{k=1}^N\rangle \otimes |\mathbf{f}\rangle$.}}
\begin{algorithmic}[1]
\Function{TranslationCutB}{$\vec{e}_i,k,\lbrace \ldots \rbrace$}

\State $n_{ccw}, n_{cw} \gets$ number of anyons with which $\alpha_k$ is braided counter-clockwise and clockwise, respectively

\State $\vec{v}\gets$ vector representing $|(\alpha_{k'},\vec{r}_{k'}+\delta_{kk'}\vec{e}_x)_{k'=1}^N, \mathbf{f}\rangle$

\For{$i\gets 1$ to $n_{ccw}$}
\State $B\gets$ matrix corresponding to $B^{ f_{k-3+i} \alpha_{k+i} \alpha_k }_{f_{k-1+i}}$, where $\lbrace f_k \rbrace$ are the fusion products associated with $\vec{v}$; $f_{-1}\equiv 1,f_0\equiv \alpha_1$
\State $\vec{v}\gets B\cdot \vec{v}$
\EndFor

\State $U_x \gets$ matrix with $(U_x)_{ij}=U_x\left( \text{\boldmath$\alpha$},\mathbf{f}_i,\mathbf{f}_j\right)$, $\text{\boldmath$\alpha$}=(\alpha_1,\ldots,\alpha_{k-1},\alpha_{k+1},\ldots,\alpha_N,\alpha_k)$
\State $\vec{v}\gets U_x\cdot \vec{v}$

\For{$i\gets -1$ to $n_{cw}-2$}
\State $B'\gets$ matrix corresponding to $B'^{ f_{i} \alpha_{2+i} \alpha_k }_{f_{2+i}}$
\State $\vec{v}\gets B'\cdot \vec{v}$
\EndFor

\State \Return $\vec{v}$
\EndFunction
\end{algorithmic}
\label{algo:Hamiltonian_x_cutB}
\end{algorithm}

\begin{algorithm}[t]
\caption{\justifyingg{Computation of the final state after translating the $k$-th anyon by one lattice spacing in positive $y$-direction across cut $A$. The initial state $|(\alpha_k,\vec{r}_k)_{k=1}^N, \mathbf{f}\rangle$ is represented by $\vec{e}_i$. The matrices corresponding to braid operators only act on the subspace defined by the fusion degrees of freedom, i.e., on the second part in the tensor product $|(\alpha_k,\vec{r}_k)_{k=1}^N\rangle \otimes |\mathbf{f}\rangle$.}}
\begin{algorithmic}[1]
\Function{TranslationCutA}{$\vec{e}_i,k,\lbrace \ldots \rbrace$}

\State $n_{cw}, n_{ccw} \gets$ number of anyons with which $\alpha_k$ is braided clockwise and counter-clockwise, respectively

\State $\vec{v}\gets$ vector representing $|(\alpha_{k'},\vec{r}_{k'}+\delta_{kk'}\vec{e}_y)_{k'=1}^N, \mathbf{f}\rangle$

\For{$i\gets 1$ to $n_{cw}$}
\State $B'\gets$ matrix corresponding to $B'^{ f_{k-3+i} \alpha_{k+i} \alpha_k }_{f_{k-1+i}}$, where $\lbrace f_k \rbrace$ are the fusion products associated with $\vec{v}$; $f_{-1}\equiv 1,f_0\equiv \alpha_1$
\State $\vec{v}\gets B'\cdot \vec{v}$
\EndFor

\State $S^{z}\gets$ matrix with $(S^{z})_{ij}=S^{(z)}_{(f_N)_i(f_N)_j}$, $(f_N)_i$ is the $N$-th entry in $\mathbf{f}_i$
\State $U_y \gets$ matrix with $(U_y)_{ij}=\sum_{k,l} S^{(z)\dagger}_{(f_N)_j,k}U_x\left( \text{\boldmath$\alpha$},\widetilde{\mathbf{f'}},\widetilde{\mathbf{f}}\right)S^{(z)}_{l,(f_N)_i}$, $(f_n)_i$: $n$-th entry in $\mathbf{f}_i$, $\widetilde{\mathbf{f}}=((f_1)_j,\ldots ,(f_{N-1})_j,k)$, $\widetilde{\mathbf{f}}=((f_1)_i,\ldots ,(f_{N-1})_i,l)$, $\text{\boldmath$\alpha$}=(\alpha_1,\ldots,\alpha_{k-1},\alpha_{k+1},\ldots,\alpha_N,\alpha_k)$
\State $\vec{v}\gets U_y\cdot \vec{v}$

\For{$i\gets -1$ to $n_{ccw}-2$}
\State $B\gets$ matrix corresponding to $B^{ f_{i} \alpha_{2+i} \alpha_k }_{f_{2+i}}$
\State $\vec{v}\gets B\cdot \vec{v}$
\EndFor

\State \Return $\vec{v}$
\EndFunction
\end{algorithmic}
\label{algo:Hamiltonian_x_cutA}
\end{algorithm}

Finally, let us briefly state the string of operators needed to compute the matrix elements $\mathcal{H}_{s's}$ for the braided fusion diagram given as example in the context of translations in $y$-direction in Fig.~\ref{eq:translate_A}. The matrix elements are
\begin{align}
\begin{split}
	\mathcal{H}_{s's}=-t\sum_{f_2'',f_3'',f_5'',f_5'''}&S^{(f_4)\dagger}_{f_5f_5''}\left[B'^{f_1dc}_{f_3} \right]_{f_2f_2''}\left[B'^{f_2''ec}_{f_4}  \right]_{f_3f_3''}\\ &\times U_x\left( \text{\boldmath$\alpha$},\widetilde{\mathbf{f'}}, \widetilde{\mathbf{f}} \right) \left[ B^{1ac}_{f_1'} \right]_{ca}S^{(f_4')}_{f_5'''f_5'},
\end{split}
\end{align}
where the intital state $|s\rangle$ and the final state $|s'\rangle$ are again associated with the tuples $(f_1,f_2,f_3,f_4,f_5)$ and $(f_1',f_2',f_3',f_4',f_5')$, respectively. Further, $\widetilde{\mathbf{f}}=(f_1,f_2'',f_3'',f_4,f_5'')$ and $\widetilde{\mathbf{f'}}=(f_1',f_2',f_3',f_4',f_5''')$. Note that despite the transformation with $S^{(z)\dagger}$ reversing the orientation of the lines in the fusion diagrams, the same braid operators as for the other scenarios can be used; there is no need for any adjustments such as replacing charges with their conjugated ones. This can also be seen by noting that $S^{(z)}$ commutes with the braid operators $B$ and $B'$ since these operators do not change the charges $f_{N-1}$ or $f_N$ that determine the action of $S^{(z)}$. We can thus transform the fusion diagrams using the punctured torus $S$-matrix directly before applying $U_x\left( \text{\boldmath$\alpha$},\mathbf{f'}, \mathbf{f} \right)$, as shown in Alg.~\ref{algo:Hamiltonian_x_cutA}, which contains a pseudoalgorithm for computing translations across cut $A$.

\section{Proof of the Number of Wave Function Components Being Unaffected by the Anyon Ordering}
\label{app_distinguishable_anyons}

Here, we show that for distinguishable anyons, the number of tuples $\mathbf{f}$, and thus the number of wave function components, does not depend on the anyon ordering. First, note that two distinct anyon orderings can be related by (multiple) permutations of neighboring charges $\alpha_n$ and $\alpha_{n+1}$ in the fusion diagrams in Eq.~(\ref{eq:general_state}). We can thus focus on this case. Due to the associativity and commutativity of fusion, the fusion products $f_{n},f_{n+1},\ldots,f_{N}$ are unaffected by permuting $\alpha_n$ and $\alpha_{n+1}$. 
This also holds for the fusion products $f_1,f_2,\ldots,f_{n-2}$, which do not involve $\alpha_n$ and $\alpha_{n+1}$.
For the unpermuted anyon ordering, the contribution of the fusion product $f_{n-1}$ to the number of tuples is $\sum_{f_{n-1}}N_{f_{n-2}\alpha_n}^{f_{n-1}}N_{f_{n-1}\alpha_{n+1}}^{f_{n}}$. For the permuted anyon ordering, the corresponding contribution is $\sum_{f'_{n-1}}N_{f_{n-2}\alpha_{n+1}}^{f'_{n-1}}N_{f'_{n-1}\alpha_{n}}^{f_{n}}$. Due to the associativity of fusion, these two sums agree with each other since $\sum_eN_{ab}^eN_{ec}^d=\sum_fN_{af}^dN_{bc}^f$~\cite{bonderson_2007}. I.e., since the contribution of the only fusion product that is changed due to the permutation of $\alpha_n$ and $\alpha_{n+1}$ to the number of tuples is the same for the permuted and the unpermuted case, the number of tuples for both anyon orderings agree, $|\lbrace \mathbf{f}\rbrace|=|\lbrace \mathbf{f'}\rbrace|$. This implies that all anyon orderings feature the same number of tuples $\mathbf{f}$ and shows that we can indeed think of the wave function components as being associated with tuples that depend on the anyon ordering.

\section{Effect of Mutual Bosonic Statistics}
\label{app_commutation}

In section~\ref{sec:algo}, we excluded hopping that leads to multiple anyons being localized at the same site. In general, such scenarios may occur if the corresponding anyons do not pick up a non-trivial phase upon being exchanged. When dealing with identical anyons, this is usually not the case. This can however be relevant for anyons of different charge if one does not explicitly add hard-core interactions. Then, as we will see, one has to take special care when constructing the basis according to Sec.~\ref{sec:basis_states}; the braids remain the same as discussed in Sec.~\ref{sec:algo}.

Suppose there are two anyons $a$ and $b$ located at positions $\vec{r}_a$ and $\vec{r}_b$, respectively, that fuse to anyon $c$. The effect of exchanging the two anyons counter-clockwise is given by the $R$-moves:
\begin{align}
	\psi_{a,b\rightarrow c}(\vec{r}_a,\vec{r}_b) = R^{ab}_c\psi_{a,b\rightarrow c}(\vec{r}_b,\vec{r}_a),
\end{align}
where the first (second) argument in $\psi_{a,b\rightarrow c}$ represents the position of anyon $a$ $(b)$. If both anyons are localized at the same site, i.e., if $\vec{r}_a=\vec{r}_b$, the wave function can only be non-zero if $R^{ab}_c=1$, that is, the two anyons are mutual bosons. Using this result, it is clear that the algorithm / braid rules for the different translations described above in Sec.~\ref{sec:algo} can also be applied if an anyon is translated onto an already occupied site, given this is allowed by the just discussed constraint. In particular, it does not matter whether anyons at the same site are braided with each other due to $R^{ab}_c=1$.

There are a few things to be considered when constructing the basis states representing such lattice configurations. Let us start with fusion diagrams where all anyons located at the same sites fuse together and their fusion product then fuses with the intermediate fusion products according to the fusion convention rather then each anyon on the same site fusing one after another with the intermediate fusion products. An example for such a fusion diagram is given by
\begin{align}
\begin{split}
\begin{tikzpicture}[line width=0.75pt]
	\draw (0,0.15) ellipse (1*0.8 and 0.75*0.8);
	\draw (0,0.15) ellipse (0.25*0.8 and 0.25*0.8);
	\node[black, anchor=north west] (a) at (0.03,0.15) {$y$};
	\draw[black] (0.177*0.8,0.177*0.8+0.15) -- (-0.177*0.8,-0.177*0.8+0.15);
	\draw[black] (0.177*0.8,-0.177*0.8+0.15) -- (-0.177*0.8,0.177*0.8+0.15);
	\node[black, anchor=south] (a) at (-1.5,3.3) {$a$};
	\node[black, anchor=south] (a) at (-1.5+0.6,3.3) {$b$};
	\node[black, anchor=south] (a) at (-1.5+0.6*2,3.3) {$c$};
	\node[black, anchor=south] (a) at (-1.5+0.6*3,3.3) {$d$};
	\node[black, anchor=south] (a) at (-1.5+0.6*4,3.3) {$e$};
	\node[black, anchor=north] (a) at (-1.5+0.3*1.5,3.25-0.5196*1.5) {$f_1$};
	\node[black, anchor=north] (a) at (-1.5+0.3*1.5+0.9,3.25-0.5196*1.5) {$f_2$};
	\node[black, anchor=north] (a) at (-1.5+0.3*3,3.25-0.5196*3) {$f_3$};
	\node[black, anchor=north] (a) at (-1.5+0.3*4-0.1*0.5,3.25-0.5196*4+0.1732*0.5) {$f_4$};
	\node[black] (a) at (0.9*0.8,0.7*0.8+0.15+0.075) {$f_5$};
	\draw[black] (0,0.75) -- (-1.5,0.75+2.598) node[sloped,pos=0.90,allow upside down]{\arrowIn}; ; 
	\draw[black] (-1.5*0.59,0.75+2.598*0.59) -- (-1.5*0.6,0.75+2.598*0.6) node[sloped,pos=1,allow upside down]{\arrowIn}; ; 
	\draw[black] (-1.5*0.29,0.75+2.598*0.29) -- (-1.5*0.3,0.75+2.598*0.3) node[sloped,pos=1,allow upside down]{\arrowIn}; ; 
	\draw[black] (-1.5*0.09,0.75+2.598*0.09) -- (-1.5*0.1,0.75+2.598*0.1) node[sloped,pos=1,allow upside down]{\arrowIn}; ; 
	\draw[black] (-1.2,0.75+0.5196*4) -- (-1.5+0.6,0.75+2.598) node[sloped,pos=0.50,allow upside down]{\arrowIn}; ; 
	\draw[black] (0,0.75+4.1568/2) -- (-1.5+0.6*2,0.75+2.598) node[sloped,pos=0.50,allow upside down]{\arrowIn}; ; 
	\draw[black] (-1.2+0.3*2,0.75+0.5196*2) -- (0,0.75+4.1568/2) node[sloped,pos=0.50,allow upside down]{\arrowIn}; ; 
	\draw[black] (0,0.75+4.1568/2) -- (-1.5+0.6*3,0.75+2.598) node[sloped,pos=0.50,allow upside down]{\arrowIn}; ; 
	\draw[black] (-1.2+0.3*3,0.75+0.5196*1) -- (-1.5+0.6*4,0.75+2.598) node[sloped,pos=0.50,allow upside down]{\arrowIn}; ; 
	\draw[black, line width=0.05pt] (0.707*0.8+0.707*0.8*0.001,0.707*0.8*0.75-0.707*0.8*0.75*0.001+0.15) -- (0.707*0.8,0.707*0.8*0.75+0.15) node[sloped,pos=0.50,allow upside down]{\arrowIn}; ; 	
	\node [black, anchor=west] (a) at (1.05,3.16/2) {
            $\begin{aligned}
                ,
            \end{aligned}$
    };
\end{tikzpicture}
\end{split}
\label{eq:same_site_example}
\end{align}
where anyons $c$ and $d$ are located at the same site. Diagrams as the one above allow to straight forwardly read off the $R$-moves corresponding to the constraint discussed above. In the given example, states corresponding to the fusion diagram in Eq.~(\ref{eq:same_site_example}) can only exist if $R^{cd}_{f_2}=1$. Having assured that the considered states can indeed exist, we can now bring the fusion diagram back to the canonical form~(\ref{eq:general_state}) using $F$-moves; for the diagram in Eq.~(\ref{eq:same_site_example}), $[F^{f_1cd}_{f_3}]^{-1}_{f_2f'_2}$ needs to be used. There are two crucial observations to be discussed. Firstly, applying $F$-moves yields a superposition of fusion diagrams in canonical form. These superpositions correspond to the basis states for this anyon configuration, which is in contrast to the basis states introduced in Sec.~\ref{sec:basis_states}. Secondly, when dealing with distinguishable anyons, we have fix an order in which anyons localized at identical sites fuse. The reason is that if two anyons of different charge located at the same site are first exchanged and then, the fusion diagram is brought to the canonical form, the intermediate fusion products of the obtained diagrams in the superposition may be different from the intermediate fusion products when not exchanging the anyons (see the discussion in Sec.~\ref{sec:anyonorderings}). As these two superpositions describe the same physical state, we have discard one of them. This can be done by introducing an ``on-site fusion order'' that fixes the order in which anyons localized at identical sites fuse. In the case depicted in Eq.~(\ref{eq:same_site_example}), we choose the on-site fusion order to be the alphabetical order.

To sum up, the first thing to do when dealing with multiple anyons at identical sites is to rewrite the fusion diagrams in such a way that anyons at the same sites directly fuse with each other. Then, using the $R$-moves, we can check whether or not the states associated with the diagrams can actually exist. If they do exist, we have to order the anyons at identical sites according to an on-site fusion order that we have to define. Finally, we can use $F$-moves to express the fusion diagrams as superpositions of fusion diagrams in canonical form, which serve as basis states. In general, there may be more than two anyons localized at a single site and multiple sites may be occupied by more than one anyon. Then, we start with fusion diagrams containing small fusion trees for each of those sites. Their fusion products fuse with the other fusion products of the anyons located at the other sites. The constraint above is changed such that at each vertex of the fusion trees for a single site, the corresponding $R$-move has to equal identity.

\section{Special Case of Abelian Anyons and Relation to Existing Algorithms}
\label{app_equi}

Here, we show that for abelian anyon models, our algorithm presented in Sec.~\ref{sec:algo} is related to the algorithm in Ref.~\cite{PhysRevB.43.10761} that was briefly sketched in Sec.~\ref{sec:partcile_types_abelian} for semions. More generally, the latter algorithm simulates $N$ abelian anyons on a torus, where all anyons possess the same charge and a phase of $\theta$ is acquired upon exchanging two of them, with $e^{i2\theta N}=1$. Translating an anyon around the torus along a non-contractible loop is modelled using two $M\times M$ matrices, where $M$ is the smallest positive integer fulfilling $e^{i2\theta M}=1$, i.e., $\theta=\pi p/M$ with $p$ and $M$ being coprime integers. To show the relation between the two algorithms, we choose the same starting point as in Ref.~\cite{PhysRevB.43.10761}, that is, we only know the exchange statistics in terms of the phase $\theta$ and first need to find an abelian anyon model that features the corresponding anyons. Then, based on this anyon model, we compute all relevant phases and matrices that are to be applied when translating the anyons
and show the relation to the corresponding quantities in the algorithm in Ref.~\cite{PhysRevB.43.10761}.

\subsection{Appropriate Abelian Anyon Models}\label{sec:appranyonmodel}

First, we have to take a look at the anyonic charges of the given anyon model: From the string rules in Ref.~\cite{PhysRevB.43.10761} that were also sketched Sec.~\ref{sec:partcile_types_abelian}, one can see that when fusing two anyons and exchanging this fusion product with another anyon, a phase of $2\theta$ is picked up since the phases associated with the fusion product's string correspond to twice the phases associated with an initial anyon's string. Thus, $M$ anyons fuse to the vacuum charge and as this is the minimal number of anyons to do so, there are overall $M$ anyonic charges in the anyon model. We will refer to these charges as $[a]_M$, $[a]_M\in \lbrace0,...,M-1\rbrace$, where $[a]_M=0$ denotes the vacuum charge (note the changed convention compared to the main text) and $[a]_M=\eta$ with $1\leq \eta \leq M-1$ the charge of the anyons being simulated.

We can see from the discussion in Sec.~\ref{sec:sheets} that due to the abelian nature, there are $M$ wave function components / tuples in total, agreeing with the dimensionality of the $M\times M$ matrices in Eq.~(2.19) in Ref.~\cite{PhysRevB.43.10761}. For the description of the anyon model, we follow the relevant parts in Bonderson's PhD thesis~\cite{bonderson_2007}. In the above notation, the fusion rules are
\begin{align}
	[a]_M\times [b]_M = [a+b]_M,
\end{align}
where anyon charges within the brackets $[\,\,]_M$ are always taken mod $M$. For convenience, the brackets $[\,\,]_M$ will only be written in the following where the corresponding arguments may not be in the set $\lbrace0,...,M-1\rbrace$. The $F$- and $R$-moves are given by
\begin{align}
\begin{split}
	\left[ F^{a,b,c}_{[a+b+c]_M} \right]_{[a+b]_M,[b+c]_M}&=e^{i\frac{\pi}{M} a\left(b+c-[b+c]_M\right)}\\ \text{and}\quad R^{a,b}_{[a+b]_M}&=e^{i\frac{\theta}{\eta^2} ab},
\end{split}
\label{eq:abelian_FR}
\end{align}
the entries of the modular $S$-matrix, which corresponds to $S^{(0)}$, are
\begin{align}
	S_{a,b} = \frac{1}{\sqrt{M}}e^{i \frac{2\theta}{\eta^2} ab}.
\end{align}
Here, it was used that $R^{\eta,\eta}_{[2\eta]_M}=e^{i\theta}$ and assumed that the anyon model containing the appropriate $R$-moves is $\mathbb{Z}_M^{n+1/2}$ with $n=\frac{p}{2\eta^2}-\frac{1}{2}\in\lbrace 0,1,...,M-1 \rbrace$ and $M$ even. I.e., $\eta$, and therefore also $n$, has to be chosen based on $p=M\theta/\pi$ in order to determine an appropriate anyon model. If there is no appropriate choice of $\eta$ such that $n$ fulfills the above constraint, the second type of abelian anyon models that includes $M$ distinct charges, $\mathbb{Z}_M^{n}$ with $n=\frac{p}{2\eta^2}\in\lbrace 0,1,...,M-1 \rbrace$ has to be utilized. These models feature trivial $F$-moves and will be considered later on. Note that we can find an appropriate anyon model for every $p$ since for $\eta=1$ and odd $p$, the condition for $\mathbb{Z}_M^{n+1/2}$ with $n=(p-1)/2$ is fulfilled, whereas for $\eta=1$ and even $p$, the condition for $\mathbb{Z}_M^{n}$ with $n=p/2$ is fulfilled. We can thus restrict ourselves to $\eta=1$ from now on.\\

\subsection{Relating the Translation Processes}

Let us now show that the matrices $\tau_i$ and $\rho_i$ (Eqs.~(2.12) and (2.13) in Ref.~\cite{PhysRevB.43.10761}) and $\sigma_j=e^{i\theta}$ with $i\in\lbrace 1,...,N\rbrace$, $j\in\lbrace 1,...,N-1\rbrace$ naturally arise from the rules in our algorithm. The operators corresponding to $\tau_i$ and $\rho_i$ move the $i$-th anyon along a non-contractible loop in positive $x$- and $y$-direction (without any translations in the perpendicular direction), respectively, $\sigma_j$ corresponds to counter-clockwise exchanging the $j$-th and $(j+1)$-th anyon. Since $\tau_i$ and $\rho_i$ are matrices rather than operators, they are defined as the action of the above operators on a specific anyon configuration. This anyon configuration contains $N$ anyons, where the coordinates of the $i$-th anyon are denoted by $x_i$ and $y_i$ and fulfill $x_{i+1}>x_i$ and $y_{i+1}>y_i$ for $i=1,...,N-1$ (Fig.~1 in Ref.~\cite{PhysRevB.43.10761}).

\subsubsection{Translations in the Bulk}

For the operations correspdoning to $\sigma_j$, no cut has to be crossed, i.e., according to the rule in Fig.~\ref{eq:rule1}, it is given by
\begin{align}
\begin{split}
	\sigma_j&= \left[ B^{[j-1]_M,1,1}_{[j+1]_M} \right]_{[j]_M,[j]_M} 
	=R^{1,1}_{[2]_M}\mathbb{1}_M=e^{i\theta}\mathbb{1}_M,
\end{split}
\end{align}
in agreement with what is found in Ref.~\cite{PhysRevB.43.10761}.

\subsubsection{Translations across Cut \texorpdfstring{$B$}{B}}

Using the above result and the rules in Figs.~\ref{eq:translate_B} and \ref{eq:translate_A}, it can be seen that $\tau_{j+1}=e^{-i2\theta}\tau_j=e^{-i2\theta j}\tau_1$ and $\rho_{j+1}=e^{i2\theta}\rho_j=e^{i2\theta j}\rho_1$, which also follows from the more general relations $\tau_{j+1}=\sigma_j^{-1}\tau_j\sigma_j^{-1}$ and $\rho_{j+1}=\sigma_j\rho_j\sigma_j$~\cite{StatisticsTorusFormulae, PhysRevB.43.10761}. It thus remains to show that our algorithm reproduces $\tau_1$ and $\rho_1$. The former can be computed using Fig.~\ref{eq:translate_B} and Eq.~(\ref{eq:sheet_hopping_def}):
\begin{widetext}
\begin{align}
\begin{split}
	\left(\tau_1\right)_{[a-1]_M,a} = &e^{i\theta(N-1)}\left[F^{M-1,1,[-a]_M}_{[-a]_M}\right]_{0,[1-a]_M}\left[F^{1,M-1,[1-a]_M}_{[1-a]_M}\right]^{-1}_{[-a]_M,0}\\
                &\times\left[F^{1,1,1}_{[3]_M}\right]^{-1}_{[2]_M,[2]_M} \prod_{j=0}^{N-4}\left[F^{1,[-2-j]_M,1}_{[-j]_M}\right]^{-1}_{[-1-j]_M,[-1-j]_M}  \\
                =&e^{i\theta(N-1)}e^{i\frac{\pi}{M} \lbrace M([-a]_M-[-a+1]_M)-(2-[2]_M)-\sum_{j=0}^{N-4}([-2-j]_M +1-[-1-j]_M)\rbrace}\\
                =&e^{i\theta(N-1)}e^{-i\pi(1+ N/M+\lfloor -1/M \rfloor)}
                \\=&e^{i\theta(N-1)}e^{-i\pi N/M},
\end{split}
\label{eq:tau1}
\end{align}
\end{widetext}
where $e^{i\theta(N-1)}$ originates from the braid operations $\sigma_j$ and $N>2$ and $M\geq 2$ was assumed. The contribution $([-a]_M-[1-a]_M)$ is either $-1$ or $M-1$ and due to $M$ being even and this contribution appearing in the exponential, both cases yield the same result. The sum in the exponent was found to equal $(N-3)+[2]_M-[-1]_M$, where it was used that $[N]_M=0$. Finally, $[x]_M=x-M\lfloor x/M \rfloor$ was utilized, where $\lfloor x \rfloor$ denotes the largest integer smaller than or equal to $x$. For $M=N=2$, the correct result for $\left(\tau_1\right)_{[a-1]_M,a}$ is given by $-e^{i\theta}$ and thus also agrees with the above result. Equation~(\ref{eq:tau1}) therefore holds for all allowed choices of $M$ and $N$; $M=1$ is not considered since it corresponds to the trivial case of simulating vacuum charges.

According to Eq.~(\ref{eq:tau1}), charge $a$ is moving along the non-contractible loop in the initial state, whereas in the final state, that charge is $[a-1]_M$, which can be seen by thinking of the wave function components as entries in a vector this matrix acts on. All other entries in $\tau_1$ are zero, i.e., $\tau_1$ is given by

\begin{align}
	\tau_1 = e^{i\theta(N-1)}e^{-i\pi N/M}
	\begin{pmatrix}
0 & 1 & &\\
\vdots & 0 & \ddots & \\
0 & & \ddots & 1\\
1 & 0 & \cdots &0\\
\end{pmatrix}.
\label{eq:tau1_matrix}
\end{align}

\subsubsection{Translations across Cut \texorpdfstring{$A$}{A}}

Using this result and Fig.~\ref{eq:translate_A}, $\rho_1$ is computed to be
\begin{widetext}
\begin{align}
\begin{split}
	(\rho_1)_{jk}&=e^{-i2\theta(N-1)}\frac{1}{M}\sum_{a,b=0}^{M-1}  S^{(0)\dagger}_{ja}  (\tau_1)_{ab} S^{(0)}_{bk}\\
	&=e^{-i\theta(N-1)}e^{-i\pi N/M}\frac{1}{M}\sum_{a=0}^{M-1}  e^{-i2\theta ja}  e^{i2\theta [a+1]_Mk}\\
	&=e^{-i\theta(N-1)}e^{-i\pi N/M}\frac{1}{M}\sum_{a=0}^{M-1}  e^{i2\theta (k-j)a} e^{i2\theta k(1-M\lfloor (a+1)/M \rfloor)}\\
	&=e^{-i\theta(N-1)}e^{-i\pi N/M} e^{i2\theta k} \delta_{jk},
\end{split}
\end{align}
\end{widetext}
where we used the fact that the translation across cut $A$ can be obtained by transforming the translation across cut $B$ via $S^{(z)}$ ($z=0$ for abelian anyons). The prefactor $e^{-i2\theta(N-1)}$ compensates the factor $e^{i\theta(N-1)}$ in Eq.~(\ref{eq:tau1}) and accounts for the other braids according to Fig.~\ref{eq:translate_A}. It was used that $e^{i2\theta M}=1$. This result is again valid for $N\geq2$ and $M\geq 2$. Written as matrix, $\rho_1$ takes the form
\begin{align}
	\rho_1= e^{-i\theta(N-1)}e^{-i\pi N/M}
	\begin{pmatrix}
1 &  & &\\
 & e^{i2\theta} &  & \\
 & & \ddots & \\
 &  &  & e^{i2\theta (M-1)}\\
 \end{pmatrix}.
 \label{eq:rho1_matrix}
\end{align}

The two $M\times M$ matrices in Eqs.~(\ref{eq:tau1_matrix}) and (\ref{eq:rho1_matrix}) agree with the respective matrices in Ref.~\cite{PhysRevB.43.10761} except for the prefactor. This is however of no concern as we can always choose to introduce fluxes in both directions that compensate the differences in phase.

In addition to the rules derived from the operators corresponding to $\tau_i$, $\rho_i$ and $\sigma_j$, for the algorithm simulating abelian anyons~\cite{PhysRevB.43.10761}, one final rule is needed. If an anyon is translated across cut $A$, the wave function picks up an additional factor of $e^{i\theta}$ for each anyon located in the same column as the anyon being translated. This additional phase is naturally contained in our algorithm, see the rules summarized in Fig.~\ref{eq:translate_A}. It thus follows that for abelian anyons, our algorithm reproduces the algorithm in Ref.~\cite{PhysRevB.43.10761}.\\

Until now, we assumed that the appropriate abelian anyon model is given by $\mathbb{Z}_M^{n+1/2}$ for $n=\frac{p}{2}-\frac{1}{2}\in\lbrace 0,1,...,M-1 \rbrace$ (with $\eta=1$) and $M$ even, see App.~\ref{sec:appranyonmodel}. However, if the appropriate anyon model turns out to be $\mathbb{Z}_M^{n}$ for $n=\frac{p}{2}\in\lbrace 0,1,...,M-1 \rbrace$, a very similar result is obtained: Due to the $F$-moves being trivial, the prefactor in Eq.~(\ref{eq:tau1_matrix}) is changed to $e^{i\theta(N-1)}$ and the prefactor in Eq.~(\ref{eq:rho1_matrix}) becomes $e^{-i\theta(N-1)}$. All other results remain unaffected. This implies that also for these models, the two matrices $\tau_1$ and $\rho_1$ can be made to agree with those in Ref.~\cite{PhysRevB.43.10761} by introducing external fluxes.\\

With this, it is shown that for abelian anyons, our algorithm is related to the algorithm discussed in Ref.~\cite{PhysRevB.43.10761}. It is to be stressed that the two algorithms use different conventions for external fluxes, which has to be kept in mind when verifying this relation numerically.

\section{Momentum States}
\label{app_mom}

Due to the periodicity in $x$- and $y$-direction, it is possible to construct momentum states which block diagonalize the Hamiltonian $\mathcal{H}$ given by Eq.~(\ref{eq:hamiltonian_square_lattice}). Before discussing the actual construction, let us first introduce a new notation for superpositions of the basis states introduced in Sec.~\ref{sec:basis_states}. We have seen that tuples, which are used to associate fusion diagrams with the basis states, correspond to wave function components. By thinking of wave function components as different components of a vector, we can alternatively associate unit vectors with the tuples / fusion diagrams, i.e., we may associate tuple $\mathbf{f}$ with the unit vector in the $i$-th direction, $\vec{e}_i$. A different tuple $\mathbf{f'}$ is then assoicated with a unit vector along another direction $j\neq i$. Using this notation, we can write superpositions of basis states as
\begin{align}
\begin{split}
	\sum_{i} a_i|(\alpha_k,\vec{r}_k)_{k=1}^N, \mathbf{f}_i\rangle &\equiv |(\alpha_k,\vec{r}_k)_{k=1}^N\rangle \otimes |\textstyle{\sum_i} a_i\vec{e}_i\rangle\\
	&\equiv |(\alpha_k,\vec{r}_k)_{k=1}^N, \textstyle{\sum_i} a_i\vec{e}_i\rangle,
\end{split}
\label{eq:basis_superpos}
\end{align}
where $a_i\in \mathbb{C}$ are arbitrary complex numbers and we associate unit vector $\vec{e}_i$ with tuple $\mathbf{f}_i$. We refer to the vector $|\sum_i a_i\vec{e}_i\rangle$ describing the superposition of fusion diagrams as ``fusion diagram amplitude vector'' (FDAV) since the entries in its components correspond to the probability amplitude of the anyon configuration featuring the fusion processes associated with the respective fusion diagrams. We will further use the more convenient notation of $|\vec{\mathcal{A}}\rangle$ for the FDAV.

There are a few remarks to be made regarding the above definition. First of all, we can use the superpositions~(\ref{eq:basis_superpos}) rather than the states in Eq.~(\ref{eq:realspacebasis}) as basis for computing the Hamiltonian $\mathcal{H}$. We only need to ensure orthonormality, i.e., $\langle \vec{\mathcal{A}}|\vec{\mathcal{A}} \rangle=1$ and $\langle \vec{\mathcal{A'}}|\vec{\mathcal{A}} \rangle=0$ for $|\vec{\mathcal{A}}\rangle \neq |\vec{\mathcal{A'}}\rangle$. Further, all the operators acting on the fusion diagrams can be interpreted as matrices that are to be multiplied onto the FDAVs. I.e., the $F$-moves, $R$-moves, braid operators $B$ and $B'$ and the punctured torus $S$-matrices $S^{(z)}$ become unitary matrices in the subspaces they act on. These subspaces are given by the components corresponding to fusion diagrams that are consistent with the indices of respective operators. The punctured torus $S$-matrices $S^{(z)}$ for example represent unitary matrices in the subspaces corresponding to fusion diagrams for which $f_{N-1}=z$ in Eq.~(\ref{eq:general_state}). The matrices corresponding to the operators $U_x\left( \text{\boldmath$\alpha$},\mathbf{f'}, \mathbf{f} \right)$ defined by Eq.~(\ref{eq:sheet_hopping_def}) in App.~\ref{app_example_computations_sheet_hopping} turn out to be of particular interest for the construction of the momentum states. We thus introduce the notation
\begin{align}
	\widetilde{U}_x\left( \text{\boldmath$\alpha$}\right)_{ij}=U_x\left( \text{\boldmath$\alpha$},\mathbf{f}_i, \mathbf{f}_j \right)
	\label{eq:sheet_hopping_matrix_def}
\end{align}
for these matrices. Multiplying $\widetilde{U}_x\left( \text{\boldmath$\alpha$}\right)$ onto $|\vec{\mathcal{A}}\rangle$ thus corresponds to moving the final anyon in the fusion order around the torus in $x$-direction for each fusion diagram. Since this operation is reasonable for every fusion diagram / tuple, the matrices $\widetilde{U}_x\left( \text{\boldmath$\alpha$}\right)$ are unitary in the space corresponding to all possible linear combinations of fusion diagrams. We can also express the property in Eq.~(\ref{eq:sheet_hopping_identity_tuples}) in terms of $\widetilde{U}_x\left( \text{\boldmath$\alpha$}\right)$:
\begin{align}
	\widetilde{U}_x\left( \sigma^{N-1}( \text{\boldmath$\alpha$})\right)\widetilde{U}_x\left( \sigma^{N-2} (\text{\boldmath$\alpha$})\right)\ldots \widetilde{U}_x\left(  \text{\boldmath$\alpha$}\right)= \mathbb{1},
	\label{eq:sheet_hopping_identity}
\end{align}
where $\sigma(1,2,\ldots,N) = (N,1,2,\ldots,N-1)$ again.

Note that if there are distinguishable anyonic excitations in the system (see Sec.~\ref{sec:anyonorderings}), the tuples corresponding to all the allowed fusion diagrams in general depend on the anyon ordering. This means that while the dimensions of the vectors in the superpostions~(\ref{eq:basis_superpos}) do not change, the fusion diagrams associated with the individual components might change when exchanging anyons of distinct charge.

Having introduced the new notation for superpositions of fusion diagrams, we can now discuss the construction of momentum states. To do this, we start in App.~\ref{sec:translationoperators} by taking a look at the action of the two operators $\mathcal{T}_x$ and $\mathcal{T}_y$ that translate all anyons in positive $x$- and $y$-direction, respectively. We then use this knowledge to construct the momentum states (App.~\ref{sec:construction}), where we first consider momentum eigenstates in $x$-direction in App.~\ref{app:momentum_x}. These states are then extended to momentum eigenstates in both $x$- and $y$-direction in App.~\ref{app:momentum_xy}. Lastly, in App.~\ref{sec:completeness}, we argue that the translation operators commute with each other and with the Hamiltonian $\mathcal{H}$, which is a requirement for them to block diagonalize $\mathcal{H}$. We also argue that both of the momentum states indeed form bases of the Hilbert space. Note that the final section does not contain information that is important for the construction of the momentum states. It can therefore be omitted.

\subsection{The Translation Operators \texorpdfstring{$\mathcal{T}_x$}{Tx} and \texorpdfstring{$\mathcal{T}_y$}{Ty}}\label{sec:translationoperators}

In this section, we introduce the two translation operators $\mathcal{T}_x$ and $\mathcal{T}_y$ and discuss some properties that are crucial for the construction of the momentum states later on.

\subsubsection{The Translation Operator \texorpdfstring{$\mathcal{T}_x$}{Tx}}
\label{sec:translationoperators_x}

Applying the translation operator $\mathcal{T}_x$ onto an arbitrary state $|s\rangle=|(\alpha_k,\vec{r}_k)_{k=1}^N,\vec{\mathcal{A}}\rangle$ does in general not only shift the anyons' positions in positive $x$-direction, but does also change the FDAV $|\vec{\mathcal{A}}\rangle$. From the braid rules discussed in the Sec.~\ref{sec:algo}, we know that braiding can only occur if anyons are translated over cut $B$ since $\mathcal{T}_x$ translates all anyons simultaneously. If only one anyon is translated across cut $B$ upon applying $\mathcal{T}_x$, the FDAV is multiplied by the matrix $\widetilde{U}_x(\text{\boldmath$\alpha$})$ introduced above.

\begin{figure*}
\begin{tikzpicture}[line width=0.75pt]
	\begin{scope}[shift={(13.7,2.61/2-2.18/2)}]
	\node [black, anchor=east] (a) at (-2.25-0.15,2.18/2) {
            $\begin{aligned}
                =
            \end{aligned}$
    };
	\draw (0,0.15) ellipse (1*0.8 and 0.75*0.8);
	\draw (0,0.15) ellipse (0.25*0.8 and 0.25*0.8);
	\node[black, anchor=north west] (a) at (0.03,0.15) {$x$};
	\draw[black] (0.177*0.8,0.177*0.8+0.15) -- (-0.177*0.8,-0.177*0.8+0.15);
	\draw[black] (0.177*0.8,-0.177*0.8+0.15) -- (-0.177*0.8,0.177*0.8+0.15);
	\draw [black,domain=0.419-0.14:1.2+0.5196/sin(60), samples=10] plot ({\x*cos(120)-0.6}, {\x*sin(120)+0.75});
	\draw [black,domain=0:1.8-0.13, samples=10] plot ({\x*cos(120)}, {\x*sin(120)+0.75});
	\draw [black,domain=1.8+0.13:2.1+0.01, samples=10] plot ({\x*cos(120)}, {\x*sin(120)+0.75});
	\draw [black,domain=2.25-0.01:2.4, samples=10] plot ({\x*cos(120)}, {\x*sin(120)+0.75});
	\draw [black,domain=0.582-0.14:1.8*0.5+0.582*0.5+0.01, samples=10] plot ({\x*cos(120)+0.6}, {\x*sin(120)+0.75});
	\draw [black,domain=0.5*(2.4-0.5196/sin(60)*2):2.4-0.5196/sin(60)*2, samples=10] plot ({\x*cos(60)-0.3*2}, {\x*sin(60)+0.75+0.5196*2});
	\draw [black,domain=-1.51:-0.29, samples=10] plot ({\x}, {2.309});
	\draw[black] (-0.9,2.309) -- (-0.9001,2.309) node[sloped,pos=0.50,allow upside down]{\arrowIn}; ; 
	\draw [black,domain=0.815:2.1+0.01, samples=10] plot ({\x*cos(120)+1.2},
	 {\x*sin(120)+0.75});
	 \draw[black] (0.47125,2.012) -- (0.47125-0.005,2.012+0.00866) node[sloped,pos=0.50,allow upside down]{\arrowIn}; ; 
	\draw [black,domain=0.582-0.22:2.25+0.01, samples=10] plot ({\x*cos(120)-1.2}, {\x*sin(120)+0.75});
	\draw[black] (-1.853,1.88) -- (-1.853+0.005,1.88-0.00866) node[sloped,pos=0.50,allow upside down]{\arrowIn}; ; 
	\draw [black,domain=-2.33:-1.13, samples=10] plot ({\x}, {2.699});
	\draw[black] (-1.73,2.699) -- (-1.7301,2.699) node[sloped,pos=0.50,allow upside down]{\arrowIn}; ; 
	\draw [black,domain=-1.05-0.01:-0.15-0.13, samples=10] plot ({\x}, {2.569});
	\draw [black,domain=-0.15+0.13:0.15, samples=10] plot ({\x}, {2.569});
	\draw[black] (-0.45,2.569) -- (-0.4501,2.569) node[sloped,pos=0.50,allow upside down]{\arrowIn}; ; 
	\node[black, anchor=south] (a) at (-1.2,2.75) {$a$};
	\node[black, anchor=south] (a) at (0,2.75) {$b$};
	\node[black, anchor=north] (a) at (-1.5+0.3*3.5-0.05,3.25-0.5196*3.5) {$f_1$};
	\node[black, anchor=west] (a) at (0.17-0.02,0.42-0.02) {$f_2$};
	\draw[black] (-1.5*0.8*0.63,0.75+2.598*0.8*0.63) -- (-1.5*0.8*0.625,0.75+2.598*0.8*0.625) node[sloped,pos=1,allow upside down]{\arrowIn}; ; 
	\draw[black] (-1.5*0.8*0.26,0.75+2.598*0.8*0.26) -- (-1.5*0.8*0.25,0.75+2.598*0.8*0.25) node[sloped,pos=1,allow upside down]{\arrowIn}; ; 
	\draw[black] (0.1568,1.518) -- (0.1568-0.005,1.518+0.00866) node[sloped,pos=1,allow upside down]{\arrowIn}; ; 
	\draw[black] (-0.6+0.9*0.666+0.01,0.75+0.5196*2) -- (-1.2+0.3*2,0.75+0.5196*2) node[sloped,pos=0.50,allow upside down]{\arrowIn}; ; 
	\draw[black] (-1.1197,1.65) -- (-1.1197+0.005,1.65-0.00866) node[sloped,pos=1,allow upside down]{\arrowIn}; ; 
	\draw[black] (-0.15,2.57) -- (-0.15-0.005,2.57-0.00866) node[sloped,pos=1,allow upside down]{\arrowIn}; ; 
	\draw[black, line width=0.05pt] (0.707*0.8,0.707*0.75*0.8+0.15) -- (0.707*0.8+0.707*0.001,0.707*0.75*0.8-0.707*0.75*0.001+0.15) node[sloped,pos=0.50,allow upside down]{\arrowIn}; ; 
	\draw[black, line width=0.05pt] (0.707*1.5*1.074*0.8+0.707*1.5*1.074*0.001,0.707*1.2*1.074*0.8-0.707*1.2*1.074*0.001+0.15) -- (0.707*1.5*1.074*0.8,0.707*1.2*1.074*0.8+0.15) node[sloped,pos=0.50,allow upside down]{\arrowIn}; ; 
	\draw[black, line width=0.05pt] (0.707*2.222*0.8+0.707*2.222*0.001,0.707*1.83*0.8-0.707*1.83*0.001+0.15) -- (0.707*2.222*0.8,0.707*1.83*0.8+0.15) node[sloped,pos=0.50,allow upside down]{\arrowIn}; ; 
	\draw [black,domain=-235.5:73.25, samples=100] plot ({1.5*cos(\x)*1.074*0.8}, {1.2*sin(\x)*1.074*0.8+0.15});
	\draw [black,domain=-219.25:63.5, samples=100] plot ({2.222*cos(\x)*0.8}, {1.83*sin(\x)*0.8+0.15});
	\end{scope}
	\begin{scope}[shift={(3.7,2.61/2-2.18/2)}]
	\draw (0,0.15) ellipse (1*0.8 and 0.75*0.8);
	\draw (0,0.15) ellipse (0.25*0.8 and 0.25*0.8);
	\node[black, anchor=north west] (a) at (0.03,0.15) {$x$};
	\draw[black] (0.177*0.8,0.177*0.8+0.15) -- (-0.177*0.8,-0.177*0.8+0.15);
	\draw[black] (0.177*0.8,-0.177*0.8+0.15) -- (-0.177*0.8,0.177*0.8+0.15);
	\draw [black,domain=0.419-0.14:1.2+0.5196/sin(60)+0.01, samples=10] plot ({\x*cos(120)-0.6}, {\x*sin(120)+0.75});
	\draw [black,domain=0:1.8-0.3+0.01, samples=10] plot ({\x*cos(120)}, {\x*sin(120)+0.75});
	\draw [black,domain=1.8-0.01:2.1+0.01-0.13, samples=10] plot ({\x*cos(120)}, {\x*sin(120)+0.75});
	\draw [black,domain=2.1+0.13-0.01:2.4, samples=10] plot ({\x*cos(120)}, {\x*sin(120)+0.75});
	\draw [black,domain=0.582-0.14:1.8-0.3+0.01, samples=10] plot ({\x*cos(120)+0.6}, {\x*sin(120)+0.75});
	\draw [black,domain=(2.25-0.5196/sin(60)*2-0.01-0.15):2.4-0.5196/sin(60)*2, samples=10] plot ({\x*cos(60)-0.3*2}, {\x*sin(60)+0.75+0.5196*2});
	\draw [black,domain=0:0.5*(2.4-0.5196/sin(60)*2), samples=10] plot ({\x*cos(60)-0.3*2}, {\x*sin(60)+0.75+0.5196*2});
	\draw [black,domain=-1.5-0.01:-0.9, samples=10] plot ({\x}, {2.309});
	\draw[black] (-1.2,2.309) -- (-1.2001,2.309) node[sloped,pos=0.50,allow upside down]{\arrowIn}; ; 
	\draw [black,domain=0.815:1.8+0.01, samples=10] plot ({\x*cos(120)+1.2},
	 {\x*sin(120)+0.75});
	\draw[black] (0.54625,1.8823) -- (0.54625-0.005,1.8823+0.00866) node[sloped,pos=0.50,allow upside down]{\arrowIn}; ; 
	\draw [black,domain=0.582-0.22:2.25+0.01-0.15, samples=10] plot ({\x*cos(120)-1.2}, {\x*sin(120)+0.75});
	\draw[black] (-1.8155,1.816) -- (-1.8155+0.005,1.816-0.00866) node[sloped,pos=0.50,allow upside down]{\arrowIn}; ; 
	\draw [black,domain=-2.33+0.075:-0.075-0.075, samples=10] plot ({\x}, {2.569});
	\draw[black] (-1.2025,2.569) -- (-1.2026,2.569) node[sloped,pos=0.50,allow upside down]{\arrowIn}; ; 
	\draw [black,domain=-0.75-0.01:-0.45-0.13, samples=10] plot ({\x}, {2.049});
	\draw [black,domain=-0.45+0.13:-0.15, samples=10] plot ({\x}, {2.049});
	\draw[black] (-0.45,2.049) -- (-0.45-0.005,2.049-0.00866) node[sloped,pos=0.50,allow upside down]{\arrowIn}; ; 
	\draw [black,domain=-0.3-0.01:0.3, samples=10] plot ({\x}, {2.309});
	\draw[black] (0,2.309) -- (-0.005,2.309) node[sloped,pos=0.50,allow upside down]{\arrowIn}; ; 
	\node[black, anchor=south] (a) at (-1.2,2.75) {$a$};
	\node[black, anchor=south] (a) at (0,2.75) {$b$};
	\node[black, anchor=north] (a) at (-1.5+0.3*3.5-0.05,3.25-0.5196*3.5) {$f_1$};
	\node[black, anchor=west] (a) at (0.17-0.02,0.42-0.02) {$f_2$};
	\draw[black] (-1.5*0.8*0.57,0.75+2.598*0.8*0.57) -- (-1.5*0.8*0.5625,0.75+2.598*0.8*0.5625) node[sloped,pos=1,allow upside down]{\arrowIn}; ; 
	\draw[black] (-1.5*0.8*0.26,0.75+2.598*0.8*0.26) -- (-1.5*0.8*0.25,0.75+2.598*0.8*0.25) node[sloped,pos=1,allow upside down]{\arrowIn}; ; 
	\draw[black] (0.1145,1.591) -- (0.1145-0.005,1.591+0.00866) node[sloped,pos=1,allow upside down]{\arrowIn}; ; 
	\draw[black] (-1.1197,1.65) -- (-1.1197+0.005,1.65-0.00866) node[sloped,pos=1,allow upside down]{\arrowIn}; ; 
	\draw[black, line width=0.05pt] (0.707*0.8,0.707*0.75*0.8+0.15) -- (0.707*0.8+0.707*0.001,0.707*0.75*0.8-0.707*0.75*0.001+0.15) node[sloped,pos=0.50,allow upside down]{\arrowIn}; ; 
	\draw[black, line width=0.05pt] (0.707*1.5*1.074*0.8+0.707*1.5*1.074*0.001,0.707*1.2*1.074*0.8-0.707*1.2*1.074*0.001+0.15) -- (0.707*1.5*1.074*0.8,0.707*1.2*1.074*0.8+0.15) node[sloped,pos=0.50,allow upside down]{\arrowIn}; ; 
	\draw[black, line width=0.05pt] (0.707*2.222*0.8+0.707*2.222*0.001,0.707*1.83*0.8-0.707*1.83*0.001+0.15) -- (0.707*2.222*0.8,0.707*1.83*0.8+0.15) node[sloped,pos=0.50,allow upside down]{\arrowIn}; ; 
	\draw [black,domain=-235.5:73.25, samples=100] plot ({1.5*cos(\x)*1.074*0.8}, {1.2*sin(\x)*1.074*0.8+0.15});
	\draw [black,domain=-219.25:63.5, samples=100] plot ({2.222*cos(\x)*0.8}, {1.83*sin(\x)*0.8+0.15});
	\end{scope}
	\begin{scope}[shift={(8.7,2.61/2-2.18/2)}]
	\node [black, anchor=east] (a) at (-2.25-0.15,2.18/2) {
            $\begin{aligned}
                =
            \end{aligned}$
    };
	\draw (0,0.15) ellipse (1*0.8 and 0.75*0.8);
	\draw (0,0.15) ellipse (0.25*0.8 and 0.25*0.8);
	\node[black, anchor=north west] (a) at (0.03,0.15) {$x$};
	\draw[black] (0.177*0.8,0.177*0.8+0.15) -- (-0.177*0.8,-0.177*0.8+0.15);
	\draw[black] (0.177*0.8,-0.177*0.8+0.15) -- (-0.177*0.8,0.177*0.8+0.15);
	\draw [black,domain=0.419-0.14:1.2+0.5196/sin(60), samples=10] plot ({\x*cos(120)-0.6}, {\x*sin(120)+0.75});
	\draw [black,domain=0:1.5, samples=10] plot ({\x*cos(120)}, {\x*sin(120)+0.75});
	\draw [black,domain=2.25-0.01:2.4, samples=10] plot ({\x*cos(120)}, {\x*sin(120)+0.75});
	\draw [black,domain=0.582-0.14:1.8*0.5+0.582*0.5+0.01, samples=10] plot ({\x*cos(120)+0.6}, {\x*sin(120)+0.75});
	\draw [black,domain=0.5*(2.4-0.5196/sin(60)*2):2.4-0.5196/sin(60)*2, samples=10] plot ({\x*cos(60)-0.3*2}, {\x*sin(60)+0.75+0.5196*2});
	\draw [black,domain=-1.51:-0.29, samples=10] plot ({\x}, {2.309});
	\draw[black] (-0.9,2.309) -- (-0.9001,2.309) node[sloped,pos=0.50,allow upside down]{\arrowIn}; ; 
	\draw [black,domain=0.815:1.5+0.01, samples=10] plot ({\x*cos(120)+1.2},
	 {\x*sin(120)+0.75});
	 \draw[black] (0.6213,1.7524) -- (0.6213-0.005,1.7524+0.00866) node[sloped,pos=0.50,allow upside down]{\arrowIn}; ; 
	\draw [black,domain=0.582-0.22:2.25+0.01, samples=10] plot ({\x*cos(120)-1.2}, {\x*sin(120)+0.75});
	\draw[black] (-1.853,1.88) -- (-1.853+0.005,1.88-0.00866) node[sloped,pos=0.50,allow upside down]{\arrowIn}; ; 
	\draw [black,domain=-2.33:-1.13, samples=10] plot ({\x}, {2.699});
	\draw[black] (-1.73,2.699) -- (-1.7301,2.699) node[sloped,pos=0.50,allow upside down]{\arrowIn}; ; 
	\node[black, anchor=south] (a) at (-1.2,2.75) {$a$};
	\node[black, anchor=south] (a) at (0,2.75) {$b$};
	\node[black, anchor=north] (a) at (-1.5+0.3*3.5-0.05,3.25-0.5196*3.5) {$f_1$};
	\node[black, anchor=west] (a) at (0.17-0.02,0.42-0.02) {$f_2$};
	\draw[black] (0.45+0.01,2.049) -- (-0.75-0.01,2.049) node[sloped,pos=0.50,allow upside down]{\arrowIn}; ; 
	\draw[black] (-1.5*0.8*0.5625,0.75+2.598*0.8*0.5625) -- (-1.5*0.8*0.562,0.75+2.598*0.8*0.562) node[sloped,pos=1,allow upside down]{\arrowIn}; ; 
	\draw[black] (-1.5*0.8*0.26,0.75+2.598*0.8*0.26) -- (-1.5*0.8*0.25,0.75+2.598*0.8*0.25) node[sloped,pos=1,allow upside down]{\arrowIn}; ; 
	\draw[black] (0.1568,1.518) -- (0.1568-0.005,1.518+0.00866) node[sloped,pos=1,allow upside down]{\arrowIn}; ; 
	\draw[black] (-0.6+0.9*0.666+0.01,0.75+0.5196*2) -- (-1.2+0.3*2,0.75+0.5196*2) node[sloped,pos=0.50,allow upside down]{\arrowIn}; ; 
	\draw[black] (-1.1197,1.65) -- (-1.1197+0.005,1.65-0.00866) node[sloped,pos=1,allow upside down]{\arrowIn}; ; 
	\draw[black] (-0.15,2.57) -- (-0.15-0.005,2.57-0.00866) node[sloped,pos=1,allow upside down]{\arrowIn}; ; 
	\draw[black, line width=0.05pt] (0.707*0.8,0.707*0.75*0.8+0.15) -- (0.707*0.8+0.707*0.001,0.707*0.75*0.8-0.707*0.75*0.001+0.15) node[sloped,pos=0.50,allow upside down]{\arrowIn}; ; 
	\draw[black, line width=0.05pt] (0.707*1.5*1.074*0.8+0.707*1.5*1.074*0.001,0.707*1.2*1.074*0.8-0.707*1.2*1.074*0.001+0.15) -- (0.707*1.5*1.074*0.8,0.707*1.2*1.074*0.8+0.15) node[sloped,pos=0.50,allow upside down]{\arrowIn}; ; 
	\draw[black, line width=0.05pt] (0.707*2.222*0.8+0.707*2.222*0.001,0.707*1.83*0.8-0.707*1.83*0.001+0.15) -- (0.707*2.222*0.8,0.707*1.83*0.8+0.15) node[sloped,pos=0.50,allow upside down]{\arrowIn}; ; 
	\draw [black,domain=-235.5:73.25, samples=100] plot ({1.5*cos(\x)*1.074*0.8}, {1.2*sin(\x)*1.074*0.8+0.15});
	\draw [black,domain=-219.25:63.5, samples=100] plot ({2.222*cos(\x)*0.8}, {1.83*sin(\x)*0.8+0.15});
	\end{scope}
\end{tikzpicture}
\caption{When translating two anyons simulateously across cut $A$, it is not important to apply the rule summarized in Fig.~\ref{eq:translate_A} in a specific order. Applying it first on anyon $a$ (left diagram) leads to the same result as applying it first to anyon $b$ (right diagram).}
\label{eq:translate_A_multiple}
\end{figure*}
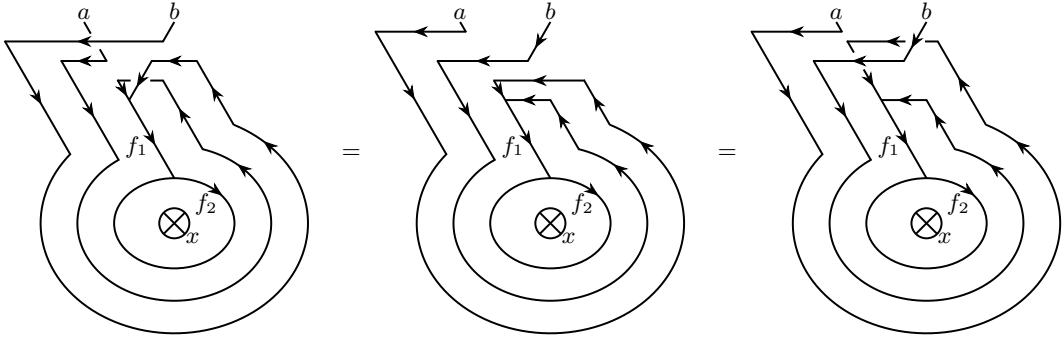

If multiple anyons are translated simultaneously over cut $B$, the resulting FDAV can be expressed in terms of multiple matrices $\widetilde{U}_x(\text{\boldmath$\alpha$})$ being multiplied onto $|\vec{\mathcal{A}}\rangle$. The reason for this can be seen when moving multiple anyons around the non-contractible loop in a fusion diagram. Successively applying $\widetilde{U}_x(\text{\boldmath$\alpha$})$ on $|\vec{\mathcal{A}}\rangle$ with the proper order of anyon charges in the arguments then takes care of the loops one after another, starting with the most inner one. In fact, such fusion diagrams are also obtained if $\mathcal{T}_x$ is applied multiple times onto a state, where for each application, no more than a single anyon crosses cut $B$. Thus, the application of $\mathcal{T}_x$ onto an arbitrary state can be written as
\begin{align}
\begin{split}
	\mathcal{T}_x|(\alpha_k,\vec{r}_k)_{k=1}^N,\vec{\mathcal{A}}\rangle = &|(\alpha_k,\vec{r}_k+\vec{e}_x)_{k=1}^N\rangle   \\&\otimes \prod_{i=0}^{n_x-1} \widetilde{U}_x (\sigma^i \left(\text{\boldmath$\alpha$} \right) ) |\vec{\mathcal{A}}\rangle .
\end{split}
	\label{eq:apply_Tx_state}
\end{align}
Here, $\vec{r}_k+\vec{e}_x$ is understood modulo the lattice size due to the PBC. The product $\prod_{i=0}^{j-1} \widetilde{U}_x\left(\sigma^i \left(\text{\boldmath$\alpha$}\right) \right)=\widetilde{U}_x\left(\sigma^{j-1} \left(\text{\boldmath$\alpha$}\right) \right)\widetilde{U}_x\left(\sigma^{j-2} \left(\text{\boldmath$\alpha$} \right)\right)...\widetilde{U}_x\left(\text{\boldmath$\alpha$} \right)$ is ordered, $\sigma(1,2,...,N)=(N,1,2,...,N-1)$ and $n_x$ is the number of anyons being translated over cut $B$ when applying $\mathcal{T}_x$.

From Eqs.~(\ref{eq:sheet_hopping_identity}) and~(\ref{eq:apply_Tx_state}), it can be seen that $\mathcal{T}_x^{L_x}=\mathbb{1}$, i.e., both the final anyon configuration and the final FDAV after applying $\mathcal{T}_x^{L_x}$ agree with the ones of the initial state. This is fairly obvious for the anyon configuration due to the PBC. That this also holds for the FDAVs can be seen when looking at an arbitrary fusion diagram as given in Eq.~(\ref{eq:general_state}): Applying $\mathcal{T}_x^{L_x}$ is in terms of the fusion diagrams equivalent to moving all $N$ anyons around the non-contractible loop, without any further braiding among them. As this is done with all anyons, the diagram is in fact equivalent to the initial one, as discussed in the context of Eq.~(\ref{eq:sheet_hopping_identity_tuples}).\\

The insight above has important consequences: If an anyon configuration on the lattice possesses a periodicity $p_x$ with $0<p_x<L_x$ such that applying $\mathcal{T}_x^{p_x}$ yields the same anyon configuration again (where one also has to consider the anyonic charges, not only the positions), the matrices $\widetilde{U}_x(\text{\boldmath$\alpha$})$ associated with these configurations have an additional property. Suppose that by applying $\mathcal{T}_x^{p_x}$, $n_x=Np_x/L_x$ anyons are translated over cut $B$. Then, it follows from Eq.~(\ref{eq:sheet_hopping_identity}) that
\begin{align}
	\left[\prod_{i=0}^{n_x-1} \widetilde{U}_x\left(\sigma^i \left(\text{\boldmath$\alpha$} \right) \right)\right]^{L_x/p_x} \mathcal{P}^x_{p_x} = \mathbb{1} \mathcal{P}^x_{p_x},
	\label{eq:Tx_id}
\end{align}
where $\mathcal{P}^x_{p_x}$ denotes the projector onto the basis states associated with anyon configurations possessing a periodicity of $p_x$ in $x$-direction. This implies that within this subspace, $\prod_{i=0}^{n_x-1} \widetilde{U}_x\left(\sigma^i \left(\text{\boldmath$\alpha$} \right)\right)$ is diagonalizable with eigenvalues $e^{2\pi i j p_x/L_x}$, $j\in\lbrace 0,1,...,L_x/p_x-1 \rbrace$. In particular, if there are $N$ \emph{identical} anyons on the lattice, the matrix $\widetilde{U}_x(\text{\boldmath$\alpha$})$ (there exists only one matrix in this case as each possible configuration of the charges $\lbrace \alpha_k  \rbrace$ is the same) is diagonalizable with eigenvalues $e^{2\pi i j /N}$, $j\in\lbrace 0,1,...,N-1 \rbrace$.

\subsubsection{The Translation Operator \texorpdfstring{$\mathcal{T}_y$}{Ty}}
\label{sec:translationoperators_y}

Applying the translation operator in $y$-direction $\mathcal{T}_y$ onto a state $|(\alpha_k,\vec{r}_k)_{k=1}^N,\vec{\mathcal{A}}\rangle$ shifts the positions of all anyons by one lattice spacing in positive $y$-direction and entails braiding in the fusion diagrams if an anyon crosses cut $A$. This braiding occurs as summarized in Fig.~\ref{eq:translate_A} in Sec.~\ref{sec:algo_cuts}. If multiple anyons are translated across cut $A$ simultaneously, this rule is applied for each of those anyons, where it is \emph{not} important to apply the rule to the relevant anyons in a specific order, as can be seen from the following argument.

Suppose that two anyons $a$ and $b$ are translated across cut $A$ simultaneously, as depicted in diagrammatic form in Fig.~\ref{eq:translate_A_multiple}. If we first apply the rule in Fig.~\ref{eq:translate_A} to anyon $a$ and then to anyon $b$, the fusion diagram on the left in Fig.~\ref{eq:translate_A_multiple} is obtained. On the other hand, if we first apply the rule to anyon $b$ and then to anyon $a$, the diagram on the right in Fig.~\ref{eq:translate_A_multiple} is obtained. In both cases, the line crossings can be resolved since the line of one anyon is always in front of / behind the line of the other one, which leads to the fusion diagram in the center of Fig.~\ref{eq:translate_A_multiple}. Note that the presence of additional anyons that are not translated across cut $A$ upon applying $\mathcal{T}_y$ does not change the above observation. It is straight forward to see from Fig.~\ref{eq:translate_A} that braids involving such anyons are independent from the order in which the rule in Fig.~\ref{eq:translate_A} is applied since the positions of these anyons in the fusion diagrams do not change.

With this argument, it becomes clear that anyons crossing cut $A$ do not braid with each other. These anyons, however, do braid with all other anyons, which is not unexpected, given the rule in Fig.~\ref{eq:translate_A}. Further, the same braided fusion diagram can be obtained if the anyons are translated over cut $A$ one after another, that is, $\mathcal{T}_y$ is applied multiple times and each time no more than a single anyon crosses cut $A$. This also implies that applying $\mathcal{T}_y^{L_y}$ moves all anyons around the non-contractible loop without braiding them with each other, as the above argument can be employed for each pair of anyons in this scenario. The resulting fusion diagram is thus equal to the one before the application of $\mathcal{T}_y^{L_y}$, implying that $\mathcal{T}_y^{L_y}=\mathbb{1}$ both in terms of the anyon configurations and the fusion diagrams.\\

In principle, one could now write the action of $\mathcal{T}_y$ onto an arbitrary state similar to Eq.~(\ref{eq:apply_Tx_state}) in terms of $B$- and $B'$-operators, the punctured torus $S$-matrix and the matrices $\widetilde{U}_x(\text{\boldmath$\alpha$})$. We will forgo this since it would not add any benefits as the operator string is already evident from the rule in Fig.~\ref{eq:translate_A} and the example in App.~\ref{app_example_computations}; it would be necessary to introduce many new auxiliary quantities in order to obtain a general, bloated expression that would have to be absorbed into another quantity anyways in order to maintain readable expressions. We will thus simply write
\begin{align}
\begin{split}
	\mathcal{T}_y|(\alpha_k,\vec{r}_k)_{k=1}^N,\vec{\mathcal{A}}\rangle =&|(\alpha_k,\vec{r}_k+\vec{e}_y)_{k=1}^N\rangle\\&\otimes \widetilde{U}_y( (\alpha_k,\vec{r}_k)_{k=1}^N)| \vec{\mathcal{A}}\rangle ,
\end{split}
	\label{eq:apply_Ty_state}
\end{align}
where $\vec{r}_k+\vec{e}_y$ is understood modulo the lattice size. $\widetilde{U}_y( (\alpha_k,\vec{r}_k)_{k=1}^N)$ are matrices that apply the rule in Fig.~\ref{eq:translate_A} to the superposition of fusion diagrams encoded in $|\vec{\mathcal{A}}\rangle$ for each anyon that is translated across cut $A$ when applying $\mathcal{T}_y$. These matrices do not only depend on the order of the anyons $\lbrace \alpha_k \rbrace$ in the initial state, but also on their positions. We generalize this to the ordered product $\prod_{i=0}^{j-1}\widetilde{U}_y( (\alpha_k,\vec{r}_k+i\vec{e}_y)_{k=1}^N )=\widetilde{U}_y( (\alpha_k,\vec{r}_k+(j-1)\vec{e}_y)_{k=1}^N)\widetilde{U}_y( (\alpha_k,\vec{r}_k+(j-2)\vec{e}_y)_{k=1}^N )\ldots \widetilde{U}_y( (\alpha_k,\vec{r}_k)_{k=1}^N )$ when applying $\mathcal{T}_y^j$.

It is to be emphasized that $\widetilde{U}_x(\text{\boldmath$\alpha$})$ as defined in Eq.~(\ref{eq:apply_Tx_state}) and $\widetilde{U}_y( (\alpha_k,\vec{r}_k)_{k=1}^N )$ as defined above are different in the sense that for each time an anyon crosses cut $B$, $\widetilde{U}_x(\text{\boldmath$\alpha$})$ is to be applied, i.e., the total number of $\widetilde{U}_x(\text{\boldmath$\alpha$})$ to be applied depends on the particle number. $\widetilde{U}_y( (\alpha_k,\vec{r}_k)_{k=1}^N)$ on the other hand is applied for each time $\mathcal{T}_y$ is acting on a state, which means the total number of $\widetilde{U}_y( (\alpha_k,\vec{r}_k)_{k=1}^N )$ to be applied does not depend on the particle number.\\

Similar to the translation in $x$-direction, an anyon configuration might be periodic in $y$-direction with a periodicity $p_y$ that is smaller than the lattice size, $0<p_y < L_y$. In this case,
\begin{align}
	\left[\prod_{i=0}^{p_y-1}\widetilde{U}_y( (\alpha_k,\vec{r}_k+i\vec{e}_y)_{k=1}^N )\right]^{L_y/p_y}\mathcal{P}^y_{p_y}=\mathbb{1}\mathcal{P}^y_{p_y}
	\label{eq:Ty_id}
\end{align}
holds with $\mathcal{P}^y_{p_y}$ being the projector onto the subspace of anyon configurations possessing a periodicity of $p_y$ in $y$-direction. This implies that $\prod_{i=0}^{p_y-1}\widetilde{U}_y( (\alpha_k,\vec{r}_k+i\vec{e}_y)_{k=1}^N )$ is diagonalizable with eigenvalues $e^{2\pi ijp_y/L_y}, j\in \lbrace 0,1,...,L_y/p_y-1 \rbrace$.

\subsubsection{Combining \texorpdfstring{$\mathcal{T}_x$}{Tx} and \texorpdfstring{$\mathcal{T}_y$}{Ty}}

We can further introduce the ``mixed'' periodicity $\mathbf{p}_{xy}\equiv (p_x',p_y')$ involving translations in both $x$- and $y$-direction, which is defined by
\begin{widetext}
\begin{align}
	(p_x',p_y')\equiv \min_i  \left(\min_j \left(\lbrace (i,j): T_y^j T_x^i| (\alpha_k,\vec{r}_k)_{k=1}^N \rangle=| (\alpha_k,\vec{r}_k)_{k=1}^N \rangle, 0<i\leq L_x,0<j\leq L_y \rbrace \right) \right),
	\label{eq:mixed_periodicity_definition}
\end{align}
\end{widetext}
where $p_y'\leq p_y$ holds. Here, the operators $T_x$ and $T_y$ correspond to the parts in the translation operators $\mathcal{T}_x$ and $\mathcal{T}_y$ that act on the anyon configurations and not on the fusion diagrams. Using $[\mathcal{T}_x,\mathcal{T}_y]=0$, which will be argued below in App.~\ref{sec:commutators}, it is evident that
\begin{align}
\begin{split}
	\Bigg[ \prod_{i=0}^{p_y'-1}\widetilde{U}_y( (&\alpha_k,\vec{r}_k+p_x'\vec{e}_x+i\vec{e}_y)_{k=1}^N ) \\& \times \prod_{i=0}^{n_x(p_x')-1} \widetilde{U}_x (\sigma^i \left(\text{\boldmath$\alpha$} \right)) \Bigg]^{\gamma}\mathcal{P}^{xy}_{p_x',p_y'}=\mathbb{1}\mathcal{P}^{xy}_{p_x',p_y'},
\end{split}
	\label{eq:mixed_p}
\end{align}
where $\gamma$ is the least common multiple of $L_x/p_x'$ and $L_y/p_y'$ and $n_x(p_x')$ again denotes the number of anyons being translated over cut $B$ when applying $\mathcal{T}_x^{p_x'}$. The projector $\mathcal{P}^{xy}_{p_x',p_y'}$ projects onto the subspace spanned by the anyon configurations possessing the mixed periodicity $(p_x',p_y')$. Similar to the other cases, it follows that $\prod_{i=0}^{p_y'-1}\widetilde{U}_y( (\alpha_k,\vec{r}_k+p_x'\vec{e}_x+i\vec{e}_y)_{k=1}^N )\prod_{i=0}^{n_x(p_x')-1} \widetilde{U}_x (\sigma^i \left(\text{\boldmath$\alpha$} \right))$ is diagonalizable with eigenvalues $e^{2\pi ij/\gamma}$, $j\in \lbrace 0,1,...,\gamma-1 \rbrace$.

\subsection{Construction of Momentum States}\label{sec:construction}

Using the above relations, we can now construct momentum states in $x$- and $y$-direction.

\subsubsection{Construction of Momentum States in \texorpdfstring{$x$}{x}-Direction}\label{app:momentum_x}

Let us start by first constructing momentum states that exploit the translational invariance in $x$-direction, for which property~(\ref{eq:Tx_id}) is crucial. These states are
\begin{widetext}
\begin{align}
\begin{split}
	|(\alpha_k,\vec{r}_k)_{k=1}^N,k_x,\vec{\mathcal{A}}\rangle &\equiv \frac{1}{\sqrt{N_x}}\sum_{j=0}^{L_x-1} e^{ijk_x}\mathcal{T}_x^j|(\alpha_k,\vec{r}_k)_{k=1}^N,\vec{\mathcal{A}}\rangle 
	\\ &= \frac{1}{\sqrt{N_x}}\sum_{j=0}^{L_x-1} e^{ijk_x}|(\alpha_k,\vec{r}_k+j \vec{e}_x)_{k=1}^N \rangle \otimes \prod_{i=0}^{n_x(j)-1} \widetilde{U}_x (\sigma^i \left(\text{\boldmath$\alpha$} \right) ) |\vec{\mathcal{A}}\rangle ,
\end{split}
\label{eq:mom_x_def}
\end{align}
\end{widetext}
where $n_x(j)$ denotes the number of anyons being translated across cut $B$ when applying $\mathcal{T}_x^j$ onto $|(\alpha_k,\vec{r}_k)_{k=1}^N,\vec{\mathcal{A}}\rangle $, $N_x=L_x^2/p_x$ the normalization constant and $k_x$ the momentum in $x$-direction with $k_x\in\lbrace 0,2\pi/L_x,...,2\pi(L_x-1)/L_x \rbrace$ if $p_x=L_x$ with the periodicity in $x$-direction $p_x$ (the case $p_x<L_x$ is considered below). $|(\alpha_k,\vec{r}_k)_{k=1}^N,\vec{\mathcal{A}}\rangle $ will be refered to as ``reference state'' of $|(\alpha_k,\vec{r}_k)_{k=1}^N,k_x,\vec{\mathcal{A}}\rangle $, as it is the state generating the momentum state by being translated. From the definition in Eq.~(\ref{eq:mom_x_def}), it can be seen that
\begin{align}
	\mathcal{T}_x|(\alpha_k,\vec{r}_k)_{k=1}^N,k_x,\vec{\mathcal{A}}\rangle  = e^{-ik_x}|(\alpha_k,\vec{r}_k)_{k=1}^N,k_x,\vec{\mathcal{A}}\rangle ,
\end{align}
as expected for a momentum state with momentum $k_x$.

If the anyon configuration of $|(\alpha_k,\vec{r}_k)_{k=1}^N,\vec{\mathcal{A}}\rangle$ in Eq.~(\ref{eq:mom_x_def}) is periodic in $x$-direction with a periodicity $p_x<L_x$, there is a constraint for the allowed momenta $k_x$. By choosing $|\vec{\mathcal{A}}\rangle$ to be an eigenvector of $\prod_{i=0}^{n_x(p_x)-1} \widetilde{U}_x\left(\sigma^i \left(\text{\boldmath$\alpha$} \right) \right)$,
\begin{align}
\begin{split}
	e^{ip_xk_x}e^{2\pi i \beta p_x/L_x} &=1\\\text{with}\quad \prod_{i=0}^{n_x(p_x)-1} \widetilde{U}_x\left(\sigma^i \left(\text{\boldmath$\alpha$} \right) \right)|\vec{\mathcal{A}}\rangle &=e^{2\pi i \beta p_x/L_x}|\vec{\mathcal{A}}\rangle
\end{split}
	\label{eq:mom_x_constraint}
\end{align}
has to be fulfilled, restricting the allowed momenta to $k_x=2\pi n/p_x-2\pi\beta /L_x$, $n\in\lbrace 0,1,...,p_x-1 \rbrace$. Here, $\beta\in\lbrace 0,1,...,L_x/p_x-1 \rbrace$ is an integer that is determined by the second equation in Eq.~(\ref{eq:mom_x_constraint}). The above constraint follows from Eq.~(\ref{eq:Tx_id}) and assures that $e^{ip_xk_x}\mathcal{T}_x^{p_x}|(\alpha_k,\vec{r}_k)_{k=1}^N,\vec{\mathcal{A}}\rangle =|(\alpha_k,\vec{r}_k)_{k=1}^N,\vec{\mathcal{A}}\rangle $. Note that the allowed values for $k_x$ are in sharp contrast to the respective values for fermionic and bosonic systems since the momenta for anyons are shifted by $2\pi\beta /L_x$, i.e., not all momenta are multiples of $2\pi /p_x$.

Thus, when constructing momentum states with momentum in $x$-direction, one first has to find the periodicity $p_x$ of each anyon configuration. Then, if $p_x<L_x$, eigenstates of $\prod_{i=0}^{n_x(p_x)-1} \widetilde{U}_x\left(\sigma^i \left(\text{\boldmath$\alpha$} \right) \right)$ have to be used as FDAVs. The momentum sectors in which such a state may exist are obtained by checking condition~(\ref{eq:mom_x_constraint}). If there are multiple eigenstates $|\vec{\mathcal{A}}\rangle$ to the same eigenvalue, it is to be assured that these eigenstates are pairwise orthonormal.

\subsubsection{Construction of Momentum States in \texorpdfstring{$x$}{x}- and \texorpdfstring{$y$}{y}-Direction}\label{app:momentum_xy}

Using the considerations for the translation operators discussed in App.~\ref{sec:translationoperators} and those in the previous section, we can now construct momentum states possessing the momentum $k_x$ in $x$-direction and $k_y$ in $y$-direction:
\begin{widetext}
\begin{align}
\begin{split}
	&| (\alpha_k,\vec{r}_k)_{k=1}^N , \vec{k},\vec{\mathcal{A}}\rangle  \equiv \frac{1}{\sqrt{N_{xy}}}\sum_{j=0}^{L_x-1}\sum_{l=0}^{L_y-1} e^{ijk_x+ilk_y}\mathcal{T}_y^l\mathcal{T}_x^j| (\alpha_k,\vec{r}_k)_{k=1}^N ,\vec{\mathcal{A}}\rangle \\ &= \frac{1}{\sqrt{N_{xy}}}\sum_{j=0}^{L_x-1} \sum_{l=0}^{L_y-1} e^{ijk_x+ilk_y}| (\alpha_k,\vec{r}_k+j\vec{e}_x+l\vec{e}_y)_{k=1}^N \rangle \otimes \prod_{i=0}^{l-1}\widetilde{U}_y( (\alpha_k,\vec{r}_k+j\vec{e}_x+i\vec{e}_y)_{k=1}^N ) \prod_{i=0}^{n_x(j)-1} \widetilde{U}_x (\sigma^i \left(\text{\boldmath$\alpha$} \right) ) |\vec{\mathcal{A}}\rangle ,
\end{split}
\label{eq:mom_xy_def}
\end{align}
\end{widetext}
where $\vec{k}=(k_x,k_y)^{\top}$ denotes the momentum vector and $N_{xy}=L_x^2L_y^2/p_xp_y'$ the normalization constant. When there are no additional constraints due to the periodicities $p_x$, $p_y$, $p_x'$ and $p_y'$, the allowed momenta are given by $k_{x(y)}\in \lbrace 0,2\pi/L_{x(y)},...,2\pi(L_{x(y)}-1)/L_{x(y)} \rbrace$. Similar to the momentum eigenstates in $x$-direction, for $| (\alpha_k,\vec{r}_k)_{k=1}^N , \vec{k}, \vec{\mathcal{A}}\rangle$,
\begin{align}
	\mathcal{T}_{x(y)}| (\alpha_k,\vec{r}_k)_{k=1}^N , \vec{k},\vec{\mathcal{A}}\rangle =e^{-ik_{x(y)}}| (\alpha_k,\vec{r}_k)_{k=1}^N, \vec{k},\vec{\mathcal{A}}\rangle 
\end{align}
holds. When constructing the momentum states for a reference state $| (\alpha_k,\vec{r}_k)_{k=1}^N ,\vec{\mathcal{A}}\rangle$ with non-trivial periodicities, there are multiple things to consider. First of all, if $p_x<L_x$, the same constraint as for momentum states in $x$-direction has to be fulfilled: $|\vec{\mathcal{A}}\rangle$ has to be chosen to be an eigenvector of $\prod_{i=0}^{n_x(p_x)-1} \widetilde{U}_x (\sigma^i \left(\text{\boldmath$\alpha$} \right) )$ and Eq.~(\ref{eq:mom_x_constraint}) has to be fulfilled, restricting the allowed momenta in $x$-direction to $k_x=2\pi n/p_x-2\pi\beta /L_x$, $n\in\lbrace 0,1,...,p_x-1 \rbrace$, where $\beta\in\lbrace 0,1,...,L_x/p_x-1 \rbrace$ is determined by the second equation in Eq.~(\ref{eq:mom_x_constraint}).

If $p_y<L_y$, $|\vec{\mathcal{A}}\rangle$ has to be chosen to be an eigenvector of $\prod_{i=0}^{p_y-1}\widetilde{U}_y( (\alpha_k,\vec{r}_k+i\vec{e}_y)_{k=1}^N )$ that fulfills
\begin{align}
\begin{split}
	e^{ip_yk_y}e^{2\pi i\beta'p_y/L_y}&=1 \\\text{with}\quad \prod_{i=0}^{p_y-1}\widetilde{U}_y((\alpha_k,\vec{r}_k+i\vec{e}_y)_{k=1}^N )|\vec{\mathcal{A}}\rangle&=e^{2\pi i \beta' p_y/L_y}|\vec{\mathcal{A}}\rangle,
\end{split}
	\label{eq:mom_xy_constraint1}
\end{align}
which restricts the allowed momenta in $y$-direction to $k_y=2\pi n/p_y-2\pi\beta' /L_y$, $n\in\lbrace 0,1,...,p_y-1 \rbrace$, where $\beta'\in\lbrace 0,1,...,L_y/p_y-1 \rbrace$ is determined by the second relation in Eq.~(\ref{eq:mom_xy_constraint1}).

Finally, if $p_y'<p_y$, one has to choose $|\vec{\mathcal{A}}\rangle$ to be an eigenvector of $\prod_{i=0}^{p_y'-1}\widetilde{U}_y( (\alpha_k,\vec{r}_k+p_x'\vec{e}_x+i\vec{e}_y)_{k=1}^N )\prod_{i=0}^{n_x(p_x')-1} \widetilde{U}_x\left(\sigma^i \left(\text{\boldmath$\alpha$} \right) \right)$ that fulfills
\begin{align}
\begin{split}
	e^{ip_x'k_x+ip_y'k_y}e^{2\pi i\beta''/\gamma}&=1 \\\text{with}\quad  \prod_{i=0}^{p_y'-1}\widetilde{U}_y( (\alpha_k,\vec{r}_k+p_x'\vec{e}_x+&i\vec{e}_y)_{k=1}^N )\\ \times\prod_{i=0}^{n_x(p_x')-1} \widetilde{U}_x\left(\sigma^i \left(\text{\boldmath$\alpha$} \right) \right)&|\vec{\mathcal{A}}\rangle=e^{2\pi i \beta''/\gamma}|\vec{\mathcal{A}}\rangle,
\end{split}
	\label{eq:mom_xy_constraint2}
\end{align}
where $\gamma$ again denotes the least common multiple of $L_x/p_x'$ and $L_y/p_y'$. This restricts the allowed momenta to $k_x=2\pi n/p_x'-2\pi \beta_x/L_x$ and $k_y=2\pi m/p_y'-2\pi \beta_y/L_y$, $n\in\lbrace 0,1,...,p_x'-1 \rbrace$, $m\in\lbrace 0,1,...,p_y'-1 \rbrace$ for all choices of $\beta_x$ and $ \beta_y $ such that $\exp(2\pi i(p_x'\beta_x/L_x+p_y'\beta_y/L_y))=\exp(2\pi i\beta''/\gamma)$, where $\beta_x$ and $\beta_y$ are integers fulfilling $0 \leq \beta_x< L_x/p_x'$ and $0 \leq \beta_y < L_y/p_y'$
and $\beta'' \in \lbrace 0,1,\ldots \gamma-1 \rbrace$ is determined by the second equation in Eq.~(\ref{eq:mom_xy_constraint2}). Note that if $p_y'=p_y$, this case is already covered by combining the constraints due to the periodicities in $x$- and $y$-direction. In this case, Eq.~(\ref{eq:mom_xy_constraint2}) does not correspond to an additonal constraint.\\

If a reference state $| (\alpha_k,\vec{r}_k)_{k=1}^N,\vec{\mathcal{A}}\rangle$ fulfills multiple of the conditions mentioned above ($p_x<L_x$, $p_y<L_y$, $p_y'<p_y$), the FDAVs $|\vec{\mathcal{A}}\rangle$ need to be eigenvectors of all the corresponding operators. The allowed momenta for the momentum states $| (\alpha_k,\vec{r}_k)_{k=1}^N, \vec{k},\vec{\mathcal{A}}\rangle $ are then restricted by all the respective constraints given by Eqs.~(\ref{eq:mom_x_constraint}), (\ref{eq:mom_xy_constraint1}) and / or (\ref{eq:mom_xy_constraint2}). Due to the commutation relation $[\mathcal{T}_x,\mathcal{T}_y]=0$ (see App.~\ref{sec:commutators} below), we can indeed choose the FDAVs to be simultaneous eigenstates of the relevant operators.

To summarize, when constructing the momentum states $| (\alpha_k,\vec{r}_k)_{k=1}^N , \vec{k},\vec{\mathcal{A}}\rangle $, we should first find the periodicities $p_x$, $p_y$, $p_x'$ and $p_y'$ of the anyon configuration and identify which of the three constraints discussed above need to be considered. Then, the FDAVs have to be chosen to be (simultaneous) eigenstates of the relevant operators, which also determines the allowed momenta. Different FDAVs to identical anyon configurations have to be orthogonal. Simultaneous eigenvectors of two operators can be found numerically using the Zassenhaus algorithm~\cite{LUKS1997335, fischer}, which computes a basis for the sum and the intersection of two subspaces of a vectorspace. In particular, we are interested in the intersection of the subspace of an operator $O_1$ corresponding to an eigenvalue $v_1$ with the subspace of another operator $O_2$ corresponding to some other eigenvalue $v_2$, where the vectorspace is spanned by the FDAVs. All the vectors in the intersection are therefore simultaneous eigenvectors of the two operators $O_1$ and $O_2$ to the associated eigenvalues $v_1$ and $v_2$. If for an anyon configuration, e.g., $p_x<L_x$, $p_y<L_y$ and $p_y'=p_y$, we can find the eigenvectors and therefore the subspaces of $\prod_{i=0}^{n_x(p_x)-1} \widetilde{U}_x\left(\sigma^i \left(\text{\boldmath$\alpha$} \right) \right)$ and $\prod_{i=0}^{p_y-1}\widetilde{U}_y( (\alpha_k,\vec{r}_k+i\vec{e}_y)_{k=1}^N )$ by diagonalization and then find the simultaneous eigenvectors in the intersections of all the combinations of subspaces to the different eigenvalues of the two operators using the Zassenhaus algorithm. When identifying the different subspaces, it is crucial to account for the numerical precision: In the diagonalization process, one might obtain two slightly different eigenvalues which in fact would agree with each other in an exact treatment. Such cases can be identified since we know the typical magnitude of numerical errors and already discussed the form of all possible eigenvalues to the relevant operators. We thus need to combine the corresponding eigenvectors in order to form to correct (higher dimensional) subspaces. Otherwise, we might not obtain the correct number of simultaneous eigenstates in the intersections.

The simultaneous eigenvectors of three operators can be found by applying the Zassenhaus algorithm a second time, where one subspace corresponds to an intersection of two subspaces of the first two operators, the other subspace corresponds to a subpace of the third operator, i.e., one subspace already corresponds to an eigenvalue pair of the first two operators, the second subspace to an eigenvalue of the third operator. With this method, it is possible to construct all allowed momentum states $| (\alpha_k,\vec{r}_k)_{k=1}^N , \vec{k},\vec{\mathcal{A}}\rangle $ to every anyon configuration.

\subsection{Existence and Completeness}\label{sec:completeness}

Here, we argue that the momentum states constructed above can indeed be used to block diagonalize the Hamiltonian $\mathcal{H}$. We start by showing the necessary commutation relations involving $\mathcal{H}$, $\mathcal{T}_x$ and $\mathcal{T}_y$ and then argue that the momentum states form bases of the Hilbert space.

\subsubsection{Commutator Relations}\label{sec:commutators}

First, let us note that the translation operator $\mathcal{T}_x$ as described in App.~\ref{sec:translationoperators_x} can be thought of as applying the part of the Hamiltonian $\mathcal{H}$ that translates anyons in positive $x$-direction in arbitrary order on each site at which an anyon is located. This point of view is straight forward to verify using the rules summarized in Figs.~\ref{eq:rule1} and \ref{eq:translate_B}. Due to the fact that \emph{all} anyons are translated, there is no braiding between different anyons involved; they can at most move acorss cut $B$, which corresponds to moving the corresponding lines around the non-contractible loop in the fusion diagrams. Similarly, we can see from App.~\ref{sec:translationoperators_y} that $\mathcal{T}_y$ can be thought of as applying the part of the Hamiltonian $\mathcal{H}$ that translates anyons in positive $y$-direction on each occupied site in arbitrary order. This time however, in the fusion diagrams, braiding occurs between anyons that cross cut $A$ and anyons that do not cross cut $A$.

From these considerations, it can be seen that $\mathcal{T}_x$ ($\mathcal{T}_y$) commutes with the part of $\mathcal{H}$ that translates anyons in $x$-direction ($y$-direction), that is, $\mathcal{T}_x T_{\vec{r}_i,\vec{r}_i+\vec{e}_x}  \mathcal{T}_x^{-1} = T_{\vec{r}_i+\vec{e}_x,\vec{r}_i+2\vec{e}_x}$ and $\mathcal{T}_y T_{\vec{r}_i,\vec{r}_i+\vec{e}_y}  \mathcal{T}_y^{-1} = T_{\vec{r}_i+\vec{e}_y,\vec{r}_i+2\vec{e}_y}$, where $T_{\vec{r}_i,\vec{r}_i+\vec{e}_x}$ and $T_{\vec{r}_i,\vec{r}_i+\vec{e}_y}$ are the translation operators in the Hamiltonian $\mathcal{H}$ in Eq.~(\ref{eq:hamiltonian_square_lattice}).

To argue that $\mathcal{T}_x T_{\vec{r}_i,\vec{r}_i+\vec{e}_y}  \mathcal{T}_x^{-1} = T_{\vec{r}_i+\vec{e}_x,\vec{r}_i+\vec{e}_x+\vec{e}_y}$, we note that the action of $T_{\vec{r}_i,\vec{r}_i+\vec{e}_y}$ is only non-trivial in terms of the fusion diagrams if the anyon located at $\vec{r}_i$ is translated across cut $A$. Even in this case, $\mathcal{T}_x T_{\vec{r}_i,\vec{r}_i+\vec{e}_y}  \mathcal{T}_x^{-1} = T_{\vec{r}_i+\vec{e}_x,\vec{r}_i+\vec{e}_x+\vec{e}_y}$ holds since $\mathcal{T}_x$ compensates all effects introduced by applying $\mathcal{T}_x^{-1}$. This can be understood intuitively by looking at Fig.~\ref{fig:intuitive_picture}. If the term $\mathcal{T}_x T_{\vec{r}_i,\vec{r}_i+\vec{e}_y}  \mathcal{T}_x^{-1}$ was not equal to $T_{\vec{r}_i+\vec{e}_x,\vec{r}_i+\vec{e}_x+\vec{e}_y}$, this inequality would need to show up in the fusion diagrams, that is, the braids would differ. This is equivalent to $\mathcal{T}_x T_{\vec{r}_i,\vec{r}_i+\vec{e}_y}  \mathcal{T}_x^{-1}T_{\vec{r}_i+\vec{e}_x,\vec{r}_i+\vec{e}_x+\vec{e}_y}^{-1} \neq \mathbb{1}$, which describes a process after which all anyons are again at their initial positions. This process can only be non-trivial in the fusion diargams if one anyon circles around another one. Since this is not possible due to $\mathcal{T}_x$ translating all anyons simultaneously, it follows that $\mathcal{T}_x T_{\vec{r}_i,\vec{r}_i+\vec{e}_y}  \mathcal{T}_x^{-1} = T_{\vec{r}_i+\vec{e}_x,\vec{r}_i+\vec{e}_x+\vec{e}_y}$.

Similarly, it follows that $\mathcal{T}_y T_{\vec{r}_i,\vec{r}_i+\vec{e}_x}  \mathcal{T}_y^{-1} = T_{\vec{r}_i+\vec{e}_y,\vec{r}_i+\vec{e}_x+\vec{e}_y}$ since after the process described by $\mathcal{T}_y T_{\vec{r}_i,\vec{r}_i+\vec{e}_x}  \mathcal{T}_y^{-1} T_{\vec{r}_i+\vec{e}_y,\vec{r}_i+\vec{e}_x+\vec{e}_y}^{-1}$, all anyons are back at their initial positions without any braids due to $\mathcal{T}_y$ translating all anyons simultaneously. Overall, it thus follows that $[\mathcal{H},\mathcal{T}_x]=[\mathcal{H},\mathcal{T}_y]=0$. Since both $\mathcal{T}_x$ and $\mathcal{T}_y$ commute with the parts in $\mathcal{H}$ that translate anyons along the different directions and can be viewed as these very parts acting on all the anyons, it is clear that $[\mathcal{T}_x,\mathcal{T}_y]=0$ holds. Together with the other commutation relations, it follows that $\mathcal{H}$ can be block diagonalized using the simultaneous eigenstates of $\mathcal{T}_x$ and $\mathcal{T}_y$ that are constructed in App.~\ref{sec:construction}.

\subsubsection{Completeness of Momentum States in \texorpdfstring{$x$}{x}-Direction}

Here, we show that the momentum states $|(\alpha_k,\vec{r}_k)_{k=1}^N,k_x,\vec{\mathcal{A}}\rangle $ introduced in Eq.~(\ref{eq:mom_x_def}) indeed form a basis of the Hilbert space, i.e., that different momentum states are orthogonal and that the number of momentum states equals the number of states in the original basis given by Eq.~(\ref{eq:realspacebasis}). As for the orthogonality, it is easy to see that two momentum states with unrelated reference states, meaning that the anyon configuration of one reference state can not be brought to the form of the anyon configuration of the other one using $\mathcal{T}_x$-operators, are orthogonal. We can thus focus on momentum states with related reference states and for those, we assume that the reference states possess identical anyon configurations. This does not further restrict the number of momentum states since a momentum state generated by a translated version of a reference state differs from the original one merely by a phase.
The overlap between two momentum states whose reference states have identical anyon configurations is given by
\begin{widetext}
\begin{align}
\begin{split}
	&\langle(\alpha_k,\vec{r}_k)_{k=1}^N,k'_x,\vec{\mathcal{A'}} |(\alpha_k,\vec{r}_k)_{k=1}^N,k_x,\vec{\mathcal{A}}\rangle \\
	&=\frac{1}{N_x}\sum_{j,j'=0}^{L_x-1} e^{i(jk_x-j'k_x')} \langle(\alpha_k,\vec{r}_k)_{k=1}^N,\vec{\mathcal{A'}}|  \mathcal{T}_x^{j-j'}|(\alpha_k,\vec{r}_k)_{k=1}^N,\vec{\mathcal{A}}\rangle \\
	&=\frac{1}{N_x}\sum_{j,j'=0}^{p_x-1}\sum_{l,l'=0}^{L_x/p_x-1} e^{i(jk_x-j'k_x')} \langle(\alpha_k,\vec{r}_k)_{k=1}^N,\vec{\mathcal{A'}}|  \mathcal{T}_x^{j-j'}|(\alpha_k,\vec{r}_k)_{k=1}^N,\vec{\mathcal{A}}\rangle =\frac{L_x^2/p_x}{N_x}\delta_{k_xk_x'}\langle \vec{\mathcal{A}'}|\vec{\mathcal{A}} \rangle.
\end{split}
\label{eq:mom_x_calc}
\end{align}
\end{widetext}
First, each sum ranging from $0$ to $L_x-1$ appearing in the momentum states can be divided into two sums, where the limits of the first one are $0$ and $p_x-1$ and the limits of the second one $0$ and $L_x/p_x-1$; both momentum states have the same periodicity since the reference states possess identical anyon configurations. Then, using that the overlap of the anyon configurations yields $\delta_{jj'}$, the sum over $j$ is $p_x\delta_{k_xk_x'}$. The latter holds because the allowed momenta for a momentum state with periodicity $p_x$ are $2\pi n/p_x-2\pi\beta /L_x$, $n\in\lbrace 0,1,...,p_x-1 \rbrace$. The relevant part in the sum is thus $\sum_{j=0}^{p_x-1}\exp{2\pi i(n-n')j/p_x}=p_x\delta_{n,n'}=p_x\delta_{k_xk_x'}$, where it was already used that the eigenvalue of $\prod_{i=0}^{n_x(p_x)-1} \widetilde{U}_x\left(\sigma^i \left(\text{\boldmath$\alpha$} \right) \right)$ for both of the two reference states is identical due to the overlap $\langle \vec{\mathcal{A}'}|\vec{\mathcal{A}}\rangle$. One finally arrives at the last expression in Eq.~(\ref{eq:mom_x_calc}), which states that both the momenta and the FDAVs have to be identical for a non-zero overlap. As we already argued that the anyon configuration also has to match, it follows that the momentum states are pairwise orthonormal. This calculation also shows that $N_x=L_x^2/p_x$.

There is a straight forward argument that the total number of momentum states $|(\alpha_k,\vec{r}_k)_{k=1}^N,k_x,\vec{\mathcal{A}}\rangle $ is the same as the total number of states $|(\alpha_k,\vec{r}_k)_{k=1}^N,\vec{\mathcal{A}}\rangle $ in the original basis: Every given anyon configuration possesses a periodicity $p_x\leq L_x$. By using a certain configuration for the reference state, one can generate $p_x$ different momentum states. At the same time, for the original basis, there are also $p_x$ states that have this exact anyon configuration with the only difference being that all anyons are translated in $x$-direction by $0,1,\ldots,p_x-1$ sites. Taking the FDAVs into account yields a similar result. Due to the fact that $\prod_{i=0}^{n_x(p_x)-1} \widetilde{U}_x\left(\sigma^i \left(\text{\boldmath$\alpha$} \right) \right)$ is diagonalizable, the number of eigenvectors coincides with the number of distinct FDAVs of the original basis and thus, it follows that the number of momentum states equals the number of states in the original basis. Therefore, the momentum states $|(\alpha_k,\vec{r}_k)_{k=1}^N,k_x,\vec{\mathcal{A}}\rangle $ indeed form a basis of the Hilbert space.

\subsubsection{Completeness of Momentum States in \texorpdfstring{$x$}{x}- and \texorpdfstring{$y$}{y}-Direction}

In order to show that the momentum states $| (\alpha_k,\vec{r}_k)_{k=1}^N , \vec{k},\vec{\mathcal{A}}\rangle $ also form a basis of the Hilbert space, we first show their orthogonality, similar to the procedure for the momentum states in $x$-direction $| (\alpha_k,\vec{r}_k)_{k=1}^N , k_x,\vec{\mathcal{A}}\rangle $. Momentum states generated by reference states whose anyon configurations can not be made to agree with each other by applying $\mathcal{T}_x$- and $\mathcal{T}_y$-operators are orthogonal. Thus, we assume in the following that the considered momentum states are generated by the same reference state. The overlap between two such momentum states with momenta $\vec{k}=(k_x,k_y)$ and $\vec{k}'=(k_x',k_y')$ is
\begin{widetext}
\begin{align}
\begin{split}
	&\langle (\alpha_k,\vec{r}_k)_{k=1}^N , \vec{k}',\vec{\mathcal{A}'} | (\alpha_k,\vec{r}_k)_{k=1}^N , \vec{k},\vec{\mathcal{A}}\rangle \\
	&=\frac{1}{N_{xy}}\sum_{j,j'=0}^{L_x-1}\sum_{l,l'=0}^{L_y-1} e^{i(jk_x-j'k_x')+i(lk_y-l'k_y')} \langle (\alpha_k,\vec{r}_k)_{k=1}^N ,\vec{\mathcal{A'}}|  \mathcal{T}_y^{l-l'}\mathcal{T}_x^{j-j'}| (\alpha_k,\vec{r}_k)_{k=1}^N ,\vec{\mathcal{A}}\rangle \\
	&=\frac{L_x^2L_y^2/p_xp_y'}{N_{xy}}\delta_{k_xk_x'}\delta_{k_yk_y'}\langle \vec{\mathcal{A}'}|\vec{\mathcal{A}} \rangle.
\end{split}
\label{eq:mom_xy_calc}
\end{align}
\end{widetext}
The calculation is similar to the one in Eq.~(\ref{eq:mom_x_calc}): Using the definition of the translation operators and the relations~(\ref{eq:mom_x_constraint}), (\ref{eq:mom_xy_constraint1}) and (\ref{eq:mom_xy_constraint2}), the sums over $j,j'$ and $l,l'$ yield $\delta_{k_xk_x'}L_x^2/p_x$ and $\delta_{k_yk_y'}L_y^2/p_y'$ and the FDAVs' scalar product. The periodicity $p_y'$ appears instead of $p_y$ since in the definition of the mixed periodicity in Eq.~(\ref{eq:mixed_periodicity_definition}), we first minimize with respect to the translations in $y$-direction. From Eq.~(\ref{eq:mom_xy_calc}), we see that the momentum states $| (\alpha_k,\vec{r}_k)_{k=1}^N , \vec{k},\vec{\mathcal{A}}\rangle $ are orthonormal.

There is also an argument that the total number of momentum states $| (\alpha_k,\vec{r}_k)_{k=1}^N , \vec{k},\vec{\mathcal{A}}\rangle $ agrees with the total number of basis states $| (\alpha_k,\vec{r}_k)_{k=1}^N ,\vec{\mathcal{A}}\rangle $. For a given anyon configuration, each combination of translations in $x$- and $y$-direction is also contained in the latter basis, where the number of translations in $x$- and $y$-direction have to be chosen such that double counting is avoided. If we now use this anyon configuration (or an arbitrary translation of it) for the reference state to generate all allowed momentum states, it can be seen from the constraints~(\ref{eq:mom_x_constraint}), (\ref{eq:mom_xy_constraint1}) and (\ref{eq:mom_xy_constraint2}) that the number of momentum states with distinct momentum quantum numbers agrees with the number of states $| (\alpha_k,\vec{r}_k)_{k=1}^N ,\vec{\mathcal{A}}\rangle $ that possess this anyon configuration modulo translations. A similar result is obtained when also considering the FDAVs. As $\mathcal{T}_x$ and $\mathcal{T}_y$ are unitary and commute, these operators can be simultaneously diagonalized. It follows that the number of eigenvectors agrees with the dimensions of the FDAVs and thus also with the number of states $| (\alpha_k,\vec{r}_k)_{k=1}^N ,\vec{\mathcal{A}}\rangle $. Using this result and the orthonormality, it is shown that the momentum states with momentum in $x$- and $y$-direction indeed form another basis of the Hilbert space.

\section{Constraints on the Phases Associated with Non-trivial Loops}\label{app:Verlinde}

As previously mentioned in Sec.~\ref{sec:algo_cuts}, there are different conventions~\cite{PhysRevB.43.10761, SemionsTorus, PhysRevB.86.155111, darragh_thesis} regarding which phase is to be picked up by the wave function when an anyon is translated along a non-trivial loop around the torus. This phase is independent of the presence of other particles, that is, it is also relevant for the single particle case and thus has to be considered in addition to the statistical effects among the anyons. It is clear that such phases cannot be model dependent if they are truly associated with the anyons themselves.

An answer to the question of which values such phases are allowed to take can be found using TQFT. The action of Wilson loop operators $W_a(C)$~\cite{doi:10.1063/1.4939783, Turaev+2016}, which act along closed loops $C$, corresponds to creating an anyon-antianyon pair ($a$ and $\overline{a}$), moving one of the anyons along $C$ and letting the anyons annihilate again. If translating anyons along non-trivial loops results in non-trivial phases, such phases must be consistent with the Wilson loop operators and their properties. In particular, the Wilson loop operators fulfil the Verlinde algebra~\cite{VERLINDE1988360}:
\begin{align}
	W_a(C)W_b(C) = \sum_c N_{ab}^c W_c(C). \label{eq:Verlinde}
\end{align}
Let us now assume that there are multiple choices for the phases we are interested in. Then, Eq.~(\ref{eq:Verlinde}) has to hold for each of these choices. We can thus replace\footnote{If we think of Wilson loop operators as consisting of local operators that translate anyons from one lattice site to a neighboring one (assuming there is one anyon localized on one of the two sites), we can interpret this replacement as introducing this phase only for the operators translating anyons over cuts $A$ and $B$ in our convention (see Fig.~\ref{fig:lattice}). In particular, this means that Wilson loop operators acting along trivial loops are not affected.} $W_a(C)\rightarrow \alpha_a W_a(C)$ ($|\alpha_a|=1$) to derive a constraint on the allowed phases. In general, this implies that $\alpha_c=\alpha_a \alpha_b$ if $N_{ab}^c > 0 $, which already fixes $\alpha_1=1$ (due to $N_{11}^1=1$) for the vacuum charge.

Let us focus on what the above implies for the models considered in this work. For abelian anyon models, using the notation introduced in Sec.~\ref{app_equi} (i.e., the vacuum charge is now denoted by $0$), we arrive at $\alpha_n=e^{2\pi i nl/M}$ with $l\in \lbrace 0,1,...,M-1 \rbrace$ for a model containing $M$ anyonic charges, where $M=2$ for the semionic anyon model. It has already been pointed out in Ref.~\cite{PhysRevB.43.10761} that all these choices for $\alpha_n$ are equivalent. For Fibonacci anyons, we obtain the equation $\alpha_\tau^2(W_1(C)+W_\tau(C))=W_1(C)+\alpha_\tau W_\tau(C)$, implying $\alpha_\tau=1$. For Ising anyons, using $N_{\sigma \psi}^\sigma = 1=N_{\sigma \sigma}^1=N_{\sigma\sigma}^\psi$, we get $\alpha_\psi = 1$ and $\alpha_\sigma = \pm 1$\footnote{Doing a quick analysis of the energy spectrum for different lattice sizes, we found no differences between the choices $\alpha_\sigma = 1$ and $\alpha_\sigma=-1$; the momenta are however shifted accordingly. This suggests that both choices are equivalent, similar to the abelian case.}.

Note that replacing the Wilson loop operators along the non-trivial loops $W_a(C)\rightarrow \alpha_a W_a(C)$ does also affect the $S$-matrix, see, e.g., the relevant equations in Ref.~\cite{VERLINDE1988360}. The $S$-matrix then includes both the mutual statistics of the anyons and these phases. Further, note that it is also possible to use different choices for $\lbrace \alpha_a \rbrace$ for the non-trivial loops along different directions, as long as consistency is ensured.\\

Having found whether or not the phases of interest are unique, we still need a method to actually determine them since the above considerations only show how one set of phases can be obtained from another one. We choose the determine the phases by considering string-net models~\cite{PhysRevB.71.045110}. For these models, acting with Wilson loop operators $W_a(C)$ onto the vacuum state only yields trivial phases, which in particular also holds for non-trivial closed loops (the action of $W_a(C)$ does have additional effects on the vacuum state, which are of no interest for this discussion). Specifically, there are string-net models that realize the doubled versions of the three anyon models we consider, meaning that we are indeed correct in setting these phase factors to one.

Note that the above discussion does \emph{not} necessarily mean that other choices for the phases, as found in, e.g., Refs.~\cite{SemionsTorus, PhysRevB.86.155111, darragh_thesis} are false. As already noted in Ref.~\cite{PhysRevB.86.155111}, the phases may actually be model-dependent. The Read-Rezayi state~\cite{PhysRevB.59.8084}, for example, exhibits a $U(1)$ symmetry that can be thought of as gauge freedom if the physical system does not choose a particular $U(1)$ sector.


\bibliography{./references.bib}

\begin{thebibliography}{87}%
\makeatletter
\providecommand \@ifxundefined [1]{%
 \@ifx{#1\undefined}
}%
\providecommand \@ifnum [1]{%
 \ifnum #1\expandafter \@firstoftwo
 \else \expandafter \@secondoftwo
 \fi
}%
\providecommand \@ifx [1]{%
 \ifx #1\expandafter \@firstoftwo
 \else \expandafter \@secondoftwo
 \fi
}%
\providecommand \natexlab [1]{#1}%
\providecommand \enquote  [1]{``#1''}%
\providecommand \bibnamefont  [1]{#1}%
\providecommand \bibfnamefont [1]{#1}%
\providecommand \citenamefont [1]{#1}%
\providecommand \href@noop [0]{\@secondoftwo}%
\providecommand \href [0]{\begingroup \@sanitize@url \@href}%
\providecommand \@href[1]{\@@startlink{#1}\@@href}%
\providecommand \@@href[1]{\endgroup#1\@@endlink}%
\providecommand \@sanitize@url [0]{\catcode `\\12\catcode `\$12\catcode
  `\&12\catcode `\#12\catcode `\^12\catcode `\_12\catcode `\%12\relax}%
\providecommand \@@startlink[1]{}%
\providecommand \@@endlink[0]{}%
\providecommand \url  [0]{\begingroup\@sanitize@url \@url }%
\providecommand \@url [1]{\endgroup\@href {#1}{\urlprefix }}%
\providecommand \urlprefix  [0]{URL }%
\providecommand \Eprint [0]{\href }%
\providecommand \doibase [0]{https://doi.org/}%
\providecommand \selectlanguage [0]{\@gobble}%
\providecommand \bibinfo  [0]{\@secondoftwo}%
\providecommand \bibfield  [0]{\@secondoftwo}%
\providecommand \translation [1]{[#1]}%
\providecommand \BibitemOpen [0]{}%
\providecommand \bibitemStop [0]{}%
\providecommand \bibitemNoStop [0]{.\EOS\space}%
\providecommand \EOS [0]{\spacefactor3000\relax}%
\providecommand \BibitemShut  [1]{\csname bibitem#1\endcsname}%
\let\auto@bib@innerbib\@empty
\bibitem [{\citenamefont {Leinaas}\ and\ \citenamefont
  {Myrheim}(1977)}]{Leinaas1977}%
  \BibitemOpen
  \bibfield  {author} {\bibinfo {author} {\bibfnamefont {J.~M.}\ \bibnamefont
  {Leinaas}}\ and\ \bibinfo {author} {\bibfnamefont {J.}~\bibnamefont
  {Myrheim}},\ }\bibfield  {title} {\bibinfo {title} {On the theory of
  identical particles},\ }\href {https://doi.org/10.1007/BF02727953} {\bibfield
   {journal} {\bibinfo  {journal} {Il Nuovo Cimento B (1971-1996)}\ }\textbf
  {\bibinfo {volume} {37}},\ \bibinfo {pages} {1} (\bibinfo {year}
  {1977})}\BibitemShut {NoStop}%
\bibitem [{\citenamefont {Wilczek}(1982{\natexlab{a}})}]{PhysRevLett.48.1144}%
  \BibitemOpen
  \bibfield  {author} {\bibinfo {author} {\bibfnamefont {F.}~\bibnamefont
  {Wilczek}},\ }\bibfield  {title} {\bibinfo {title} {{Magnetic Flux, Angular
  Momentum, and Statistics}},\ }\href
  {https://doi.org/10.1103/PhysRevLett.48.1144} {\bibfield  {journal} {\bibinfo
   {journal} {Phys. Rev. Lett.}\ }\textbf {\bibinfo {volume} {48}},\ \bibinfo
  {pages} {1144} (\bibinfo {year} {1982}{\natexlab{a}})}\BibitemShut {NoStop}%
\bibitem [{\citenamefont {Wilczek}(1982{\natexlab{b}})}]{PhysRevLett.49.957}%
  \BibitemOpen
  \bibfield  {author} {\bibinfo {author} {\bibfnamefont {F.}~\bibnamefont
  {Wilczek}},\ }\bibfield  {title} {\bibinfo {title} {{Quantum Mechanics of
  Fractional-Spin Particles}},\ }\href
  {https://doi.org/10.1103/PhysRevLett.49.957} {\bibfield  {journal} {\bibinfo
  {journal} {Phys. Rev. Lett.}\ }\textbf {\bibinfo {volume} {49}},\ \bibinfo
  {pages} {957} (\bibinfo {year} {1982}{\natexlab{b}})}\BibitemShut {NoStop}%
\bibitem [{\citenamefont {Nayak}\ \emph {et~al.}(2008)\citenamefont {Nayak},
  \citenamefont {Simon}, \citenamefont {Stern}, \citenamefont {Freedman},\ and\
  \citenamefont {Das~Sarma}}]{einleitung}%
  \BibitemOpen
  \bibfield  {author} {\bibinfo {author} {\bibfnamefont {C.}~\bibnamefont
  {Nayak}}, \bibinfo {author} {\bibfnamefont {S.~H.}\ \bibnamefont {Simon}},
  \bibinfo {author} {\bibfnamefont {A.}~\bibnamefont {Stern}}, \bibinfo
  {author} {\bibfnamefont {M.}~\bibnamefont {Freedman}},\ and\ \bibinfo
  {author} {\bibfnamefont {S.}~\bibnamefont {Das~Sarma}},\ }\bibfield  {title}
  {\bibinfo {title} {{Non-Abelian anyons and topological quantum
  computation}},\ }\href {https://doi.org/10.1103/RevModPhys.80.1083}
  {\bibfield  {journal} {\bibinfo  {journal} {Rev. Mod. Phys.}\ }\textbf
  {\bibinfo {volume} {80}},\ \bibinfo {pages} {1083} (\bibinfo {year}
  {2008})}\BibitemShut {NoStop}%
\bibitem [{\citenamefont {Field}\ and\ \citenamefont
  {Simula}(2018)}]{1802.06176}%
  \BibitemOpen
  \bibfield  {author} {\bibinfo {author} {\bibfnamefont {B.}~\bibnamefont
  {Field}}\ and\ \bibinfo {author} {\bibfnamefont {T.}~\bibnamefont {Simula}},\
  }\bibfield  {title} {\bibinfo {title} {{Introduction to topological quantum
  computation with non-Abelian anyons}},\ }\href
  {https://doi.org/10.1088/2058-9565/aacad2} {\bibfield  {journal} {\bibinfo
  {journal} {Quantum Science and Technology}\ }\textbf {\bibinfo {volume}
  {3}},\ \bibinfo {pages} {045004} (\bibinfo {year} {2018})}\BibitemShut
  {NoStop}%
\bibitem [{\citenamefont {Mong}\ \emph {et~al.}(2014)\citenamefont {Mong},
  \citenamefont {Clarke}, \citenamefont {Alicea}, \citenamefont {Lindner},
  \citenamefont {Fendley}, \citenamefont {Nayak}, \citenamefont {Oreg},
  \citenamefont {Stern}, \citenamefont {Berg}, \citenamefont {Shtengel},\ and\
  \citenamefont {Fisher}}]{PhysRevX.4.011036}%
  \BibitemOpen
  \bibfield  {author} {\bibinfo {author} {\bibfnamefont {R.~S.~K.}\
  \bibnamefont {Mong}}, \bibinfo {author} {\bibfnamefont {D.~J.}\ \bibnamefont
  {Clarke}}, \bibinfo {author} {\bibfnamefont {J.}~\bibnamefont {Alicea}},
  \bibinfo {author} {\bibfnamefont {N.~H.}\ \bibnamefont {Lindner}}, \bibinfo
  {author} {\bibfnamefont {P.}~\bibnamefont {Fendley}}, \bibinfo {author}
  {\bibfnamefont {C.}~\bibnamefont {Nayak}}, \bibinfo {author} {\bibfnamefont
  {Y.}~\bibnamefont {Oreg}}, \bibinfo {author} {\bibfnamefont {A.}~\bibnamefont
  {Stern}}, \bibinfo {author} {\bibfnamefont {E.}~\bibnamefont {Berg}},
  \bibinfo {author} {\bibfnamefont {K.}~\bibnamefont {Shtengel}},\ and\
  \bibinfo {author} {\bibfnamefont {M.~P.~A.}\ \bibnamefont {Fisher}},\
  }\bibfield  {title} {\bibinfo {title} {{Universal Topological Quantum
  Computation from a Superconductor-Abelian Quantum Hall Heterostructure}},\
  }\href {https://doi.org/10.1103/PhysRevX.4.011036} {\bibfield  {journal}
  {\bibinfo  {journal} {Phys. Rev. X}\ }\textbf {\bibinfo {volume} {4}},\
  \bibinfo {pages} {011036} (\bibinfo {year} {2014})}\BibitemShut {NoStop}%
\bibitem [{\citenamefont {Kitaev}(2003)}]{KITAEV20032}%
  \BibitemOpen
  \bibfield  {author} {\bibinfo {author} {\bibfnamefont {A.}~\bibnamefont
  {Kitaev}},\ }\bibfield  {title} {\bibinfo {title} {Fault-tolerant quantum
  computation by anyons},\ }\href
  {https://doi.org/https://doi.org/10.1016/S0003-4916(02)00018-0} {\bibfield
  {journal} {\bibinfo  {journal} {Annals of Physics}\ }\textbf {\bibinfo
  {volume} {303}},\ \bibinfo {pages} {2} (\bibinfo {year} {2003})}\BibitemShut
  {NoStop}%
\bibitem [{\citenamefont {Freedman}\ \emph {et~al.}(2002)\citenamefont
  {Freedman}, \citenamefont {Larsen},\ and\ \citenamefont
  {Wang}}]{freedman2002modular}%
  \BibitemOpen
  \bibfield  {author} {\bibinfo {author} {\bibfnamefont {M.~H.}\ \bibnamefont
  {Freedman}}, \bibinfo {author} {\bibfnamefont {M.}~\bibnamefont {Larsen}},\
  and\ \bibinfo {author} {\bibfnamefont {Z.}~\bibnamefont {Wang}},\ }\bibfield
  {title} {\bibinfo {title} {A modular functor which is universal for quantum
  computation},\ }\href {https://doi.org/https://doi.org/10.1007/s002200200645}
  {\bibfield  {journal} {\bibinfo  {journal} {Communications in Mathematical
  Physics}\ }\textbf {\bibinfo {volume} {227}},\ \bibinfo {pages} {605}
  (\bibinfo {year} {2002})}\BibitemShut {NoStop}%
\bibitem [{\citenamefont {Savary}\ and\ \citenamefont
  {Balents}(2016)}]{Savary2016}%
  \BibitemOpen
  \bibfield  {author} {\bibinfo {author} {\bibfnamefont {L.}~\bibnamefont
  {Savary}}\ and\ \bibinfo {author} {\bibfnamefont {L.}~\bibnamefont
  {Balents}},\ }\bibfield  {title} {\bibinfo {title} {Quantum spin liquids: a
  review},\ }\href {https://doi.org/10.1088/0034-4885/80/1/016502} {\bibfield
  {journal} {\bibinfo  {journal} {Reports on Progress in Physics}\ }\textbf
  {\bibinfo {volume} {80}},\ \bibinfo {pages} {016502} (\bibinfo {year}
  {2016})}\BibitemShut {NoStop}%
\bibitem [{\citenamefont {Levin}\ and\ \citenamefont
  {Wen}(2005)}]{PhysRevB.71.045110}%
  \BibitemOpen
  \bibfield  {author} {\bibinfo {author} {\bibfnamefont {M.~A.}\ \bibnamefont
  {Levin}}\ and\ \bibinfo {author} {\bibfnamefont {X.-G.}\ \bibnamefont
  {Wen}},\ }\bibfield  {title} {\bibinfo {title} {{String-net condensation: A
  physical mechanism for topological phases}},\ }\href
  {https://doi.org/10.1103/PhysRevB.71.045110} {\bibfield  {journal} {\bibinfo
  {journal} {Phys. Rev. B}\ }\textbf {\bibinfo {volume} {71}},\ \bibinfo
  {pages} {045110} (\bibinfo {year} {2005})}\BibitemShut {NoStop}%
\bibitem [{\citenamefont {Kitaev}\ and\ \citenamefont
  {Laumann}(2009)}]{0904.2771}%
  \BibitemOpen
  \bibfield  {author} {\bibinfo {author} {\bibfnamefont {A.}~\bibnamefont
  {Kitaev}}\ and\ \bibinfo {author} {\bibfnamefont {C.}~\bibnamefont
  {Laumann}},\ }\href@noop {} {\bibinfo {title} {Topological phases and quantum
  computation}} (\bibinfo {year} {2009}),\ \Eprint
  {https://arxiv.org/abs/arXiv:0904.2771} {arXiv:0904.2771} \BibitemShut
  {NoStop}%
\bibitem [{\citenamefont {Balents}\ \emph {et~al.}(2002)\citenamefont
  {Balents}, \citenamefont {Fisher},\ and\ \citenamefont
  {Girvin}}]{PhysRevB.65.224412}%
  \BibitemOpen
  \bibfield  {author} {\bibinfo {author} {\bibfnamefont {L.}~\bibnamefont
  {Balents}}, \bibinfo {author} {\bibfnamefont {M.~P.~A.}\ \bibnamefont
  {Fisher}},\ and\ \bibinfo {author} {\bibfnamefont {S.~M.}\ \bibnamefont
  {Girvin}},\ }\bibfield  {title} {\bibinfo {title} {Fractionalization in an
  easy-axis kagome antiferromagnet},\ }\href
  {https://doi.org/10.1103/PhysRevB.65.224412} {\bibfield  {journal} {\bibinfo
  {journal} {Phys. Rev. B}\ }\textbf {\bibinfo {volume} {65}},\ \bibinfo
  {pages} {224412} (\bibinfo {year} {2002})}\BibitemShut {NoStop}%
\bibitem [{\citenamefont {Hermele}\ \emph {et~al.}(2004)\citenamefont
  {Hermele}, \citenamefont {Fisher},\ and\ \citenamefont
  {Balents}}]{PhysRevB.69.064404}%
  \BibitemOpen
  \bibfield  {author} {\bibinfo {author} {\bibfnamefont {M.}~\bibnamefont
  {Hermele}}, \bibinfo {author} {\bibfnamefont {M.~P.~A.}\ \bibnamefont
  {Fisher}},\ and\ \bibinfo {author} {\bibfnamefont {L.}~\bibnamefont
  {Balents}},\ }\bibfield  {title} {\bibinfo {title} {{Pyrochlore photons: The
  $U(1)$ spin liquid in a $S=\frac{1}{2}$ three-dimensional frustrated
  magnet}},\ }\href {https://doi.org/10.1103/PhysRevB.69.064404} {\bibfield
  {journal} {\bibinfo  {journal} {Phys. Rev. B}\ }\textbf {\bibinfo {volume}
  {69}},\ \bibinfo {pages} {064404} (\bibinfo {year} {2004})}\BibitemShut
  {NoStop}%
\bibitem [{\citenamefont {Laughlin}(1983)}]{PhysRevLett.50.1395}%
  \BibitemOpen
  \bibfield  {author} {\bibinfo {author} {\bibfnamefont {R.~B.}\ \bibnamefont
  {Laughlin}},\ }\bibfield  {title} {\bibinfo {title} {{Anomalous Quantum Hall
  Effect: An Incompressible Quantum Fluid with Fractionally Charged
  Excitations}},\ }\href {https://doi.org/10.1103/PhysRevLett.50.1395}
  {\bibfield  {journal} {\bibinfo  {journal} {Phys. Rev. Lett.}\ }\textbf
  {\bibinfo {volume} {50}},\ \bibinfo {pages} {1395} (\bibinfo {year}
  {1983})}\BibitemShut {NoStop}%
\bibitem [{\citenamefont {Halperin}(1984)}]{PhysRevLett.52.1583}%
  \BibitemOpen
  \bibfield  {author} {\bibinfo {author} {\bibfnamefont {B.~I.}\ \bibnamefont
  {Halperin}},\ }\bibfield  {title} {\bibinfo {title} {{Statistics of
  Quasiparticles and the Hierarchy of Fractional Quantized Hall States}},\
  }\href {https://doi.org/10.1103/PhysRevLett.52.1583} {\bibfield  {journal}
  {\bibinfo  {journal} {Phys. Rev. Lett.}\ }\textbf {\bibinfo {volume} {52}},\
  \bibinfo {pages} {1583} (\bibinfo {year} {1984})}\BibitemShut {NoStop}%
\bibitem [{\citenamefont {Read}\ and\ \citenamefont
  {Rezayi}(1999)}]{PhysRevB.59.8084}%
  \BibitemOpen
  \bibfield  {author} {\bibinfo {author} {\bibfnamefont {N.}~\bibnamefont
  {Read}}\ and\ \bibinfo {author} {\bibfnamefont {E.}~\bibnamefont {Rezayi}},\
  }\bibfield  {title} {\bibinfo {title} {{Beyond paired quantum Hall states:
  Parafermions and incompressible states in the first excited Landau level}},\
  }\href {https://doi.org/10.1103/PhysRevB.59.8084} {\bibfield  {journal}
  {\bibinfo  {journal} {Phys. Rev. B}\ }\textbf {\bibinfo {volume} {59}},\
  \bibinfo {pages} {8084} (\bibinfo {year} {1999})}\BibitemShut {NoStop}%
\bibitem [{\citenamefont {Stern}(2008)}]{STERN2008204}%
  \BibitemOpen
  \bibfield  {author} {\bibinfo {author} {\bibfnamefont {A.}~\bibnamefont
  {Stern}},\ }\bibfield  {title} {\bibinfo {title} {{Anyons and the quantum
  Hall effect—A pedagogical review}},\ }\href
  {https://doi.org/https://doi.org/10.1016/j.aop.2007.10.008} {\bibfield
  {journal} {\bibinfo  {journal} {Annals of Physics}\ }\textbf {\bibinfo
  {volume} {323}},\ \bibinfo {pages} {204} (\bibinfo {year} {2008})},\ \bibinfo
  {note} {january Special Issue 2008}\BibitemShut {NoStop}%
\bibitem [{\citenamefont {Bartolomei}\ \emph {et~al.}(2020)\citenamefont
  {Bartolomei}, \citenamefont {Kumar}, \citenamefont {Bisognin}, \citenamefont
  {Marguerite}, \citenamefont {Berroir}, \citenamefont {Bocquillon},
  \citenamefont {Plaçais}, \citenamefont {Cavanna}, \citenamefont {Dong},
  \citenamefont {Gennser}, \citenamefont {Jin},\ and\ \citenamefont
  {Fève}}]{doi:10.1126/science.aaz5601}%
  \BibitemOpen
  \bibfield  {author} {\bibinfo {author} {\bibfnamefont {H.}~\bibnamefont
  {Bartolomei}}, \bibinfo {author} {\bibfnamefont {M.}~\bibnamefont {Kumar}},
  \bibinfo {author} {\bibfnamefont {R.}~\bibnamefont {Bisognin}}, \bibinfo
  {author} {\bibfnamefont {A.}~\bibnamefont {Marguerite}}, \bibinfo {author}
  {\bibfnamefont {J.-M.}\ \bibnamefont {Berroir}}, \bibinfo {author}
  {\bibfnamefont {E.}~\bibnamefont {Bocquillon}}, \bibinfo {author}
  {\bibfnamefont {B.}~\bibnamefont {Plaçais}}, \bibinfo {author}
  {\bibfnamefont {A.}~\bibnamefont {Cavanna}}, \bibinfo {author} {\bibfnamefont
  {Q.}~\bibnamefont {Dong}}, \bibinfo {author} {\bibfnamefont {U.}~\bibnamefont
  {Gennser}}, \bibinfo {author} {\bibfnamefont {Y.}~\bibnamefont {Jin}},\ and\
  \bibinfo {author} {\bibfnamefont {G.}~\bibnamefont {Fève}},\ }\bibfield
  {title} {\bibinfo {title} {Fractional statistics in anyon collisions},\
  }\href {https://doi.org/10.1126/science.aaz5601} {\bibfield  {journal}
  {\bibinfo  {journal} {Science}\ }\textbf {\bibinfo {volume} {368}},\ \bibinfo
  {pages} {173} (\bibinfo {year} {2020})}\BibitemShut {NoStop}%
\bibitem [{\citenamefont {Kasahara}\ \emph {et~al.}(2018)\citenamefont
  {Kasahara}, \citenamefont {Ohnishi}, \citenamefont {Mizukami}, \citenamefont
  {Tanaka}, \citenamefont {Ma}, \citenamefont {Sugii}, \citenamefont {Kurita},
  \citenamefont {Tanaka}, \citenamefont {Nasu}, \citenamefont {Motome},
  \citenamefont {Shibauchi},\ and\ \citenamefont {Matsuda}}]{Kasahara2018}%
  \BibitemOpen
  \bibfield  {author} {\bibinfo {author} {\bibfnamefont {Y.}~\bibnamefont
  {Kasahara}}, \bibinfo {author} {\bibfnamefont {T.}~\bibnamefont {Ohnishi}},
  \bibinfo {author} {\bibfnamefont {Y.}~\bibnamefont {Mizukami}}, \bibinfo
  {author} {\bibfnamefont {O.}~\bibnamefont {Tanaka}}, \bibinfo {author}
  {\bibfnamefont {S.}~\bibnamefont {Ma}}, \bibinfo {author} {\bibfnamefont
  {K.}~\bibnamefont {Sugii}}, \bibinfo {author} {\bibfnamefont
  {N.}~\bibnamefont {Kurita}}, \bibinfo {author} {\bibfnamefont
  {H.}~\bibnamefont {Tanaka}}, \bibinfo {author} {\bibfnamefont
  {J.}~\bibnamefont {Nasu}}, \bibinfo {author} {\bibfnamefont {Y.}~\bibnamefont
  {Motome}}, \bibinfo {author} {\bibfnamefont {T.}~\bibnamefont {Shibauchi}},\
  and\ \bibinfo {author} {\bibfnamefont {Y.}~\bibnamefont {Matsuda}},\
  }\bibfield  {title} {\bibinfo {title} {{Majorana quantization and
  half-integer thermal quantum Hall effect in a Kitaev spin liquid}},\ }\href
  {https://doi.org/10.1038/s41586-018-0274-0} {\bibfield  {journal} {\bibinfo
  {journal} {Nature}\ }\textbf {\bibinfo {volume} {559}},\ \bibinfo {pages}
  {227} (\bibinfo {year} {2018})}\BibitemShut {NoStop}%
\bibitem [{\citenamefont {Yokoi}\ \emph {et~al.}(2021)\citenamefont {Yokoi},
  \citenamefont {Ma}, \citenamefont {Kasahara}, \citenamefont {Kasahara},
  \citenamefont {Shibauchi}, \citenamefont {Kurita}, \citenamefont {Tanaka},
  \citenamefont {Nasu}, \citenamefont {Motome}, \citenamefont {Hickey},
  \citenamefont {Trebst},\ and\ \citenamefont
  {Matsuda}}]{doi:10.1126/science.aay5551}%
  \BibitemOpen
  \bibfield  {author} {\bibinfo {author} {\bibfnamefont {T.}~\bibnamefont
  {Yokoi}}, \bibinfo {author} {\bibfnamefont {S.}~\bibnamefont {Ma}}, \bibinfo
  {author} {\bibfnamefont {Y.}~\bibnamefont {Kasahara}}, \bibinfo {author}
  {\bibfnamefont {S.}~\bibnamefont {Kasahara}}, \bibinfo {author}
  {\bibfnamefont {T.}~\bibnamefont {Shibauchi}}, \bibinfo {author}
  {\bibfnamefont {N.}~\bibnamefont {Kurita}}, \bibinfo {author} {\bibfnamefont
  {H.}~\bibnamefont {Tanaka}}, \bibinfo {author} {\bibfnamefont
  {J.}~\bibnamefont {Nasu}}, \bibinfo {author} {\bibfnamefont {Y.}~\bibnamefont
  {Motome}}, \bibinfo {author} {\bibfnamefont {C.}~\bibnamefont {Hickey}},
  \bibinfo {author} {\bibfnamefont {S.}~\bibnamefont {Trebst}},\ and\ \bibinfo
  {author} {\bibfnamefont {Y.}~\bibnamefont {Matsuda}},\ }\bibfield  {title}
  {\bibinfo {title} {Half-integer quantized anomalous thermal {H}all effect in
  the {K}itaev material candidate $\alpha$-$\mathrm{RuCl}_3$},\ }\href
  {https://doi.org/10.1126/science.aay5551} {\bibfield  {journal} {\bibinfo
  {journal} {Science}\ }\textbf {\bibinfo {volume} {373}},\ \bibinfo {pages}
  {568} (\bibinfo {year} {2021})}\BibitemShut {NoStop}%
\bibitem [{\citenamefont {Bruin}\ \emph {et~al.}(2022)\citenamefont {Bruin},
  \citenamefont {Claus}, \citenamefont {Matsumoto}, \citenamefont {Kurita},
  \citenamefont {Tanaka},\ and\ \citenamefont {Takagi}}]{2104.12184}%
  \BibitemOpen
  \bibfield  {author} {\bibinfo {author} {\bibfnamefont {J.~A.~N.}\
  \bibnamefont {Bruin}}, \bibinfo {author} {\bibfnamefont {R.~R.}\ \bibnamefont
  {Claus}}, \bibinfo {author} {\bibfnamefont {Y.}~\bibnamefont {Matsumoto}},
  \bibinfo {author} {\bibfnamefont {N.}~\bibnamefont {Kurita}}, \bibinfo
  {author} {\bibfnamefont {H.}~\bibnamefont {Tanaka}},\ and\ \bibinfo {author}
  {\bibfnamefont {H.}~\bibnamefont {Takagi}},\ }\bibfield  {title} {\bibinfo
  {title} {Robustness of the thermal {H}all effect close to half-quantization
  in $\alpha$-$\mathrm{RuCl}_3$},\ }\href
  {https://doi.org/10.1038/s41567-021-01501-y} {\bibfield  {journal} {\bibinfo
  {journal} {Nature Physics}\ }\textbf {\bibinfo {volume} {18}},\ \bibinfo
  {pages} {401} (\bibinfo {year} {2022})}\BibitemShut {NoStop}%
\bibitem [{\citenamefont {Satzinger}\ \emph {et~al.}(2021)\citenamefont
  {Satzinger}, \citenamefont {Liu}, \citenamefont {Smith}, \citenamefont
  {Knapp}, \citenamefont {Newman}, \citenamefont {Jones}, \citenamefont {Chen},
  \citenamefont {Quintana}, \citenamefont {Mi}, \citenamefont {Dunsworth},
  \citenamefont {Gidney}, \citenamefont {Aleiner}, \citenamefont {Arute},
  \citenamefont {Arya}, \citenamefont {Atalaya}, \citenamefont {Babbush},
  \citenamefont {Bardin}, \citenamefont {Barends}, \citenamefont {Basso},
  \citenamefont {Bengtsson}, \citenamefont {Bilmes}, \citenamefont {Broughton},
  \citenamefont {Buckley}, \citenamefont {Buell}, \citenamefont {Burkett},
  \citenamefont {Bushnell}, \citenamefont {Chiaro}, \citenamefont {Collins},
  \citenamefont {Courtney}, \citenamefont {Demura}, \citenamefont {Derk},
  \citenamefont {Eppens}, \citenamefont {Erickson}, \citenamefont {Faoro},
  \citenamefont {Farhi}, \citenamefont {Fowler}, \citenamefont {Foxen},
  \citenamefont {Giustina}, \citenamefont {Greene}, \citenamefont {Gross},
  \citenamefont {Harrigan}, \citenamefont {Harrington}, \citenamefont {Hilton},
  \citenamefont {Hong}, \citenamefont {Huang}, \citenamefont {Huggins},
  \citenamefont {Ioffe}, \citenamefont {Isakov}, \citenamefont {Jeffrey},
  \citenamefont {Jiang}, \citenamefont {Kafri}, \citenamefont {Kechedzhi},
  \citenamefont {Khattar}, \citenamefont {Kim}, \citenamefont {Klimov},
  \citenamefont {Korotkov}, \citenamefont {Kostritsa}, \citenamefont
  {Landhuis}, \citenamefont {Laptev}, \citenamefont {Locharla}, \citenamefont
  {Lucero}, \citenamefont {Martin}, \citenamefont {McClean}, \citenamefont
  {McEwen}, \citenamefont {Miao}, \citenamefont {Mohseni}, \citenamefont
  {Montazeri}, \citenamefont {Mruczkiewicz}, \citenamefont {Mutus},
  \citenamefont {Naaman}, \citenamefont {Neeley}, \citenamefont {Neill},
  \citenamefont {Niu}, \citenamefont {O’Brien}, \citenamefont {Opremcak},
  \citenamefont {Pató}, \citenamefont {Petukhov}, \citenamefont {Rubin},
  \citenamefont {Sank}, \citenamefont {Shvarts}, \citenamefont {Strain},
  \citenamefont {Szalay}, \citenamefont {Villalonga}, \citenamefont {White},
  \citenamefont {Yao}, \citenamefont {Yeh}, \citenamefont {Yoo}, \citenamefont
  {Zalcman}, \citenamefont {Neven}, \citenamefont {Boixo}, \citenamefont
  {Megrant}, \citenamefont {Chen}, \citenamefont {Kelly}, \citenamefont
  {Smelyanskiy}, \citenamefont {Kitaev}, \citenamefont {Knap}, \citenamefont
  {Pollmann},\ and\ \citenamefont {Roushan}}]{doi:10.1126/science.abi8378}%
  \BibitemOpen
  \bibfield  {author} {\bibinfo {author} {\bibfnamefont {K.~J.}\ \bibnamefont
  {Satzinger}}, \bibinfo {author} {\bibfnamefont {Y.-J.}\ \bibnamefont {Liu}},
  \bibinfo {author} {\bibfnamefont {A.}~\bibnamefont {Smith}}, \bibinfo
  {author} {\bibfnamefont {C.}~\bibnamefont {Knapp}}, \bibinfo {author}
  {\bibfnamefont {M.}~\bibnamefont {Newman}}, \bibinfo {author} {\bibfnamefont
  {C.}~\bibnamefont {Jones}}, \bibinfo {author} {\bibfnamefont
  {Z.}~\bibnamefont {Chen}}, \bibinfo {author} {\bibfnamefont {C.}~\bibnamefont
  {Quintana}}, \bibinfo {author} {\bibfnamefont {X.}~\bibnamefont {Mi}},
  \bibinfo {author} {\bibfnamefont {A.}~\bibnamefont {Dunsworth}}, \bibinfo
  {author} {\bibfnamefont {C.}~\bibnamefont {Gidney}}, \bibinfo {author}
  {\bibfnamefont {I.}~\bibnamefont {Aleiner}}, \bibinfo {author} {\bibfnamefont
  {F.}~\bibnamefont {Arute}}, \bibinfo {author} {\bibfnamefont
  {K.}~\bibnamefont {Arya}}, \bibinfo {author} {\bibfnamefont {J.}~\bibnamefont
  {Atalaya}}, \bibinfo {author} {\bibfnamefont {R.}~\bibnamefont {Babbush}},
  \bibinfo {author} {\bibfnamefont {J.~C.}\ \bibnamefont {Bardin}}, \bibinfo
  {author} {\bibfnamefont {R.}~\bibnamefont {Barends}}, \bibinfo {author}
  {\bibfnamefont {J.}~\bibnamefont {Basso}}, \bibinfo {author} {\bibfnamefont
  {A.}~\bibnamefont {Bengtsson}}, \bibinfo {author} {\bibfnamefont
  {A.}~\bibnamefont {Bilmes}}, \bibinfo {author} {\bibfnamefont
  {M.}~\bibnamefont {Broughton}}, \bibinfo {author} {\bibfnamefont {B.~B.}\
  \bibnamefont {Buckley}}, \bibinfo {author} {\bibfnamefont {D.~A.}\
  \bibnamefont {Buell}}, \bibinfo {author} {\bibfnamefont {B.}~\bibnamefont
  {Burkett}}, \bibinfo {author} {\bibfnamefont {N.}~\bibnamefont {Bushnell}},
  \bibinfo {author} {\bibfnamefont {B.}~\bibnamefont {Chiaro}}, \bibinfo
  {author} {\bibfnamefont {R.}~\bibnamefont {Collins}}, \bibinfo {author}
  {\bibfnamefont {W.}~\bibnamefont {Courtney}}, \bibinfo {author}
  {\bibfnamefont {S.}~\bibnamefont {Demura}}, \bibinfo {author} {\bibfnamefont
  {A.~R.}\ \bibnamefont {Derk}}, \bibinfo {author} {\bibfnamefont
  {D.}~\bibnamefont {Eppens}}, \bibinfo {author} {\bibfnamefont
  {C.}~\bibnamefont {Erickson}}, \bibinfo {author} {\bibfnamefont
  {L.}~\bibnamefont {Faoro}}, \bibinfo {author} {\bibfnamefont
  {E.}~\bibnamefont {Farhi}}, \bibinfo {author} {\bibfnamefont {A.~G.}\
  \bibnamefont {Fowler}}, \bibinfo {author} {\bibfnamefont {B.}~\bibnamefont
  {Foxen}}, \bibinfo {author} {\bibfnamefont {M.}~\bibnamefont {Giustina}},
  \bibinfo {author} {\bibfnamefont {A.}~\bibnamefont {Greene}}, \bibinfo
  {author} {\bibfnamefont {J.~A.}\ \bibnamefont {Gross}}, \bibinfo {author}
  {\bibfnamefont {M.~P.}\ \bibnamefont {Harrigan}}, \bibinfo {author}
  {\bibfnamefont {S.~D.}\ \bibnamefont {Harrington}}, \bibinfo {author}
  {\bibfnamefont {J.}~\bibnamefont {Hilton}}, \bibinfo {author} {\bibfnamefont
  {S.}~\bibnamefont {Hong}}, \bibinfo {author} {\bibfnamefont {T.}~\bibnamefont
  {Huang}}, \bibinfo {author} {\bibfnamefont {W.~J.}\ \bibnamefont {Huggins}},
  \bibinfo {author} {\bibfnamefont {L.~B.}\ \bibnamefont {Ioffe}}, \bibinfo
  {author} {\bibfnamefont {S.~V.}\ \bibnamefont {Isakov}}, \bibinfo {author}
  {\bibfnamefont {E.}~\bibnamefont {Jeffrey}}, \bibinfo {author} {\bibfnamefont
  {Z.}~\bibnamefont {Jiang}}, \bibinfo {author} {\bibfnamefont
  {D.}~\bibnamefont {Kafri}}, \bibinfo {author} {\bibfnamefont
  {K.}~\bibnamefont {Kechedzhi}}, \bibinfo {author} {\bibfnamefont
  {T.}~\bibnamefont {Khattar}}, \bibinfo {author} {\bibfnamefont
  {S.}~\bibnamefont {Kim}}, \bibinfo {author} {\bibfnamefont {P.~V.}\
  \bibnamefont {Klimov}}, \bibinfo {author} {\bibfnamefont {A.~N.}\
  \bibnamefont {Korotkov}}, \bibinfo {author} {\bibfnamefont {F.}~\bibnamefont
  {Kostritsa}}, \bibinfo {author} {\bibfnamefont {D.}~\bibnamefont {Landhuis}},
  \bibinfo {author} {\bibfnamefont {P.}~\bibnamefont {Laptev}}, \bibinfo
  {author} {\bibfnamefont {A.}~\bibnamefont {Locharla}}, \bibinfo {author}
  {\bibfnamefont {E.}~\bibnamefont {Lucero}}, \bibinfo {author} {\bibfnamefont
  {O.}~\bibnamefont {Martin}}, \bibinfo {author} {\bibfnamefont {J.~R.}\
  \bibnamefont {McClean}}, \bibinfo {author} {\bibfnamefont {M.}~\bibnamefont
  {McEwen}}, \bibinfo {author} {\bibfnamefont {K.~C.}\ \bibnamefont {Miao}},
  \bibinfo {author} {\bibfnamefont {M.}~\bibnamefont {Mohseni}}, \bibinfo
  {author} {\bibfnamefont {S.}~\bibnamefont {Montazeri}}, \bibinfo {author}
  {\bibfnamefont {W.}~\bibnamefont {Mruczkiewicz}}, \bibinfo {author}
  {\bibfnamefont {J.}~\bibnamefont {Mutus}}, \bibinfo {author} {\bibfnamefont
  {O.}~\bibnamefont {Naaman}}, \bibinfo {author} {\bibfnamefont
  {M.}~\bibnamefont {Neeley}}, \bibinfo {author} {\bibfnamefont
  {C.}~\bibnamefont {Neill}}, \bibinfo {author} {\bibfnamefont {M.~Y.}\
  \bibnamefont {Niu}}, \bibinfo {author} {\bibfnamefont {T.~E.}\ \bibnamefont
  {O’Brien}}, \bibinfo {author} {\bibfnamefont {A.}~\bibnamefont {Opremcak}},
  \bibinfo {author} {\bibfnamefont {B.}~\bibnamefont {Pató}}, \bibinfo
  {author} {\bibfnamefont {A.}~\bibnamefont {Petukhov}}, \bibinfo {author}
  {\bibfnamefont {N.~C.}\ \bibnamefont {Rubin}}, \bibinfo {author}
  {\bibfnamefont {D.}~\bibnamefont {Sank}}, \bibinfo {author} {\bibfnamefont
  {V.}~\bibnamefont {Shvarts}}, \bibinfo {author} {\bibfnamefont
  {D.}~\bibnamefont {Strain}}, \bibinfo {author} {\bibfnamefont
  {M.}~\bibnamefont {Szalay}}, \bibinfo {author} {\bibfnamefont
  {B.}~\bibnamefont {Villalonga}}, \bibinfo {author} {\bibfnamefont {T.~C.}\
  \bibnamefont {White}}, \bibinfo {author} {\bibfnamefont {Z.}~\bibnamefont
  {Yao}}, \bibinfo {author} {\bibfnamefont {P.}~\bibnamefont {Yeh}}, \bibinfo
  {author} {\bibfnamefont {J.}~\bibnamefont {Yoo}}, \bibinfo {author}
  {\bibfnamefont {A.}~\bibnamefont {Zalcman}}, \bibinfo {author} {\bibfnamefont
  {H.}~\bibnamefont {Neven}}, \bibinfo {author} {\bibfnamefont
  {S.}~\bibnamefont {Boixo}}, \bibinfo {author} {\bibfnamefont
  {A.}~\bibnamefont {Megrant}}, \bibinfo {author} {\bibfnamefont
  {Y.}~\bibnamefont {Chen}}, \bibinfo {author} {\bibfnamefont {J.}~\bibnamefont
  {Kelly}}, \bibinfo {author} {\bibfnamefont {V.}~\bibnamefont {Smelyanskiy}},
  \bibinfo {author} {\bibfnamefont {A.}~\bibnamefont {Kitaev}}, \bibinfo
  {author} {\bibfnamefont {M.}~\bibnamefont {Knap}}, \bibinfo {author}
  {\bibfnamefont {F.}~\bibnamefont {Pollmann}},\ and\ \bibinfo {author}
  {\bibfnamefont {P.}~\bibnamefont {Roushan}},\ }\bibfield  {title} {\bibinfo
  {title} {Realizing topologically ordered states on a quantum processor},\
  }\href {https://doi.org/10.1126/science.abi8378} {\bibfield  {journal}
  {\bibinfo  {journal} {Science}\ }\textbf {\bibinfo {volume} {374}},\ \bibinfo
  {pages} {1237} (\bibinfo {year} {2021})}\BibitemShut {NoStop}%
\bibitem [{\citenamefont {Kitaev}\ and\ \citenamefont
  {Preskill}(2006)}]{PhysRevLett.96.110404}%
  \BibitemOpen
  \bibfield  {author} {\bibinfo {author} {\bibfnamefont {A.}~\bibnamefont
  {Kitaev}}\ and\ \bibinfo {author} {\bibfnamefont {J.}~\bibnamefont
  {Preskill}},\ }\bibfield  {title} {\bibinfo {title} {{Topological
  Entanglement Entropy}},\ }\href
  {https://doi.org/10.1103/PhysRevLett.96.110404} {\bibfield  {journal}
  {\bibinfo  {journal} {Phys. Rev. Lett.}\ }\textbf {\bibinfo {volume} {96}},\
  \bibinfo {pages} {110404} (\bibinfo {year} {2006})}\BibitemShut {NoStop}%
\bibitem [{\citenamefont {Semeghini}\ \emph {et~al.}(2021)\citenamefont
  {Semeghini}, \citenamefont {Levine}, \citenamefont {Keesling}, \citenamefont
  {Ebadi}, \citenamefont {Wang}, \citenamefont {Bluvstein}, \citenamefont
  {Verresen}, \citenamefont {Pichler}, \citenamefont {Kalinowski},
  \citenamefont {Samajdar}, \citenamefont {Omran}, \citenamefont {Sachdev},
  \citenamefont {Vishwanath}, \citenamefont {Greiner}, \citenamefont
  {Vuletić},\ and\ \citenamefont {Lukin}}]{doi:10.1126/science.abi8794}%
  \BibitemOpen
  \bibfield  {author} {\bibinfo {author} {\bibfnamefont {G.}~\bibnamefont
  {Semeghini}}, \bibinfo {author} {\bibfnamefont {H.}~\bibnamefont {Levine}},
  \bibinfo {author} {\bibfnamefont {A.}~\bibnamefont {Keesling}}, \bibinfo
  {author} {\bibfnamefont {S.}~\bibnamefont {Ebadi}}, \bibinfo {author}
  {\bibfnamefont {T.~T.}\ \bibnamefont {Wang}}, \bibinfo {author}
  {\bibfnamefont {D.}~\bibnamefont {Bluvstein}}, \bibinfo {author}
  {\bibfnamefont {R.}~\bibnamefont {Verresen}}, \bibinfo {author}
  {\bibfnamefont {H.}~\bibnamefont {Pichler}}, \bibinfo {author} {\bibfnamefont
  {M.}~\bibnamefont {Kalinowski}}, \bibinfo {author} {\bibfnamefont
  {R.}~\bibnamefont {Samajdar}}, \bibinfo {author} {\bibfnamefont
  {A.}~\bibnamefont {Omran}}, \bibinfo {author} {\bibfnamefont
  {S.}~\bibnamefont {Sachdev}}, \bibinfo {author} {\bibfnamefont
  {A.}~\bibnamefont {Vishwanath}}, \bibinfo {author} {\bibfnamefont
  {M.}~\bibnamefont {Greiner}}, \bibinfo {author} {\bibfnamefont
  {V.}~\bibnamefont {Vuletić}},\ and\ \bibinfo {author} {\bibfnamefont
  {M.~D.}\ \bibnamefont {Lukin}},\ }\bibfield  {title} {\bibinfo {title}
  {Probing topological spin liquids on a programmable quantum simulator},\
  }\href {https://doi.org/10.1126/science.abi8794} {\bibfield  {journal}
  {\bibinfo  {journal} {Science}\ }\textbf {\bibinfo {volume} {374}},\ \bibinfo
  {pages} {1242} (\bibinfo {year} {2021})}\BibitemShut {NoStop}%
\bibitem [{\citenamefont {Witten}(1989)}]{Witten1989}%
  \BibitemOpen
  \bibfield  {author} {\bibinfo {author} {\bibfnamefont {E.}~\bibnamefont
  {Witten}},\ }\bibfield  {title} {\bibinfo {title} {{Quantum field theory and
  the Jones polynomial}},\ }\href {https://doi.org/10.1007/BF01217730}
  {\bibfield  {journal} {\bibinfo  {journal} {Communications in Mathematical
  Physics}\ }\textbf {\bibinfo {volume} {121}},\ \bibinfo {pages} {351}
  (\bibinfo {year} {1989})}\BibitemShut {NoStop}%
\bibitem [{\citenamefont {Atiyah}(1988)}]{Atiyah1988}%
  \BibitemOpen
  \bibfield  {author} {\bibinfo {author} {\bibfnamefont {M.}~\bibnamefont
  {Atiyah}},\ }\bibfield  {title} {\bibinfo {title} {Topological quantum field
  theories},\ }\href {https://doi.org/10.1007/BF02698547} {\bibfield  {journal}
  {\bibinfo  {journal} {Publications Math{\'e}matiques de l'Institut des Hautes
  {\'E}tudes Scientifiques}\ }\textbf {\bibinfo {volume} {68}},\ \bibinfo
  {pages} {175} (\bibinfo {year} {1988})}\BibitemShut {NoStop}%
\bibitem [{\citenamefont {Turaev}(2016)}]{Turaev+2016}%
  \BibitemOpen
  \bibfield  {author} {\bibinfo {author} {\bibfnamefont {V.~G.}\ \bibnamefont
  {Turaev}},\ }\href {https://doi.org/doi:10.1515/9783110435221} {\emph
  {\bibinfo {title} {{Quantum Invariants of Knots and 3-Manifolds}}}}\
  (\bibinfo  {publisher} {De Gruyter},\ \bibinfo {year} {2016})\BibitemShut
  {NoStop}%
\bibitem [{\citenamefont {Kitaev}(2006)}]{KitaevFormulae}%
  \BibitemOpen
  \bibfield  {author} {\bibinfo {author} {\bibfnamefont {A.}~\bibnamefont
  {Kitaev}},\ }\bibfield  {title} {\bibinfo {title} {Anyons in an exactly
  solved model and beyond},\ }\href
  {https://doi.org/https://doi.org/10.1016/j.aop.2005.10.005} {\bibfield
  {journal} {\bibinfo  {journal} {Annals of Physics}\ }\textbf {\bibinfo
  {volume} {321}},\ \bibinfo {pages} {2} (\bibinfo {year} {2006})},\ \bibinfo
  {note} {january Special Issue}\BibitemShut {NoStop}%
\bibitem [{\citenamefont {Heinrich}\ \emph {et~al.}(2016)\citenamefont
  {Heinrich}, \citenamefont {Burnell}, \citenamefont {Fidkowski},\ and\
  \citenamefont {Levin}}]{PhysRevB.94.235136}%
  \BibitemOpen
  \bibfield  {author} {\bibinfo {author} {\bibfnamefont {C.}~\bibnamefont
  {Heinrich}}, \bibinfo {author} {\bibfnamefont {F.}~\bibnamefont {Burnell}},
  \bibinfo {author} {\bibfnamefont {L.}~\bibnamefont {Fidkowski}},\ and\
  \bibinfo {author} {\bibfnamefont {M.}~\bibnamefont {Levin}},\ }\bibfield
  {title} {\bibinfo {title} {{Symmetry-enriched string nets: Exactly solvable
  models for SET phases}},\ }\href {https://doi.org/10.1103/PhysRevB.94.235136}
  {\bibfield  {journal} {\bibinfo  {journal} {Phys. Rev. B}\ }\textbf {\bibinfo
  {volume} {94}},\ \bibinfo {pages} {235136} (\bibinfo {year}
  {2016})}\BibitemShut {NoStop}%
\bibitem [{\citenamefont {Morampudi}\ \emph {et~al.}(2017)\citenamefont
  {Morampudi}, \citenamefont {Turner}, \citenamefont {Pollmann},\ and\
  \citenamefont {Wilczek}}]{SpectralFunction}%
  \BibitemOpen
  \bibfield  {author} {\bibinfo {author} {\bibfnamefont {S.~C.}\ \bibnamefont
  {Morampudi}}, \bibinfo {author} {\bibfnamefont {A.~M.}\ \bibnamefont
  {Turner}}, \bibinfo {author} {\bibfnamefont {F.}~\bibnamefont {Pollmann}},\
  and\ \bibinfo {author} {\bibfnamefont {F.}~\bibnamefont {Wilczek}},\
  }\bibfield  {title} {\bibinfo {title} {{Statistics of Fractionalized
  Excitations through Threshold Spectroscopy}},\ }\href
  {https://doi.org/10.1103/PhysRevLett.118.227201} {\bibfield  {journal}
  {\bibinfo  {journal} {Phys. Rev. Lett.}\ }\textbf {\bibinfo {volume} {118}},\
  \bibinfo {pages} {227201} (\bibinfo {year} {2017})}\BibitemShut {NoStop}%
\bibitem [{\citenamefont {Trebst}\ \emph
  {et~al.}(2008{\natexlab{a}})\citenamefont {Trebst}, \citenamefont {Troyer},
  \citenamefont {Wang},\ and\ \citenamefont {Ludwig}}]{0902.3275}%
  \BibitemOpen
  \bibfield  {author} {\bibinfo {author} {\bibfnamefont {S.}~\bibnamefont
  {Trebst}}, \bibinfo {author} {\bibfnamefont {M.}~\bibnamefont {Troyer}},
  \bibinfo {author} {\bibfnamefont {Z.}~\bibnamefont {Wang}},\ and\ \bibinfo
  {author} {\bibfnamefont {A.~W.~W.}\ \bibnamefont {Ludwig}},\ }\bibfield
  {title} {\bibinfo {title} {{A Short Introduction to Fibonacci Anyon
  Models}},\ }\href {https://doi.org/10.1143/PTPS.176.384} {\bibfield
  {journal} {\bibinfo  {journal} {Progress of Theoretical Physics Supplement}\
  }\textbf {\bibinfo {volume} {176}},\ \bibinfo {pages} {384} (\bibinfo {year}
  {2008}{\natexlab{a}})}\BibitemShut {NoStop}%
\bibitem [{\citenamefont {Feiguin}\ \emph {et~al.}(2007)\citenamefont
  {Feiguin}, \citenamefont {Trebst}, \citenamefont {Ludwig}, \citenamefont
  {Troyer}, \citenamefont {Kitaev}, \citenamefont {Wang},\ and\ \citenamefont
  {Freedman}}]{PhysRevLett.98.160409}%
  \BibitemOpen
  \bibfield  {author} {\bibinfo {author} {\bibfnamefont {A.}~\bibnamefont
  {Feiguin}}, \bibinfo {author} {\bibfnamefont {S.}~\bibnamefont {Trebst}},
  \bibinfo {author} {\bibfnamefont {A.~W.~W.}\ \bibnamefont {Ludwig}}, \bibinfo
  {author} {\bibfnamefont {M.}~\bibnamefont {Troyer}}, \bibinfo {author}
  {\bibfnamefont {A.}~\bibnamefont {Kitaev}}, \bibinfo {author} {\bibfnamefont
  {Z.}~\bibnamefont {Wang}},\ and\ \bibinfo {author} {\bibfnamefont {M.~H.}\
  \bibnamefont {Freedman}},\ }\bibfield  {title} {\bibinfo {title}
  {{Interacting Anyons in Topological Quantum Liquids: The Golden Chain}},\
  }\href {https://doi.org/10.1103/PhysRevLett.98.160409} {\bibfield  {journal}
  {\bibinfo  {journal} {Phys. Rev. Lett.}\ }\textbf {\bibinfo {volume} {98}},\
  \bibinfo {pages} {160409} (\bibinfo {year} {2007})}\BibitemShut {NoStop}%
\bibitem [{\citenamefont {Poilblanc}\ \emph {et~al.}(2011)\citenamefont
  {Poilblanc}, \citenamefont {Ludwig}, \citenamefont {Trebst},\ and\
  \citenamefont {Troyer}}]{PhysRevB.83.134439}%
  \BibitemOpen
  \bibfield  {author} {\bibinfo {author} {\bibfnamefont {D.}~\bibnamefont
  {Poilblanc}}, \bibinfo {author} {\bibfnamefont {A.~W.~W.}\ \bibnamefont
  {Ludwig}}, \bibinfo {author} {\bibfnamefont {S.}~\bibnamefont {Trebst}},\
  and\ \bibinfo {author} {\bibfnamefont {M.}~\bibnamefont {Troyer}},\
  }\bibfield  {title} {\bibinfo {title} {{Quantum spin ladders of non-Abelian
  anyons}},\ }\href {https://doi.org/10.1103/PhysRevB.83.134439} {\bibfield
  {journal} {\bibinfo  {journal} {Phys. Rev. B}\ }\textbf {\bibinfo {volume}
  {83}},\ \bibinfo {pages} {134439} (\bibinfo {year} {2011})}\BibitemShut
  {NoStop}%
\bibitem [{\citenamefont {Trebst}\ \emph
  {et~al.}(2008{\natexlab{b}})\citenamefont {Trebst}, \citenamefont {Ardonne},
  \citenamefont {Feiguin}, \citenamefont {Huse}, \citenamefont {Ludwig},\ and\
  \citenamefont {Troyer}}]{PhysRevLett.101.050401}%
  \BibitemOpen
  \bibfield  {author} {\bibinfo {author} {\bibfnamefont {S.}~\bibnamefont
  {Trebst}}, \bibinfo {author} {\bibfnamefont {E.}~\bibnamefont {Ardonne}},
  \bibinfo {author} {\bibfnamefont {A.}~\bibnamefont {Feiguin}}, \bibinfo
  {author} {\bibfnamefont {D.~A.}\ \bibnamefont {Huse}}, \bibinfo {author}
  {\bibfnamefont {A.~W.~W.}\ \bibnamefont {Ludwig}},\ and\ \bibinfo {author}
  {\bibfnamefont {M.}~\bibnamefont {Troyer}},\ }\bibfield  {title} {\bibinfo
  {title} {{Collective States of Interacting Fibonacci Anyons}},\ }\href
  {https://doi.org/10.1103/PhysRevLett.101.050401} {\bibfield  {journal}
  {\bibinfo  {journal} {Phys. Rev. Lett.}\ }\textbf {\bibinfo {volume} {101}},\
  \bibinfo {pages} {050401} (\bibinfo {year} {2008}{\natexlab{b}})}\BibitemShut
  {NoStop}%
\bibitem [{\citenamefont {Ludwig}\ \emph {et~al.}(2011)\citenamefont {Ludwig},
  \citenamefont {Poilblanc}, \citenamefont {Trebst},\ and\ \citenamefont
  {Troyer}}]{10.1088/1367-2630/13/4/045014}%
  \BibitemOpen
  \bibfield  {author} {\bibinfo {author} {\bibfnamefont {A.~W.~W.}\
  \bibnamefont {Ludwig}}, \bibinfo {author} {\bibfnamefont {D.}~\bibnamefont
  {Poilblanc}}, \bibinfo {author} {\bibfnamefont {S.}~\bibnamefont {Trebst}},\
  and\ \bibinfo {author} {\bibfnamefont {M.}~\bibnamefont {Troyer}},\
  }\bibfield  {title} {\bibinfo {title} {{Two-dimensional quantum liquids from
  interacting non-Abelian anyons}},\ }\href
  {https://doi.org/10.1088/1367-2630/13/4/045014} {\bibfield  {journal}
  {\bibinfo  {journal} {New Journal of Physics}\ }\textbf {\bibinfo {volume}
  {13}},\ \bibinfo {pages} {045014} (\bibinfo {year} {2011})}\BibitemShut
  {NoStop}%
\bibitem [{\citenamefont {Wright}\ \emph {et~al.}(2014)\citenamefont {Wright},
  \citenamefont {Rigol}, \citenamefont {Davis},\ and\ \citenamefont
  {Kheruntsyan}}]{1312.4657}%
  \BibitemOpen
  \bibfield  {author} {\bibinfo {author} {\bibfnamefont {T.~M.}\ \bibnamefont
  {Wright}}, \bibinfo {author} {\bibfnamefont {M.}~\bibnamefont {Rigol}},
  \bibinfo {author} {\bibfnamefont {M.~J.}\ \bibnamefont {Davis}},\ and\
  \bibinfo {author} {\bibfnamefont {K.~V.}\ \bibnamefont {Kheruntsyan}},\
  }\bibfield  {title} {\bibinfo {title} {{Nonequilibrium Dynamics of
  One-Dimensional Hard-Core Anyons Following a Quench: Complete Relaxation of
  One-Body Observables}},\ }\href
  {https://doi.org/10.1103/PhysRevLett.113.050601} {\bibfield  {journal}
  {\bibinfo  {journal} {Phys. Rev. Lett.}\ }\textbf {\bibinfo {volume} {113}},\
  \bibinfo {pages} {050601} (\bibinfo {year} {2014})}\BibitemShut {NoStop}%
\bibitem [{\citenamefont {Girardeau}(2006)}]{PhysRevLett.97.100402}%
  \BibitemOpen
  \bibfield  {author} {\bibinfo {author} {\bibfnamefont {M.~D.}\ \bibnamefont
  {Girardeau}},\ }\bibfield  {title} {\bibinfo {title} {{Anyon-Fermion Mapping
  and Applications to Ultracold Gases in Tight Waveguides}},\ }\href
  {https://doi.org/10.1103/PhysRevLett.97.100402} {\bibfield  {journal}
  {\bibinfo  {journal} {Phys. Rev. Lett.}\ }\textbf {\bibinfo {volume} {97}},\
  \bibinfo {pages} {100402} (\bibinfo {year} {2006})}\BibitemShut {NoStop}%
\bibitem [{\citenamefont {Hao}\ and\ \citenamefont
  {Chen}(2012)}]{PhysRevA.86.043631}%
  \BibitemOpen
  \bibfield  {author} {\bibinfo {author} {\bibfnamefont {Y.}~\bibnamefont
  {Hao}}\ and\ \bibinfo {author} {\bibfnamefont {S.}~\bibnamefont {Chen}},\
  }\bibfield  {title} {\bibinfo {title} {Dynamical properties of hard-core
  anyons in one-dimensional optical lattices},\ }\href
  {https://doi.org/10.1103/PhysRevA.86.043631} {\bibfield  {journal} {\bibinfo
  {journal} {Phys. Rev. A}\ }\textbf {\bibinfo {volume} {86}},\ \bibinfo
  {pages} {043631} (\bibinfo {year} {2012})}\BibitemShut {NoStop}%
\bibitem [{\citenamefont {Piroli}\ and\ \citenamefont
  {Calabrese}(2017)}]{PhysRevA.96.023611}%
  \BibitemOpen
  \bibfield  {author} {\bibinfo {author} {\bibfnamefont {L.}~\bibnamefont
  {Piroli}}\ and\ \bibinfo {author} {\bibfnamefont {P.}~\bibnamefont
  {Calabrese}},\ }\bibfield  {title} {\bibinfo {title} {Exact dynamics
  following an interaction quench in a one-dimensional anyonic gas},\ }\href
  {https://doi.org/10.1103/PhysRevA.96.023611} {\bibfield  {journal} {\bibinfo
  {journal} {Phys. Rev. A}\ }\textbf {\bibinfo {volume} {96}},\ \bibinfo
  {pages} {023611} (\bibinfo {year} {2017})}\BibitemShut {NoStop}%
\bibitem [{\citenamefont {Li}(2013)}]{Li2013}%
  \BibitemOpen
  \bibfield  {author} {\bibinfo {author} {\bibfnamefont {Y.}~\bibnamefont
  {Li}},\ }\bibfield  {title} {\bibinfo {title} {Ground-state properties of
  hard-core anyons in one-dimensional periodic lattices},\ }\href
  {https://doi.org/10.1140/epjp/i2013-13094-0} {\bibfield  {journal} {\bibinfo
  {journal} {The European Physical Journal Plus}\ }\textbf {\bibinfo {volume}
  {128}},\ \bibinfo {pages} {94} (\bibinfo {year} {2013})}\BibitemShut
  {NoStop}%
\bibitem [{\citenamefont {Tang}\ \emph {et~al.}(2015)\citenamefont {Tang},
  \citenamefont {Eggert},\ and\ \citenamefont {Pelster}}]{Tang_2015}%
  \BibitemOpen
  \bibfield  {author} {\bibinfo {author} {\bibfnamefont {G.}~\bibnamefont
  {Tang}}, \bibinfo {author} {\bibfnamefont {S.}~\bibnamefont {Eggert}},\ and\
  \bibinfo {author} {\bibfnamefont {A.}~\bibnamefont {Pelster}},\ }\bibfield
  {title} {\bibinfo {title} {Ground-state properties of anyons in a
  one-dimensional lattice},\ }\href
  {https://doi.org/10.1088/1367-2630/17/12/123016} {\bibfield  {journal}
  {\bibinfo  {journal} {New Journal of Physics}\ }\textbf {\bibinfo {volume}
  {17}},\ \bibinfo {pages} {123016} (\bibinfo {year} {2015})}\BibitemShut
  {NoStop}%
\bibitem [{\citenamefont {del Campo}(2008)}]{PhysRevA.78.045602}%
  \BibitemOpen
  \bibfield  {author} {\bibinfo {author} {\bibfnamefont {A.}~\bibnamefont {del
  Campo}},\ }\bibfield  {title} {\bibinfo {title} {Fermionization and
  bosonization of expanding one-dimensional anyonic fluids},\ }\href
  {https://doi.org/10.1103/PhysRevA.78.045602} {\bibfield  {journal} {\bibinfo
  {journal} {Phys. Rev. A}\ }\textbf {\bibinfo {volume} {78}},\ \bibinfo
  {pages} {045602} (\bibinfo {year} {2008})}\BibitemShut {NoStop}%
\bibitem [{\citenamefont {Wang}(2022)}]{2202.06543}%
  \BibitemOpen
  \bibfield  {author} {\bibinfo {author} {\bibfnamefont {Q.-W.}\ \bibnamefont
  {Wang}},\ }\bibfield  {title} {\bibinfo {title} {Exact dynamical correlations
  of hard-core anyons in one-dimensional lattices},\ }\href
  {https://doi.org/10.1103/PhysRevB.105.205143} {\bibfield  {journal} {\bibinfo
   {journal} {Phys. Rev. B}\ }\textbf {\bibinfo {volume} {105}},\ \bibinfo
  {pages} {205143} (\bibinfo {year} {2022})}\BibitemShut {NoStop}%
\bibitem [{\citenamefont {Hatsugai}\ \emph
  {et~al.}(1991{\natexlab{a}})\citenamefont {Hatsugai}, \citenamefont
  {Kohmoto},\ and\ \citenamefont {Wu}}]{PhysRevB.43.2661_fluxconvention}%
  \BibitemOpen
  \bibfield  {author} {\bibinfo {author} {\bibfnamefont {Y.}~\bibnamefont
  {Hatsugai}}, \bibinfo {author} {\bibfnamefont {M.}~\bibnamefont {Kohmoto}},\
  and\ \bibinfo {author} {\bibfnamefont {Y.-S.}\ \bibnamefont {Wu}},\
  }\bibfield  {title} {\bibinfo {title} {Braid group and anyons on a
  cylinder},\ }\href {https://doi.org/10.1103/PhysRevB.43.2661} {\bibfield
  {journal} {\bibinfo  {journal} {Phys. Rev. B}\ }\textbf {\bibinfo {volume}
  {43}},\ \bibinfo {pages} {2661} (\bibinfo {year}
  {1991}{\natexlab{a}})}\BibitemShut {NoStop}%
\bibitem [{\citenamefont {Hatsugai}\ \emph
  {et~al.}(1991{\natexlab{b}})\citenamefont {Hatsugai}, \citenamefont
  {Kohmoto},\ and\ \citenamefont {Wu}}]{PhysRevB.43.10761}%
  \BibitemOpen
  \bibfield  {author} {\bibinfo {author} {\bibfnamefont {Y.}~\bibnamefont
  {Hatsugai}}, \bibinfo {author} {\bibfnamefont {M.}~\bibnamefont {Kohmoto}},\
  and\ \bibinfo {author} {\bibfnamefont {Y.-S.}\ \bibnamefont {Wu}},\
  }\bibfield  {title} {\bibinfo {title} {{Anyons on a torus: Braid group,
  Aharonov-Bohm period, and numerical study}},\ }\href
  {https://doi.org/10.1103/PhysRevB.43.10761} {\bibfield  {journal} {\bibinfo
  {journal} {Phys. Rev. B}\ }\textbf {\bibinfo {volume} {43}},\ \bibinfo
  {pages} {10761} (\bibinfo {year} {1991}{\natexlab{b}})}\BibitemShut {NoStop}%
\bibitem [{\citenamefont {Kallin}(1993)}]{SemionsTorus}%
  \BibitemOpen
  \bibfield  {author} {\bibinfo {author} {\bibfnamefont {C.}~\bibnamefont
  {Kallin}},\ }\bibfield  {title} {\bibinfo {title} {Flux quantization for
  semions on a torus},\ }\href {https://doi.org/10.1103/PhysRevB.48.13742}
  {\bibfield  {journal} {\bibinfo  {journal} {Phys. Rev. B}\ }\textbf {\bibinfo
  {volume} {48}},\ \bibinfo {pages} {13742} (\bibinfo {year}
  {1993})}\BibitemShut {NoStop}%
\bibitem [{\citenamefont {Zatloukal}\ \emph {et~al.}(2014)\citenamefont
  {Zatloukal}, \citenamefont {Lehman}, \citenamefont {Singh}, \citenamefont
  {Pachos},\ and\ \citenamefont {Brennen}}]{PhysRevB.90.134201}%
  \BibitemOpen
  \bibfield  {author} {\bibinfo {author} {\bibfnamefont {V.}~\bibnamefont
  {Zatloukal}}, \bibinfo {author} {\bibfnamefont {L.}~\bibnamefont {Lehman}},
  \bibinfo {author} {\bibfnamefont {S.}~\bibnamefont {Singh}}, \bibinfo
  {author} {\bibfnamefont {J.~K.}\ \bibnamefont {Pachos}},\ and\ \bibinfo
  {author} {\bibfnamefont {G.~K.}\ \bibnamefont {Brennen}},\ }\bibfield
  {title} {\bibinfo {title} {Transport properties of anyons in random
  topological environments},\ }\href
  {https://doi.org/10.1103/PhysRevB.90.134201} {\bibfield  {journal} {\bibinfo
  {journal} {Phys. Rev. B}\ }\textbf {\bibinfo {volume} {90}},\ \bibinfo
  {pages} {134201} (\bibinfo {year} {2014})}\BibitemShut {NoStop}%
\bibitem [{\citenamefont {Singh}\ \emph {et~al.}(2014)\citenamefont {Singh},
  \citenamefont {Pfeifer}, \citenamefont {Vidal},\ and\ \citenamefont
  {Brennen}}]{PhysRevB.89.075112}%
  \BibitemOpen
  \bibfield  {author} {\bibinfo {author} {\bibfnamefont {S.}~\bibnamefont
  {Singh}}, \bibinfo {author} {\bibfnamefont {R.~N.~C.}\ \bibnamefont
  {Pfeifer}}, \bibinfo {author} {\bibfnamefont {G.}~\bibnamefont {Vidal}},\
  and\ \bibinfo {author} {\bibfnamefont {G.~K.}\ \bibnamefont {Brennen}},\
  }\bibfield  {title} {\bibinfo {title} {Matrix product states for anyonic
  systems and efficient simulation of dynamics},\ }\href
  {https://doi.org/10.1103/PhysRevB.89.075112} {\bibfield  {journal} {\bibinfo
  {journal} {Phys. Rev. B}\ }\textbf {\bibinfo {volume} {89}},\ \bibinfo
  {pages} {075112} (\bibinfo {year} {2014})}\BibitemShut {NoStop}%
\bibitem [{\citenamefont {Poilblanc}\ \emph {et~al.}(2013)\citenamefont
  {Poilblanc}, \citenamefont {Feiguin}, \citenamefont {Troyer}, \citenamefont
  {Ardonne},\ and\ \citenamefont {Bonderson}}]{poilblanc2013one}%
  \BibitemOpen
  \bibfield  {author} {\bibinfo {author} {\bibfnamefont {D.}~\bibnamefont
  {Poilblanc}}, \bibinfo {author} {\bibfnamefont {A.}~\bibnamefont {Feiguin}},
  \bibinfo {author} {\bibfnamefont {M.}~\bibnamefont {Troyer}}, \bibinfo
  {author} {\bibfnamefont {E.}~\bibnamefont {Ardonne}},\ and\ \bibinfo {author}
  {\bibfnamefont {P.}~\bibnamefont {Bonderson}},\ }\bibfield  {title} {\bibinfo
  {title} {One-dimensional itinerant interacting non-abelian anyons},\ }\href
  {https://doi.org/10.1103/PhysRevB.87.085106} {\bibfield  {journal} {\bibinfo
  {journal} {Phys. Rev. B}\ }\textbf {\bibinfo {volume} {87}},\ \bibinfo
  {pages} {085106} (\bibinfo {year} {2013})}\BibitemShut {NoStop}%
\bibitem [{\citenamefont {Poilblanc}\ \emph {et~al.}(2012)\citenamefont
  {Poilblanc}, \citenamefont {Troyer}, \citenamefont {Ardonne},\ and\
  \citenamefont {Bonderson}}]{poilblanc2012fractionalization}%
  \BibitemOpen
  \bibfield  {author} {\bibinfo {author} {\bibfnamefont {D.}~\bibnamefont
  {Poilblanc}}, \bibinfo {author} {\bibfnamefont {M.}~\bibnamefont {Troyer}},
  \bibinfo {author} {\bibfnamefont {E.}~\bibnamefont {Ardonne}},\ and\ \bibinfo
  {author} {\bibfnamefont {P.}~\bibnamefont {Bonderson}},\ }\bibfield  {title}
  {\bibinfo {title} {Fractionalization of itinerant anyons in one-dimensional
  chains},\ }\href {https://doi.org/10.1103/PhysRevLett.108.207201} {\bibfield
  {journal} {\bibinfo  {journal} {Phys. Rev. Lett.}\ }\textbf {\bibinfo
  {volume} {108}},\ \bibinfo {pages} {207201} (\bibinfo {year}
  {2012})}\BibitemShut {NoStop}%
\bibitem [{\citenamefont {Soni}\ \emph {et~al.}(2016)\citenamefont {Soni},
  \citenamefont {Troyer},\ and\ \citenamefont {Poilblanc}}]{soni2016effective}%
  \BibitemOpen
  \bibfield  {author} {\bibinfo {author} {\bibfnamefont {M.}~\bibnamefont
  {Soni}}, \bibinfo {author} {\bibfnamefont {M.}~\bibnamefont {Troyer}},\ and\
  \bibinfo {author} {\bibfnamefont {D.}~\bibnamefont {Poilblanc}},\ }\bibfield
  {title} {\bibinfo {title} {Effective models of doped quantum ladders of
  non-abelian anyons},\ }\href {https://doi.org/10.1103/PhysRevB.93.035124}
  {\bibfield  {journal} {\bibinfo  {journal} {Phys. Rev. B}\ }\textbf {\bibinfo
  {volume} {93}},\ \bibinfo {pages} {035124} (\bibinfo {year}
  {2016})}\BibitemShut {NoStop}%
\bibitem [{\citenamefont {Ayeni}\ \emph {et~al.}(2018)\citenamefont {Ayeni},
  \citenamefont {Pfeifer},\ and\ \citenamefont {Brennen}}]{ayeni2018phase}%
  \BibitemOpen
  \bibfield  {author} {\bibinfo {author} {\bibfnamefont {B.~M.}\ \bibnamefont
  {Ayeni}}, \bibinfo {author} {\bibfnamefont {R.~N.~C.}\ \bibnamefont
  {Pfeifer}},\ and\ \bibinfo {author} {\bibfnamefont {G.~K.}\ \bibnamefont
  {Brennen}},\ }\bibfield  {title} {\bibinfo {title} {Phase transitions on a
  ladder of braided non-abelian anyons},\ }\href
  {https://doi.org/10.1103/PhysRevB.98.045432} {\bibfield  {journal} {\bibinfo
  {journal} {Phys. Rev. B}\ }\textbf {\bibinfo {volume} {98}},\ \bibinfo
  {pages} {045432} (\bibinfo {year} {2018})}\BibitemShut {NoStop}%
\bibitem [{\citenamefont {Pfeifer}\ \emph {et~al.}(2010)\citenamefont
  {Pfeifer}, \citenamefont {Corboz}, \citenamefont {Buerschaper}, \citenamefont
  {Aguado}, \citenamefont {Troyer},\ and\ \citenamefont
  {Vidal}}]{PhysRevB.82.115126}%
  \BibitemOpen
  \bibfield  {author} {\bibinfo {author} {\bibfnamefont {R.~N.~C.}\
  \bibnamefont {Pfeifer}}, \bibinfo {author} {\bibfnamefont {P.}~\bibnamefont
  {Corboz}}, \bibinfo {author} {\bibfnamefont {O.}~\bibnamefont {Buerschaper}},
  \bibinfo {author} {\bibfnamefont {M.}~\bibnamefont {Aguado}}, \bibinfo
  {author} {\bibfnamefont {M.}~\bibnamefont {Troyer}},\ and\ \bibinfo {author}
  {\bibfnamefont {G.}~\bibnamefont {Vidal}},\ }\bibfield  {title} {\bibinfo
  {title} {Simulation of anyons with tensor network algorithms},\ }\href
  {https://doi.org/10.1103/PhysRevB.82.115126} {\bibfield  {journal} {\bibinfo
  {journal} {Phys. Rev. B}\ }\textbf {\bibinfo {volume} {82}},\ \bibinfo
  {pages} {115126} (\bibinfo {year} {2010})}\BibitemShut {NoStop}%
\bibitem [{\citenamefont {K{\"o}nig}\ and\ \citenamefont
  {Bilgin}(2010)}]{konig2010anyonic}%
  \BibitemOpen
  \bibfield  {author} {\bibinfo {author} {\bibfnamefont {R.}~\bibnamefont
  {K{\"o}nig}}\ and\ \bibinfo {author} {\bibfnamefont {E.}~\bibnamefont
  {Bilgin}},\ }\bibfield  {title} {\bibinfo {title} {Anyonic entanglement
  renormalization},\ }\href {https://doi.org/10.1103/PhysRevB.82.125118}
  {\bibfield  {journal} {\bibinfo  {journal} {Phys. Rev. B}\ }\textbf {\bibinfo
  {volume} {82}},\ \bibinfo {pages} {125118} (\bibinfo {year}
  {2010})}\BibitemShut {NoStop}%
\bibitem [{\citenamefont {Ayeni}\ \emph {et~al.}(2016)\citenamefont {Ayeni},
  \citenamefont {Singh}, \citenamefont {Pfeifer},\ and\ \citenamefont
  {Brennen}}]{PhysRevB.93.165128}%
  \BibitemOpen
  \bibfield  {author} {\bibinfo {author} {\bibfnamefont {B.~M.}\ \bibnamefont
  {Ayeni}}, \bibinfo {author} {\bibfnamefont {S.}~\bibnamefont {Singh}},
  \bibinfo {author} {\bibfnamefont {R.~N.~C.}\ \bibnamefont {Pfeifer}},\ and\
  \bibinfo {author} {\bibfnamefont {G.~K.}\ \bibnamefont {Brennen}},\
  }\bibfield  {title} {\bibinfo {title} {Simulation of braiding anyons using
  matrix product states},\ }\href {https://doi.org/10.1103/PhysRevB.93.165128}
  {\bibfield  {journal} {\bibinfo  {journal} {Phys. Rev. B}\ }\textbf {\bibinfo
  {volume} {93}},\ \bibinfo {pages} {165128} (\bibinfo {year}
  {2016})}\BibitemShut {NoStop}%
\bibitem [{\citenamefont {{Derzhko}}(2001)}]{0101188}%
  \BibitemOpen
  \bibfield  {author} {\bibinfo {author} {\bibfnamefont {O.}~\bibnamefont
  {{Derzhko}}},\ }\bibfield  {title} {\bibinfo {title} {{Jordan-Wigner
  fermionization for spin-1/2 systems in two dimensions: A brief review}},\
  }\href {https://ui.adsabs.harvard.edu/abs/2001JPhSt...5...49D} {\bibfield
  {journal} {\bibinfo  {journal} {Journal of Physical Studies}\ }\textbf
  {\bibinfo {volume} {5}},\ \bibinfo {pages} {49} (\bibinfo {year}
  {2001})}\BibitemShut {NoStop}%
\bibitem [{\citenamefont {Beverland}\ \emph {et~al.}(2016)\citenamefont
  {Beverland}, \citenamefont {Buerschaper}, \citenamefont {Koenig},
  \citenamefont {Pastawski}, \citenamefont {Preskill},\ and\ \citenamefont
  {Sijher}}]{doi:10.1063/1.4939783}%
  \BibitemOpen
  \bibfield  {author} {\bibinfo {author} {\bibfnamefont {M.~E.}\ \bibnamefont
  {Beverland}}, \bibinfo {author} {\bibfnamefont {O.}~\bibnamefont
  {Buerschaper}}, \bibinfo {author} {\bibfnamefont {R.}~\bibnamefont {Koenig}},
  \bibinfo {author} {\bibfnamefont {F.}~\bibnamefont {Pastawski}}, \bibinfo
  {author} {\bibfnamefont {J.}~\bibnamefont {Preskill}},\ and\ \bibinfo
  {author} {\bibfnamefont {S.}~\bibnamefont {Sijher}},\ }\bibfield  {title}
  {\bibinfo {title} {Protected gates for topological quantum field theories},\
  }\href {https://doi.org/10.1063/1.4939783} {\bibfield  {journal} {\bibinfo
  {journal} {Journal of Mathematical Physics}\ }\textbf {\bibinfo {volume}
  {57}},\ \bibinfo {pages} {022201} (\bibinfo {year} {2016})}\BibitemShut
  {NoStop}%
\bibitem [{\citenamefont {Jordan}\ and\ \citenamefont
  {Wigner}(1928)}]{Jordan1928}%
  \BibitemOpen
  \bibfield  {author} {\bibinfo {author} {\bibfnamefont {P.}~\bibnamefont
  {Jordan}}\ and\ \bibinfo {author} {\bibfnamefont {E.}~\bibnamefont
  {Wigner}},\ }\bibfield  {title} {\bibinfo {title} {{{\"U}ber das Paulische
  {\"A}quivalenzverbot}},\ }\href {https://doi.org/10.1007/BF01331938}
  {\bibfield  {journal} {\bibinfo  {journal} {{Zeitschrift f{\"u}r Physik}}\
  }\textbf {\bibinfo {volume} {47}},\ \bibinfo {pages} {631} (\bibinfo {year}
  {1928})}\BibitemShut {NoStop}%
\bibitem [{\citenamefont {Simon}(2021)}]{simon2020topological}%
  \BibitemOpen
  \bibfield  {author} {\bibinfo {author} {\bibfnamefont {S.~H.}\ \bibnamefont
  {Simon}},\ }\href
  {{https://www-thphys.physics.ox.ac.uk/people/SteveSimon/topological2021/TopoBook-Sep1-2021.pdf}}
  {\bibinfo {title} {{Topological Quantum: Lecture Notes and Proto-Book}}}
  (\bibinfo {year} {2021})\BibitemShut {NoStop}%
\bibitem [{\citenamefont {Einarsson}(1990)}]{StatisticsTorusFormulae}%
  \BibitemOpen
  \bibfield  {author} {\bibinfo {author} {\bibfnamefont {T.}~\bibnamefont
  {Einarsson}},\ }\bibfield  {title} {\bibinfo {title} {Fractional statistics
  on a torus},\ }\href {https://doi.org/10.1103/PhysRevLett.64.1995} {\bibfield
   {journal} {\bibinfo  {journal} {Phys. Rev. Lett.}\ }\textbf {\bibinfo
  {volume} {64}},\ \bibinfo {pages} {1995} (\bibinfo {year}
  {1990})}\BibitemShut {NoStop}%
\bibitem [{\citenamefont {Batista}\ and\ \citenamefont
  {Ortiz}(2001)}]{PhysRevLett.86.1082}%
  \BibitemOpen
  \bibfield  {author} {\bibinfo {author} {\bibfnamefont {C.~D.}\ \bibnamefont
  {Batista}}\ and\ \bibinfo {author} {\bibfnamefont {G.}~\bibnamefont
  {Ortiz}},\ }\bibfield  {title} {\bibinfo {title} {{Generalized Jordan-Wigner
  Transformations}},\ }\href {https://doi.org/10.1103/PhysRevLett.86.1082}
  {\bibfield  {journal} {\bibinfo  {journal} {Phys. Rev. Lett.}\ }\textbf
  {\bibinfo {volume} {86}},\ \bibinfo {pages} {1082} (\bibinfo {year}
  {2001})}\BibitemShut {NoStop}%
\bibitem [{\citenamefont {Bernevig}\ and\ \citenamefont
  {Neupert}(2015)}]{1506.05805}%
  \BibitemOpen
  \bibfield  {author} {\bibinfo {author} {\bibfnamefont {A.}~\bibnamefont
  {Bernevig}}\ and\ \bibinfo {author} {\bibfnamefont {T.}~\bibnamefont
  {Neupert}},\ }\href@noop {} {\bibinfo {title} {{Topological Superconductors
  and Category Theory}}} (\bibinfo {year} {2015}),\ \Eprint
  {https://arxiv.org/abs/arXiv:1506.05805} {arXiv:1506.05805} \BibitemShut
  {NoStop}%
\bibitem [{\citenamefont {Preskill}(1999)}]{preskill1999lecture}%
  \BibitemOpen
  \bibfield  {author} {\bibinfo {author} {\bibfnamefont {J.}~\bibnamefont
  {Preskill}},\ }\href@noop {} {\bibinfo {title} {{Lecture Notes for Physics
  219: Quantum computation}}} (\bibinfo {year} {1999})\BibitemShut {NoStop}%
\bibitem [{\citenamefont {Bonderson}\ \emph {et~al.}(2008)\citenamefont
  {Bonderson}, \citenamefont {Shtengel},\ and\ \citenamefont
  {Slingerland}}]{0707.4206}%
  \BibitemOpen
  \bibfield  {author} {\bibinfo {author} {\bibfnamefont {P.}~\bibnamefont
  {Bonderson}}, \bibinfo {author} {\bibfnamefont {K.}~\bibnamefont
  {Shtengel}},\ and\ \bibinfo {author} {\bibfnamefont {J.}~\bibnamefont
  {Slingerland}},\ }\bibfield  {title} {\bibinfo {title} {{Interferometry of
  non-Abelian anyons}},\ }\href
  {https://doi.org/https://doi.org/10.1016/j.aop.2008.01.012} {\bibfield
  {journal} {\bibinfo  {journal} {Annals of Physics}\ }\textbf {\bibinfo
  {volume} {323}},\ \bibinfo {pages} {2709} (\bibinfo {year}
  {2008})}\BibitemShut {NoStop}%
\bibitem [{\citenamefont {Bonderson}(2007)}]{bonderson_2007}%
  \BibitemOpen
  \bibfield  {author} {\bibinfo {author} {\bibfnamefont {P.~H.}\ \bibnamefont
  {Bonderson}},\ }\emph {\bibinfo {title} {{Non-Abelian anyons and
  interferometry}}},\ \href@noop {} {Ph.D. thesis} (\bibinfo {year}
  {2007})\BibitemShut {NoStop}%
\bibitem [{\citenamefont {Bonderson}(2021)}]{2102.05677}%
  \BibitemOpen
  \bibfield  {author} {\bibinfo {author} {\bibfnamefont {P.}~\bibnamefont
  {Bonderson}},\ }\bibfield  {title} {\bibinfo {title} {{Measuring Topological
  Order}},\ }\href {https://doi.org/10.1103/PhysRevResearch.3.033110}
  {\bibfield  {journal} {\bibinfo  {journal} {Physical Review Research}\
  }\textbf {\bibinfo {volume} {3}},\ \bibinfo {pages} {033110} (\bibinfo {year}
  {2021})}\BibitemShut {NoStop}%
\bibitem [{\citenamefont {Bonderson}\ \emph {et~al.}(2013)\citenamefont
  {Bonderson}, \citenamefont {Fidkowski}, \citenamefont {Freedman},\ and\
  \citenamefont {Walker}}]{arxiv.1306.2379}%
  \BibitemOpen
  \bibfield  {author} {\bibinfo {author} {\bibfnamefont {P.}~\bibnamefont
  {Bonderson}}, \bibinfo {author} {\bibfnamefont {L.}~\bibnamefont
  {Fidkowski}}, \bibinfo {author} {\bibfnamefont {M.}~\bibnamefont
  {Freedman}},\ and\ \bibinfo {author} {\bibfnamefont {K.}~\bibnamefont
  {Walker}},\ }\href@noop {} {\bibinfo {title} {Twisted interferometry}}
  (\bibinfo {year} {2013}),\ \Eprint {https://arxiv.org/abs/arXiv:1306.2379}
  {arXiv:1306.2379} \BibitemShut {NoStop}%
\bibitem [{\citenamefont {Bonderson}\ \emph {et~al.}(2017)\citenamefont
  {Bonderson}, \citenamefont {Knapp},\ and\ \citenamefont
  {Patel}}]{InsideOutsideBases}%
  \BibitemOpen
  \bibfield  {author} {\bibinfo {author} {\bibfnamefont {P.}~\bibnamefont
  {Bonderson}}, \bibinfo {author} {\bibfnamefont {C.}~\bibnamefont {Knapp}},\
  and\ \bibinfo {author} {\bibfnamefont {K.}~\bibnamefont {Patel}},\ }\bibfield
   {title} {\bibinfo {title} {Anyonic entanglement and topological entanglement
  entropy},\ }\href {https://doi.org/https://doi.org/10.1016/j.aop.2017.07.018}
  {\bibfield  {journal} {\bibinfo  {journal} {Annals of Physics}\ }\textbf
  {\bibinfo {volume} {385}},\ \bibinfo {pages} {399} (\bibinfo {year}
  {2017})}\BibitemShut {NoStop}%
\bibitem [{\citenamefont {Kirillov~Jr}(1996)}]{kirillov1996inner}%
  \BibitemOpen
  \bibfield  {author} {\bibinfo {author} {\bibfnamefont {A.}~\bibnamefont
  {Kirillov~Jr}},\ }\bibfield  {title} {\bibinfo {title} {On an inner product
  in modular tensor categories},\ }\href
  {https://doi.org/10.1090/S0894-0347-96-00210-X} {\bibfield  {journal}
  {\bibinfo  {journal} {Journal of the American Mathematical Society}\ }\textbf
  {\bibinfo {volume} {9}},\ \bibinfo {pages} {1135} (\bibinfo {year}
  {1996})}\BibitemShut {NoStop}%
\bibitem [{\citenamefont {Pfeifer}\ \emph {et~al.}(2012)\citenamefont
  {Pfeifer}, \citenamefont {Buerschaper}, \citenamefont {Trebst}, \citenamefont
  {Ludwig}, \citenamefont {Troyer},\ and\ \citenamefont
  {Vidal}}]{PhysRevB.86.155111}%
  \BibitemOpen
  \bibfield  {author} {\bibinfo {author} {\bibfnamefont {R.~N.~C.}\
  \bibnamefont {Pfeifer}}, \bibinfo {author} {\bibfnamefont {O.}~\bibnamefont
  {Buerschaper}}, \bibinfo {author} {\bibfnamefont {S.}~\bibnamefont {Trebst}},
  \bibinfo {author} {\bibfnamefont {A.~W.~W.}\ \bibnamefont {Ludwig}}, \bibinfo
  {author} {\bibfnamefont {M.}~\bibnamefont {Troyer}},\ and\ \bibinfo {author}
  {\bibfnamefont {G.}~\bibnamefont {Vidal}},\ }\bibfield  {title} {\bibinfo
  {title} {Translation invariance, topology, and protection of criticality in
  chains of interacting anyons},\ }\href
  {https://doi.org/10.1103/PhysRevB.86.155111} {\bibfield  {journal} {\bibinfo
  {journal} {Phys. Rev. B}\ }\textbf {\bibinfo {volume} {86}},\ \bibinfo
  {pages} {155111} (\bibinfo {year} {2012})}\BibitemShut {NoStop}%
\bibitem [{\citenamefont {Oganesyan}\ and\ \citenamefont
  {Huse}(2007)}]{PhysRevB.75.155111}%
  \BibitemOpen
  \bibfield  {author} {\bibinfo {author} {\bibfnamefont {V.}~\bibnamefont
  {Oganesyan}}\ and\ \bibinfo {author} {\bibfnamefont {D.~A.}\ \bibnamefont
  {Huse}},\ }\bibfield  {title} {\bibinfo {title} {Localization of interacting
  fermions at high temperature},\ }\href
  {https://doi.org/10.1103/PhysRevB.75.155111} {\bibfield  {journal} {\bibinfo
  {journal} {Phys. Rev. B}\ }\textbf {\bibinfo {volume} {75}},\ \bibinfo
  {pages} {155111} (\bibinfo {year} {2007})}\BibitemShut {NoStop}%
\bibitem [{\citenamefont {Kollath}\ \emph {et~al.}(2010)\citenamefont
  {Kollath}, \citenamefont {Roux}, \citenamefont {Biroli},\ and\ \citenamefont
  {Läuchli}}]{Kollath_2010}%
  \BibitemOpen
  \bibfield  {author} {\bibinfo {author} {\bibfnamefont {C.}~\bibnamefont
  {Kollath}}, \bibinfo {author} {\bibfnamefont {G.}~\bibnamefont {Roux}},
  \bibinfo {author} {\bibfnamefont {G.}~\bibnamefont {Biroli}},\ and\ \bibinfo
  {author} {\bibfnamefont {A.~M.}\ \bibnamefont {Läuchli}},\ }\bibfield
  {title} {\bibinfo {title} {Statistical properties of the spectrum of the
  extended {B}ose{\textendash}{H}ubbard model},\ }\href
  {https://doi.org/10.1088/1742-5468/2010/08/p08011} {\bibfield  {journal}
  {\bibinfo  {journal} {Journal of Statistical Mechanics: Theory and
  Experiment}\ }\textbf {\bibinfo {volume} {2010}},\ \bibinfo {pages} {P08011}
  (\bibinfo {year} {2010})}\BibitemShut {NoStop}%
\bibitem [{\citenamefont {D'Alessio}\ \emph {et~al.}(2016)\citenamefont
  {D'Alessio}, \citenamefont {Kafri}, \citenamefont {Polkovnikov},\ and\
  \citenamefont {Rigol}}]{doi:10.1080/00018732.2016.1198134}%
  \BibitemOpen
  \bibfield  {author} {\bibinfo {author} {\bibfnamefont {L.}~\bibnamefont
  {D'Alessio}}, \bibinfo {author} {\bibfnamefont {Y.}~\bibnamefont {Kafri}},
  \bibinfo {author} {\bibfnamefont {A.}~\bibnamefont {Polkovnikov}},\ and\
  \bibinfo {author} {\bibfnamefont {M.}~\bibnamefont {Rigol}},\ }\bibfield
  {title} {\bibinfo {title} {From quantum chaos and eigenstate thermalization
  to statistical mechanics and thermodynamics},\ }\href
  {https://doi.org/10.1080/00018732.2016.1198134} {\bibfield  {journal}
  {\bibinfo  {journal} {Advances in Physics}\ }\textbf {\bibinfo {volume}
  {65}},\ \bibinfo {pages} {239} (\bibinfo {year} {2016})}\BibitemShut
  {NoStop}%
\bibitem [{\citenamefont {Atas}\ \emph {et~al.}(2013)\citenamefont {Atas},
  \citenamefont {Bogomolny}, \citenamefont {Giraud},\ and\ \citenamefont
  {Roux}}]{PhysRevLett.110.084101}%
  \BibitemOpen
  \bibfield  {author} {\bibinfo {author} {\bibfnamefont {Y.~Y.}\ \bibnamefont
  {Atas}}, \bibinfo {author} {\bibfnamefont {E.}~\bibnamefont {Bogomolny}},
  \bibinfo {author} {\bibfnamefont {O.}~\bibnamefont {Giraud}},\ and\ \bibinfo
  {author} {\bibfnamefont {G.}~\bibnamefont {Roux}},\ }\bibfield  {title}
  {\bibinfo {title} {{Distribution of the Ratio of Consecutive Level Spacings
  in Random Matrix Ensembles}},\ }\href
  {https://doi.org/10.1103/PhysRevLett.110.084101} {\bibfield  {journal}
  {\bibinfo  {journal} {Phys. Rev. Lett.}\ }\textbf {\bibinfo {volume} {110}},\
  \bibinfo {pages} {084101} (\bibinfo {year} {2013})}\BibitemShut {NoStop}%
\bibitem [{\citenamefont {Lin}\ \emph {et~al.}(1993)\citenamefont {Lin},
  \citenamefont {Gubernatis}, \citenamefont {Gould},\ and\ \citenamefont
  {Tobochnik}}]{lin1993}%
  \BibitemOpen
  \bibfield  {author} {\bibinfo {author} {\bibfnamefont {H.}~\bibnamefont
  {Lin}}, \bibinfo {author} {\bibfnamefont {J.}~\bibnamefont {Gubernatis}},
  \bibinfo {author} {\bibfnamefont {H.}~\bibnamefont {Gould}},\ and\ \bibinfo
  {author} {\bibfnamefont {J.}~\bibnamefont {Tobochnik}},\ }\bibfield  {title}
  {\bibinfo {title} {Exact diagonalization methods for quantum systems},\
  }\href {https://doi.org/10.1063/1.4823192} {\bibfield  {journal} {\bibinfo
  {journal} {Computers in Physics}\ }\textbf {\bibinfo {volume} {7}},\ \bibinfo
  {pages} {400} (\bibinfo {year} {1993})}\BibitemShut {NoStop}%
\bibitem [{\citenamefont {Wei{\ss}e}\ and\ \citenamefont
  {Fehske}(2008)}]{Weisse2008}%
  \BibitemOpen
  \bibfield  {author} {\bibinfo {author} {\bibfnamefont {A.}~\bibnamefont
  {Wei{\ss}e}}\ and\ \bibinfo {author} {\bibfnamefont {H.}~\bibnamefont
  {Fehske}},\ }\bibinfo {title} {Exact {D}iagonalization {T}echniques},\ in\
  \href {https://doi.org/10.1007/978-3-540-74686-7_18} {\emph {\bibinfo
  {booktitle} {Computational Many-Particle Physics}}},\ \bibinfo {editor}
  {edited by\ \bibinfo {editor} {\bibfnamefont {H.}~\bibnamefont {Fehske}},
  \bibinfo {editor} {\bibfnamefont {R.}~\bibnamefont {Schneider}},\ and\
  \bibinfo {editor} {\bibfnamefont {A.}~\bibnamefont {Wei{\ss}e}}}\ (\bibinfo
  {publisher} {Springer Berlin Heidelberg},\ \bibinfo {address} {Berlin,
  Heidelberg},\ \bibinfo {year} {2008})\ pp.\ \bibinfo {pages}
  {529--544}\BibitemShut {NoStop}%
\bibitem [{\citenamefont {Zhang}\ and\ \citenamefont
  {Dong}(2010)}]{Zhang_2010}%
  \BibitemOpen
  \bibfield  {author} {\bibinfo {author} {\bibfnamefont {J.~M.}\ \bibnamefont
  {Zhang}}\ and\ \bibinfo {author} {\bibfnamefont {R.~X.}\ \bibnamefont
  {Dong}},\ }\bibfield  {title} {\bibinfo {title} {Exact diagonalization: the
  {B}ose{\textendash}{H}ubbard model as an example},\ }\href
  {https://doi.org/10.1088/0143-0807/31/3/016} {\bibfield  {journal} {\bibinfo
  {journal} {European Journal of Physics}\ }\textbf {\bibinfo {volume} {31}},\
  \bibinfo {pages} {591} (\bibinfo {year} {2010})}\BibitemShut {NoStop}%
\bibitem [{\citenamefont {Giraud}\ \emph {et~al.}(2022)\citenamefont {Giraud},
  \citenamefont {Mac\'e}, \citenamefont {Vernier},\ and\ \citenamefont
  {Alet}}]{PhysRevX.12.011006}%
  \BibitemOpen
  \bibfield  {author} {\bibinfo {author} {\bibfnamefont {O.}~\bibnamefont
  {Giraud}}, \bibinfo {author} {\bibfnamefont {N.}~\bibnamefont {Mac\'e}},
  \bibinfo {author} {\bibfnamefont {E.}~\bibnamefont {Vernier}},\ and\ \bibinfo
  {author} {\bibfnamefont {F.}~\bibnamefont {Alet}},\ }\bibfield  {title}
  {\bibinfo {title} {Probing symmetries of quantum many-body systems through
  gap ratio statistics},\ }\href {https://doi.org/10.1103/PhysRevX.12.011006}
  {\bibfield  {journal} {\bibinfo  {journal} {Phys. Rev. X}\ }\textbf {\bibinfo
  {volume} {12}},\ \bibinfo {pages} {011006} (\bibinfo {year}
  {2022})}\BibitemShut {NoStop}%
\bibitem [{\citenamefont {Lake}\ and\ \citenamefont
  {Wu}(2016)}]{PhysRevB.94.115139}%
  \BibitemOpen
  \bibfield  {author} {\bibinfo {author} {\bibfnamefont {E.}~\bibnamefont
  {Lake}}\ and\ \bibinfo {author} {\bibfnamefont {Y.-S.}\ \bibnamefont {Wu}},\
  }\bibfield  {title} {\bibinfo {title} {Signatures of broken parity and
  time-reversal symmetry in generalized string-net models},\ }\href
  {https://doi.org/10.1103/PhysRevB.94.115139} {\bibfield  {journal} {\bibinfo
  {journal} {Phys. Rev. B}\ }\textbf {\bibinfo {volume} {94}},\ \bibinfo
  {pages} {115139} (\bibinfo {year} {2016})}\BibitemShut {NoStop}%
\bibitem [{\citenamefont {Robnik}\ and\ \citenamefont
  {Berry}(1986)}]{MRobnik_1986}%
  \BibitemOpen
  \bibfield  {author} {\bibinfo {author} {\bibfnamefont {M.}~\bibnamefont
  {Robnik}}\ and\ \bibinfo {author} {\bibfnamefont {M.~V.}\ \bibnamefont
  {Berry}},\ }\bibfield  {title} {\bibinfo {title} {False time-reversal
  violation and energy level statistics: the role of anti-unitary symmetry},\
  }\href {https://doi.org/10.1088/0305-4470/19/5/020} {\bibfield  {journal}
  {\bibinfo  {journal} {Journal of Physics A: Mathematical and General}\
  }\textbf {\bibinfo {volume} {19}},\ \bibinfo {pages} {669} (\bibinfo {year}
  {1986})}\BibitemShut {NoStop}%
\bibitem [{\citenamefont {Pfeifer}\ and\ \citenamefont
  {Singh}(2015)}]{PhysRevB.92.115135}%
  \BibitemOpen
  \bibfield  {author} {\bibinfo {author} {\bibfnamefont {R.~N.~C.}\
  \bibnamefont {Pfeifer}}\ and\ \bibinfo {author} {\bibfnamefont
  {S.}~\bibnamefont {Singh}},\ }\bibfield  {title} {\bibinfo {title} {Finite
  density matrix renormalization group algorithm for anyonic systems},\ }\href
  {https://doi.org/10.1103/PhysRevB.92.115135} {\bibfield  {journal} {\bibinfo
  {journal} {Phys. Rev. B}\ }\textbf {\bibinfo {volume} {92}},\ \bibinfo
  {pages} {115135} (\bibinfo {year} {2015})}\BibitemShut {NoStop}%
\bibitem [{\citenamefont {Kirchner}(2021)}]{nico_kirchner_firstVersion}%
  \BibitemOpen
  \bibfield  {author} {\bibinfo {author} {\bibfnamefont {N.}~\bibnamefont
  {Kirchner}},\ }\emph {\bibinfo {title} {Simulation of Anyon Dynamics}},\
  \href {https://doi.org/10.5281/zenodo.7566432} {Master's thesis},\ \bibinfo
  {school} {Technische Universit{\"a}t M{\"u}nchen} (\bibinfo {year}
  {2021})\BibitemShut {NoStop}%
\bibitem [{\citenamefont {Millar}(2021)}]{darragh_thesis}%
  \BibitemOpen
  \bibfield  {author} {\bibinfo {author} {\bibfnamefont {D.}~\bibnamefont
  {Millar}},\ }\emph {\bibinfo {title} {{Interacting Anyons in One and Two
  Dimensions: Strong Zero Modes in Anyon Chains and Non-Abelian Anyons on a
  Torus}}},\ \href {https://mural.maynoothuniversity.ie/14865/} {Ph.D. thesis}
  (\bibinfo {year} {2021})\BibitemShut {NoStop}%
\bibitem [{\citenamefont {Kirchner}\ \emph {et~al.}(2022)\citenamefont
  {Kirchner}, \citenamefont {Smith},\ and\ \citenamefont
  {Pollmann}}]{nico_kirchner_2022_6777951}%
  \BibitemOpen
  \bibfield  {author} {\bibinfo {author} {\bibfnamefont {N.}~\bibnamefont
  {Kirchner}}, \bibinfo {author} {\bibfnamefont {A.}~\bibnamefont {Smith}},\
  and\ \bibinfo {author} {\bibfnamefont {F.}~\bibnamefont {Pollmann}},\
  }\bibfield  {title} {\bibinfo {title} {Numerical simulation of non-abelian
  anyons [{D}ata set]},\ }\href {https://doi.org/10.5281/zenodo.6777951}
  {10.5281/zenodo.6777951} (\bibinfo {year} {2022}),\ \bibinfo {note}
  {{Zenodo}}\BibitemShut {NoStop}%
\bibitem [{\citenamefont {Luks}\ \emph {et~al.}(1997)\citenamefont {Luks},
  \citenamefont {Rákóczi},\ and\ \citenamefont {Wright}}]{LUKS1997335}%
  \BibitemOpen
  \bibfield  {author} {\bibinfo {author} {\bibfnamefont {E.~M.}\ \bibnamefont
  {Luks}}, \bibinfo {author} {\bibfnamefont {F.}~\bibnamefont {Rákóczi}},\
  and\ \bibinfo {author} {\bibfnamefont {C.~R.}\ \bibnamefont {Wright}},\
  }\bibfield  {title} {\bibinfo {title} {{Some Algorithms for Nilpotent
  Permutation Groups}},\ }\href
  {https://doi.org/https://doi.org/10.1006/jsco.1996.0092} {\bibfield
  {journal} {\bibinfo  {journal} {Journal of Symbolic Computation}\ }\textbf
  {\bibinfo {volume} {23}},\ \bibinfo {pages} {335} (\bibinfo {year}
  {1997})}\BibitemShut {NoStop}%
\bibitem [{\citenamefont {Fischer}(2019)}]{fischer}%
  \BibitemOpen
  \bibfield  {author} {\bibinfo {author} {\bibfnamefont {G.}~\bibnamefont
  {Fischer}},\ }\href
  {https://doi.org/https://doi.org/10.1007/978-3-658-27343-9} {\emph {\bibinfo
  {title} {Lernbuch {Lineare} {Algebra} und {Analytische} {Geometrie}}}},\
  \bibinfo {edition} {4th}\ ed.\ (\bibinfo  {publisher} {Springer-Verlag},\
  \bibinfo {year} {2019})\ pp.\ \bibinfo {pages} {211--214}\BibitemShut
  {NoStop}%
\bibitem [{\citenamefont {Verlinde}(1988)}]{VERLINDE1988360}%
  \BibitemOpen
  \bibfield  {author} {\bibinfo {author} {\bibfnamefont {E.}~\bibnamefont
  {Verlinde}},\ }\bibfield  {title} {\bibinfo {title} {Fusion rules and modular
  transformations in 2d conformal field theory},\ }\href
  {https://doi.org/https://doi.org/10.1016/0550-3213(88)90603-7} {\bibfield
  {journal} {\bibinfo  {journal} {Nuclear Physics B}\ }\textbf {\bibinfo
  {volume} {300}},\ \bibinfo {pages} {360} (\bibinfo {year}
  {1988})}\BibitemShut {NoStop}%
\end{thebibliography}%

\end{document}